\documentclass[a4paper, 12pt, twoside, notitlepage]{report}      

\usepackage[square]{natbib}                    
\usepackage{color}
\usepackage{epsfig}
\usepackage{epstopdf}
\usepackage{amsmath, amssymb}
\usepackage{hyperref}
\usepackage{wasysym} 
\usepackage{subfigure}
\usepackage{amsfonts}
\usepackage{amssymb}
\usepackage{empheq}
\usepackage{enumerate}
\usepackage[toc,page]{appendix}
\usepackage{abstract}
\usepackage{setspace}
\usepackage{rotating}
\usepackage{sidecap}

\pdfoutput=1

\usepackage{fancyhdr}
\renewcommand\appendix{\par
  \setcounter{section}{0}
  \setcounter{subsection}{0}
  \setcounter{figure}{0}
  \setcounter{table}{0}
  \renewcommand\thesection{Appendix \Alph{section}}
  \renewcommand\thefigure{\Alph{section}\arabic{figure}}
  \renewcommand\thetable{\Alph{section}\arabic{table}}
}

\numberwithin{equation}{chapter}

\allowdisplaybreaks

\begin{document}
\setlength{\parskip}{4pt} 
\setlength{\parindent}{22pt}

\fancyhf{}

\fancyhead[RO]{\slshape \leftmark}
\fancyhead[LE]{\slshape \rightmark}
\fancyfoot[C]{\thepage}
\renewcommand{\headrulewidth}{0.4pt} 
\renewcommand{\footrulewidth}{0pt}

\unitlength = 1mm

\titlepage

{\centering

\Large
\textsc{Cosmology with the 6-degree Field Galaxy Survey}\\[30pt]
by\\[25pt]
\textbf{Florian Beutler}\\[10pt]
\normalsize

\vspace{3.5cm}

\normalsize

Submitted to The University of Western Australia\\[0pt]
for the degree of\\[0pt]
\textbf{Doctor of Philosophy}\\ [0pt]
\vspace{0.4cm}
Supervisors: Chris Blake, Lister Staveley-Smith, Heath Jones, Peter Quinn\\[10pt]
\vspace{0.8cm}
School of Physics\\[10pt]
Perth, 2012\\[10pt]

\vspace{1.5cm}
\includegraphics[width=6cm]{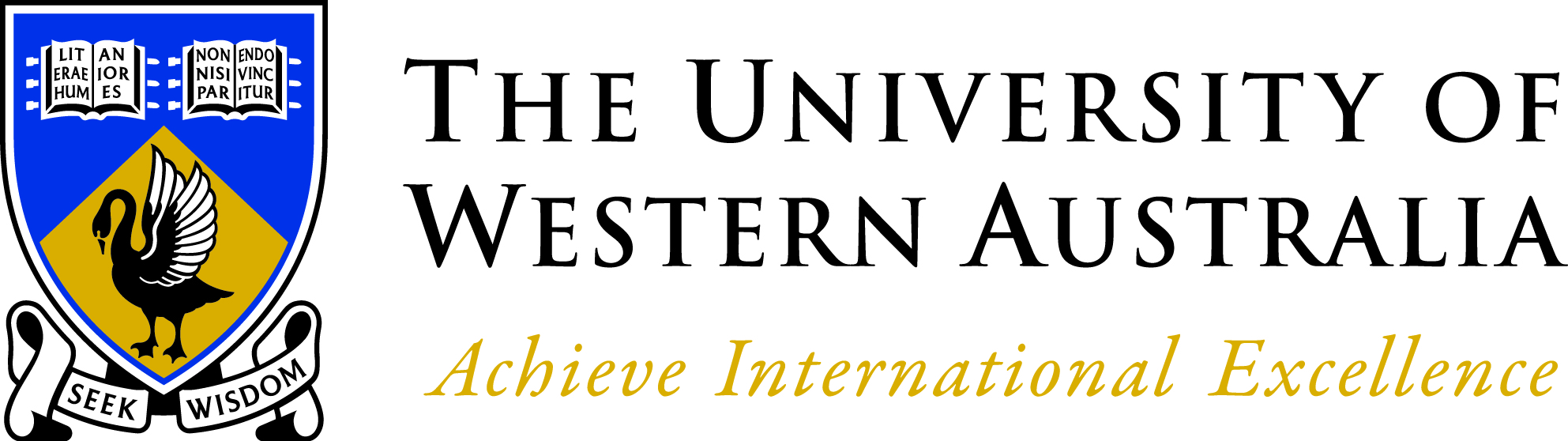}

}

\cleardoublepage

\renewcommand*\abstractname{\Large Declaration\hfill}
\begin{abstract}
This thesis is my own work, and no part of it has been submitted for a degree at this, or at any other, university.
\vspace{5cm}\\
\makebox[2.1in]{\hrulefill}\\
Florian Beutler
\end{abstract}

\cleardoublepage

\renewcommand*\abstractname{\Large Abstract\hfill}
\begin{abstract}
This thesis presents the analysis of the clustering of galaxies in the 6dF Galaxy Survey (6dFGS). At large separation scales the baryon acoustic oscillation (BAO) signal is detected which allows a measurement of the distance ratio, $r_s(z_d)/D_V(z_{\rm eff}) = 0.336\pm0.015$ ($4.5\%$ precision), where $r_s(z_d)$ is the sound horizon at the drag epoch $z_d$ and $D_V(z_{\rm eff})$ is the absolute distance to the effective redshift of the survey, given by $z_{\rm eff} = 0.106$. The low effective redshift of 6dFGS makes it a competitive and independent alternative to Cepheids and low-$z$ supernovae in constraining the Hubble constant. The value of the Hubble constant reported in this work is $H_0 = 67.0\pm3.2\;$km s$^{-1}\;$Mpc$^{-1}$ ($4.8\%$ precision) which depends only on the WMAP-7 calibration of the sound horizon and on the galaxy clustering in 6dFGS. Compared to earlier BAO studies at higher redshift, this analysis is less dependent on other cosmological parameters. 
This thesis also includes forecasts for the proposed TAIPAN all-southern-sky optical galaxy survey and the radio WALLABY survey. TAIPAN has the potential to constrain the Hubble constant with $3\%$ precision using the BAO technique.

Modelling the 2D galaxy correlation function of 6dFGS, $\xi(r_p,\pi)$, allows a measure of the parameter combination $f(z_{\rm eff})\sigma_8(z_{\rm eff}) = 0.423 \pm 0.055$, where $f \simeq \Omega_m^{\gamma}(z)$ is the growth rate of cosmic structure and $\sigma_8$ is the r.m.s. of matter fluctuations in $8h^{-1}\,$Mpc spheres. The effective redshift of this analysis is $z_{\rm eff} = 0.067$.
Such a measurement allows to test the relationship between matter and gravity on cosmic scales by constraining the growth index of density fluctuations, $\gamma$. The 6dFGS measurement of $f\sigma_8$ combined with WMAP-7, results in $\gamma = 0.547 \pm 0.088$, consistent with the prediction of General Relativity ($\gamma_{\rm GR} \approx 0.55$). Because of the low effective redshift of 6dFGS this measurement of the growth rate is independent of the fiducial cosmological model (Alcock-Paczynski effect). Using a Fisher matrix analysis it can be predicted, that the WALLABY survey will be able to measure $f\sigma_8$ with a precision of $4$ - $10\%$, depending on the modelling of non-linear structure formation. This is comparable to the predicted precision for the best redshift bins of the Baryon Oscillation Spectroscopic Survey (BOSS), demonstrating that low-redshift surveys have a significant role to play in future tests of dark energy and modified gravity.

The last chapter of this thesis studies the stellar-mass dependence of galaxy clustering in the 6dF Galaxy Survey. The near-infrared selection of 6dFGS allows more reliable stellar mass estimates compared to optical bands used in other galaxy surveys. Using the Halo Occupation Distribution (HOD) model, this analysis investigates the trend of dark matter halo mass and satellite fraction with stellar mass by measuring the projected correlation function, $w_p(r_p)$. This is the first rigorous study of halo occupation as a function of stellar mass at low redshift using galaxy clustering. The findings of this analysis are, that the typical halo mass ($M_1$) as well as the satellite power law index ($\alpha$) increase with stellar mass. This indicates, (1) that galaxies with higher stellar mass sit in more massive dark matter halos and (2) that these more massive dark matter halos accumulate satellites faster with growing mass compared to halos occupied by low stellar mass galaxies. 
Furthermore there seems to be a relation between $M_1$ and the minimum dark matter halo mass ($M_{\rm min}$) of $M_1 \approx 22\,M_{\rm min}$, in agreement with similar findings for SDSS galaxies, but explored for the first time as a function of stellar mass. The satellite fraction of 6dFGS galaxies declines with increasing stellar mass from $21\%$ at $M_{\rm stellar} = 2.6\times10^{10}h^{-2}\,M_{\odot}$ to $12\%$ at $M_{\rm stellar} = 5.4\times10^{10}h^{-2}\,M_{\odot}$ indicating that high stellar mass galaxies are more likely to be central galaxies. Finally the 6dFGS results are compared to two different semi-analytic models derived from the Millennium Simulation, finding some disagreement. The 6dFGS constraints on the satellite fraction as a function of stellar mass can be used to place new constraints on semi-analytic models in the future, particularly the behaviour of luminous red satellites. 

\end{abstract}

\cleardoublepage

\renewcommand*\abstractname{\Large Preface\hfill}
\begin{abstract}
This thesis is constructed as a $``$series of papers$"$ in compliance with the rules for PhD thesis submission from the graduate research school at the University of Western Australia.\\

Publications arisen from this thesis:
\begin{enumerate}
\item \textbf{The 6dF Galaxy Survey: Baryon Acoustic Oscillations and the Local Hubble Constant} (Chapter 2)\\
F. Beutler, C. Blake, M. Colless, H. Jones, L. Staveley-Smith, L. Campbell, Q. Parker, W. Saunders, F. Watson\\
MNRAS 416, 3017, 2011\\
arXiv:1106.3366
\item \textbf{The 6dF Galaxy Survey: $z \approx 0$ measurements of the growth of structure and $\sigma_8$} (Chapter 3)\\
F. Beutler, C. Blake, M. Colless, H. Jones, L. Staveley-Smith, G. Pool, L. Campbell, Q. Parker, W. Saunders, F.Watson\\
MNRAS 423, 3430B, 2012\\
arXiv:1204.4725
\item \textbf{The 6dF Galaxy Survey: Dependence of halo occupation on stellar mass} (Chapter 4)\\
F. Beutler, C. Blake, M. Colless, H. Jones, L. Staveley-Smith, L. Campbell, Q. Parker, W. Saunders, F.Watson\\
MNRAS 429, 3604B, 2013\\
arXiv:1212.3610
\end{enumerate}
\end{abstract}

\cleardoublepage

\renewcommand{\abstractname}{Acknowledgements}
\begin{abstract}
First I would like to thank Chris Blake for his patience and support during the last three years. His dedication to my PhD project was invaluable and I will always be grateful for having had him as a supervisor. Thank you!

Furthermore I would like to thank Lister Staveley-Smith and Peter Quinn who gave me the oportunity to do my PhD at the University of Western Australia. I would also like to thank Matthew Colless for his significant role in shaping my PhD project and for his advice and support during the last three years. Thanks also to Heath Jones for his patience with the floods of emails I send towards him. 

Furthermore I would like to thank Martin Meyer who patiently listened to my numerous questions. I always cherished your advice.

A big thanks goes to my girlfriend Morag Scrimgeour, who managed to make the last three years in Australia the best of my life. You make me question whether it is a justified scarifies to academia to be so far apart from you for the next year. 

Thanks also to Jacinta Delhaize, Toby Potter, Alan Duffy, Stefan Westerlund, Shaun Hooper, Le Kelvin and all the other students and postdocs who made ICRAR a lively and fun environment to work in.

\emph{Ich m\"ochte auch meinen Eltern und meinem Bruder danken. Danke dass Ihr mich in Australien besucht habt und danke f\"ur euere Unterst\"utzung and euer Verst\"andnis.}

I would also like to thank the University of Western Australia, which supported me through several travel funding schemes and allowed me to bridge the distance gap between Perth and the rest of the world.

I am very grateful for the supported by the Australian Government through the International Postgraduate Research Scholarship (IPRS) and by additional scholarships from ICRAR and the AAO.
\end{abstract}

\cleardoublepage

\pagenumbering{roman} \tableofcontents    

\cleardoublepage
\setcounter{page}{1}

\pagenumbering{arabic} 

\pagestyle{fancy}


  \chapter{Introduction}

With the introduction of General Relativity in 1915 by Albert Einstein~\citep{Einstein:1915}, it became possible for the first time to describe the gravitational Universe with one set of equations. Nevertheless it was not until later in the century that the modern model of cosmology emerged. Although complemented by many theoretical achievements like inflation~\citep{Guth:1980zm} and the explanation of the origin of heavy elements by Big Bang nucleosynthesis~\citep{Alpher:1948ve}, it were largely observational breakthroughs which led to what we now call the standard model of cosmology. The most important of these were the discovery of the Hubble expansion~\citep{Hubble:1929ig}, the discovery of dark matter~\citep{Zwicky:1937zza,Kahn1959,Freeman:1970mx,Rubin:1970zza}, the discovery of the Cosmic Microwave Background (CMB)~\citep{Penzias:1965}, the discovery of fluctuations in the CMB~\citep{Smoot:1992, Bennett:1996ce} and the discovery of the accelerating expansion of the Universe~\citep{Perlmutter:1998np,Riess:1998} as well as the transition to precision cosmology~\citep{Spergel:2003cb}.

In this chapter we give a brief introduction to the theoretical basis of modern cosmology. We also discuss the tools of observational cosmology which are used in this work to test the current standard model, namely baryon acoustic oscillations (BAO) and redshift space distortions. For a more detailed discussion we refer to~\citet{Kolb:1994,Peacock:1999,Dodelson:2003,Carroll:2004,Lyth:2010}.

\begin{figure}[tb]
\begin{center}
\epsfig{file=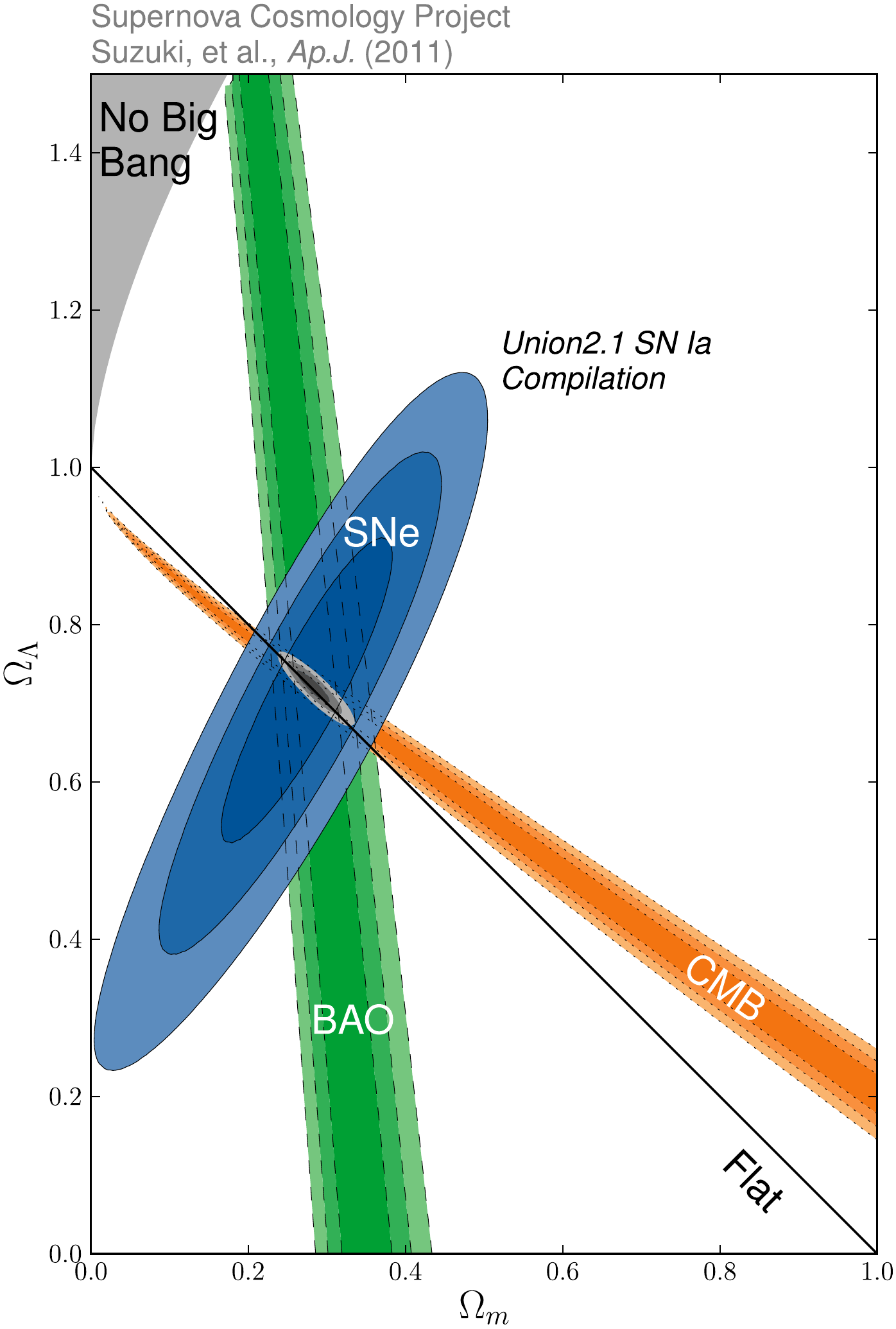,width=8cm}
\caption{Cosmological constraints indicating dark energy ($\Omega_{\Lambda}$) comprising $73\%$ of the total energy density and (dark) matter ($\Omega_m$) making up $27\%$. (credit:~\citealt{Suzuki:2011hu})}
\label{fig:first_plot}
\end{center}
\end{figure}

According to current best estimates (see e.g.~\citealt{Komatsu:2010fb, Suzuki:2011hu, Sanchez:2012sg} and Figure~\ref{fig:first_plot} for the latest results), approximately $73\%$ of our Universe is made up of dark energy, about $27\%$ is in the form of matter, but only $20\%$ of this matter fraction is in the form of baryonic matter, while the rest is so called cold dark matter (CDM). Recent constraints on dark energy as a cosmological constant ($\Omega_{\Lambda}$) and matter ($\Omega_m =$ cold dark matter + baryonic matter) are shown in Figure~\ref{fig:first_plot}. We will discuss the cosmological probes used in this diagram within this introduction, especially the CMB and BAO techniques.

It will be crucial for the understanding of the concepts used in this work to be familiar with the basic ideas of General Relativity and the expanding Universe. Hence we start with a brief introduction to General Relativity in section~\ref{sec:GR} including a discussion of the solutions of the Einstein equations that are believed to describe our Universe. In section~\ref{sec:statistics} we will introduce the statistical tools used in this analysis, namely the correlation function and power spectrum. In section~\ref{sec:relics} we introduce inflation followed by a discussion of the fundamental equations for the evolution of the initial matter fluctuations. In section~\ref{sec:pecvel} we discuss the impact of peculiar velocities on the distribution of galaxies, and how we can detect this signal with galaxy redshift surveys. In section~\ref{sec:halo} we introduce the halo model, which connects the matter and galaxy clustering. In section~\ref{sec:obs} we discuss techniques of modern observational cosmology, focusing on Baryon Acoustic Oscillations and their signature in the CMB and the local galaxy distribution. In section~\ref{sec:models} we discuss alternative models for dark energy, some of which will be tested in this analysis. In section~\ref{sec:6dFGS} we introduce the 6dFGS dataset which is used in this work. Section~\ref{sec:overview} will give a overview and a motivation for this thesis and introduce the following three chapters.

\section{General Relativity}
\label{sec:GR}

Since gravity is the dominant force on cosmic scales, the evolution of the Universe is described by the Einstein field equations, which relate the mass-energy content of the Universe to the geometry of space-time, including a geometrical cosmological constant (we use units where the speed of light is $c = 1$):
\begin{equation}
G_{\mu\nu} = R_{\mu\nu} - \frac{1}{2}g_{\mu\nu}R = 8\pi GT_{\mu\nu} + g_{\mu\nu}\Lambda,
\end{equation}
where $G_{\mu\nu}$ is the Einstein tensor describing the geometry of the Universe, $R_{\mu \nu}$ is the Ricci tensor, $R$ is the Ricci scalar, $g_{\mu \nu}$ is the metric tensor, $\Lambda$ is the cosmological constant and $G$ is Newton's gravitational constant. The matter distribution is characterised by the energy-momentum tensor $T_{\mu\nu}$, which in the case of a perfect fluid\footnote{A perfect fluid is one that can be completely specified by two quantities, the rest-frame energy density $\rho$, and an isotropic rest-frame pressure $p$ that specifies the pressure in every direction.} leads to $T_{00} = \rho$ which is the energy density and $T_{ii} = p_i$ which are the pressure components for $i = 1,2,3$. The indices $\mu = 0, 1, 2, 3$ correspond to the space-time coordinates $x^{\mu} = (x^0, x^1, x^2, x^3) = (ct, x, y, z)$.

In order to find solutions for these equations we need several simplifying assumptions. The most important of these is the Cosmological Principle, which states that on the largest scales our Universe is homogeneous and isotropic, i.e. there are no special locations or directions. Modern cosmology provides very strong support for this assumption from the observation of the almost constant CMB temperature~\citep{Smoot:1991,Bennett:1996ce} as well as large scale structure measurements at low redshift~\citep{Hogg:2004vw, Scrimgeour:2012wt}. Using these assumptions, the most general space-time interval can be written as:
\begin{equation}
ds^2 = dt^2 - a^2(t)[d\chi^2 + S^2_k(\chi)(d\theta^2 + \sin^2\theta d\phi^2)].
\label{eq:FRWmetric}
\end{equation}
This is the Friedmann-Lemaitre-Robertson-Walker (FLRW) metric~\citep{Friedman1922,Lemaitre:1931,Robertson:1935,Walker:1937} with $t$ being the proper time measured by a co-moving observer, $a(t)$ is the scale factor, $\theta$ and $\phi$ are the spherical coordinates, and $k$ is known as the curvature parameter. Furthermore
\begin{equation}
S_k(\chi) = \begin{cases} \sin \chi&\text{ if }k = +1\hspace{0.3cm}\text{  (closed)}\cr
\chi &\text{ if } k = 0\hspace{0.6cm}\text{  (flat)}\cr
\sinh \chi &\text{ if } k = -1\hspace{0.3cm}\text{  (open)},\end{cases}
\label{eq:Sk}
\end{equation}
with $\chi$ being a dimensionless radial coordinate related to physical coordinate distance by $r = a\chi$. We call a Universe open if $k = -1$, closed if $k = +1$ and flat (Euclidean) if $k = 0$. As we will see later, $k=0$ seems to be the case favoured by current observational data.
Using the scale factor $a(t)$ we can define the cosmological redshift as $z = a_0/a(t) - 1$, where $a_0$ is the scale factor today. 

We can now move on and solve the Einstein field equations to obtain a set of relations which describe the dynamics of the scale factor $a(t)$. Einstein's equations relate the evolution of this scale factor, $a(t)$, to the pressure and energy of the matter in the Universe
\begin{equation}
H^2 = \left(\frac{\dot{a}}{a}\right)^2 = \frac{8 \pi G}{3}\rho - \frac{k}{a^2} + \frac{\Lambda}{3},
\label{eq:friedman1}
\end{equation}
which is derived from the $00$-component of Einstein's field equations, and
\begin{equation}
    \frac{\ddot{a}}{a} = -\frac{4 \pi G}{3}\left(\rho+3p\right) + \frac{\Lambda}{3},
\label{eq:friedman2}
\end{equation}
which is derived from the trace of Einstein's field equations. The energy density $\rho$ and pressure $p$ correspond to all particle species, relativistic or not. The equations above also include the definition of the Hubble parameter
\begin{equation}
H(t) \equiv \frac{\dot{a}}{a},
\end{equation}
where the dot denotes a derivative with respect to time $t$. 

Whether the curvature of the Universe is closed ($k=+1$), flat ($k=0$) or open ($k=-1$) depends on the energy density of the components it contains. The density required to make the Universe flat is called the critical density $\rho_c$ and is found by using the first of the Friedmann equations and setting $\Lambda$ and the normalised spatial curvature, $k$, to zero:
\begin{equation}
    \rho_c = \frac{3 H(t)^2}{8 \pi G} = 2.78h^{-1}10^{11}\text{M}_{\odot}h^3\text{Mpc}^{-3},
\end{equation}
where $h$ is defined as
\begin{equation}
h = \frac{H_0}{100\,\text{km/s/Mpc}}.
\end{equation}
The density parameter is then defined as:
\begin{equation}
    \Omega(t) \equiv \frac{\rho}{\rho_c} = \frac{8 \pi G\rho}{3 H(t)^2}. 
\end{equation}
The evolution of the density as the Universe expands can be studied by the adiabatic equation
\begin{equation}
\frac{d(a^3\rho c^2)}{dt} + p\frac{da^3}{dt} = 0,
\end{equation}
which can be derived from the Friedmann equations above. Characterising each component by its equation of state
\begin{equation}
w = \frac{p}{\rho}
\end{equation}
and making the ansatz $\rho \propto a^n$, one obtains
\begin{equation}
\rho(a) = \rho_0 a^{-3(w+1)} \propto \begin{cases}
a^{-4}&\text{ if }w = \frac{1}{3}\hspace{0.7cm} \text{radiation \& relativistic species}\\
a^{-3}&\text{ if }w = 0\hspace{0.7cm} \text{dust (or pressure-less matter)}\\
a^{-2}&\text{ if }w = -\frac{1}{3}\hspace{0.4cm} \text{curvature}\\
\text{const.}&\text{ if }w = -1\hspace{0.4cm} \Lambda\text{ \& inflation},
\end{cases}
\label{eq:rhow}
\end{equation}
where $\rho_0$ denotes the density value at the present time.

Using the density parameter and their evolution discussed above, and substituting them into the Friedmann equation (eq.~\ref{eq:friedman1}) we can write
\begin{equation}
\frac{H(a)^2}{H_0^2} = \Omega_{\gamma} a^{-4} + \Omega_M a^{-3} + \Omega_k a^{-2} + \Omega_{\Lambda},
\label{eq:friedmann}
\end{equation}
where we treat curvature also as a density defined by
\begin{equation}
\rho_k = - \frac{3 k}{8\pi Ga^2}.
\end{equation}
Eq.~\ref{eq:friedmann} is the version of the Friedmann equation which is most useful for the study of cosmology, because the different $\Omega_i$'s are directly related to observational quantities. We will use this equation frequently in this thesis.

Since radiation density (including neutrinos, if they are relativistic) follows $\rho \simeq a^{-4}$ and matter density $\rho \sim a^{-3}$, we can infer that there must have been a time when the Universe was dominated by radiation and the transition from radiation domination to matter domination is given by
\begin{equation}
a_{eq} = \frac{\rho_r}{\rho_m} \simeq 3\times 10^{-4}.
\end{equation}
For $a > a_{eq}$ the expansion was matter-dominated. Only very recently ($a\gtrsim 0.5$) dark energy became the dominating energy component of the Universe and began to play a major role in the expansion of the Universe. In fact, since dark energy does not appear to depend on time, while all other energy components decrease with time, dark energy will only become more dominant in the future.

\subsection{Cosmic distances}
\label{sec:dist}

Within this formalism it is easy to understand how the Hubble law follows directly from the cosmological principle and the FLRW-metric. Following our nomenclature, the proper distance is defined by $r = a(t)\chi$, whereas its velocity is given by $v = dr/ dt = \dot{a}\chi = \dot{a}r/a$. Putting this in the definition of the Hubble parameter we find 
\begin{equation}
H_0 = \frac{\dot{a}}{a}\Big|_0 = v\chi,
\end{equation}
which is Hubble's law and $H_0$ is the value of the Hubble parameter today. 

The co-moving distance $D_{C}$ of an object that emits light at conformal time $t_{\rm e}$ and is observed today ($t_{\rm o}$) is
\begin{equation}
D_{C} =  \int^{t_{\rm o}}_{t_{\rm e}} \frac{c\,dt}{a} =  \int^{1}_{a_{\rm e}} \frac{c\,da}{a^2H(a)} =  \int^{z}_{0}\frac{c\,dz}{H(z)},
\label{eq:59}
\end{equation}
where we used $dt = da/(aH(a))$. Using the combined density parameter
\begin{equation}
E(z) = \left[\Omega_{\gamma}a^{-4} + \Omega_ma^{-3} + \Omega_ka^{-2} + \Omega_{\Lambda}\right]^{1/2},
\end{equation}
the distance equation can be re-written as
\begin{equation}
D_{C} = \int^{z}_{0}\frac{c\,dz}{H(z)} = \frac{c}{H_0}  \int^{z}_{0} \frac{dz}{E(z)}.
\end{equation}
The co-moving distance $D_C$ is the most common distance used in this analysis. However we also employ the transverse co-moving distance:
\begin{equation}
D_M(z) 
= \frac{c}{H_0}\frac{\chi}{\sqrt{\Omega_k}}S_{k}\left(\sqrt{\Omega_k}\right),
\end{equation}
where we used eq.~\ref{eq:Sk}, $\chi = H_0D_C(z)/c$ and $k = +1$, $k = 0$ and $k=-1$ for the cases of positive curvature ($\Omega_k > 0$), no curvature ($\Omega_k = 0$) and negative curvature ($\Omega_k < 0$), respectively. From this we can directly derive the angular diameter distance:
\begin{equation}
D_A(z) = \frac{D_M(z)}{1+z}
\end{equation}
and the luminosity distance is defined as 
\begin{equation}
D_L(z) = (1+z) D_M(z).
\end{equation}
The angular diameter distance does not increase indefinitely as $z\rightarrow \infty$, instead it turns over at $z\approx 1$. Consequently, objects with higher redshift appear larger in angular size.

\section{Clustering statistics}
\label{sec:statistics}

The Universe we live in was seeded by quantum fluctuations, which were pushed to cosmic scales during inflation and grew by the subsequent evolution to form the highly nonlinear structures we observe today~\citep{Kolb:1994,Peacock:1999}. As a consequence of this quantum mechanical origin, the structures are stochastic with random initial conditions. Due to this fact we cannot hope to develop a theory that exactly reproduces the Universe we observe today. Rather, we should consider our Universe as one representation of an ensemble of possible Universes. Therefore, we need to introduce statistical quantities, which can be used to compare theoretical predictions with the observed data. 

\subsection{Two-point correlation function and power spectrum}

We can define a dimensionless over-density or density contrast
\begin{equation}
\delta(r) = \frac{\rho(r) - \overline{\rho}}{\overline{\rho}},
\end{equation}
where $\rho(r)$ is the density at a specific position $r$ and $\overline{\rho}$ is the average density. The Two point correlation function is now defined as
\begin{equation}
\xi(r) = \langle \delta(r+x)\delta(r)\rangle.
\end{equation}
The correlation function determines the probability of finding two objects at separation $r$ in excess of the probability one would expect for a random distribution.

It will prove convenient to build up the actual density field from a superposition of modes that describe the behaviour on a certain scale. We therefore consider a finite box of volume $V$ with periodic boundary conditions and expand the density contrast in terms of its Fourier components, allowing one to write the correlation function as 
\begin{equation}
\xi(r) = \frac{V}{(2\pi)^3} \int d^3k|\delta(k)|^2\exp[-ikr],
\label{eq:deltas}
\end{equation}
with the wavenumber $k$. The ensemble average of the squared amplitude of the Fourier component of a field 
\begin{equation}
P(k) = \langle |\delta(k)|^2\rangle,
\end{equation}
is called a power spectrum. Using spherical symmetry the relation between the correlation function and the power spectrum  reduces to a Hankel transform
\begin{equation}
\xi(r) = \frac{V}{(2\pi)^3} \int d^3kP(k)k^2j_0(kr),
\end{equation}
where $j_0(kr) = \sin(kr)/kr$ refers to the spherical Bessel function of order $0$.

\subsection{Filtering of the density field and the linear bias scheme}

As we are not only interested in the local properties of perturbations, but also in averages over a certain volume, we can convolve the density field with a filter $W_R$ of scale $R$ (often called window function). This convolution in real space translates into a simple multiplication in Fourier space. The variance of the smoothed density field is given by
\begin{equation}
\sigma_R^2 = \frac{V}{(2\pi)^3}\int d^3kP(k)|W_R(k)|^2 = \langle \delta^2(x)\rangle,
\end{equation}
where we use the Fourier transform of the top hat filter with radius $R$
\begin{equation}
W_R(k) = \frac{3}{(kR)^3}\left[\sin(kR) - \cos(kR)\right].
\end{equation}
The scale $R$ of the filter is related to a typical mass by the relation
\begin{equation}
M = \frac{4\pi}{3}R^3\overline{\rho}.
\end{equation}
Historically $R = 8h^{-1}$Mpc has been chosen as the smoothing scale at which to quantify the variance of the density field and is denoted by $\sigma_8$. This parameter normalises the matter power spectrum ($P(k) \propto \sigma_8^2$) and we could expect to derive this parameter from the amplitude of the measured power spectrum. However a galaxy redshift survey actually measures the galaxy power spectrum $P_g(k)$, which is not necessarily identical to the matter power spectrum $P_m(k)$. On linear scales we can expect that the galaxy power spectrum and the matter power spectrum are related by a linear factor 
\begin{equation}
P_g(k) = b^2P_m(k),
\end{equation}
where $b$ is the linear galaxy bias. This model has been verified using N-body simulations (e.g.~\citealt{Coles:1993, Scherrer:1997hp}). A more physically motivated picture of the relation between matter and galaxies is the halo model, which will be discussed in section~\ref{sec:halo}. 

While the assumption of a linear bias will break down for small scales, it also means that the galaxy bias $b$ and $\sigma_8$ are degenerate and the normalisation of the galaxy power spectrum alone cannot constrain both of these two parameters. We will discuss this further in chapter~\ref{ch:RSD} and we will show possibilities for breaking this degeneracy using extra information in the galaxy clustering.

\section{Initial conditions and the evolution of the primordial power spectrum}
\label{sec:relics}

Temperature fluctuations in the Universe today are of the order of $10^8\,$Kelvin\footnote{Approximate temperature in the core of the sun}, while the CMB shows fluctuations of the order of $10^{-5}\,$Kelvin. The growth from these very small fluctuations to today's fluctuations can (to some extent) be calculated by perturbation theory. To do this we need to solve the coupled Einstein and Boltzmann equations. The Einstein equations tell us how perturbations in the metric influence the different energy components, while the Boltzmann equations tell us how these quantities evolve with time.
The Boltzmann equation is (see e.g.~\citealt{Kolb:1994})
\begin{equation}
\frac{df(x,p,t)}{dt} = C(f),
\label{eq:boltz}
\end{equation}
where $f(x,p,t)$ is the phase-space density of the different components of the Universe (photons, neutrinos, electrons, collision-less dark matter and protons).
The left hand side of this equation is the collision-less part, while the right hand side, $C(f)$, has to contain all the appropriate scattering physics (in the early Universe, the photons are coupled to the electrons, which in turn are coupled to the protons). To solve this equation we need to perturb the distribution function $f$, using the perturbed metric coefficients, together with their equation of motion derived from the Einstein equations (see e.g.~\citealt{Peebles:1970ag,Wilson:1981yi, Bond:1984fp, Zlosnik:2011iu}). The result is a system of coupled differential equations for the distribution functions of the non-fluid components (photons, neutrinos and collision-less dark matter) together with the density and pressure of the collisional baryon fluid and gravitational waves. Given a set of initial conditions, these equations are sufficient to describe the evolution of cosmological perturbations. To do this we have to integrate the whole system numerically and evolve it to the present. Conveniently this can be done using a publicly available code, e.g. CMBeasy~\citep{Seljak:1996is} or CAMB~\citep{Lewis:2002ah}. Some alternative methods using analytic solutions and fitting functions have been suggested~\citep{Bardeen:1985tr,Hu:1995en,Eisenstein:1997ik,Montanari:2011nz}, but modern cosmology depends on such high precision that most of the time the full approach is needed. In this analysis we use the software package CAMB and refer the interested reader to~\citet{Zlosnik:2011iu} for a detailed derivation and theoretical discussion.

\subsection{Inflation}
\label{sec:inflation}

The inflation theory was primarily introduced as an explanation of the two main problems of the standard cosmological model, the flatness problem and the horizon problem. The \emph{horizon problem} describes the observational fact that CMB photons observed from two opposite directions of the sky have the same mean temperature (with overall fluctuations of the order of $10^{-5}\,$Kelvin), although those regions were never in casual contact.

The \emph{flatness problem} can easily be discussed by looking at a modified version of the Friedmann equation
\begin{equation}
(\Omega_{\rm tot}^{-1} - 1)\rho a^2 = -\frac{3kc^2}{8 \pi G}.
\end{equation}
Flatness is expressed by $(\Omega_{\rm tot}^{-1} - 1) = 0$. The right hand side of this expression contains only constants, and therefore the left hand side must remain constant throughout the evolution of the Universe.
As the Universe expands the scale factor $a$ increases and the density $\rho$ decreases (see eq.~\ref{eq:rhow}). For most of the lifetime of the Universe it is matter or radiation dominated and hence $\rho$ decreases with $a^{-3}$ and $a^{-4}$ respectively.  Therefore the factor $\rho a^2$ will decrease. Since the time of the Planck era, shortly after the Big Bang, this term has decreased by a factor of around $10^{60}$, and so $(\Omega_{\rm tot}^{-1} - 1)$ must have increased by a similar amount to retain the constant value of their product. Nevertheless all measurements are still consistent with $(\Omega_{\rm tot}^{-1} - 1) = 0$. Why did the Universe pick flatness with such high precision?

To avoid the flatness problem, inflation introduces a period in which the term $\rho$ decreases more slowly than $a^2$ and hence $(\Omega_{\rm tot}^{-1} - 1)$ moves towards zero. This also means that regions which are not causally connected today could have been connected before the epoch of inflation. It is interesting to note that the future of the Universe, which will be dark energy dominated, will show a similar behaviour.

For a FLRW Universe a rapid expansion follows the criteria
\begin{equation}
a^{-3(w+1)} > a^{-2} \;\;\rightarrow\;\; w < -\frac{1}{3}.
\label{eq:infcondition}
\end{equation}
This condition can be obtained by introducing a scalar field $\phi$ minimally coupled to gravity. The equation of motion for such a field is given by the Klein-Gordon equation
\begin{equation}
\ddot{\phi} + 3H\dot{\phi} + \frac{dV(\phi)}{d\phi} = 0,
\end{equation}
where $V$ is the potential of the scalar field. The energy density and the pressure of the scalar field are
\begin{equation}
\begin{split}
p &= \frac{1}{2}\dot{\phi}^2 - V(\phi),\\
\rho &= \frac{1}{2}\dot{\phi}^2 + V(\phi).
\end{split}
\label{eq:inf}
\end{equation}
Using the inflation condition $w=p/\rho < -1/3$ we find that the potential $V(\phi)$ has to satisfy
\begin{equation}
\dot{\phi}^2 < V(\phi).
\end{equation}
The Universe during inflation is composed of this uniform scalar field. The scalar field energy decays in a quantum process into the energy components we see today and since it decays differently in different parts of the Universe, it seeds the density fluctuations which then gravitationally grow over time.

\begin{figure}[tb]
\begin{center}
\epsfig{file=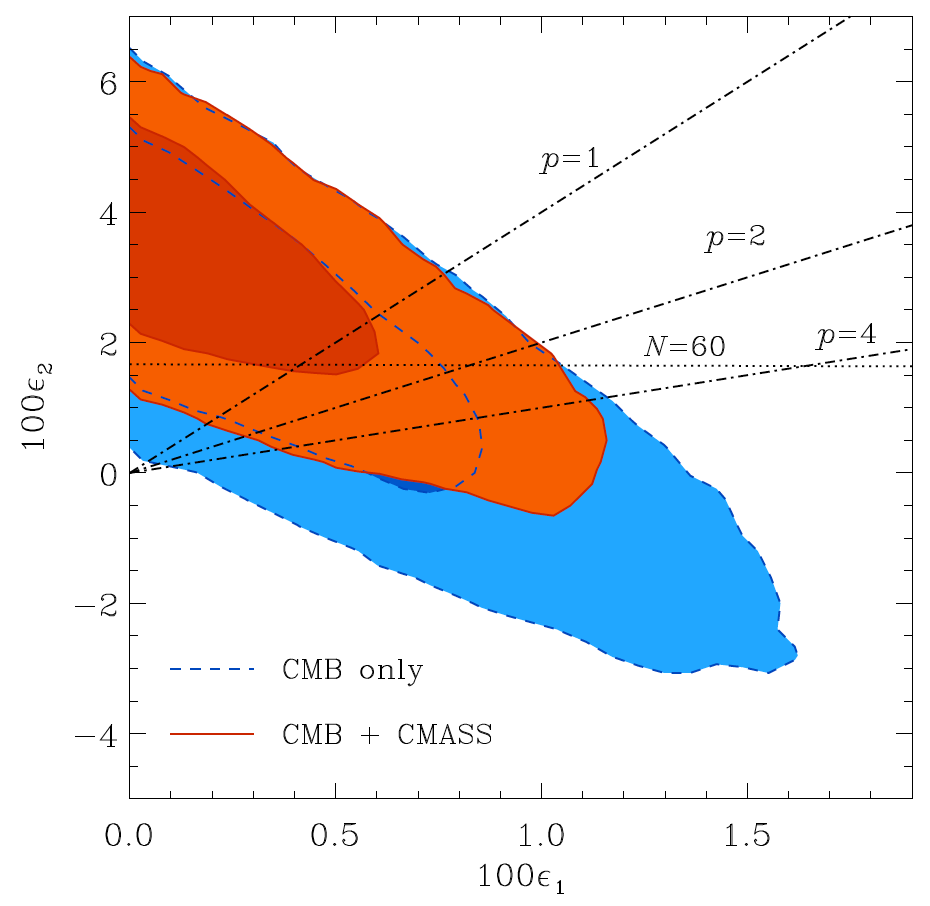,width=10cm}
\caption{Constraints on the horizon flow parameters $\epsilon_1$ and $\epsilon_2$ (see text) from WMAP7 combined with the BOSS-CMASS sample (red contours) and WMAP7 alone (blue contours). This analysis assumes a power law inflational potential ($V(\phi) \propto \phi^p$). (credit:~\cite{Sanchez:2012sg})}
\label{fig:inflation}
\end{center}
\end{figure}

Observations of the Cosmic Microwave Background are now good enough to begin to constrain the type of inflational potential (see e.g.~\citealt{Leach:2002dw,Kinney:2008wy,Finelli:2009bs}). Here we are going to use the horizon flow parameters $\epsilon_1 = -d \ln H(N)/dN$ and $\epsilon_2 = d \ln |\epsilon_1|/dN$~\citep{Schwarz:2001vv}, where $N$ is the number of e-foldings before the end of inflation. Assuming that the inflationary phase is driven by a potential of the form $V(\phi) \propto \phi^p$ (chaotic inflation) we can find the following relation between the horizon flow parameters, the power-law index, $p$, and the number of e-folds, $N$~\citep{Leach:2002dw}
\begin{align}
\epsilon_2 &= \frac{4}{p}\epsilon_1,\\
N &= \frac{p}{4}\left(\frac{1}{\epsilon_1} - 1\right).
\end{align}
Figure~\ref{fig:inflation} shows one of the latest constraints on these parameters using data from WMAP7~\citep{Komatsu:2010fb} and large scale structure data from the BOSS-CMASS sample~\citep{Sanchez:2012sg}. The dot-dashed lines correspond to inflationary models with $p = 1$, $2$ and $4$. The blue contours represent the constraints from WMAP7 alone. Large scale structure data alone cannot constrain $\epsilon_1$ or $\epsilon_2$ but such data can help to break degeneracies in the CMB. Therefore adding data from the BOSS-CMASS sample improves the constraints by about $30\%$. Using this data together with an assumed upper limit for the number of e-folds of $N \leq 60$~\citep{Dodelson:2003vq,Liddle:2003as},~\citet{Sanchez:2012sg} report a limit of $p < 1.2$ at the $95\%$ confidence level, imposing a constraint on the inflationary potential used in this analysis (see also~\citealt{Komatsu:2008hk,Komatsu:2010fb}). Future CMB experiments will improve these constraints and hence we can expect to learn much more about inflation in the coming decade.

\subsection{The Universe dominated by radiation}
\label{sec:rad}

Directly after inflation, the Universe entered a phase dominated by the radiation density which lasted untill $z\approx 3400$. Even though the matter energy density surpassed the radiation density at this point, the Universe remained optically thick to radiation until a cosmological redshift of $z\approx 1100$, when the Universe was about $380\,000$ years old. The events during this short period are essential for this thesis and will be discussed in detail (see also~\citealt{Hu:1995en,Eisenstein:2006nj}). 

The small over-densities seeded by inflation represent not only an excess of matter but also an excess of photons and thus such regions are subject to photon pressure~\citep{Peebles:1970ag}. In the radiation dominated era, the photons are tightly coupled to baryons via Thomson scattering. As a result, these over-pressured density peaks initiate sound waves, travelling with the sound speed 
\begin{equation}
c^2_s = \frac{\partial p}{\partial \rho} = \frac{c^2}{3(1 + R)},
\end{equation}
with $R\equiv 3\rho_{b}/4\rho_{\gamma}$, where $\rho_{\gamma}$ is the photon density and $\rho_b$ is the baryon density.

The cold dark matter does not interact with photons and hence its perturbations grow via gravitational collapse (see Figure~\ref{fig:Ma}). Therefore CDM perturbations begin to dominate the total density perturbations connected to the gravitational potential $\phi$. The baryon-photon fluid oscillates within these potential wells caused by the CDM.

The evolution of the baryon-photon sound wave can be expressed as a plane wave perturbation of wavenumber $k$. Ignoring the expansion of the Universe and assuming that the gravitational potential $\phi$ is caused by CDM only, the evolution of the perturbation $\delta_b$ in mode $k$ in the tight-coupling limit follows~\citep{Peebles:1970ag,Doroshkevich:1978,Hu:1996mn}
\begin{equation}
(1 + R)\ddot{\delta}_b + \frac{k^2c^2}{3}\delta_b = (1+R)(-k^2c^2\phi).
\end{equation}
This equation represents a harmonic oscillator with mass $(1+R)$ and frequency $\omega = kc/\sqrt{3(1 + R)} = kc_s$~\citep{Hu:1995en}. The solution is of the form
\begin{equation}
\delta_b + (1+R)\phi \propto cos(kc_s\tau)
\end{equation}
with $\tau$ being the conformal time $\tau = \int dt/a$. We define
\begin{equation}
r_s(t) \equiv \int^{\eta}_0 c_sd\eta = a_0\int^t_0\frac{c_s(t)}{a(t)}dt,
\end{equation}
which we call the sound horizon at time $t$, since it represents the co-moving distance sound has traveled by time $t$. This leads to 
\begin{equation}
\delta_b + (1+R)\phi \propto cos(kr_s(t)).
\end{equation}
For photons we have 
\begin{equation}
\Theta + \phi + R\phi \propto cos(kc_s\tau),
\end{equation} 
where $\Theta = \delta T/T$ and $\Theta + \phi$ is the temperature fluctuation of photons after they climb up the gravitational potential.

At redshifts around $z\sim 1100$ the tight coupling between the baryons and photons breaks down as the Universe starts to recombine and the cross section for Thomson scattering drops dramatically. At this point the sound-wave stops and the physical structure becomes imprinted into the distribution of matter as well as into the distribution at photons. From now on the photons travel through the cosmos with little interaction and form the Cosmic Microwave Background. 

The oscillation pattern present at decoupling is imprinted in the temperature and density distribution of the photons and baryons. However, since the number density of photons remains higher after decoupling, than the number density of baryons, the small percentage of photons that happen to interact with baryons remains sufficient to drive the pressure wave in the baryons until $z \approx 1030$. The epoch of baryon decoupling is called the drag epoch, and the sound horizon $r_s(t)$ for baryons is $\sim 4\%$ larger than for photons.

The sound horizon for photons is given by $r_s(t_{\rm dec})$ and marks the wavenumber of the first compression at $k = \pi/r_s(t_{\rm dec})$. Subsequent compressions are given by
\begin{equation}
kr_s(t_{\rm dec}) = m\pi
\end{equation}
with $m = 1,2,3,...$. Therefore we see strong structure in the CMB anisotropy at the multipoles
\begin{equation}
\ell = kd_A(t_{\rm dec}) = m\pi\frac{d_A(t_{\rm dec})}{r_s(t_{\rm dec})} \equiv m\ell_A,
\end{equation}
corresponding to these scales. Here
\begin{equation}
\ell_A \equiv \pi \frac{d_A(t_{\rm dec})}{r_s(t_{\rm dec})} \equiv \frac{\pi}{\theta_s}
\end{equation}
is the acoustic scale in multipole space and
\begin{equation}
\theta_s \equiv \frac{r_s(t_{\rm dec})}{d_A(t_{\rm dec})}
\end{equation}
is the sound horizon angle, i.e. the angle at which we see the sound horizon on the last scattering surface. Because of these acoustic oscillations, the CMB angular power spectrum $\mathcal{C}_{\ell}$ has a structure of acoustic peaks at sub-horizon scales. We will discuss these features further in section~\ref{sec:obs} when we discuss the CMB observations.

While photon fluctuations before the last scattering surface are dominated by density fluctuations, the growing mode of the baryon fluctuations after the drag epoch is dominated by velocity fluctuations, i.e. velocity overshoot, not by the density term~\citep{Sunyaev:1970eu,Press:1980is}. As a result, the acoustic oscillations in the baryon component are displaced from those in the CMB by a phase shift of $\pi/2$ on large scales. 

After the drag epoch, pressure-less baryons quickly fall into the potential wells of cold dark matter fluctuations due to gravity. The oscillatory features are diluted relative to the CMB because of the small abundance of baryons relative to dark matter.

Whilst the baryon acoustic oscillations are the dominant feature in the CMB power spectrum, they are difficult to detect in the matter power spectrum. While in Fourier space we have an oscillatory feature, in real-space we expect a single peak at the sound horizon scale~\citep{Eisenstein:2006nj}. 

\subsection{Growth of perturbations}

After inflation has seeded perturbations in the distribution of matter, the dark matter perturbations start to grow through gravitational collapse, while the baryonic matter is dominated by its interaction with photons as discussed above. On scales larger than the horizon, matter perturbations are driven by the evolution of the perturbations in the dominant component of the Universe, which is $\propto a^2$ in the radiation-dominated era and $\propto a$ in the matter-dominated era. As the Hubble radius grows in the expanding Universe, it encompasses larger and larger perturbations. Within the horizon, perturbations are subject to different effects which determine their growth rate.

\begin{figure}[tb]
\begin{center}
\epsfig{file=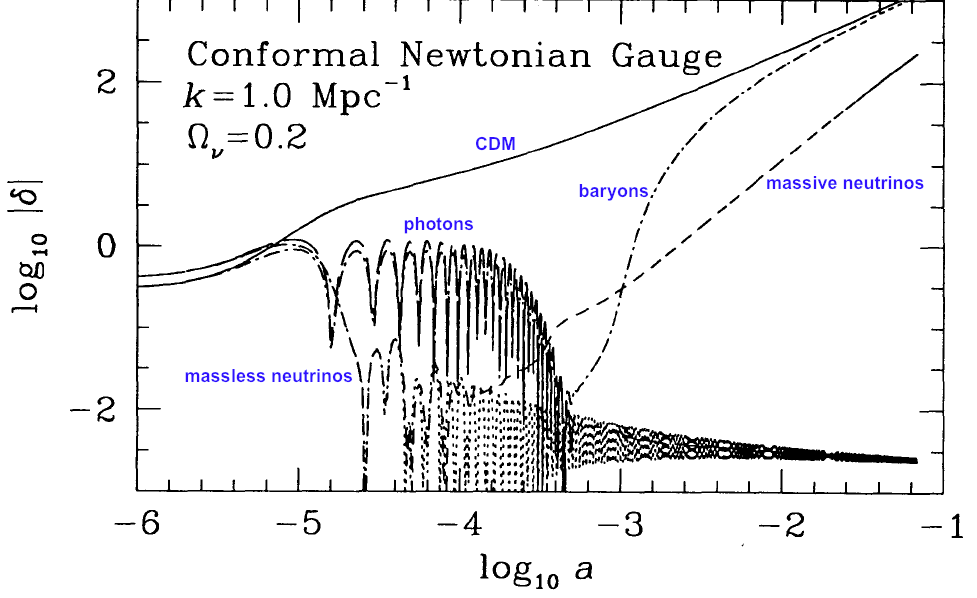,width=12cm}
\caption{This plot shows the evolution of a $k=1\,$Mpc$^{-1}$ mode for different energy components. This example represents a cold + hot dark matter model (see text for details). (credit:~\citep{Ma:1995ey} with additional labels)}
\label{fig:Ma}
\end{center}
\end{figure}

\begin{figure}[tb]
\begin{center}
\epsfig{file=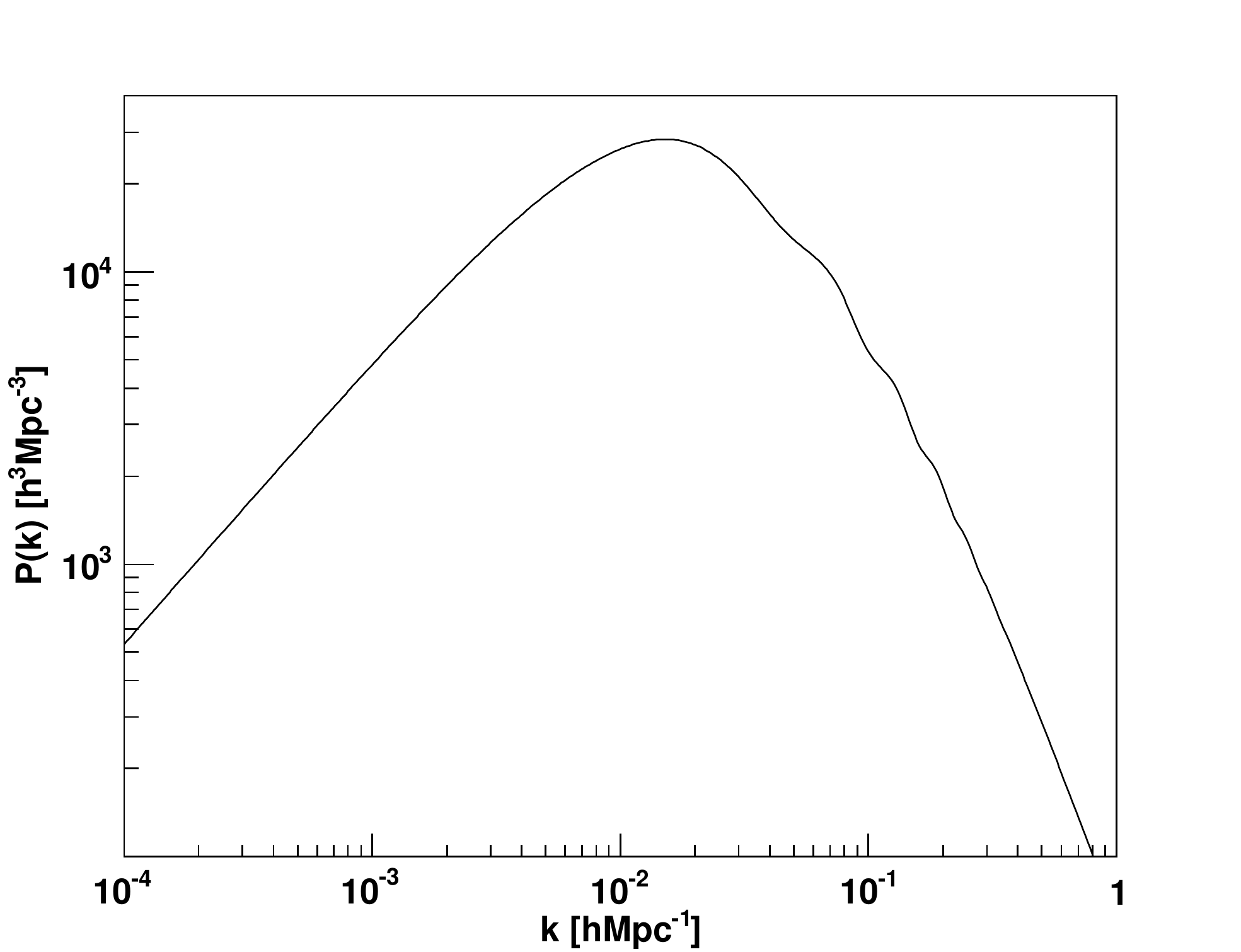,width=7.5cm}\hspace{-0.5cm}
\epsfig{file=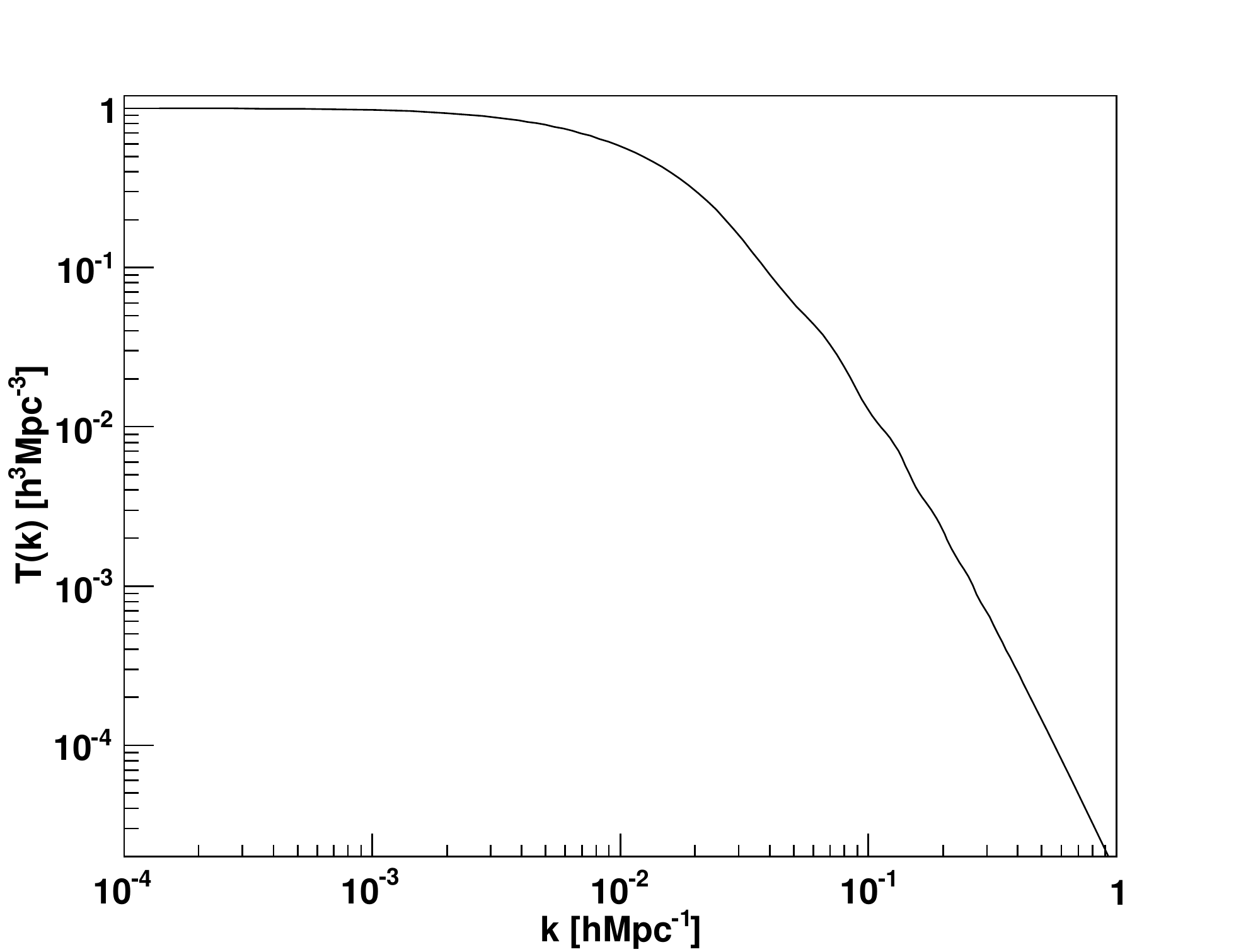,width=7.5cm}
\caption{Matter power spectrum and transfer function at redshift zero. The power spectrum follows the initial power spectrum $k^{n_s}$ untill $k_{eq}$ where its power is suppressed because such modes enter the horizon during the radiation-dominated era. The transfer function shows the suppression and is related to the power spectrum by $P(k) \propto k^{n_s}T(k)$.}
\label{fig:ps}
\end{center}
\end{figure}

Figure~\ref{fig:Ma} shows how the clustering of the different energy components evolves with time for a mode at $k=1\,$Mpc$^{-1}$~\citep{Ma:1995ey}. The solid line shows the cold dark matter which grows as $\log(a)$ in the radiation-dominated area ($-5 \lesssim \log_{10}(a) \lesssim -4.5$) and as $a$ in the matter-dominated area $\log_{10}(a) \gtrsim -4.5$. Very small wavenumbers $k$ (large scales) enter the horizon in the matter-dominated area and never encounter the suppression of growth during the radiation-dominated time. The transition of scales which enter the horizon during radiation domination to the scales which enter in the matter-dominated era is given by $k_{\rm eq}\approx 0.01h\,$Mpc$^{-1}$ and marks a peak in the power spectrum (see Figure~\ref{fig:ps}).

The radiation components show strong oscillations caused by the photon-baryon fluid. After the decoupling, the baryons (dashed-dotted line in Figure~\ref{fig:Ma}) fall into the potential wells of cold dark matter and soon have a similar clustering amplitude. The photons however do not participate in the clustering after decoupling. 

Massless neutrinos behave similarly to photons and do not participate in clustering, while massive neutrinos (dashed line) can cluster, depending on their momentum. In Figure~\ref{fig:Ma}, the neutrinos are at first, relativistic and behave like massless neutrinos, but at some point they become non-relativistic and start to cluster. The smaller clustering amplitude of massive neutrinos, which is in conflict with observations, is the strongest argument against hot dark matter. 

Furthermore the growth of cold dark matter perturbations during the radiation-dominated era is essential for the perturbation amplitude we see today. Without dark matter, the epoch of galaxy formation would occur substantially later in the Universe than is observed.

A convenient way to express how the amplitude of a certain mode changes over time is the transfer function
\begin{equation}
T(k,a) = \frac{\delta(k,a)}{\delta(k,a_e)}.
\end{equation}
Here we use $a_e$ to denote the scale factor at horizon entry, given by $a_e \propto k^{-1}$. Using the transfer function we can express the power spectrum at any time $a$ as a product of the initial power spectrum and the transfer function:
\begin{equation}
P(k,a) = T^2(k,a)P(k,a_e).
\end{equation}
The transfer function has an asymptotic behaviour of $T(k) \simeq 1$ for small $k$ and $T(k) \simeq k^{-2}$ for large $k$, with a turning point at 
\begin{equation}
\frac{1}{k} \simeq r_h(a_{eq}),
\end{equation}
where $r_h(a_{eq})$ is the Hubble radius at the time of radiation-matter equality. The behaviour $T(k) \simeq k^{-2}$ at large $k$ is because the large $k$ modes correspond to small-scale fluctuations which entered the horizon at the radiation domination era. At that time, the horizon size grows proportionally to the scale factor $r_h(a)\simeq a$. Since a fluctuation enters the horizon when $k\,r_h\simeq 1$, one has $a_{e}\propto k^{-1}$. Therefore the small-scale fluctuations are suppressed by $(a_{e}/a_{eq})^2\propto k^{-2}$.

Figure~\ref{fig:ps} shows a matter power spectrum (left) and transfer function (right) at redshift zero. On large scales (small $k$) we have the primordial power spectrum which behaves as $k^{n_s}$ since the transfer function is $1$. The power spectrum peaks at matter-radiation equality. The position of the peak (at $k_{\rm eq}$) depends on the horizon size at matter radiation equality and follows $k_{\rm eq}\propto \Omega_mh^2$. This means that all modes with $k > k_{\rm eq}$ enter the horizon when the Universe was still radiation-dominated, where the perturbations can only grow logarithmically, while the modes with $k < k_{\rm eq}$ enter the horizon when the Universe is matter dominated, where the perturbations are able to grow linearly. Hence the small-scale modes are suppressed relative to the larger-scale modes. Decreasing $\Omega_m$ will cause the matter radiation equality to happen later, the horizon size at matter radiation equality to be larger and the corresponding $k_{eq}$ to be smaller (the power spectrum peak moves to the left).

\section{Redshift space distortions}
\label{sec:pecvel}

Although it is possible to measure true distances for galaxies using standard candles, this is very time consuming. Redshifts are much easier to acquire than distances and the distortion in the mapping of galaxies in redshift space compared to real space allows a statistical measurement of peculiar velocities~\citep{Kaiser:1987qv,Hamilton:1992,Fisher:1993ye}.

The redshift-space position, $\vec{s}$, of a galaxy, is related to its real-space position, $\vec{r}$, by 
\begin{equation}
\vec{s} = \vec{r} + v_z(\vec{r})\hat{z},
\end{equation}
where $v_z(\vec{r}) \equiv u(\vec{r})/(aH(z))$ is the line-of-sight peculiar velocity with the $z$-axis being the line of sight. For a uniform z-independent mean galaxy density, the exact Jacobian for the transformation from real-space to redshift-space is
\begin{equation}
\frac{d^3s}{d^3r} = \left[1 + \frac{v_z(\vec{r})}{z}\right]^2\left[1 + \frac{dv_z(\vec{r})}{dz}\right].
\end{equation}
The galaxy over-density field in redshift-space can be obtained by imposing mass conservation, $(1+\delta_s)d^3s = (1+\delta_r)d^3r$. Furthermore we assume that $v_z$ is much smaller than the distance to the pair (plane parallel approximation; but see~\citep{Papai:2008bd} and the analysis in chapter~\ref{ch:BAO} and~\ref{ch:RSD}) and hence the term $v_z/z$ can be neglected. We get
\begin{equation}
\delta_s = \delta_r\left[1 + \frac{dv_z(\vec{r})}{dz}\right]^{-1}.
\end{equation}
If we assume an irrotational velocity field, we can write $dv_z/dz = -(d/dz)^2\nabla^{-2}\theta$, where $\theta \equiv -\nabla\cdot v$, and $\nabla$ is the Laplacian operator. In Fourier space, $(d/dz)^2\nabla^{-2} = (k_z/k)^2 = \mu_k^2$, where $\mu_k$ is the cosine of the line of sight angle, so we have
\begin{equation}
\delta_s(k) =\delta_r(k) + \mu^2\theta(k).
\end{equation}
Assuming the velocity field comes from linear perturbation theory, we have
\begin{equation}
\theta(k) = -\nabla\cdot v = \dot{\delta} = \delta \frac{d\ln \delta}{d\ln a} = \delta f,
\end{equation}
where $f = d\ln \delta/d\ln a$ is called the growth rate. We now have a linear relation between the redshift-space density field and the real space density field
\begin{equation}
\delta_s(k) = (1 + \mu^2f)\delta_r(k).
\end{equation}
For a population of galaxies with a linear bias $b$, the linear redshift-space power spectrum, $P_g(k)$ is then proportional to the linear, real-space matter power spectrum $P_m(k)$
\begin{equation}
P^s_g(k,\mu) = (b + f\mu^2)^2P_m(k) = b^2(1 + \beta\mu^2)^2P^r_m(k),
\end{equation}
where we have defined $\beta = f/b$. Although the real space power spectrum is isotropic, this is not true of the power spectrum in redshift-space. Plane waves along the line of sight are compressed in redshift-space and so their amplitudes are increased relative to those perpendicular to the line of sight. To see this effect in real-space, we can define a 2-dimensional correlation function $\xi(r_p,\pi)$ with the two separations
\begin{align}
\pi &= \frac{|\vec{s}\cdot \vec{h}|}{|\vec{s}|},\\
r_p &= \sqrt{|\vec{h}|^2 - \pi^2},
\end{align}
where we have defined the two vectors $\vec{s} = (\vec{s}_1 + \vec{s}_2)/2$ and $\vec{h} = \vec{s}_1 - \vec{s}_2$ through the positions of two galaxies $\vec{s}_1$ and $\vec{s}_2$. Now we can write~\citep{Hamilton:1992}
\begin{equation}
\xi(r_p,\pi) = \xi_0(s)\mathcal{P}_0(\mu) + \xi_2(s)\mathcal{P}_2(\mu) + \xi_4(s)\mathcal{P}_4(\mu),
\end{equation}
where the $\mathcal{P}_{\ell}$ are Legendre polynomials, $\mu = \cos(\theta)$ and $\theta$ is the angle between $r$ and the line of sight direction $\pi$. The correlation function moments $\xi_{\ell}$ are given by 
\begin{align}
\xi_{0}(s) &= \left(1 + \frac{2\beta}{3} + \frac{\beta^2}{5}\right)\xi(r),\\
\xi_{2}(s) &= \left(\frac{4\beta}{3} + \frac{4\beta^2}{7}\right)\left[\xi(r) + \overline{\xi}(r)\right],\\
\xi_{4}(s) &= \frac{8\beta^2}{35}\left[\xi(r) + \frac{5}{2}\overline{\xi}(r) + \frac{7}{2}\overline{\overline{\xi}}(r)\right]
\end{align}
and
\begin{align}
\overline{\xi}(r) &= \frac{3}{r^3}\int^r_0\xi(r')r'^2dr',\\
\overline{\overline{\xi}}(r) &= \frac{5}{r^5}\int^r_0\xi(r')r'^4dr'.
\end{align}
We will discuss more sophisticated models for the 2-dimensional correlation function in chapter~\ref{ch:RSD} including wide-angle effects.

\begin{figure}[tb]
\begin{center}
\epsfig{file=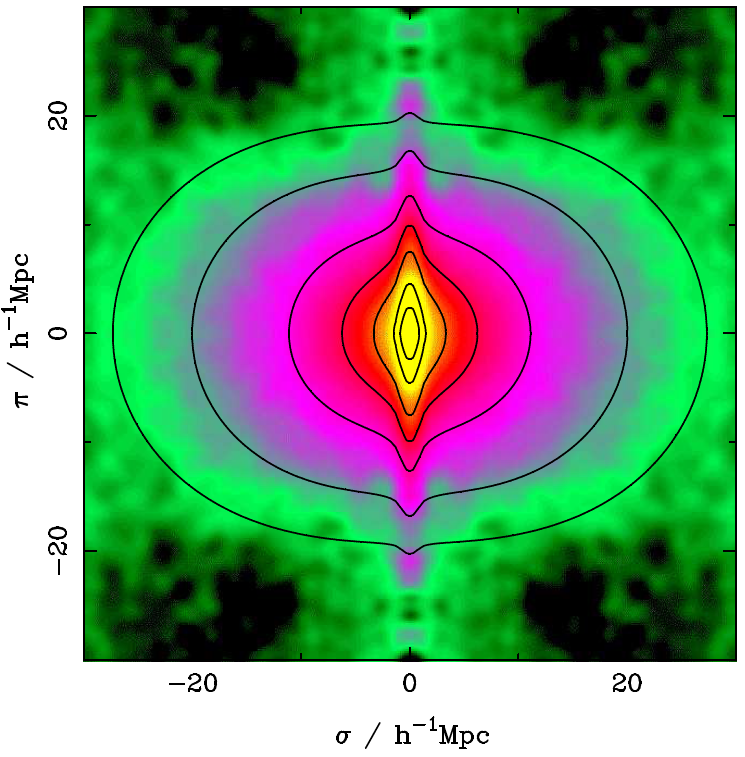,width=10cm}
\caption{The 2D correlation function of the 2-degree Field Galaxy Redshift Survey. The non-circular shape allows a measurement of the infall parameter $\beta$, which can be predicted by linear theory. While the plot uses $\sigma$ to denote the separation perpendicular to the line-of-sight, we used $r_p$ in this introduction. (credit:~\citealt{Peacock:2001gs})}
\label{fig:peacock}
\end{center}
\end{figure}

Figure~\ref{fig:peacock} shows one of the first detailed measurements of the two-dimensional correlation function, $\xi(\pi,r_p)$, using the 2-degree Field Galaxy Redshift Survey~\citep{Peacock:2001gs}. The overall non-circular structure of the signal is caused by the linear redshift space distortion effect, which we just discussed. The enhanced signal at small $r_p$ is the so called $``$finger of God$"$ effect, which is non-linear in nature and hence difficult to model.

\subsection{Testing models of gravity}
\label{sec:gravity}

While galaxy surveys measure $\beta$, the parameter interesting for cosmology is the growth rate $f = b\beta$. In linear theory, the growth rate can be predicted for a given set of cosmological parameters and allows tests of General Relativity on cosmic scales~\citep{Guzzo:2008ac}. 

\begin{figure}[tb]
\begin{center}
\epsfig{file=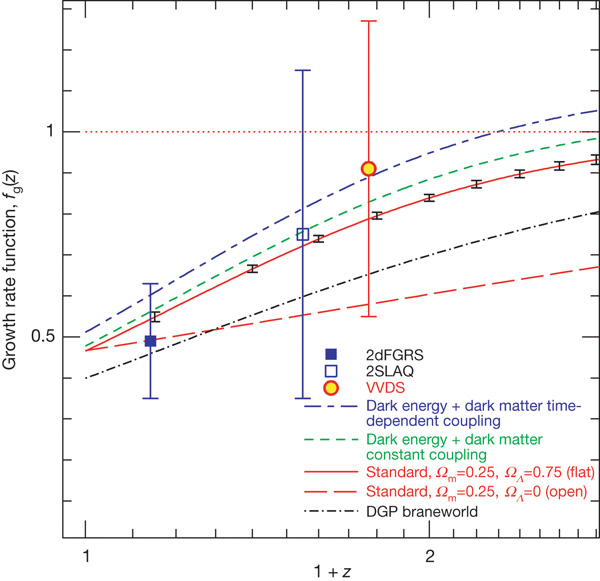,width=10cm}
\caption{The growth rate $f$ as a function of redshift $z$. The plot includes measurements from 2dFGRS, VVDS and 2SLAQ together with predictions $\Lambda$CDM (solid red line) and several alternative models. (credit:~\citealt{Guzzo:2008ac}).}
\label{fig:guzzo}
\end{center}
\end{figure}

\citet{Peebles:1980} showed, that $f = f(\Omega_m)$ and is well approximated by $f \approx \Omega_m^{0.6}(z)$ neglecting dark energy. Including dark energy as a cosmological constant, a small correction is needed~\citep{Lahav:1991wc}
\begin{equation}
f(\Omega_m,\Omega_{\Lambda}) \approx \Omega_m^{0.6}(z) + \frac{\Omega_{\Lambda}(z)}{70}\left(1 + \frac{\Omega_m(z)}{2}\right).
\end{equation}
This clearly shows that the growth rate $f$ mainly depends on the matter density $\Omega_m(z)$ and is only weakly dependent on the cosmological constant. This makes sense, since the uniform distribution of dark energy does not participate itself in the growth of perturbations and influences structure growth mainly through its impact on the expansion of the background Universe.

\citet{Linder:2005in} and \citet{Linder:2007hg} generalised this relation to $f \approx \Omega_m^{\gamma}(z)$ with the growth index $\gamma$, which for a $\Lambda$CDM Universe including General Relativity is predicted to be $\gamma_{\rm GR}\approx 0.55$. This allows us to directly test the standard model through the measurement of redshift space distortions in the galaxy distribution. Geometrical probes such as supernovae and baryon acoustic oscillations cannot alone distinguish between dark energy as a new negative pressure component of the Universe or a modification of laws of gravity. Redshift space distortions therefore offer a powerful way to complement geometrical probes by directly testing the laws of gravity on the largest scales possible. 

Figure~\ref{fig:guzzo} shows the first test of General Relativity using the growth rate $f$ from~\citet{Guzzo:2008ac}. The plot includes three data points from different galaxy surveys which all seem to be in agreement with the prediction of $\Lambda$CDM. Since this first analysis, new data have been collected, but up to now all such measurements seem to be in agreement with the standard model (see e.g.~\cite{Rapetti:2009ri,Samushia:2011cs,Blake:2011rj,Reid:2012sw,Hudson:2012gt,Rapetti:2012bu, Basilakos:2012ws} and this thesis).

\section{The halo model}
\label{sec:halo}

\begin{figure}[tb]
\begin{center}
\epsfig{file=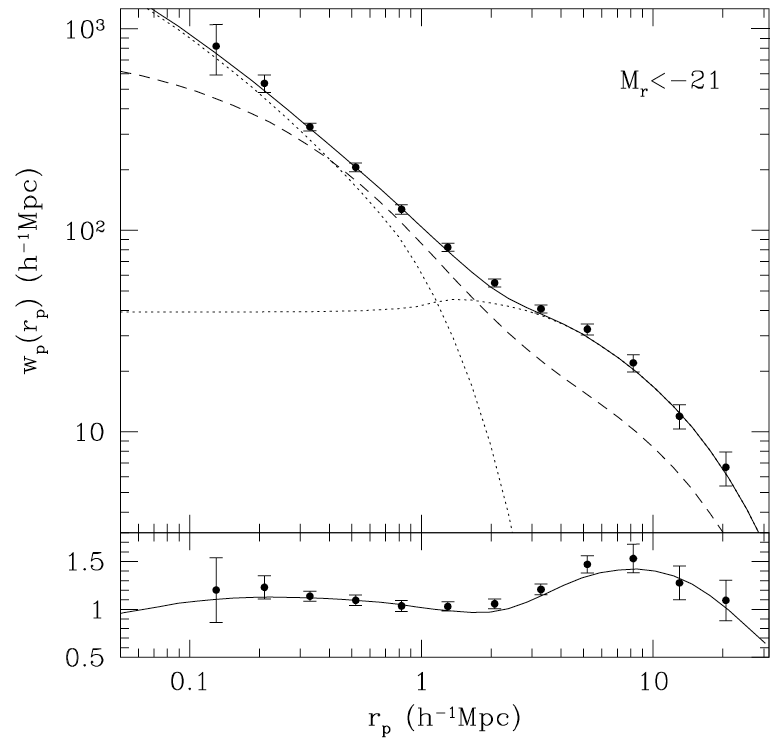,width=10cm}
\caption{The projected correlation function of a volume limited sample from SDSS together with the best fitting HOD model. (credit:~\citealt{Zehavi:2003ta})}
\label{fig:zehavi}
\end{center}
\end{figure}

The halo model approach (see e.g.~\citealt{Jing:1997nb,Ma:2000ik,Peacock:2000qk,Seljak:2000gq,Berlind:2001xk,Cooray:2002dia}) to galaxy clustering assumes that large-scale structure can be described by a distribution of dark matter halos while galaxies themselves form within these halos. Such models have been tested against n-body simulations and seem to be able to capture the galaxy-mass relation.

The contribution to the power spectrum (or correlation function) can be written as a sum of correlations of galaxies that occupy two different halos and the correlation of galaxies within a single halo
\begin{equation}
P_g(k) = P^{1h}_g(k) + P^{2h}_g(k).
\end{equation}
The exact definition of the one-halo and two-halo terms will be discussed in chapter~\ref{ch:HOD}.

The halo model is based on external inputs, such as the dark matter halo profile and the halo mass function, which come from n-body simulations. The halo model then needs to provide a strategy for how to populate dark matter halos. Usually the assumption is that above a certain mass threshold the halo will have a central galaxy, and the probability to have a satellite galaxy grows with mass. 

The first measurements of galaxy clustering in the correlation function showed that it follows a power law~\citep{Totsuji:1969,Peebles:1973,Hauser:1973,Hauser:1974,Peebles:1974}. The halo occupation distribution (HOD) model however predicts clear deviations from a power law, especially at the transition scale, between the one-halo and two halo terms, above which galaxy pairs sit in different dark matter halos. Figure~\ref{fig:zehavi} shows the correlation function of a volume limited sample of the SDSS survey together with the best fitting HOD model. The HOD model is able to explain the structure seen in the correlation function, most obviously the transition from the one-halo term to the two halo term at around $2h^{-1}\,$Mpc. We will discuss such an analysis using data from the 6dFGS in chapter~\ref{ch:HOD}.

\section{Observational cosmology}
\label{sec:obs}

In order to test the geometry of the Universe we have to measure the redshift-distance relation, since this relation is directly connected to the different components of the Universe. For this purpose one needs objects whose intrinsic luminosity or size is known. The first class of objects are called $``$standard candles$"$, while the second are known as $``$standard rulers$"$. The currently best known objects to qualify as standard candles are Type Ia supernovae, while a very good standard ruler is provided by the sound horizon, i.e. the distance a sound wave travelled between the Big Bang and the baryon drag epoch. The standard ruler technique is an important part of this thesis and will be discussed in more detail in the following sections. The standard candle technique is only used to compare cosmological constraints from different models and we refer the reader to the literature for a more detailed discussion of this method (e.g.~\citealt{Perlmutter:2003}).

\subsection{The Cosmic Microwave Background}
\label{sec:CMB}

\begin{figure}[tb]
\begin{center}
\epsfig{file=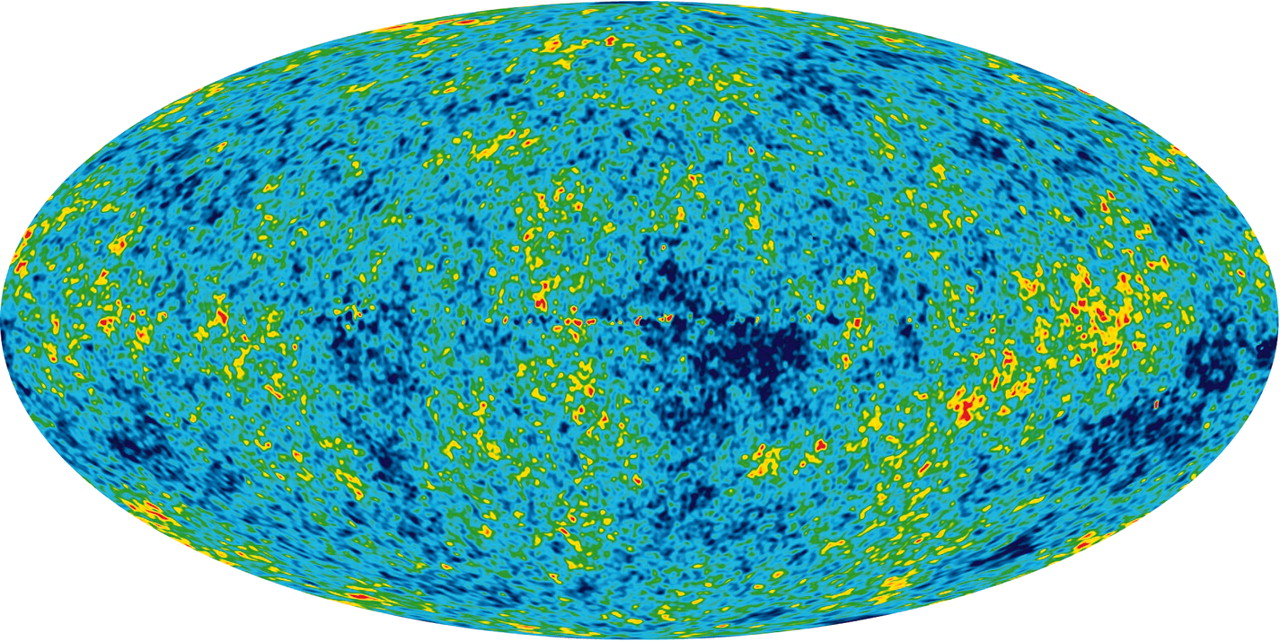,width=10cm}
\caption{The CMB temperature distribution as measured by WMAP7. To make the very small temperature fluctuations visible, the mean CMB temperature of $T_{\rm CMB} = 2.75\,$Kelvin as well as the CMB dipole and the Galactic emission have been subtracted. (credit: NASA)}
\label{fig:WMAP7}
\end{center}
\end{figure}

We have already discussed the physics before recombination, after which the photons travel with minimal interaction until today. When the photons decoupled from the baryons, they travelled out of the small matter perturbations which existed at that time. Although these perturbations were small, they caused a gravitational redshift or blueshift of the photons~\citep{Sachs:1967er}, which is still visible in the distribution of photons today (see Figure~\ref{fig:WMAP7}). By observing the photon distribution we can therefore learn about the very early matter distribution, which includes baryon acoustic oscillations.

\begin{figure}[tb]
\begin{center}
\epsfig{file=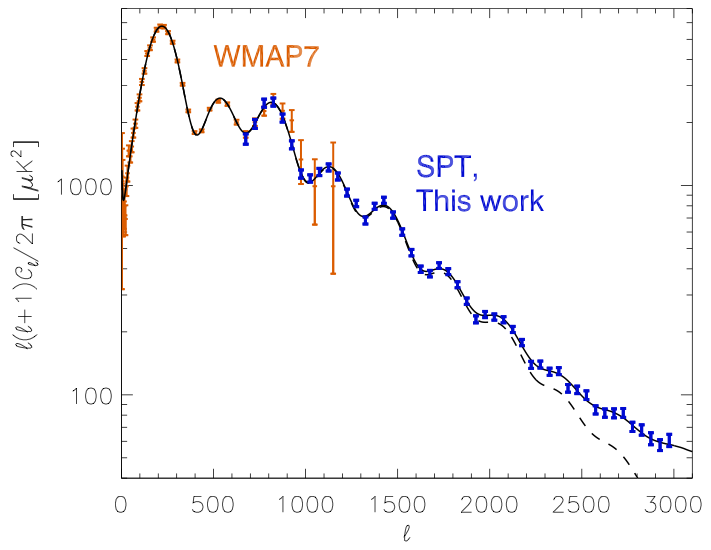,width=10cm}
\caption{The CMB angular power spectrum from WMAP7~\citep{Komatsu:2010fb} and SPT~\citep{Keisler:2011aw}. These two measurements together allow the detection of up to 8 peaks as well as CMB lensing. At large $\ell$ the data deviate from the simple model (dashed line) because of emission from background sources like clusters and starburst galaxies. 
(credit:~\citealt{Keisler:2011aw})}
\label{fig:CMB_ps}
\end{center}
\end{figure}

The CMB temperature anisotropy is a function over a sphere. We can separate out the contributions of different angular scales by doing a multipole expansion, 
\begin{equation}
\delta T(\theta,\phi) = \sum_{\ell m} Y_{\ell m}(\theta,\phi)a_{\ell m},
\end{equation}
where the sum runs over the multipole $\ell = 1, 2,... \infty$ and $m=-\ell,...,\ell$, giving $2\ell + 1$ values of $m$ for each $\ell$. The functions $Y_{\ell m}(\theta,\phi)$ are the spherical harmonics, which form an orthonormal set of functions over the sphere, so that we can calculate the multipole coefficients $a_{\ell m}$ from
\begin{equation}
a_{\ell m} = \int Y^*_{\ell m}(\theta,\phi)\delta T(\theta,\phi)d\Omega.
\end{equation}
Since each $a_{\ell m}$ represents a deviation from the average temperature, their expectation value is zero,
\begin{equation}
\langle a_{\ell m} \rangle = 0
\end{equation}
and the quantity we are interested in is the variance $\langle |a_{\ell m}|^2\rangle$ which gives the typical size of the $a_{\ell m}$. The $a_{\ell m}$ are independent of $m$ (in an isotropic Universe), since $m$ represents just the orientation, which then allows us to define
\begin{equation}
\mathcal{C}_{\ell} \equiv \langle |a_{\ell m}|^2\rangle = \frac{1}{2\ell + 1} \sum_m \langle |a_{\ell m}|^2\rangle.
\end{equation}
For Gaussian perturbations, the values of $\mathcal{C}_{\ell}$ contain all the statistical information about the CMB temperature anisotropy. The function $\mathcal{C}_{\ell}$ (of integers $\ell \geq 1$) is called the angular power spectrum, and this is the quantity which usually is compared to observations (analogous to the power spectrum $P(k)$ of density perturbations). 

The angular power spectrum $\ell(\ell + 1)\mathcal{C}_{\ell}/2\pi$ as measured by WMAP7 and SPT together with the best fitting $\Lambda$CDM model is shown in Figure~\ref{fig:CMB_ps} as a function of multipole $\ell$. 

\subsection{Baryon Acoustic Oscillations as a standard ruler}

The important consequence of the linear gravitational instability theory is that all the features present in the initial matter fluctuation spectrum should survive throughout cosmic evolution. The baryon acoustic oscillations which are responsible for the characteristic features in the CMB power spectrum should be visible in the low redshift matter distribution. 

The co-moving size of an object or a feature at redshift $z$ in the line-of-sight ($\pi$) and transverse ($r_p$) direction is related to the observed redshift size $\Delta z$ and angular size $\Delta \theta$ by the Hubble parameter $H(z)$ and the angular diameter $D_A(z)$:
\begin{equation}
\begin{split}
\pi &= \frac{c\Delta z}{H(z)},\\
r_p &= (1 + z)D_A(z)\Delta\theta.
\end{split}
\end{equation}
Thus the measurement of the observed dimensions, $\Delta z$ along the line-of-sight direction and $\Delta\theta$ in the transverse direction, give $\pi H(z)$ and $r_p/D_A(z)$. When the true physical scale of the object or feature, $\pi$ and $r_p$, are known, measurements of the observables, $\Delta z$ and $\Delta\theta$, give estimates of $H(z)$ and $D_A(z)$. Such an object is called a $``$standard ruler$"$.

In the case that the signal-to-noise is not sufficient to separate the two signals parallel and perpendicular to the line-of-sight, we can constrain a combination of both given by~\citep{Eisenstein:2005su,Padmanabhan:2008ag}
\begin{equation}
D_V(z) = \left[(1+z)^2D^2_A(z)\frac{cz}{H(z)}\right]^{1/3}.
\end{equation}
Currently we don't know of any process which can alter the sound horizon scale since decoupling by more than $1\%$ and hence such a measurement has a low systematic uncertainty~\citep{Eisenstein:2006nk,Crocce:2007dt,Sanchez:2008iw,Matsubara:2008wx,Seo:2008yx,Smith:2007gi,Padmanabhan:2009yr}.

\begin{figure}[tb]
\begin{center}
\epsfig{file=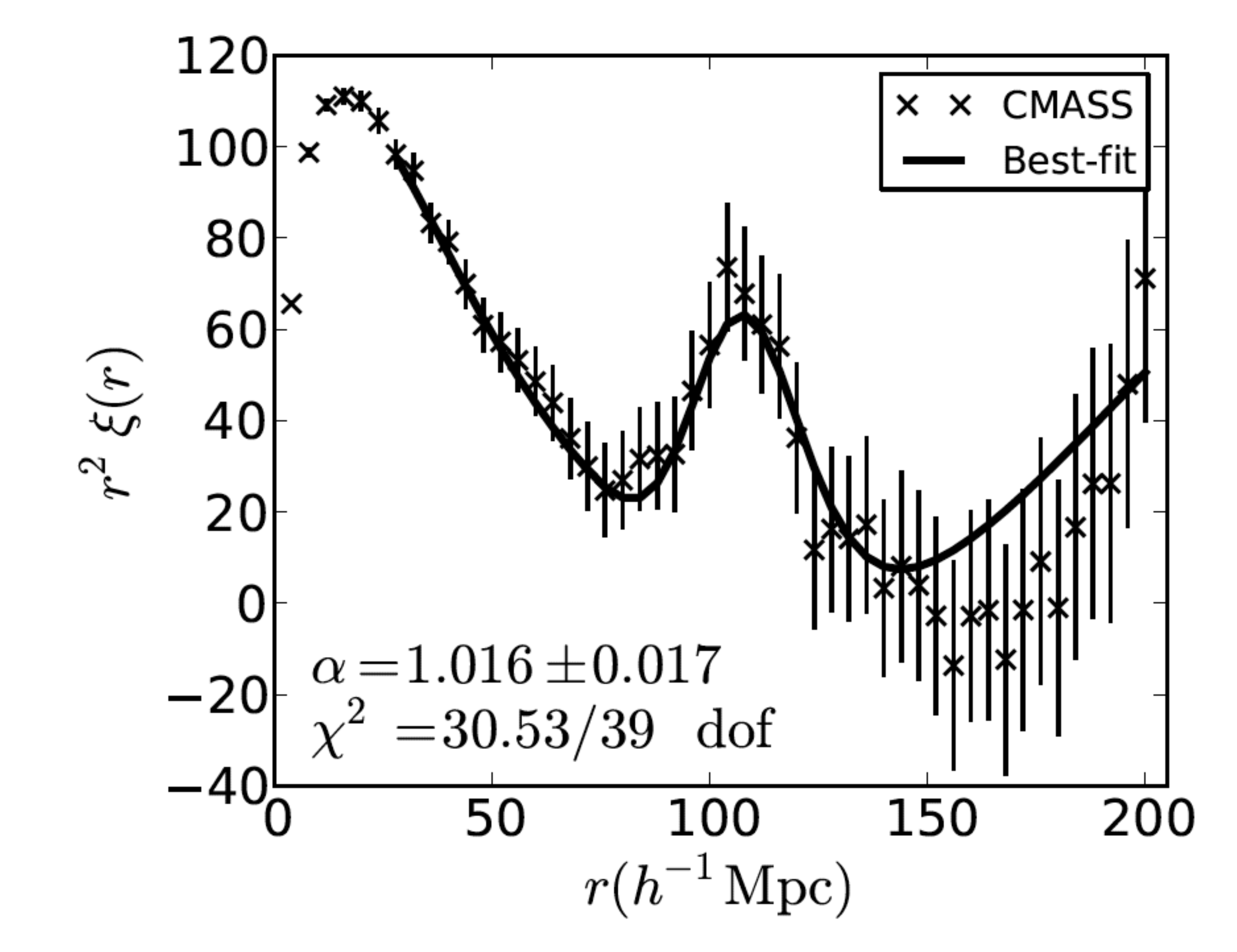,width=12cm}
\caption{The large scale correlation function of the BOSS-CMASS sample. (credit:~\citealt{Anderson:2012sa})}
\label{fig:BOSS_BAO}
\end{center}
\end{figure}

The first measurements of the baryon acoustic peak were performed using the SDSS-LRG sample~\citep{Eisenstein:2005su} and the 2dFGRS sample~\citep{Cole:2005sx}. More recently similar measurements haven been made in other galaxy surveys like 6dFGS (this thesis), WiggleZ~\citep{Blake:2011en} and BOSS~\citep{Anderson:2012sa}. The correlation function of the BOSS-CMASS sample is shown in Figure~\ref{fig:BOSS_BAO}. The BAO peak is clearly visible at around $r \approx 100h^{-1}\,$Mpc.

Currently this technique is limited by the statistical precision of the BAO signal detection in the galaxy distribution, but ongoing surveys like BOSS~\citep{Schlegel:2009hj} aim for $1\%$ constraints on $H(z)$ and $D_A(z)$. 

We will discuss the 6dFGS BAO analysis in chapter~\ref{ch:BAO}. 

\subsection{Combining different cosmological probes}

\begin{figure}[tb]
\begin{center}
\epsfig{file=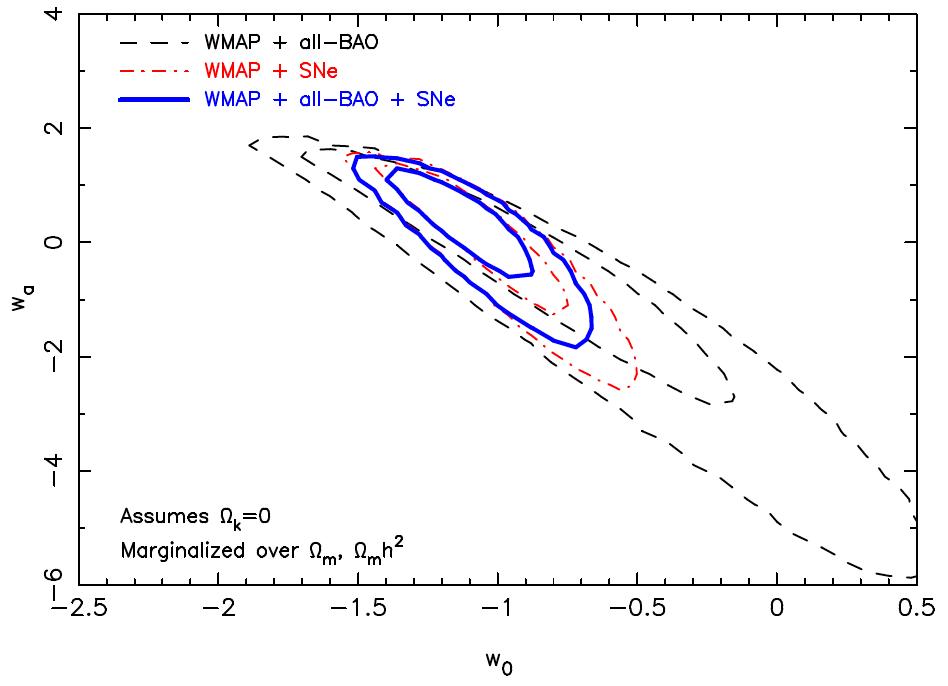,width=12cm}
\caption{Constraints in $w(z)$ from BAO, CMB and SN data. Because SN data sample the redshift distribution with more data-points, the SN data contribute strongly to this constraint. BAO data are expected to become dominant in the future, because SN samples have reached strong systematic limitations, which are difficult to overcome, while BAO data are still statistically limited. (credit:~\citealt{Blake:2011en})}
\label{fig:chris}
\end{center}
\end{figure}

Different cosmological probes are sensitive to different cosmological parameters. For example, supernova (SN) data are particularly good for constraints on the dark energy equation of state parameter especially if this parameter is varied as a function of redshift (see Figure~\ref{fig:chris}). The reason for this is that SN Ia data sample many different redshifts and hence can detect variations in $w(z)$ more precisely. However SN data are now limited by systematic uncertainties, which are difficult to overcome, while BAO data are still statistically limited. With BAO data rapidly improving, we expect the BAO data to eventually dominate over SN data, while the Cosmic Microwave Background will remain unchallenged as the most powerful cosmological probe.
Compared to SN, BAO data are more sensitive to the curvature parameter, because they have a very large lever arm from low redshift to the era of decoupling. Different cosmological probes can easily be combined using the product of the likelihoods or the sum of the $\chi^2$
\begin{equation}
\chi^2_{tot} = \chi^2_{CMB} + \chi^2_{BAO} + \chi^2_{SN}.
\end{equation}
with
\begin{equation}
\chi^2 = \sum_i \frac{\left(x^{\rm data}_i - x^{\rm model}_i\right)^2}{\sigma_i^2}.
\end{equation}
The likelihood contours of such a combination of different probes are shown in Figure~\ref{fig:chris} for $w(z)$ and in Figure~\ref{fig:first_plot} for $\Omega_{\Lambda}$ and $\Omega_m$.

It is reassuring that the standard model of cosmology has now been confirmed by several different independent probes. For example  BAO data alone now require dark energy~\citep{Blake:2011wn} and since CMB lensing has been detected, the CMB data alone also points towards dark energy~\citep{vanEngelen:2012va}. High precision constraints can be obtained by combining any two of the three main cosmological probes, SN Ia, BAO or CMB. Hence, although the CMB remains the most powerful cosmological probe, even excluding the CMB data, the remaining observational probes still give strong evidence for the standard model of cosmology~\citep{Percival:2009xn}.

\section{Alternative models of dark energy}
\label{sec:models}

The interpretation of the accelerated expansion of the Universe as vacuum energy has several theoretical weaknesses. For example, it is not clear why we happen to live at a time when dark matter and dark energy have approximately the same energy density, while the Universe was dominated by only one energy component during most of its lifetime and will be in its future (coincidence problem).

It is also not clear why the vacuum energy is so small. Quantum field theory predicts a level of vacuum energy $120$ orders of magnitude larger than observed. It was always clear that the Quantum field theoretical prediction would not lead to a Universe which would allow life to develop and hence some unknown process must cancel out vacuum energy. 
While this situation should have been already unsettling, the discovery of the accelerated expansion of the Universe in 1998~\citep{Perlmutter:1998np,Riess:1998} and its consistency with a cosmological constant implied that vacuum energy did not cancel out completely, but an imbalance of one part in $10^{120}$ remains. 

This situation has led to the suggestion of many alternative models. In this section we will outline the three main categories: modified gravity models, scalar field models and inhomogeneous models, grouped by which part of the standard model they propose to modify. We will focus on the basic idea of these models and refer to recent reviews of this topic for a more comprehensive discussion~\citep{Copeland:2006wr,Durrer:2008in,Frieman:2008sn,Tsujikawa:2010sc,Li:2011sd,Clifton:2011jh}.

\subsection{Scalar Fields: Quintessence}

Scalar fields naturally arise in particle physics and can act as candidates for dark energy. Quintessence represents a scalar field that is coupled with gravity. Given a particular potential, quintessence can describe the late-time acceleration of the expansion of the Universe. The action for quintessence is given by
\begin{equation}
S = \int d^4x\sqrt{-g}\left[\frac{1}{2}(\nabla\phi)^2 - V(\phi)\right],
\end{equation}
where $g = \det(g_{\mu\nu})$, with the metric tensor $g_{\mu\nu}$, $(\nabla\phi)^2 = g^{\mu\nu}\partial_{\mu}\phi\partial_{\nu}\phi$ and $V(\phi)$ is the potential of the field. In a flat FLRW Universe
this action varies as
\begin{equation}
\ddot{\phi} + 3H\dot{\phi} + \frac{dV}{d\phi} = 0
\end{equation}
with respect to $\phi$, where homogeneity of the scalar field is assumed. The stress-energy tensor has a form identical to that of an ideal fluid with pressure and density identical to the one we found for the case of inflation (see eq.~\ref{eq:inf})
\begin{equation}
\begin{split}
p = \frac{1}{2}\dot{\phi}^2 + V(\phi),\\
\rho = \frac{1}{2}\dot{\phi}^2 - V(\phi).
\end{split}
\end{equation}
The equation of state parameter for the field $\phi$ is given by
\begin{equation}
w_{Q} = \frac{p}{\rho} = \frac{\frac{1}{2}\dot{\phi}^2 + V(\phi)}{\frac{1}{2}\dot{\phi}^2 - V(\phi)},
\end{equation}
which suggests that $-1 \leq w \leq 1$. If the time evolution is slow $w \simeq -1$, and the field behaves like a slowly varying vacuum energy.

Because of the lack of observational constraints, there is no reason to favour one form of the potential with respect to others. For example, the original scenario proposed by~\citet{Ratra:1987rm} was a potential of the form $V(\phi) = M^{4+\alpha}/\phi^{\alpha}$, whereas~\citet{Caldwell:2005tm} proposed a quintessence theory consisting of two classes of thawing and freezing models, depending on whether the field accelerates or decelerates with time. Another approach involves a modification of the kinetic term, as in k-essence models~\citep{ArmendarizPicon:2000ah}.

Dynamical models can generally deliver some answers to the question of the nature of dark energy, but do not yet have the capability to provide a complete solution. Some classes of models have the capability to solve the coincidence problem, but the smallness of the cosmological constant (or minimum of the potential in this case) means that the fine tuning problem remains.

\subsection{Modified Gravity: f(R)}

Here we discuss models which modify the geometrical part of the Einstein equation. The $f(R)$ modified gravity theories replace the standard Einstein-Hilbert action by an arbitrary function of the Ricci scalar $R$. In Riemannian geometry, $R$ describes the simplest curvature invariant of a Riemannian manifold. It assigns a single real number to each point on the manifold which is determined by the intrinsic geometry of the manifold near that point~\citep{Lobo:2008sg}. 

The general action for a modified gravity field is given by
\begin{equation}
S = \frac{1}{16\pi G}\int d^4x\sqrt{-g}[f(R) + \mathcal{L}_m],
\end{equation}
where $\mathcal{L}_m$ is the matter Lagrangian density. Variation of this action with respect to the metric $g^{\mu\nu}$ yields the field equation
\begin{equation}
FR_{\mu\nu} - \frac{1}{2}fg_{\mu\nu} - \nabla_{\mu}\nabla_{\nu}F + g_{\mu\nu}\square F = 8\pi GT^m_{\mu\nu},
\end{equation}
where $F\equiv df(R)/dR$, $\square = \partial^{\mu}\partial_{\mu}=\partial^2/\partial t^2 - \nabla^2$ is the d'Alembert operator and $T^m_{\mu\nu}$ is the matter stress-energy tensor. This equation can be written as
\begin{equation}
G_{\mu\nu} \equiv R_{\mu\nu} - \frac{1}{2}Rg_{\mu\nu} = 8\pi GT^{\rm eff}_{\mu\nu},
\end{equation}
where the new term $T^{\rm eff}_{\mu\nu} = T^c_{\mu\nu} + T^m_{\mu\nu}/F$ describes the effective stress-energy tensor. All of the effects of the modification of gravity are now contained within the curvature part of the tensor $T^c_{\mu\nu}$.

For a FLRW metric, the generalised Friedmann equations have an identical form to eqs.~\ref{eq:friedman1} and~\ref{eq:friedman2} with $p_{eff} = p_m + p_c$ and $\rho_{eff} = \rho_m + \rho_c$. The pressure and density of the curvature field are given by:
\begin{equation}
\begin{split}
p_c &= \frac{1}{8\pi GF}\left[2\frac{\dot{a}}{a}\dot{R}f' + \ddot{R}F' + \dot{R}^2F'' - \frac{1}{2}[f - RF]\right],\\
\rho_c &= \frac{1}{8\pi GF}\left[\frac{1}{2}[f - Rf] - 3\frac{\dot{a}}{a}\dot{R}F'\right].
\end{split}
\end{equation}
Given the right choice of $f(R)$, these models can reproduce the late-time acceleration of the Universe.

While $f(R)$ models are very general, they can provide viable models to compare to observations, and many $f(R)$ theories have already been ruled out~\citep{Bertotti:2003rm,Amendola:2007nt,Hu:2007pj}.

\subsection{Inhomogeneous Models}

Using the Friedmann equations for a Universe which is clearly not homogeneous on small scales could introduce a bias into cosmological parameter constraints. We have to consider that non-linear processes could have a back-reaction effect on the background cosmology. Furthermore it is not obvious how we have to perform a covariant and gauge invariant averaging over the inhomogeneous Universe to arrive at the correct FLRW background.

It has been claimed in recent years that for certain inhomogeneous models, such effects are large enough to mimic an accelerating Universe (see e.g.~\citealt{Buchert:1999er,Buchert:2001sa,Maartens:2011yx,Ellis:2011hk}). This would be a satisfying resolution of the coincidence problem without the need for any dark energy field. However in many cases it also means that the Cosmological Principle is not correct. 

In perturbation theory, we introduce small perturbations to a homogeneous metric. In an inhomogeneous model, it is assumed that the perturbations are too big to just treat them as linear perturbations and hence the exact solutions of Einstein's field equations need to be used. The best known example of such an exact solution is the Lemaitre-Tolman-Bondi metric~\citep{Lemaitre:1933gd,Tolman:1934za,Bondi:1947av}, which describes a spherical cloud of dust (finite or infinite) that is expanding or collapsing under gravity. A spherical under-density of sufficient size could explain the discrepancy between nearby and distant supernovae luminosities without dark energy~\citep{PascualSanchez:1999zr,Vanderveld:2006rb,Enqvist:2007vb}.

These proposals have been disputed, and so far we can not say that there is a convincing demonstration that acceleration could emerge naturally from nonlinear effects during structure formation. Even if such effects are not the cause of the accelerating expansion of the Universe, we should note that back-reaction effects could significantly affect our estimations of cosmological parameters, even if they do not lead to acceleration~\citep{Li:2007ci}.

\section{The 6\lowercase{d}F Galaxy Survey}
\label{sec:6dFGS}

\begin{figure*}[tbp]
\begin{center}
\epsfig{file=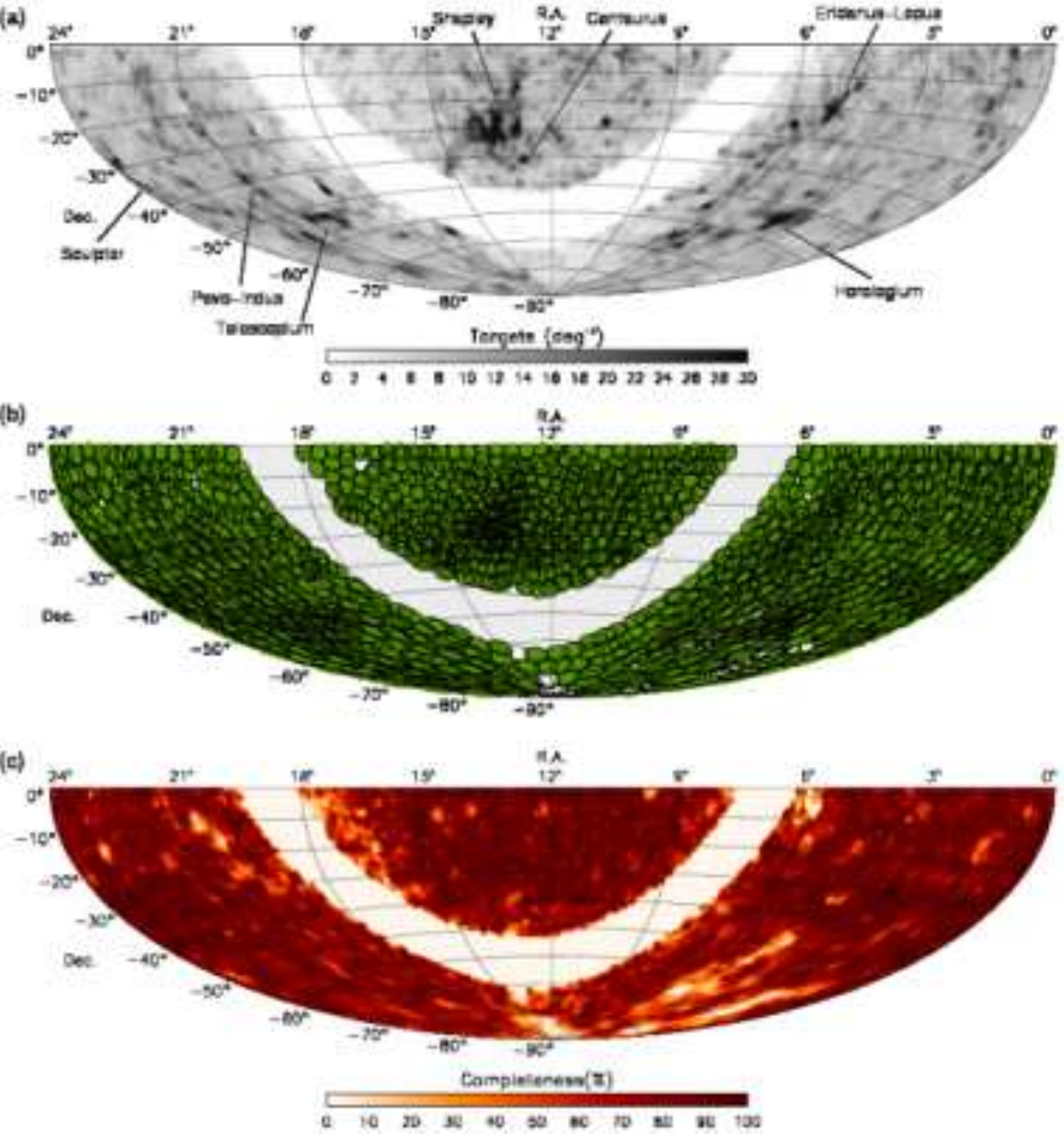,width=15.2cm}
\caption{Density of 6dFGS target sources (per square degree) on the sky; key supercluster over-densities are labelled. (b) Full
6dFGS field coverage (filled discs) and unobserved target fields (open circles). (c) Redshift completeness for $K \leq 12.65$. All panels
show equal-area Aitoff projections. (credit:~\citealt{Jones:2009yz})}
\label{fig:heath}
\end{center}
\end{figure*}

This thesis presents new constraints on the cosmological model by analysing data from the 6-degree Field Galaxy Survey, a near-infrared selected ($JHK$) redshift survey of about $125\,000$ galaxies across almost the complete southern sky, with secondary samples selected in $b_J$ and $r_F$. The $J$,$H$ and $K$ bands exclude a $\pm 10^{\circ}$ band along the Galactic plane to minimise extinction and foreground source confusion, while $b_J$ and $r_F$ exclude a $20^{\circ}$ band. 

The total sky coverage is about $17\,000\,$deg$^2$ and the median redshift of the sample is $z = 0.05$. The galaxies were selected from the Two-Micron All-Sky Survey - Extended Source Catalog~\citep[2MASS XSC;][]{Jarrett:2000me}. The spectra were taken with the 6-degree Field (6dF) multi-fibre instrument on the UK Schmidt Telescope from 2001 to 2006. The mean completeness of 6dFGS is $92\%$ but varies with sky position and magnitude (see~\citet{Jones:2006xy} for details). 

For our analysis we mostly used the $K$-band selected sample, since it is by far the largest sample in 6dFGS. For the BAO analysis in chapter~\ref{ch:BAO} we used a faint magnitude limit of $K=12.95$, while for the redshift space distortion analysis in chapter~\ref{ch:RSD} we used a limit of $K=12.75$. The motivation for the higher limit in the BAO analysis was to maximise the effective volume by selecting more faint, high redshift galaxies. For the HOD analysis in chapter~\ref{ch:HOD} we used the $J$-band selected sample, because this band is particularly robust against background noise and hence provides the best stellar mass estimates. The details of the samples used for the different analysis techniques will be explained at the beginning of each of the following chapters.

A subset of early-type 6dFGS galaxies (approximately $10\,000$) have measured line-widths that will be used to derive Fundamental Plane distances and peculiar velocities. This will lead to the largest peculiar velocity dataset ever produced and will allow bulk flow measurements as well as tests of General Relativity.

\subsection{Comparison to Other Surveys}

The 6dF Galaxy Survey covers approximately ten times the area of the 2dF Galaxy Redshift Survey (2dFGRS;~\citealt{Colless:2001gk}) and  has more than twice the areal coverage of the Sloan Digital Sky Survey seventh data release (SDSS DR7;~\citealt{Abazajian:2008wr}).
In terms of secure redshifts, 6dFGS has around half the number of 2dFGRS and one-sixth those of SDSS DR7 ($r < 17.77$). The co-moving volume covered by 6dFGS is about the same as 2dFGRS at their respective median redshifts, and around $30\%$ that of SDSS DR7. A sub-sample of about $60\,000$ luminous red galaxies (LRGs)~\citep{Eisenstein:2001cq} in SDSS covers a volume of about $1h^{-3}\,$Gpc$^3$ at redshift $z\approx 0.35$, with the main focus on cosmological measurements like baryon acoustic oscillations.

In terms of fibre aperture size, the larger apertures of 6dFGS ($6.7''$) give a projected diameter of $4.8h^{-1}\,$kpc at the median redshift of the survey, covering $40\%$ more projected area than SDSS at its median redshift, and more than three times the area of 2dFGRS. 

The WiggleZ Dark Energy Survey~\citep{Drinkwater:2009sd} measured the redshifts of about $250\,000$ galaxies over a volume comparable to the SDSS-LRG sample but at higher redshift of around $z\approx 0.6$. This survey focuses on the measurement of dark energy through the detection of the baryon acoustic oscillations, similar to the still ongoing Baryon Oscillation Spectroscopic Survey (BOSS)~\citep{Schlegel:2009hj} which is part of the Sloan Digital Sky Survey III (SDSS-III)~\citep{Eisenstein:2011sa}. BOSS has a mean redshift of $z\approx 0.5$ and will collect $1.5\,$million galaxy redshifts over a volume of about $6h^{-3}\,$Gpc$^3$. It will also measure baryon acoustic oscillations in the distribution of $Ly$-alpha absorption from the spectra of $\sim 150\,000$ quasars at redshifts $z > 2.2$~\citep{Ross:2011ky}.

\section{Overview and motivation for this thesis}
\label{sec:overview}

In the next chapter (chapter~\ref{ch:BAO}) we will discuss the baryon acoustic oscillation analysis of 6dFGS. The detection of the acoustic feature allows an absolute distance measurement to the effective redshift of the survey $z_{\rm eff} = 0.1$. At such a low redshift, the distance is independent of most cosmological parameters, except the Hubble constant $H_0$. Therefore our analysis allows a robust measurement of this parameter similar to the distance ladder technique. The systematic uncertainties of the BAO technique to derive the Hubble constant are very different to the distance ladder technique and hence allow an independent crosscheck. It is also possible to break degeneracies in the WMAP7 dataset by combining it with our measurement. 

The redshift of 6dFGS is much lower than other galaxy surveys which reported BAO detections so far and hence delivers a new data point on the BAO Hubble diagram. Combining BAO measurements at different redshifts allows us to detect dark energy by just using BAO data, without combining it with the CMB~\citep{Blake:2011wn}.

The error in our BAO measurement is sample variance limited and hence better constraints are possible with larger surveys. At the end of chapter~\ref{ch:BAO} we include predictions for two future low redshift surveys, the radio galaxy survey, WALLABY, which will be conducted using the Australian SKA pathfinder (ASKAP) telescope in Western Australia and the TAIPAN survey, which is planned for the UK Schmidt telescope at Siding Spring Observatory. With an effective volume three times that of 6dFGS, TAIPAN will be able to constrain the Hubble constant to $3\%$ using the BAO technique, significantly improving upon the 6dFGS result.\\

In chapter~\ref{ch:RSD} we use the amplitude of redshift-space distortions to measure the parameter combination $f\sigma_8$, where $f = d\ln D/d\ln a$ is the linear growth rate and $\sigma_8$ is the r.m.s. of matter fluctuations in $8h^{-1}\,$Mpc spheres. The expected growth rate can be derived from General Relativity, allowing a test of this theory on cosmic scales. As we explained in section~\ref{sec:gravity}, such tests might reveal more information on the nature of dark energy. With 6dFGS, we can probe a new redshift range, which has not been investigated by other galaxy surveys so far. The simplest $\Lambda$CDM parameterisation of the growth rate $f$ is given by $f = \Omega_m^{\gamma}(z)$, where the growth index, $\gamma$, is the parameter we are interested in. Within this simple parameterisation we can see that we are most sensitive to $\gamma$ when $\Omega_m(z)$ is small. The minimum of $\Omega_m(z)$ is at $z=0$ with $\Omega_m(z) \approx 1$ at redshift $z > 1$. This means that low redshift constraints on the growth rate are the most powerful in measuring $\gamma$. This is not necessarily true for other parameterisations of the growth rate, but other parameterisations usually depend on more parameters and as long as we haven't detected any deviation from $\Lambda$CDM, and without a convincing alternative model of gravity, it is practical to follow the simplest parameterisation. 

A further advantage of low redshift constraints on the growth rate is their relative independence from the Alcock-Paczynski (AP) effect. The AP effect originates from the assumption of a fiducial cosmological model to turn redshifts into distances. If this initial assumption is wrong, this will introduce a distortion into the 2D correlation function, which is very similar to the linear redshift-space distortion signal. While the AP effect is interesting as an additional tool to test cosmological models, it leads to a degeneracy with $f\sigma_8$. In 6dFGS the redshift distance relation is almost completely independent of the fiducial cosmological model and hence the AP effect is negligible.

Galaxy redshift surveys alone cannot constrain the growth rate $f$ because there is a degeneracy between the galaxy bias, $b$, and $\sigma_8$. This is the reason why most galaxy surveys report constraints on $f\sigma_8$ or $\beta$ instead of the actual parameter of interest, $f$. In our analysis we will combine our measurement of $f\sigma_8$ with WMAP7, but we also investigate alternative methods to directly constrain $f$ without additional datasets.

Again we include forecasts for constraints on $f\sigma_8$ expected from the WALLABY and TAIPAN surveys. Although WALLABY has a smaller effective volume, compared to TAIPAN, it will deliver better constraint on $f\sigma_8$, since it has a much smaller galaxy bias.\\

Chapter~\ref{ch:HOD} uses the halo model to investigate the relation between dark matter clustering and galaxy clustering. The question is how can we relate the observable properties of galaxies to the not-observable properties of dark matter halos? Studies have shown that stellar mass is tightly related to dark matter halo mass (e.g.~\citet{Wang:2007pa}). Hence we chose to investigate galaxy clustering as a function of this observable in 6dFGS. The near-infrared selection of 6dFGS allows very reliable estimates of stellar masses, more robust against background noise~\citep{Bell:2000jt,Drory:2004ib,Kannappan:2007ys,Longhetti:2008gv,Grillo:2007kg,Gallazzi:2009aj} compared to other galaxy surveys which rely on optical bands. We will discuss this aspect in more detail in chapter~\ref{ch:HOD}.
We divide 6dFGS into four volume-limited sub-samples with thresholds in stellar mass and redshift and calculate the projected correlation functions $w_p(r_p)$. We then fit these correlation functions with our HOD model. Our analysis allows us to investigate the effective dark matter halo mass for a given stellar mass as well as the satellite fraction. 

We compare our results to semi-analytic catalogues derived from the Millennium Simulation. Semi-analytic catalogues attempt to model baryonic effects, such as gas cooling and supernova feedback, with analytical descriptions placed on top of a dark matter only N-body simulation. The analytical descriptions are calibrated to reproduce different observables, like the galaxy luminosity function or stellar mass function. By comparing observational results with semi-analytic models we can investigate whether the different properties of these models are a satisfying description of all processes influencing the formation and evolution of galaxies. Such information can than be used to improve upon these semi-analytic models, with the goal of furthering our understanding of galaxy formation.

We summarise and conclude in chapter~\ref{ch:conc}
 
 \cleardoublepage

\renewcommand*\abstractname{\flushleft \Large Abstract\hfill}

\chapter{Baryon Acoustic Oscillations and the Local Hubble Constant}
\label{ch:BAO}
\begin{center}
\emph{\textbf{\citeauthor{Beutler:2011hx}}}\\
\emph{\textbf{MNRAS 416, 3017B (2011)}}
\end{center}

\begin{abstract}
We analyse the large-scale correlation function of the 6dF Galaxy Survey (6dFGS) and detect a Baryon Acoustic Oscillation (BAO) signal. The 6dFGS BAO detection allows us to constrain the distance-redshift relation at $z_{\rm eff} = 0.106$. We achieve a distance measure of $D_V(z_{\rm eff}) = 456\pm27\;$Mpc and a measurement of the distance ratio, $r_s(z_d)/D_V(z_{\rm eff}) = 0.336\pm0.015$ ($4.5\%$ precision), where $r_s(z_d)$ is the sound horizon at the drag epoch $z_d$. The low effective redshift of 6dFGS makes it a competitive and independent alternative to Cepheids and low-$z$ supernovae in constraining the Hubble constant. We find a Hubble constant of $H_0 = 67.0\pm3.2\;$km s$^{-1}\;$Mpc$^{-1}$ ($4.8\%$ precision) that depends only on the WMAP-7 calibration of the sound horizon and on the galaxy clustering in 6dFGS. Compared to earlier BAO studies at higher redshift, our analysis is less dependent on other cosmological parameters. The sensitivity to $H_0$ can be used to break the degeneracy between the dark energy equation of state parameter $w$ and $H_0$ in the CMB data. We determine that $w = -0.97\pm0.13$, using only WMAP-7 and BAO data from both 6dFGS and~\citet{Percival:2009xn}. 

We also discuss predictions for the large scale correlation function of two future wide-angle surveys: the WALLABY blind H{\sc I} survey (with the Australian SKA Pathfinder, ASKAP), and the proposed TAIPAN all-southern-sky optical galaxy survey with the UK Schmidt Telescope (UKST). We find that both surveys are very likely to yield detections of the BAO peak, making WALLABY the first radio galaxy survey to do so. We also predict that TAIPAN has the potential to constrain the Hubble constant with $3\%$ precision.
\end{abstract}

\section{Introduction}

The current standard cosmological model, $\Lambda$CDM, assumes that the initial fluctuations in the distribution of matter were seeded by quantum fluctuations pushed to cosmological scales by inflation. Directly after inflation, the Universe is radiation dominated and the baryonic matter is ionised and coupled to radiation through Thomson scattering. The radiation pressure drives sound-waves originating from over-densities in the matter distribution~\citep{Peebles:1970ag,Sunyaev:1970eu,Bond:1987ub}. At the time of recombination ($z_* \approx 1090$) the photons decouple from the baryons and shortly after that (at the baryon drag epoch $z_d \approx 1020$) the sound wave stalls. Through this process each over-density of the original density perturbation field has evolved to become a centrally peaked perturbation surrounded by a spherical shell~\citep{Bashinsky:2000uh,Bashinsky:2002vx, Eisenstein:2006nj}. The radius of these shells is called the sound horizon $r_s$. Both over-dense regions attract baryons and dark matter and will be preferred regions of galaxy formation. This process can equivalently be described in Fourier space, where during the photon-baryon coupling phase, the amplitude of the baryon perturbations cannot grow and instead undergo harmonic motion leading to an oscillation pattern in the power spectrum.

After the time of recombination, the mean free path of photons increases and becomes larger than the Hubble distance. Hence from now on the radiation remains almost undisturbed, eventually becoming the Cosmic Microwave Background (CMB).

The CMB is a powerful probe of cosmology due to the good theoretical understanding of the physical processes described above. The size of the sound horizon depends (to first order) only on the sound speed in the early Universe and the age of the Universe at recombination, both set by the physical matter and baryon densities, $\Omega_mh^2$ and $\Omega_bh^2$~\citep{Eisenstein:1997ik}. Hence, measuring the sound horizon in the CMB gives extremely accurate constraints on these quantities~\citep{Komatsu:2010fb}. Measurements of other cosmological parameters often show degeneracies in the CMB data alone~\citep{Efstathiou:1998xx}, especially in models with extra parameters beyond flat $\Lambda$CDM. Combining low redshift data with the CMB can break these degeneracies.

Within galaxy redshift surveys we can use the correlation function, $\xi$, to quantify the clustering on different scales. The sound horizon marks a preferred separation of galaxies and hence predicts a peak in the correlation function at the corresponding scale. The expected enhancement at $s = r_s$ is only $\Delta\xi \approx 10^{-3}b^2(1 + 2\beta/3 + \beta^2/5)$ in the galaxy correlation function, where $b$ is the galaxy bias compared to the matter correlation function and $\beta$ accounts for linear redshift space distortions. Since the signal appears at very large scales, it is necessary to probe a large volume of the Universe to decrease sample variance, which dominates the error on these scales~\citep{Tegmark:1997rp,Goldberg:1997kv,Eisenstein:1998tu}.

Very interesting for cosmology is the idea of using the sound horizon scale as a standard ruler~\citep{Eisenstein:1998tu,Cooray:2001av, Seo:2003pu, Blake:2003rh}. A standard ruler is a feature whose absolute size is known. By measuring its apparent size, one can determine its distance from the observer. The BAO signal can be measured in the matter distribution at low redshift, with the CMB calibrating the absolute size, and hence the distance-redshift relation can be mapped (see e.g.~\cite{Bassett:2009mm} for a summary).

The Sloan Digital Sky Survey (SDSS; \citealt{York:2000gk}), and the 2dF Galaxy Redshift Survey (2dFGRS; \citealt{Colless:2001gk}) were the first redshift surveys which have directly detected the BAO signal. Recently the WiggleZ Dark Energy Survey has reported a BAO measurement at redshift $z=0.6$~\citep{Blake:2011wn}.

\cite{Eisenstein:2005su} were able to constrain the distance-redshift relation to $5\%$ accuracy at an effective redshift of $z_{\rm eff} = 0.35$ using an early data release of the SDSS-LRG sample containing $\approx 47\,000$ galaxies. Subsequent studies using the final SDSS-LRG sample and combining it with the SDSS-main and the 2dFGRS sample were able to improve on this measurement and constrain the distance-redshift relation at $z_{\rm eff} = 0.2$ and $z_{\rm eff} = 0.35$ with $3\%$ accuracy~\citep{Percival:2009xn}. Other studies of the same data found similar results using the correlation function $\xi(s)$ \citep{Martinez:2009,Gaztanaga:2008xz,Labini:2009ke,Sanchez:2009jq,Kazin:2009cj}, the power spectrum $P(k)$ \citep{Cole:2005sx,Tegmark:2006az,Huetsi:2006gu,Reid:2009xm}, the projected correlation function $w(r_p)$ of photometric redshift samples \citep{Padmanabhan:2006ku,Blake:2006kv} and a cluster sample based on the SDSS photometric data~\citep{Huetsi:2009zq}. Several years earlier a study by~\cite{Miller:2001cf} found first hints of the BAO feature in a combination of smaller datasets.

Low redshift distance measurements can directly measure the Hubble constant $H_0$ with a relatively weak dependence on other cosmological parameters such as the dark energy equation of state parameter $w$. The 6dF Galaxy Survey is the biggest galaxy survey in the local Universe, covering almost half the sky. If 6dFGS could be used to constrain the redshift-distance relation through baryon acoustic oscillations, such a measurement could directly determine the Hubble constant, depending only on the calibration of the sound horizon through the matter and baryon density. The objective of the present paper is to measure the two-point correlation function on large scales for the 6dF Galaxy Survey and extract the BAO signal.

Many cosmological parameter studies add a prior on $H_0$ to help break degeneracies. The 6dFGS derivation of $H_0$ can provide an alternative source of that prior. The 6dFGS $H_0$-measurement can also be used as a consistency check of other low redshift distance calibrators such as Cepheid variables and Type Ia supernovae (through the so called distance ladder technique; see e.g.~\citealp{Freedman:2000cf,Riess:2011yx}). Compared to these more classical probes of the Hubble constant, the BAO analysis has an advantage of simplicity, depending only on $\Omega_mh^2$ and $\Omega_bh^2$ from the CMB and the sound horizon measurement in the correlation function, with small systematic uncertainties. 

Another motivation for our study is that the SDSS data after data release 3 (DR3) show more correlation on large scales than expected by $\Lambda$CDM and have no sign of a cross-over to negative $\xi$ up to $200h^{-1}\;$Mpc (the $\Lambda$CDM prediction is $140h^{-1}\;$Mpc)~\citep{Kazin:2009cj}. It could be that the LRG sample is a rather unusual realisation, and the additional power just reflects sample variance. It is interesting to test the location of the cross-over scale in another red galaxy sample at a different redshift.

This paper is organised as follows. In Section~\ref{sec:6dF} we introduce the 6dFGS survey and the $K$-band selected sub-sample used in this analysis. In Section~\ref{sec:analysis2} we explain the technique we apply to derive the correlation function and summarise our error estimate, which is based on log-normal realisations. In Section~\ref{sec:model} we discuss the need for wide angle corrections and several linear and non-linear effects which influence our measurement. Based on this discussion we introduce our correlation function model. In Section~\ref{sec:DV} we fit the data and derive the distance estimate $D_V(z_{\rm eff})$. In Section~\ref{sec:results2} we derive the Hubble constant and constraints on dark energy. In Section~\ref{sec:sig} we discuss the significance of the BAO detection of 6dFGS. In Section~\ref{sec:future} we give a short overview of future all-sky surveys and their power to measure the Hubble constant. We conclude and summarise our results in Section~\ref{sec:conclusion}.\\


\section{The 6dF galaxy survey}
\label{sec:6dF}

\subsection{Targets and Selection Function}
\label{sec:selectionfn}

The galaxies used in this analysis were selected to $K \leq 12.9$ from the 2MASS Extended Source Catalog~\citep[2MASS XSC;][]{Jarrett:2000me} and combined with redshift data from the 6dF Galaxy Survey \citep[6dFGS;][]{Jones:2009yz}. The 6dF Galaxy Survey is a combined redshift and peculiar velocity survey covering nearly the entire southern sky with $|b| < 10^\circ$. It was undertaken with the Six-Degree Field (6dF) multi-fibre instrument on the UK Schmidt Telescope from 2001 to 2006. The median redshift of the survey is $z = 0.052$ and the $25\%:50\%:75\%$ percentile completeness values are $0.61:0.79:0.92$. Papers by~\citet{Jones:2004zy, Jones:2006xy, Jones:2009yz} describe 6dFGS in full detail, including comparisons between 6dFGS, 2dFGRS and SDSS.

Galaxies were excluded from our sample if they resided in sky regions with completeness lower than 60 percent. After applying these cuts our sample contains $75\,117$ galaxies. The selection function was derived by scaling the survey completeness as a function of magnitude to match the integrated on-sky completeness, using mean galaxy counts. This method is the same adopted by~\citet{Colless:2001gk} for 2dFGRS and is explained in~\citet{Jones:2006xy} in detail. The redshift of each object was checked visually and care was taken to exclude foreground Galactic sources. The derived completeness function was used in the real galaxy catalogue to weight each galaxy by its inverse completeness. The completeness function was also applied to the mock galaxy catalogues to mimic the selection characteristics of the survey. Jones et al. (in preparation) describe the derivation of the 6dFGS selection function, and interested readers are referred to this paper for a more comprehensive treatment.

\subsection{Survey volume}

\begin{figure}[tb]
\begin{center}
\epsfig{file=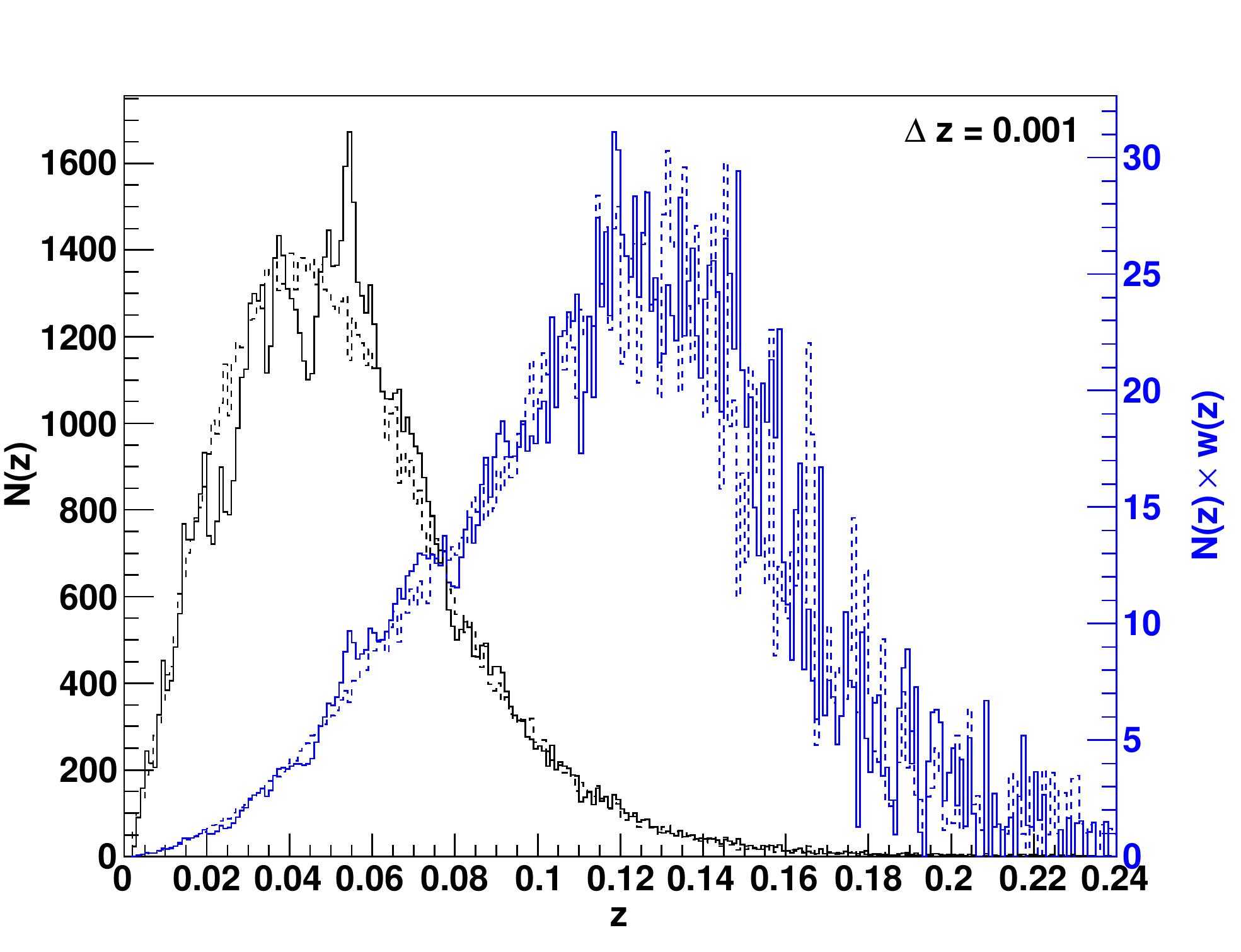,width=10cm}
\caption{Redshift distribution of the data (black solid line) and the random catalogue (black dashed line). The weighted 
distribution (using weights from eq.~\ref{eq:weight}) is shifted to higher redshift and has increased shot noise 
but a smaller error due to sample variance (blue solid and dashed lines).}
\label{fig:wred}
\end{center}
\end{figure}

We calculated the effective volume of the survey using the estimate of~\citet{Tegmark:1997rp}
\begin{equation}
V_{\rm eff} = \int d^3\vec{x} \left[\frac{n(\vec{x})P_0}{1 + n(\vec{x})P_0}\right]^2
\label{eq:veff}
\end{equation}
where $n(\vec{x})$ is the mean galaxy density at position $\vec{x}$, determined from the data, and $P_0$ is the characteristic power spectrum amplitude of the BAO signal. The parameter $P_0$ is crucial for the weighting scheme introduced later. We find that the value of $P_0 = 40\,000h^{-3}\;$Mpc$^3$ (corresponding to the value of the galaxy power spectrum at $k \approx 0.06h\;$Mpc$^{-1}$ in 6dFGS) minimises the error of the correlation function near the BAO peak.

Using $P_0 = 40\,000h^{-3}\;$Mpc$^3$ yields an effective volume of $0.08h^{-3}\;$Gpc$^3$, while using instead $P_0 = 10\,000h^{-3}\;$Mpc$^3$ (corresponding to $k \approx 0.15h\;$Mpc$^{-1}$) gives an effective volume of $0.045h^{-3}\;$Gpc$^3$.

The volume of the 6dF Galaxy Survey is approximately as large as the volume covered by the 2dF Galaxy Redshift Survey, with a sample density similar to SDSS-DR7~\citep{Abazajian:2008wr}. \citet{Percival:2009xn} reported successful BAO detections in several samples obtained from a combination of SDSS DR7, SDSS-LRG and 2dFGRS with effective volumes in the range $0.15$ - $0.45h^{-3}\;$Gpc$^3$ (using $P_0 = 10\,000h^{-3}\;$Mpc$^3$), while the original detection by~\citet{Eisenstein:2005su} used a sample with $V_{\rm eff} = 0.38h^{-3}\;$Gpc$^3$ (using $P_0 = 40\,000h^{-3}\;$Mpc$^3$).

\section{Clustering measurement}
\label{sec:analysis2}

We focus our analysis on the two-point correlation function. In the following sub-sections we introduce the technique used to estimate the correlation function and outline the method of log-normal realisations, which we employed to derive a covariance matrix for our measurement.

\subsection{Random catalogues}

To calculate the correlation function we need a random sample of galaxies which follows the same angular and redshift selection function as the 6dFGS sample. We base our random catalogue generation on the 6dFGS luminosity function of Jones et al. (in preparation), where we use random numbers to pick volume-weighted redshifts and luminosity function-weighted absolute magnitudes. We then test whether the redshift-magnitude combination falls within the 6dFGS $K$-band faint and bright apparent magnitude limits ($8.75 \leq K \leq 12.9$).

Figure~\ref{fig:wred} shows the redshift distribution of the 6dFGS $K$-selected sample (black solid line) compared to a random catalogue with the same number of galaxies (black dashed line). The random catalogue is a good description of the 6dFGS redshift distribution in both the weighted and unweighted case.

\subsection{The correlation function}
\label{sec:cor}

\begin{figure}[tb]
\begin{center}
\epsfig{file=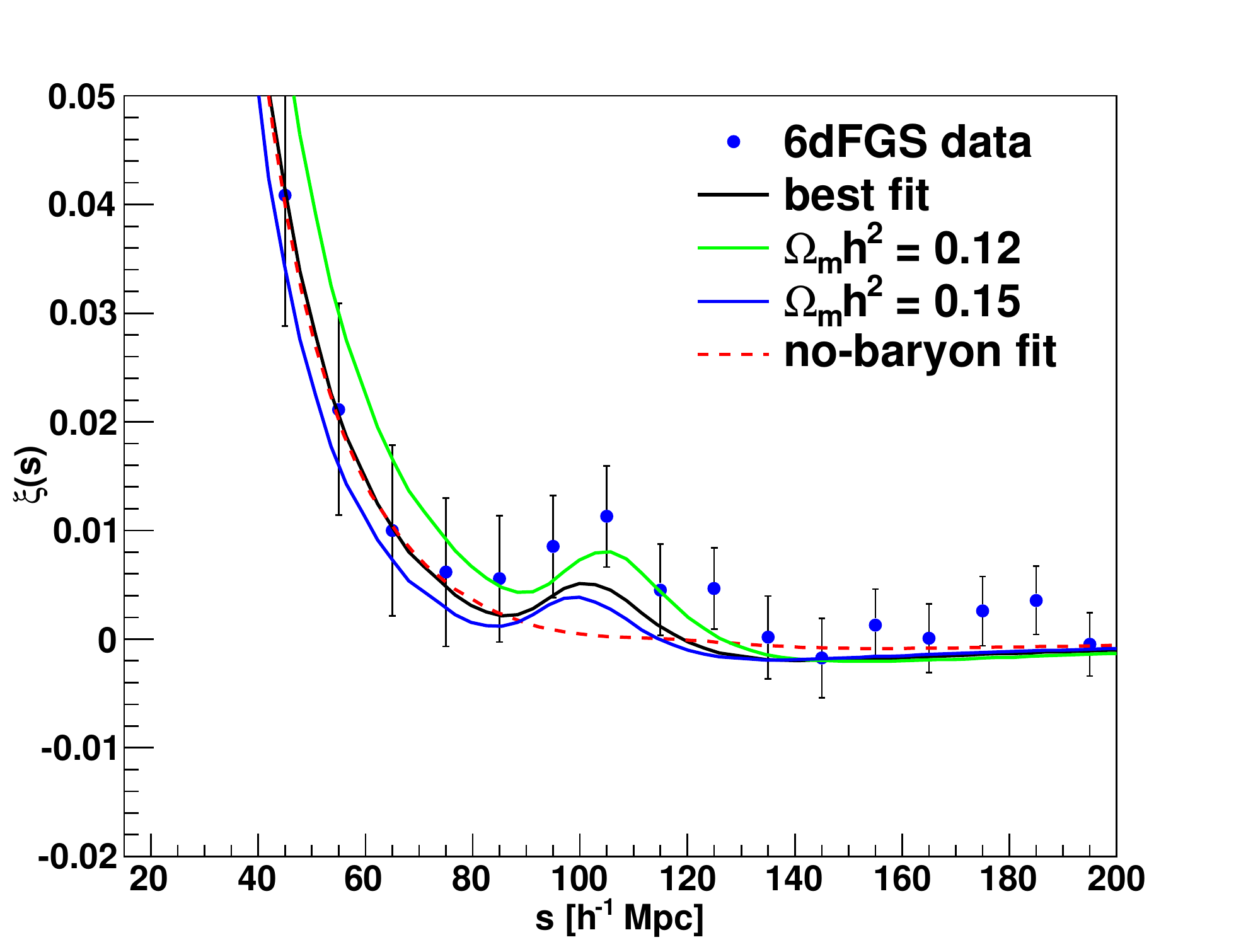,width=10cm}
\caption{The large scale correlation function of 6dFGS. The best fit model is shown by the black line with the best fit value of $\Omega_mh^2 = 0.138\pm0.020$. Models with different $\Omega_mh^2$ are shown by the green line ($\Omega_mh^2 = 0.12$) and the blue line ($\Omega_mh^2 = 0.15$). The red dashed line is a linear CDM model with $\Omega_bh^2 = 0$ (and $\Omega_mh^2 = 0.1$), while all other models use the WMAP-7 best fit value of $\Omega_bh^2 = 0.02227$~{\protect \citep{Komatsu:2010fb}}. The significance of the BAO detection in the black line relative to the red dashed line is $2.4\sigma$ (see Section~\ref{sec:sig}). The error-bars at the data points are the diagonal elements of the covariance matrix derived using
log-normal mock catalogues.}
\label{fig:bao}
\end{center}
\end{figure}

We turn the measured redshift into co-moving distance via
\begin{equation}
D_C(z) = \frac{c}{H_0}\int^z_0\frac{dz'}{E(z')}
\end{equation} 
with
\begin{align}
E(z) = \big[&\Omega^{\rm fid}_m(1+z)^3 + \Omega^{\rm fid}_k(1+z)^2\cr
             &+ \Omega^{\rm fid}_{\Lambda}(1+z)^{3(1+w^{\rm fid})})\big]^{1/2},
\end{align}
where the curvature $\Omega^{\rm fid}_k$ is set to zero, the dark energy density is given by $\Omega^{\rm fid}_{\Lambda} = 1-\Omega^{\rm fid}_m$ and the equation of state for dark energy is $w^{\rm fid} = -1$. Because of the very low redshift of 6dFGS, our data are not very sensitive to $\Omega_k$, $w$ or any other higher dimensional parameter which influences the expansion history of the Universe. We will discuss this further in Section~\ref{sec:AandR}.

Now we measure the separation between all galaxy pairs in our survey and count the number of such pairs in each separation bin. We do this for the 6dFGS data catalogue, a random catalogue with the same selection function and a combination of data-random pairs. We call the pair-separation distributions obtained from this analysis $DD(s), RR(s)$ and $DR(s)$, respectively. The binning is chosen to be from $10h^{-1}$\;Mpc up to $190h^{-1}$\;Mpc, in $10h^{-1}$\;Mpc steps. In the analysis we used $30$ random catalogues with the same size as the data catalogue. The redshift correlation function itself is given by~\citet{Landy:1993yu}:
\begin{equation}
\xi'_{\rm data}(s) = 1 + \frac{DD(s)}{RR(s)} \left(\frac{n_r}{n_d} \right)^2 - 2\frac{DR(s)}{RR(s)} \left(\frac{n_r}{n_d} \right),
\end{equation}
where the ratio $n_r/n_d$ is given by
\begin{equation}
\frac{n_r}{n_d} = \frac{\sum^{N_r}_iw_i(\vec{x})}{\sum^{N_d}_jw_j(\vec{x})}
\end{equation}
and the sums go over all random ($N_r$) and data ($N_d$) galaxies. We use the inverse density weighting of~\citet{Feldman:1993ky}:
\begin{equation}
w_i(\vec{x}) = \frac{C_i}{1 + n(\vec{x})P_0},
\label{eq:weight}
\end{equation}
with $P_0 = 40\,000h^{3}$\;Mpc$^{-3}$ and $C_i$ being the inverse completeness weighting for 6dFGS (see Section~\ref{sec:selectionfn} and Jones et al., in preparation). This weighting is designed to minimise the error on the BAO measurement, and since our sample is strongly limited by sample variance on large scales this weighting results in a significant improvement to the analysis. The effect of the weighting on the redshift distribution is illustrated in Figure~\ref{fig:wred}.

Other authors have used the so called $J_3$-weighting which optimises the error over all scales by weighting each scale differently~\citep[e.g.][]{Efstathiou:1988, Loveday:1994dx}. In a magnitude limited sample there is a correlation between luminosity and redshift, which establishes a correlation between bias and redshift~\citep{Zehavi:2004zn}. A scale-dependent weighting would imply a different effective redshift for each scale, causing a scale dependent bias. 

Finally we considered a luminosity dependent weighting as suggested by~\citet{Percival:2003pi}. However the same authors found that explicitly accounting for the luminosity-redshift relation has a negligible effect for 2dFGRS. We found that the effect to the 6dFGS correlation function is $\ll 1\sigma$ for all bins. Hence the static weighting of eq.~\ref{eq:weight} is sufficient for our dataset.

We also include an integral constraint correction in the form of 
\begin{equation}
\xi_{\rm data}(s) = \xi'_{\rm data}(s) + ic,
\end{equation}
where $ic$ is defined as
\begin{equation}
ic = \frac{\sum_s RR(s)\xi_{\rm model}(s)}{\sum_s RR(s)}.
\end{equation}
The function $RR(s)$ is calculated from our mock catalogue and $\xi_{\rm model}(s)$ is a correlation function model. Since $ic$ depends on the model of the correlation function we have to re-calculate it at each step during the fitting procedure. However we note that $ic$ has no significant impact to the final result.

Figure~\ref{fig:bao} shows the correlation function of 6dFGS at large scales. The BAO peak at $\approx 105h^{-1}\;$Mpc is clearly visible. The plot includes model predictions of different cosmological parameter sets. We will discuss these models in Section~\ref{sec:DV2}. 

\subsection{Log-normal error estimate}
\label{sec:log}

\begin{figure}[tb]
\begin{center}
\epsfig{file=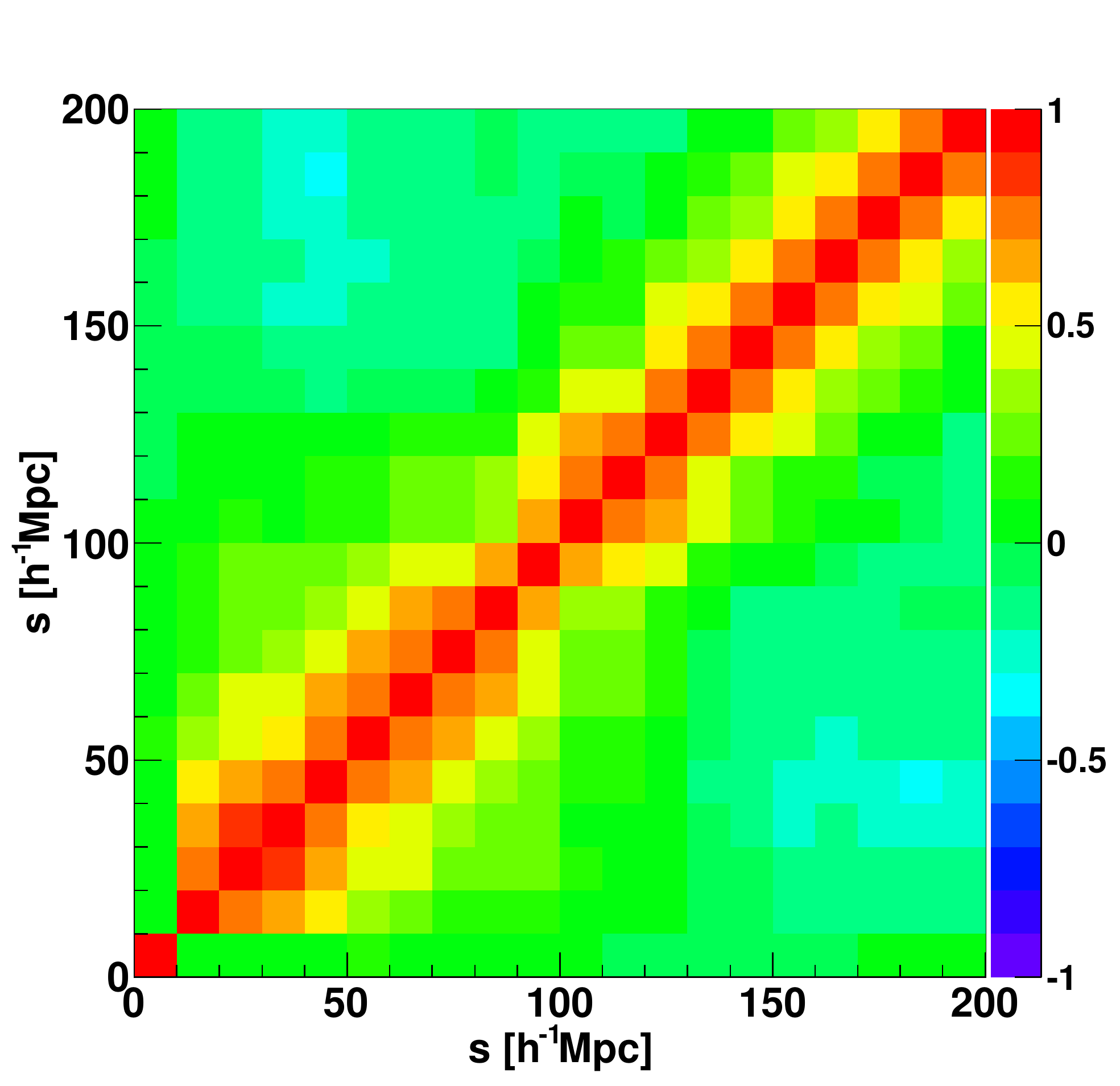,width=10cm}
\caption{Correlation matrix derived from a covariance matrix calculated from $200$ log-normal realisations. }
\label{fig:matrix_bao}
\end{center}
\end{figure}

To obtain reliable error-bars for the correlation function we use log-normal realisations~\citep{Coles:1991if, Cole:2005sx, Kitaura:2009jc}. In what follows we summarise the main steps, but refer the interested reader to Appendix~\ref{ap:log} 
in which we give a detailed explanation of how we generate the log-normal mock catalogues. 
In Appendix~\ref{ap:jk_comp} we compare the log-normal errors with jack-knife estimates.\\

Log-normal realisations of a galaxy survey are usually obtained by deriving a density field from a model power spectrum, $P(k)$, assuming Gaussian fluctuations. This density field is then Poisson sampled, taking into account the window function and the total number of galaxies. The assumption that the input power spectrum has Gaussian fluctuations can only be used in a model for a density field with over-densities $\ll 1$. As soon as we start to deal with finite r.m.s. fluctuations, the Gaussian model assigns a non-zero probability to regions of negative density. A log-normal random field $LN(\vec{x})$, can avoid this unphysical behaviour. It is obtained from a Gaussian field $G(\vec{x})$ by
\begin{equation}
LN(\vec{x}) = \exp[G(\vec{x}) - \sigma_G^2/2] - 1
\end{equation}
where $\sigma_G$ is the variance of the field. $LN(\vec{x})$ is  positive-definite but approaches $1+G(\vec{x})$ whenever the perturbations are small (e.g. at large scales). Calculating the power spectrum of a Poisson sampled density field with such a distribution will reproduce the input power spectrum convolved with the window function. As an input power spectrum for the log-normal field we use
\begin{equation}
P_{\rm nl}(k) = A P_{\rm lin}(k)\exp[-(k/k_*)^2]
\end{equation}
where $A = b^2(1 + 2\beta/3 + \beta^2/5)$ accounts for the linear bias and the linear redshift space distortions. $P_{\rm lin}(k)$ is a linear model power spectrum in real space obtained from CAMB~\citep{Lewis:1999bs} and $P_{\rm nl}(k)$ is the non-linear power spectrum in redshift space. Comparing the model above with the 6dFGS data gives $A = 4$. The damping parameter $k_*$ is set to $k_* = 0.33h$\;Mpc$^{-1}$, as found in 6dFGS (see fitting results later). How well this input model matches the 6dFGS data can be seen in Figure~\ref{fig:log_6df}.

We produce $200$ such realisations and calculate the correlation function for each of them, deriving a covariance matrix
\begin{equation}
C_{ij} = \sum^N_{n=1}\frac{\left[\xi_n(s_i) - \overline{\xi}(s_i)\right]\left[\xi_n(s_j) - \overline{\xi}(s_j)\right]}{N-1} .
\end{equation}
Here, $\xi_n(s_i)$ is the correlation function estimate at separation $s_i$ and the sum goes over all $N$ log-normal realisations. The mean value is defined as
\begin{equation}
\overline{\xi}(s_i) = \frac{1}{N}\sum^N_{n=1}\xi_n(s_i) .
\end{equation}
The case $i = j$ gives the error (ignoring correlations between bins, $\sigma_i^2 = C_{ii}$). In the following we will use this uncertainty in all diagrams, while the fitting procedures use the full covariance matrix.

The distribution of recovered correlation functions includes the effects of sample variance and shot noise. Non-linearities are also approximately included since the distribution of over-densities is skewed.

In Figure~\ref{fig:matrix_bao} we show the log-normal correlation matrix $r_{ij}$ calculated from the covariance matrix. The correlation matrix is defined as
\begin{equation}
r_{ij} = \frac{C_{ij}}{\sqrt{C_{ii}C_{jj}}},
\end{equation}
where $C$ is the covariance matrix (for a comparison to jack-knife errors see appendix~\ref{ap:jk_comp}).

\section{Modelling the BAO signal}
\label{sec:model}

In this section we will discuss wide-angle effects and non-linearities. We also introduce a model for the large scale correlation function, which we later use to fit our data.

\subsection{Wide angle formalism}

The model of linear redshift space distortions introduced by~\citet{Kaiser:1987qv} is based on the plane parallel approximation. Earlier surveys such as SDSS and 2dFGRS are at sufficiently high redshift that the maximum opening angle between a galaxy pair remains small enough to ensure the plane parallel approximation is valid. However,
the 6dF Galaxy Survey has a maximum opening angle of $180^{\circ}$ and a lower mean redshift of $z\approx 0.1$ (for our weighted sample) and so it is necessary to test the validity of  the plane parallel approximation. 
The wide angle description of redshift space distortions has been laid out in several papers~\citep{Szalay:1997cc,Szapudi:2004gh,Matsubara:2004fr,Papai:2008bd,Raccanelli:2010hk}, which we summarise in Appendix~\ref{ap:wide}.

We find that the wide-angle corrections have only a very minor effect on our sample. For our fiducial model we found a correction of $\Delta\xi = 4\cdot10^{-4} $ in amplitude at $s = 100h^{-1}$\;Mpc and $\Delta\xi = 4.5\cdot10^{-4}$ at $s = 200h^{-1}$\;Mpc, (Figure~\ref{fig:wide2} in the appendix). This is much smaller than the error bars on these scales. Despite the small size of the effect, we nevertheless include all first order correction terms in our correlation function model. It is important to note that wide angle corrections affect the correlation function amplitude only and do not cause any shift in the BAO scale. The effect of the wide-angle correction on the unweighted sample is much greater and is already noticeable on scales of $20h^{-1}$\;Mpc. Weighting to higher redshifts mitigates the effect because it reduces the average opening angle between galaxy pairs, by giving less weight to wide angle pairs (on average).

\subsection{Non-linear effects}
\label{sec:nlin}

There are a number of non-linear effects which can potentially influence a measurement of the BAO signal. These include scale-dependent bias, the non-linear growth of structure on smaller scales, and redshift-space distortions. We discuss each of these in the context of our 6dFGS sample.

As the Universe evolves, the acoustic signature in the correlation function is broadened by non-linear gravitational structure formation. Equivalently we can say that the higher harmonics in the power spectrum, which represent smaller scales, are erased~\citep{Eisenstein:2006nj}.

The early Universe physics, which we discussed briefly in the introduction, is well understood and several authors have produced software packages (e.g. CMBFAST and CAMB) and published fitting functions \citep[e.g ][]{Eisenstein:1997ik} to make predictions for the correlation function and power spectrum using thermodynamical models of the early Universe. These models already include the basic linear physics most relevant for the BAO peak. In our analysis we use the CAMB software package~\citep{Lewis:1999bs}. The non-linear evolution of the power spectrum in CAMB is calculated using the halofit code~\citep{Smith:2002dz}. This code is calibrated by $n$-body simulations and can describe non-linear effects in the shape of the matter power spectrum for pure CDM models to an accuracy of around $5$ - $10\%$~\citep{Heitmann:2008eq}. However, it has previously been shown that this non-linear model is a poor description of the non-linear effects around the BAO peak~\citep{Crocce:2007dt}. We therefore decided to use the linear model output from CAMB and incorporate the non-linear effects separately.

All non-linear effects influencing the correlation function can be approximated by a convolution with a Gaussian damping factor $\exp[-(rk_*/2)^2]$~\citep{Eisenstein:2006nk, Eisenstein:2006nj}, where $k_*$ is the damping scale. We will use this factor in our correlation function model introduced in the next section. The convolution with a Gaussian causes a shift of the peak position to larger scales, since the correlation function is not symmetric around the peak. However this shift is usually very small.

All of the non-linear processes discussed so far are not at the fundamental scale of $105h^{-1}$\;Mpc but are instead at the cluster-formation scale of up to $10h^{-1}$Mpc. The scale of $105h^{-1}$\;Mpc is far larger than any known non-linear effect in cosmology. This has led some authors to the conclusion that the peak will not be shifted significantly, but rather only blurred out. For example,~\citet{Eisenstein:2006nj} have argued that any systematic shift of the acoustic scale in real space must be small ($\apprle 0.5\%$), even at $z = 0$. 

However, several authors report possible shifts of up to $1\%$~\citep{Guzik:2006bu,Smith:2007gi,Smith:2006ne,Angulo:2007fw}. \citet{Crocce:2007dt} used re-normalised perturbation theory (RPT) and found percent-level shifts in the BAO peak. In addition to non-linear evolution, they found that mode-coupling generates additional oscillations in the power spectrum, which are out of phase with the BAO oscillations predicted by linear theory. This leads to shifts in the scale of oscillation nodes with respect to a smooth spectrum. In real space this corresponds to a peak shift towards smaller scales.
Based on their results,~\citet{Crocce:2007dt} propose a model to be used for the correlation function analysis at large scales. We will introduce this model in the next section.

\subsection{Large-scale correlation function}
\label{sec:lss}

To model the correlation function on large scales, we follow~\citet{Crocce:2007dt} and~\citet{Sanchez:2008iw} and adopt the following parametrisation\footnote{note that $r=s$, the different letters just specify whether the function is evaluated in redshift space or real space.}: 
\begin{equation}
\xi'_{\rm model}(s) = B(s)b^2\left[\xi(s) * G(r) + \xi^1_1(r)\frac{\partial \xi(s)}{\partial s}\right].
\label{eq:scocci1}
\end{equation}
Here, we decouple the scale dependency of the bias $B(s)$ and the linear bias $b$. $G(r)$ is a Gaussian damping term, accounting for non-linear suppression of the BAO signal. $\xi(s)$ is the linear correlation function (including wide angle description of redshift space distortions; eq.~\ref{eq:mom} in the appendix). The second term in eq~\ref{eq:scocci1} accounts for the mode-coupling of different Fourier modes. It contains $\partial \xi(s)/\partial s$, which is the first derivative of the redshift space correlation function, and $\xi^1_1(r)$, which is defined as 
\begin{equation}
\xi^1_1(r) = \frac{1}{2\pi^2}\int^{\infty}_0 dk\;kP_{\rm lin}(k)j_1(rk) ,
\end{equation}
with $j_1(x)$ being the spherical Bessel function of order $1$. 
\citet{Sanchez:2008iw} used an additional parameter $A_{\rm MC}$ which multiplies the mode coupling term in equation~\ref{eq:scocci1}. We found that our data are not good enough to constrain this parameter, and hence adopted $A_{\rm MC} = 1$ as in the original model by~\citet{Crocce:2007dt}.

In practice we generate linear model power spectra $P_{\rm lin}(k)$ from CAMB and convert them into a correlation function using a Hankel transform
\begin{equation}
\xi(r) = \frac{1}{2\pi^2}\int^{\infty}_0 dk\;k^2P_{\rm lin}(k)j_0(rk),
\end{equation}
where $j_0(x) = \sin(x)/x$ is the spherical Bessel function of order $0$.

The $*$-symbol in eq.~\ref{eq:scocci1} is only equivalent to a convolution in the case of a 3D correlation function, where we have the Fourier theorem relating the 3D power spectrum to the correlation function. In case of the spherically averaged quantities this is not true. Hence, the $*$-symbol in our equation stands for the multiplication of the power spectrum with $\tilde{G}(k)$ before transforming it into a correlation function. $\tilde{G}(k)$ is defined as 
\begin{equation}
\tilde{G}(k) = \exp\left[-(k/k_*)^2\right],
\label{eq:damp}
\end{equation}
with the property
\begin{equation}
\tilde{G}(k) \rightarrow 0 \;\text{ as }\; k \rightarrow \infty.
\end{equation}
The damping scale $k_*$ can be calculated from linear theory~\citep{Crocce:2005xy,Matsubara:2007wj} by
\begin{equation}
k_* = \left[\frac{1}{6\pi^2}\int^{\infty}_0 dk\;P_{\rm lin}(k)\right]^{-1/2},
\end{equation}
where $P_{\rm lin}(k)$ is again the linear power spectrum. $\Lambda$CDM predicts a value of $k_* \simeq 0.17h\;$Mpc$^{-1}$. However, we will include $k_*$ as a free fitting parameter.

The scale dependance of the 6dFGS bias, $B(s)$, is derived from the GiggleZ simulation (Poole et al., in preparation); a dark matter simulation containing $2160^3$ particles in a $1h^{-1}\;$Gpc box.  We rank-order the halos of this simulation by $V_{\rm max}$ and choose a contiguous set of $250\,000$ of them, selected to have the same clustering amplitude of 6dFGS as quantified by the separation scale $r_0$, where $\xi(r_0) = 1$. In the case of 6dFGS we found $r_0 = 9.3h^{-1}\;$Mpc. Using the redshift space correlation function of these halos and of a randomly subsampled set of $\sim {10^6}$ dark matter particles, we obtain
\begin{equation}
B(s) = 1 + \left(s/0.474h^{-1}\text{Mpc}\right)^{-1.332},
\end{equation}
which describes a $1.7\%$ correction of the correlation function amplitude at separation scales of $10h^{-1}\;$Mpc. To derive this function, the GiggleZ correlation function (snapshot $z = 0$) has been fitted down to $6h^{-1}\;$Mpc, well below the smallest scales we are interested in.

\section{Extracting the BAO signal}
\label{sec:DV}

\begin{table}[tb]
\begin{center}
\caption{This table contains all parameter constraints from 6dFGS obtained in this paper. The priors used to derive these parameters are listed in square brackets. All parameters assume $\Omega_bh^2 = 0.02227$ and in cases where a prior on $\Omega_mh^2$ is used, we adopt the WMAP-7 Markov chain probability distribution~{\protect \citep{Komatsu:2010fb}}. $A(z_{\rm eff})$ is the acoustic parameter defined by~{\protect \citet{Eisenstein:2005su}} (see equation~\ref{eq:A} in the text) and $R(z_{\rm eff})$ is the distance ratio of the 6dFGS BAO measurement to the last-scattering surface. The most sensible value for cosmological parameter constraints is $r_s(z_d)/D_V(z_{\rm eff})$, since this measurement is uncorrelated with $\Omega_mh^2$. The effective redshift of 6dFGS is $z_{\rm eff} = 0.106$ and the fitting range is from $10-190h^{-1}$\;Mpc.}
\vspace{0.4cm}
	\begin{tabular}{rll}
		\hline
  		\multicolumn{3}{c}{Summary of parameter constraints from 6dFGS}\\
		\hline
		$\Omega_mh^2$ & $0.138\pm 0.020$ ($14.5\%$)& \\
     		$D_V(z_{\rm eff})$ & $456\pm27\;$Mpc ($5.9\%$)& \\
     		$D_V(z_{\rm eff})$ & $459\pm18\;$Mpc ($3.9\%$)& $[\Omega_mh^2$ prior]\\
		$\mathbf{r_s(z_d)/D_V(z_{\rm \textbf{eff}})}$ & $\mathbf{0.336\pm0.015}$ ($\mathbf{4.5\%}$) & \\
		$R(z_{\rm eff})$ & $0.0324\pm 0.0015$ ($4.6\%$) & \\
		$A(z_{\rm eff})$ & $0.526\pm0.028$ ($5.3\%$) & \\
		\hline
		$\Omega_m$ & $0.296\pm0.028$ ($9.5\%$) & $[\Omega_mh^2$ prior]\\
		$\mathbf{H_0}$ & $\mathbf{67.0\pm3.2}$ $\mathbf{(4.8\%)}$ & $\mathbf{[\Omega_mh^2}$ \textbf{prior]}\\
		\hline
	  \end{tabular}
\label{tab:para}
\end{center}
\end{table}

In this section we fit the model correlation function developed in the previous section to our data. Such a fit can be used to derive the distance scale $D_V(z_{\rm eff})$ at the effective redshift of the survey.

\subsection{Fitting preparation}
\label{sec:fitprep}

The effective redshift of our sample is determined by
\begin{equation}
z_{\rm eff} = \sum^{N_b}_{i}\sum^{N_b}_{j}\frac{w_iw_j}{2N_b^2}(z_i + z_j),
\end{equation}
where $N_{b}$ is the number of galaxies in a particular separation bin and $w_i$ and $w_j$ are the weights for those galaxies 
from eq.~\ref{eq:weight}. We choose $z_{\rm eff}$ from bin $10$ which has the limits $100h^{-1}$\;Mpc and $110h^{-1}$\;Mpc and which gave $z_{\rm eff} = 0.106$. Other bins show values very similar to this, with a standard deviation of $\pm0.001$. The final result does not depend on a very precise determination of $z_{\rm eff}$, since we are not constraining a distance to the mean redshift, but a distance ratio (see equation~\ref{eq:alpha}, later). In fact, if the fiducial model is correct, the result is completely independent of $z_{\rm eff}$. Only if there is a $z$-dependent deviation from the fiducial model do we need $z_{\rm eff}$ to quantify this deviation at a specific redshift.\\

Along the line-of-sight, the BAO signal directly constrains the Hubble constant $H(z)$ at redshift $z$. When measured in a redshift shell, it constrains the angular diameter distance $D_A(z)$~\citep{Matsubara:2004fr}. In order to separately measure $D_A(z)$ and $H(z)$ we require a BAO detection in the 2D correlation function, where it will appear as a ring at around $105 h^{-1}$\;Mpc. Extremely large volumes are necessary for such a measurement. While there are studies that report a successful (but very low signal-to-noise) detection in the 2D correlation function using the SDSS-LRG data~\citep[e.g. ][but see also~\citealt{Kazin:2010nd}]{Gaztanaga:2008xz,Chuang:2011fy}, our sample does not allow this kind of analysis. Hence we restrict ourself to the 1D correlation function, where we measure a combination of $D_A(z)$ and $H(z)$. What we actually measure is a superposition of two angular measurements (R.A. and Dec.) and one line-of-sight measurement (redshift). To account for this mixture of measurements it is common to report the BAO distance constraints as~\citep{Eisenstein:2005su, Padmanabhan:2008ag}
\begin{equation}
D_V(z) = \left[(1+z)^2D_A^2(z)\frac{cz}{H_0E(z)}\right]^{1/3},
\label{eq:dv}
\end{equation}
where $D_A$ is the angular distance, which in the case of $\Omega_k = 0$ is given by $D_A(z) = D_C(z)/(1+z)$.

To derive model power spectra from CAMB we have to specify a complete cosmological model, which in the case of the simplest $\Lambda$CDM model ($\Omega_k = 0, w = -1$), is specified by six parameters: $\omega_{c}, \omega_{b}, n_s, \tau$, $A_s$ and $h$. These parameters are: the physical cold dark matter and baryon density, ($\omega_{c} = \Omega_ch^2,\; \omega_{b} = \Omega_bh^2$), the scalar spectral index, ($n_s$), the optical depth at recombination, ($\tau$), the scalar amplitude of the CMB temperature fluctuation, ($A_s$), and the Hubble constant in units of $100\;$km\;s$^{-1}$Mpc$^{-1}$ ($h$).

Our fit uses the parameter values from WMAP-7~\citep{Komatsu:2010fb}:  $\Omega_bh^2 = 0.02227$, $\tau = 0.085$ and $n_s = 0.966$ (maximum likelihood values). The scalar amplitude $A_s$ is set so that it results in $\sigma_8 = 0.8$, which depends on $\Omega_mh^2$. However $\sigma_8$ is degenerated with the bias parameter $b$ which is a free parameter in our fit. Furthermore, $h$ is set to $0.7$ in the fiducial model, but can vary freely in our fit through a scale distortion parameter $\alpha$, which enters the model as
\begin{equation}
\xi_{\rm model}(s) = \xi'_{\rm model}(\alpha s).
\end{equation}
This parameter accounts for deviations from the fiducial cosmological model, which we use to derive distances from the measured redshift. It is defined as~\citep{Eisenstein:2005su, Padmanabhan:2008ag}
\begin{equation}
\alpha = \frac{D_V(z_{\rm eff})}{D_V^{\rm fid}(z_{\rm eff})}.
\label{eq:alpha}
\end{equation}
The parameter $\alpha$ enables us to fit the correlation function derived with the fiducial model, without the need to re-calculate the correlation function for every new cosmological parameter set. 

At low redshift we can approximate $H(z) \approx H_0$, which results in
\begin{equation}
\alpha \approx \frac{H^{\rm fid}_0}{H_0} .
\label{eq:H0}
\end{equation}
Compared to the correct equation~\ref{eq:alpha} this approximation has an error of about $3\%$ at redshift $z = 0.1$ for our fiducial model. Since this is a significant systematic bias, we do not use this approximation at any point in our analysis.

\subsection{Extracting $D_V(z_{\rm eff})$ and $r_s(z_d)/D_V(z_{\rm eff})$}
\label{sec:DV2}

\begin{figure}[tb]
\begin{center}
\epsfig{file=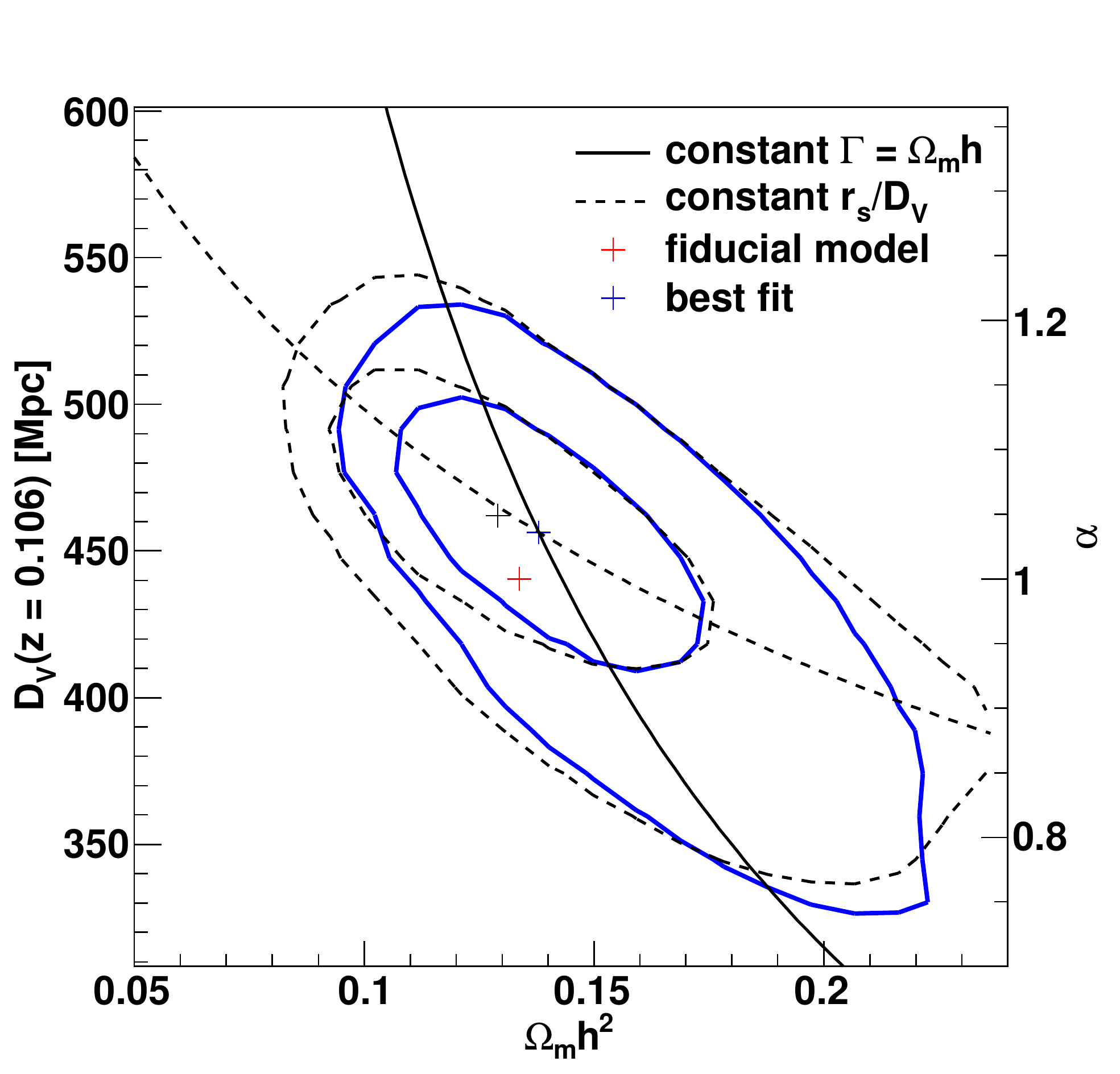,width=9.5cm}
\caption{Likelihood contours of the distance $D_V(z_{\rm eff})$ against $\Omega_mh^2$. The corresponding values of $\alpha$ are given on the right-hand axis. The contours show $1$ and  $2\sigma$ errors for both a full fit (blue solid contours) and a fit over $20-190h^{-1}$\;Mpc (black dashed contours) excluding the first data point. The black cross marks the best fitting values corresponding to the dashed black contours with $(D_V,\Omega_mh^2) = (462,0.129)$, while the blue cross marks the best fitting values for the blue contours. The black solid curve corresponds to a constant $\Omega_m h^2D_V(z_{\rm eff})$ ($D_V \sim h^{-1}$), while the dashed line corresponds to a constant angular size of the sound horizon, as described in the text.}
\label{fig:chi2}
\end{center}
\end{figure}

\begin{figure}[tb]
\begin{center}
\epsfig{file=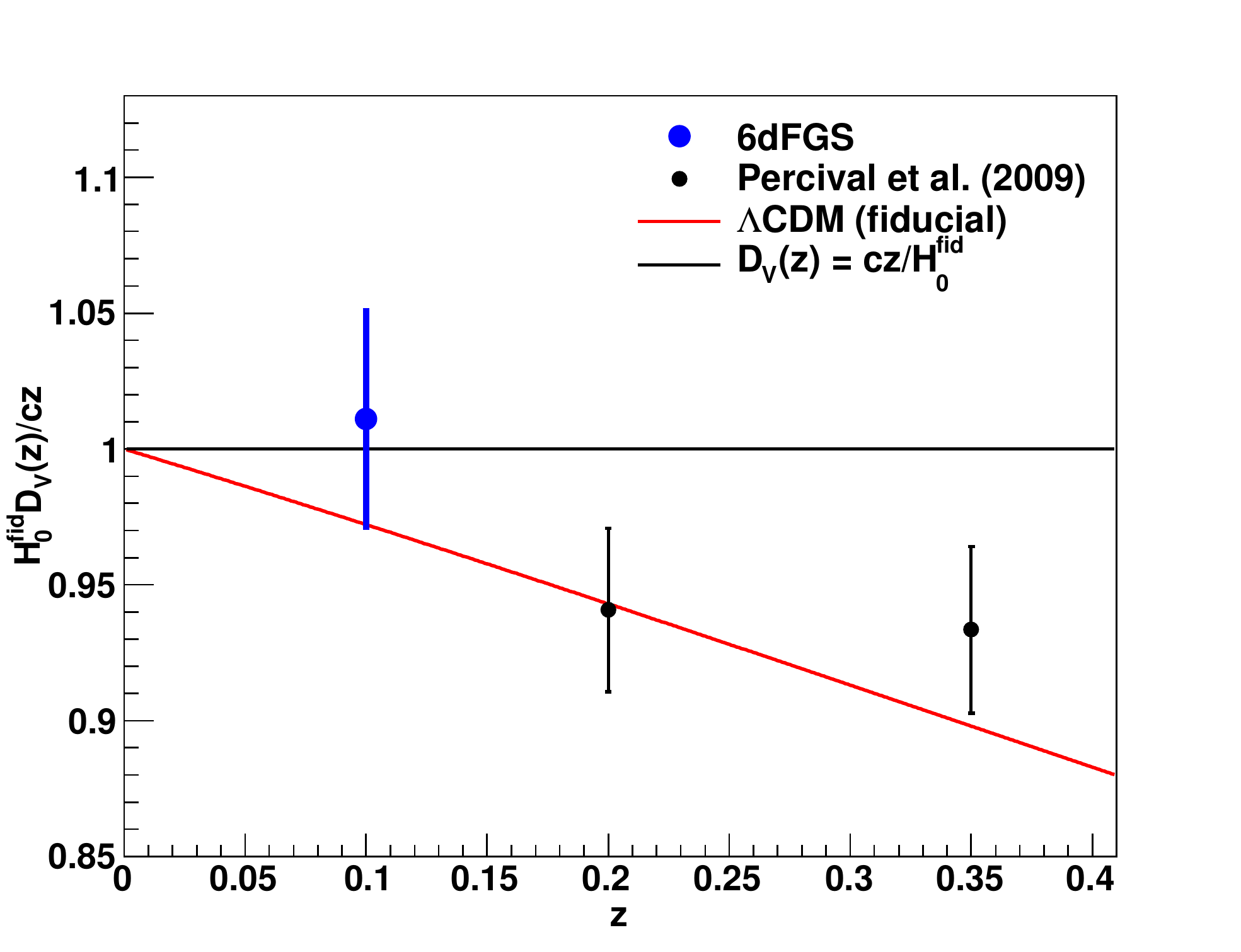,width=10cm}
\caption{The distance measurement $D_V(z)$ relative to a low redshift approximation. The points show 6dFGS data and those of~\citet{Percival:2009xn}.}
\label{fig:plot5_sn}
\end{center}
\end{figure}

Using the model introduced above we performed fits to $18$ data points between $10h^{-1}\;$Mpc and $190h^{-1}\;$Mpc. We excluded the data below $10h^{-1}\;$Mpc, since our model for non-linearities is not good enough to capture the effects on such scales. The upper limit is chosen to be well above the BAO scale, although the constraining contribution of the bins above $130h^{-1}\;$Mpc is very small. Our final model has $4$ free parameters:  $\Omega_mh^2$, $b$, $\alpha$ and $k_*$.

The best fit corresponds to a minimum $\chi^2$ of $15.7$ with $14$ degrees of freedom ($18$ data-points and $4$ free parameters). The best fitting model is included in Figure~\ref{fig:bao} (black line). The parameter values are $\Omega_m h^2 = 0.138\pm 0.020$, $b = 1.81\pm0.13$ and $\alpha = 1.036\pm0.062$, where the errors are derived for each parameter by marginalising over all other parameters. For $k_*$ we can give a lower limit of $k_* = 0.19h\;$Mpc$^{-1}$ (with $95\%$ confidence level).

We can use eq.~\ref{eq:alpha} to turn the measurement of $\alpha$ into a measurement of the distance to the effective redshift $D_V(z_{\rm eff}) = \alpha D_V^{\rm fid}(z_{\rm eff}) = 456\pm27\;$Mpc, with a precision of $5.9\%$. Our fiducial model gives $D_V^{\rm fid}(z_{\rm eff}) = 440.5\;$Mpc, where we have followed the distance definitions of~\citet{Wright:2006up} throughout. For each fit we derive the parameter $\beta = \Omega_m(z)^{0.545}/b$, which we need to calculate the wide angle corrections for the correlation function. 

The maximum likelihood distribution of $k_*$ seems to prefer smaller values than predicted by $\Lambda$CDM, although we are not able to constrain this parameter very well. This is connected to the high significance of the BAO peak in the 6dFGS data (see Section~\ref{sec:sig}). A smaller value of $k_*$ damps the BAO peak and weakens the distance constraint. For comparison we also performed a fit fixing $k_*$ to the $\Lambda$CDM prediction of $k_* \simeq 0.17h\;$Mpc$^{-1}$. We found that the error on the distance $D_V(z_{\rm eff})$ increases from $5.9\%$ to $8\%$. However since the data do not seem to support such a small value of $k_*$ we prefer to marginalise over this parameter.

The contours of $D_V(z_{\rm eff}) - \Omega_mh^2$ are shown in Figure~\ref{fig:chi2}, together with two degeneracy predictions~\citep{Eisenstein:2005su}. The solid line is that of constant $\Omega_mh^2D_V(z_{\rm eff})$, which gives the direction of degeneracy for a pure CDM model, where only the shape of the correlation function contributes to the fit, without a BAO peak. The dashed line corresponds to a constant $r_s(z_d)/D_V(z_{\rm eff})$, which is the degeneracy if only the position of the acoustic scale contributes to the fit. The dashed contours exclude the first data point, fitting from $20$ - $190 h^{-1}$\;Mpc only, with the best fitting values $\alpha = 1.049\pm 0.071$ (corresponding to $D_V(z_{\rm eff}) = 462\pm31\;$Mpc), $\Omega_mh^2 = 0.129\pm0.025$ and $b=1.72\pm0.17$. The contours of this fit are tilted towards the dashed line, which means that the fit is now driven by the BAO peak, while the general fit (solid contours) seems to have some contribution from the shape of the correlation function. Excluding the first data point increases the error on the distance constraint only slightly from $5.9\%$ to $6.8\%$. The value of $\Omega_mh^2$ tends to be smaller, but agrees within $1\sigma$ with the former value.

Going back to the complete fit from $10$ - $190h^{-1}\;$Mpc, we can include an external prior on $\Omega_mh^2$ from WMAP-7, which carries an error of only $4\%$ (compared to the $\approx 15\%$ we obtain by fitting our data). Marginalising over $\Omega_mh^2$ now gives $D_V(z_{\rm eff}) = 459\pm18\;$Mpc, which reduces the error from $5.9\%$ to $3.9\%$. The uncertainty in $\Omega_mh^2$ from WMAP-7 contributes only about $5\%$ of the error in $D_V$ (assuming no error in the WMAP-7 value of $\Omega_mh^2$ results in $D_V(z_{\rm eff}) = 459\pm17\;$Mpc).

In Figure~\ref{fig:plot5_sn} we plot the ratio $D_V(z)/D^{low-z}_V(z)$ as a function of redshift, where $D^{low-z}_V(z) = cz/H_0$. At sufficiently low redshift the approximation $H(z) \approx H_0$ is valid and the measurement is independent of any cosmological parameter except the Hubble constant. This figure also contains the results from~\citet{Percival:2009xn}.

Rather than including the WMAP-7 prior on $\Omega_mh^2$ to break the degeneracy between $\Omega_mh^2$ and the distance constraint, we can fit the ratio $r_s(z_d)/D_V(z_{\rm eff})$, where $r_s(z_d)$ is the sound horizon at the baryon drag epoch $z_d$. In principle, this is rotating Figure~\ref{fig:chi2} so that the dashed black line is parallel to the x-axis and hence breaks the degeneracy if the fit is driven by the BAO peak; it will be less efficient if the fit is driven by the shape of the correlation function. During the fit we calculate $r_s(z_d)$ using the fitting formula of~\citet{Eisenstein:1997ik}.

The best fit results in $r_s(z_d)/D_V(z_{\rm eff}) = 0.336\pm0.015$, which has an error of $4.5\%$,  smaller than the $5.9\%$ found for $D_V$ but larger than the error in $D_V$ when adding the WMAP-7 prior on $\Omega_mh^2$. This is caused by the small disagreement in the $D_V-\Omega_mh^2$ degeneracy and the line of constant sound horizon in Figure~\ref{fig:chi2}. The $\chi^2$ is $15.7$, similar to the previous fit with the same number of degrees of freedom.

\subsection{Extracting $A(z_{\rm eff})$ and $R(z_{\rm eff})$}
\label{sec:AandR}

We can also fit for the ratio of the distance between the effective redshift, $z_{\rm eff}$, and the redshift of decoupling \citep[$z_* = 1091$; ][]{Eisenstein:2005su}; 
\begin{equation}
R(z_{\rm eff}) = \frac{D_V(z_{\rm eff})}{(1+z_*)D_A(z_*)} , 
\end{equation}
with $(1+z_*)D_A(z_*)$ being the CMB angular comoving distance. Beside the fact that the Hubble constant $H_0$ cancels out in the determination of $R$, this ratio is also more robust against effects caused by  possible extra relativistic species~\citep{Eisenstein:2004an}. We calculate $D_A(z_*)$ for each $\Omega_mh^2$ during the fit and then marginalise over $\Omega_mh^2$. The best fit results in $R = 0.0324\pm 0.0015$, with $\chi^2 = 15.7$ and the same $14$ degrees of freedom.

Focusing on the path from $z=0$ to $z_{\rm eff}=0.106$, our dataset can give interesting constraints on $\Omega_m$. We derive the parameter~\citep{Eisenstein:2005su}
\begin{equation}
A(z_{\rm eff}) = 100D_V(z_{\rm eff})\frac{\sqrt{\Omega_mh^2}}{cz_{\rm eff}},
\label{eq:A}
\end{equation}
which has no dependence on the Hubble constant since $D_V \propto h^{-1}$. We obtain $A(z_{\rm eff}) = 0.526\pm0.028$ with $\chi^2/\rm d.o.f. = 15.7/14$. The value of $A$ would be identical to $\sqrt{\Omega_m}$ if measured at redshift $z=0$. At redshift $z_{\rm eff} = 0.106$ we obtain a deviation from this approximation of $6\%$ for our fiducial model, which is small but systematic. We can express $A$, including the curvature term $\Omega_k$ and the dark energy equation of state parameter $w$,  as
\begin{equation}
A(z) = \frac{\sqrt{\Omega_m}}{E(z)^{1/3}}
\begin{cases}
\left[\frac{\sinh\left(\sqrt{\Omega_k}\chi(z)\right)}{\sqrt{\Omega_k}z}\right]^{2/3} & \Omega_k > 0\cr
\left[\frac{\chi(z)}{z}\right]^{2/3} & \Omega_k = 0\cr
\left[\frac{\sin\left(\sqrt{|\Omega_k|}\chi(z)\right)}{\sqrt{|\Omega_k|}z}\right]^{2/3} & \Omega_k < 0
 \end{cases}
\end{equation}
with 
\begin{equation}
\chi(z) = D_C(z)\frac{H_0}{c} = \int^{z}_0\frac{dz'}{E(z')}
\end{equation}
and
\begin{align}
E(z) = \big[&\Omega_m(1+z)^3 + \Omega_k(1+z)^2\cr
             &+ \Omega_{\Lambda}(1+z)^{3(1+w)})\big]^{1/2} .
\end{align}
Using this equation we now linearise our result for $\Omega_m$ in $\Omega_k$ and $w$ and get
\begin{equation}
\Omega_m = 0.287 + 0.039(1+w) + 0.039\Omega_k \pm 0.027.
\end{equation}
For comparison, ~\citet{Eisenstein:2005su} found 
\begin{equation}
\Omega_m = 0.273 + 0.123(1+w) + 0.137\Omega_k \pm 0.025
\end{equation}
based on the SDSS LRG DR3 sample. This result shows the reduced sensitivity of the 6dFGS measurement to $w$ and $\Omega_k$.

\section{Cosmological implications}
\label{sec:results2}

In this section we compare our results to other studies and discuss the implications for constraints on cosmological parameters. We first note that we do not see any excess correlation on large scales as found in the SDSS-LRG sample. Our correlation function is in agreement with a crossover to negative scales at $140h^{-1}$\;Mpc, as predicted from $\Lambda$CDM. 

\begin{table*}[tb]
\begin{center}
\caption{$w$CDM constraints from different datasets. Comparing the two columns shows the influence of the 6dFGS data point. The 6dFGS data point reduces the error on $w$ by $24\%$ compared to WMAP-7+LRG which contains only the BAO data points of~\citet{Percival:2009xn}. We assume flat priors of $0.11 < \Omega_mh^2 < 0.16$ and marginalise over $\Omega_mh^2$. The asterisks denote the free parameters in each fit.}
\vspace{0.4cm}
	\begin{tabular}{llll}
     		\hline
		parameter & WMAP-7+LRG & WMAP-7+LRG+6dFGS\\
		\hline
		$H_0$ & $69.9\pm3.8$(*) & $68.7\pm2.8$(*)\\
		$\Omega_m$ & $0.283\pm0.033$ & $0.293\pm0.027$\\
		$\Omega_{\Lambda}$ & $0.717\pm0.033$ & $0.707\pm0.027$\\
		$w$ & $\llap{-}1.01\pm0.17$(*) & $\llap{-}0.97\pm0.13$(*)\\
		\hline
	  \end{tabular}
	  \label{tab:comp}
\end{center}
\end{table*}

\begin{table*}[tb]
\begin{center}
\caption{Parameter constraints from WMAP7+BAO for (i) a flat $\Lambda$CDM model, (ii) an open $\Lambda$CDM (o$\Lambda$CDM), (iii) a flat model with $w = \rm const.$ ($w$CDM),  and (iv) an open model with $w = \rm constant$ (o$w$CDM). We assume flat priors of $0.11 < \Omega_mh^2 < 0.16$ and marginalise over $\Omega_mh^2$. The asterisks denote the free parameters in each fit.}
	\begin{tabular}{lllll}
     		\hline
		 parameter & $\Lambda$CDM & o$\Lambda$CDM & $w$CDM & o$w$CDM\\
		\hline
		$H_0$ & $69.2\pm1.1$(*) & $68.3\pm1.7$(*) & $68.7\pm2.8$(*) & $70.4\pm4.3$(*)\\
		$\Omega_m$ & $0.288\pm0.011$ & $0.290\pm0.019$ & $0.293\pm0.027$ & $0.274\pm0.035$\\
		$\Omega_{k}$ & ($0$) & $\llap{-}0.0036\pm0.0060$(*) & ($0$) & $\llap{-}0.013\pm0.010$(*)\\
		$\Omega_{\Lambda}$ & 0.$712\pm0.011$ & $0.714\pm0.020$ & $0.707\pm0.027$ & $0.726\pm0.036$\\
		$w$ & ($\;\llap{-}1$) & ($\;\llap{-}1$) & $\llap{-}0.97\pm0.13$(*) & $\llap{-}1.24\pm0.39$(*)\\
		\hline
	  \end{tabular}
	  \label{tab:models}
\end{center}
\end{table*}

\subsection{Constraining the Hubble constant, $H_0$}

\begin{figure}[tb]
\begin{center}
\epsfig{file=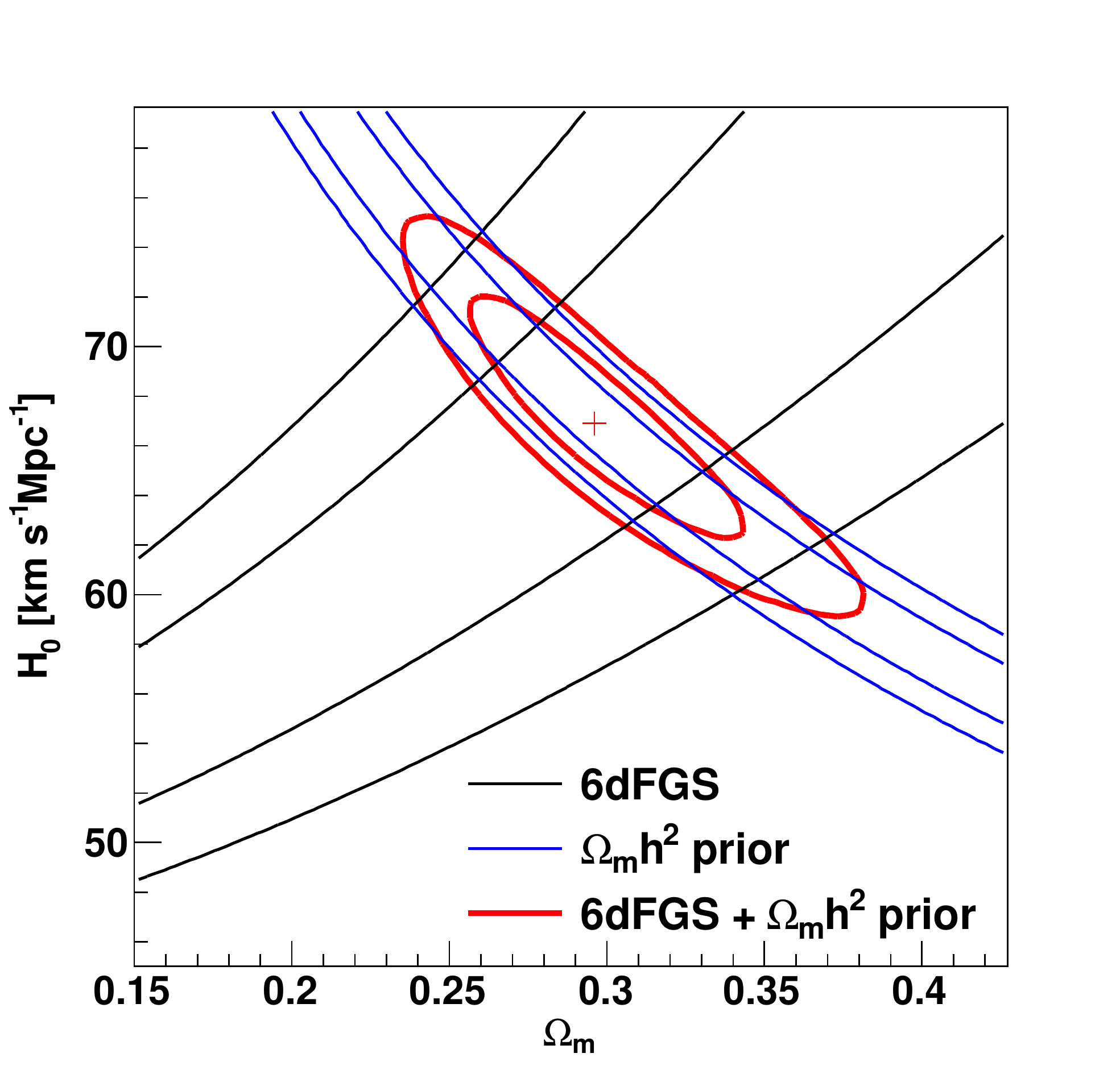,width=10cm}
\caption{The blue contours show the WMAP-7 $\Omega_mh^2$ prior~{\protect \citep{Komatsu:2010fb}}. The black contour shows constraints from 6dFGS derived by fitting to the measurement of $r_s(z_d)/D_V(z_{\rm eff})$. The solid red contours show the combined constraints resulting in $H_0 = 67.0\pm3.2\;$km\;s$^{-1}$Mpc$^{-1}$ and $\Omega_m = 0.296\pm0.028$. Combining the clustering measurement with $\Omega_mh^2$ from the CMB corresponds to the calibration of the standard ruler.}
\label{fig:chi2_h}
\end{center}
\end{figure}

\begin{figure}[tb]
\begin{center}
\epsfig{file=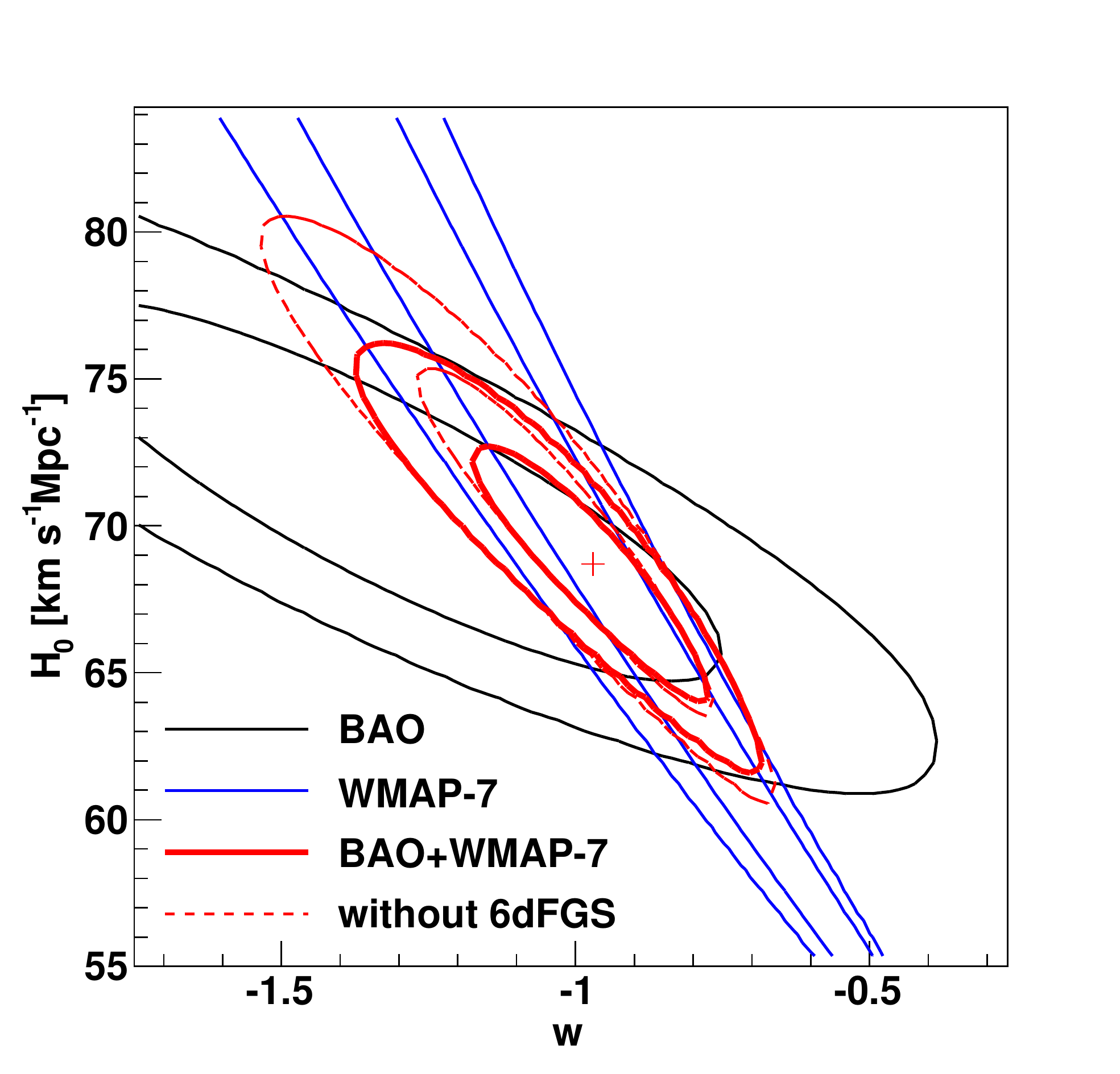,width=10cm}
\caption{The blue contours shows the WMAP-7 degeneracy in $H_0$ and $w$~{\protect \citep{Komatsu:2010fb}},
highlighting the need for a second dataset to break the degeneracy. The black contours show constraints from BAO data incorporating the $r_s(z_d)/D_V(z_{\rm eff})$ measurements of~{\protect \cite{Percival:2009xn}} and 6dFGS. The solid red contours show the combined constraints resulting in $w = -0.97\pm0.13$. Excluding the 6dFGS data point widens the constraints to the dashed red line with $w = -1.01\pm0.17$.}
\label{fig:chi2_w}
\end{center}
\end{figure}

We now use the 6dFGS data to derive an estimate of the Hubble constant. We use the 6dFGS measurement of $r_s(z_d)/D_V(0.106) = 0.336\pm0.015$ and fit directly for the Hubble constant and $\Omega_m$. We combine our measurement with a prior on $\Omega_mh^2$ coming from the WMAP-7 Markov chain results~\citep{Komatsu:2010fb}. Combining the clustering measurement with $\Omega_mh^2$ from the CMB corresponds to the calibration of the standard ruler.

We obtain values of $H_0 = 67.0\pm3.2\;$km s$^{-1}$Mpc$^{-1}$ (which has an uncertainty of only $4.8\%$) and $\Omega_m = 0.296\pm0.028$. Table~\ref{tab:para} and Figure~\ref{fig:chi2_h} summarise the results. The value of $\Omega_m$ agrees with the value we derived earlier (Section~\ref{sec:AandR}). 

To combine our measurement with the latest CMB data we use the WMAP-7 distance priors, namely the acoustic scale
\begin{equation}
\ell_A = (1+z_*)\frac{\pi D_A(z_*)}{r_s(z_*)},
\end{equation}
the shift parameter
\begin{equation}
R = 100\frac{\sqrt{\Omega_mh^2}}{c}(1+z_*)D_A(z_*)
\end{equation}
and the redshift of decoupling $z_*$ (Tables 9 and 10 in~\citealt{Komatsu:2010fb}). This combined analysis reduces the error further and yields $H_0 = 68.7\pm1.5\;$km s$^{-1}$Mpc$^{-1}$ ($2.2\%$) and $\Omega_m = 0.29\pm0.022$ ($7.6\%$).

\citet{Percival:2009xn} determine a value of $H_0 = 68.6\pm2.2$\;km s$^{-1}$Mpc$^{-1}$ using SDSS-DR7, SDSS-LRG and 2dFGRS, while \cite{Reid:2009xm} found $H_0 = 69.4\pm1.6$\;km\;s$^{-1}$Mpc$^{-1}$ using the SDSS-LRG sample and WMAP-5. In contrast to these results, 6dFGS is less affected by parameters like $\Omega_k$ and $w$ because of its lower redshift. In any case, our result of the Hubble constant agrees very well with earlier BAO analyses. Furthermore our result agrees with the latest CMB measurement of $H_0 = 70.3\pm2.5$\;km\;s$^{-1}$Mpc$^{-1}$~\citep{Komatsu:2010fb}.

The SH0ES program~\citep{Riess:2011yx} determined the Hubble constant using the distance ladder method. They used about $600$ near-IR observations of Cepheids in eight galaxies to improve the calibration of $240$ low redshift ($z < 0.1$) SN Ia, and  calibrated the Cepheid distances using the geometric distance to the maser galaxy NGC 4258. They found $H_0 = 73.8\pm2.4$\;km\;s$^{-1}$Mpc$^{-1}$, a value consistent with the initial results of the Hubble Key project \citet[$H_0 = 72\pm8$\;km\;s$^{-1}$Mpc$^{-1}$; ][]{Freedman:2000cf} but $1.7\sigma$ higher than our value (and $1.8\sigma$ higher when we combine our dataset with WMAP-7). While this could point toward unaccounted or under-estimated systematic errors in either one of the methods, the likelihood of such a deviation by chance is about $10\%$ and hence is not enough to represent a significant discrepancy. Possible systematic errors affecting the BAO measurements are the modelling of non-linearities, bias and redshift-space distortions, although these systematics are not expected to be significant at the large scales relevant to our analysis.

To summarise the finding of this section we can state that our measurement of the Hubble constant is competitive with the latest result of the  distance ladder method. The different techniques employed to derive these results have very different potential systematic errors. Furthermore we found that BAO studies provide the most accurate measurement of $H_0$ that exists, when combined with the CMB distance priors.

\subsection{Constraining dark energy}
\label{sec:de}

One key problem driving current cosmology is the determination of the dark energy equation of state parameter, $w$. When adding additional parameters like $w$ to $\Lambda$CDM we find large degeneracies in the WMAP-7-only data. One example is shown in Figure~\ref{fig:chi2_w}. WMAP-7 alone can not constrain $H_0$ or $w$ within sensible physical boundaries (e.g. $w < -1/3$). As we are sensitive to $H_0$, we can break the degeneracy between $w$ and $H_0$ inherent in the CMB-only data. Our assumption of a fiducial cosmology with $w=-1$ does not introduce a bias, since our data are not sensitive to this parameter and any deviation from this assumption is modelled within the shift parameter $\alpha$.

We again use the WMAP-7 distance priors introduced in the last section. In addition to our value of $r_s(z_d)/D_V(0.106) = 0.336\pm0.015$ we use the results of~\citet{Percival:2009xn}, who found $r_s(z_d)/D_V(0.2) = 0.1905\pm0.0061$ and $r_s(z_d)/D_V(0.35) = 0.1097\pm0.0036$. To account for the correlation between the two latter data points we employ the covariance matrix reported in their paper. Our fit has $3$ free parameters, $\Omega_mh^2$, $H_0$ and $w$.

The best fit gives $w = -0.97\pm0.13$, $H_0 = 68.7\pm 2.8$\;km\;s$^{-1}$Mpc$^{-1}$ and $\Omega_mh^2 = 0.1380\pm 0.0055$, with a $\chi^2/\rm d.o.f. = 1.3/3$. Table~\ref{tab:comp} and Figure~\ref{fig:chi2_w} summarise the results.
To illustrate the importance of the 6dFGS result to the overall fit we also show how the results change if 6dFGS is omitted. The 6dFGS data improve the constraint on $w$ by $24\%$.

Finally we show the best fitting cosmological parameters for different cosmological models using WMAP-7 and BAO results in Table~\ref{tab:models}.

\section{Significance of the BAO detection}
\label{sec:sig}

\begin{figure}[tb]
\begin{center}
\epsfig{file=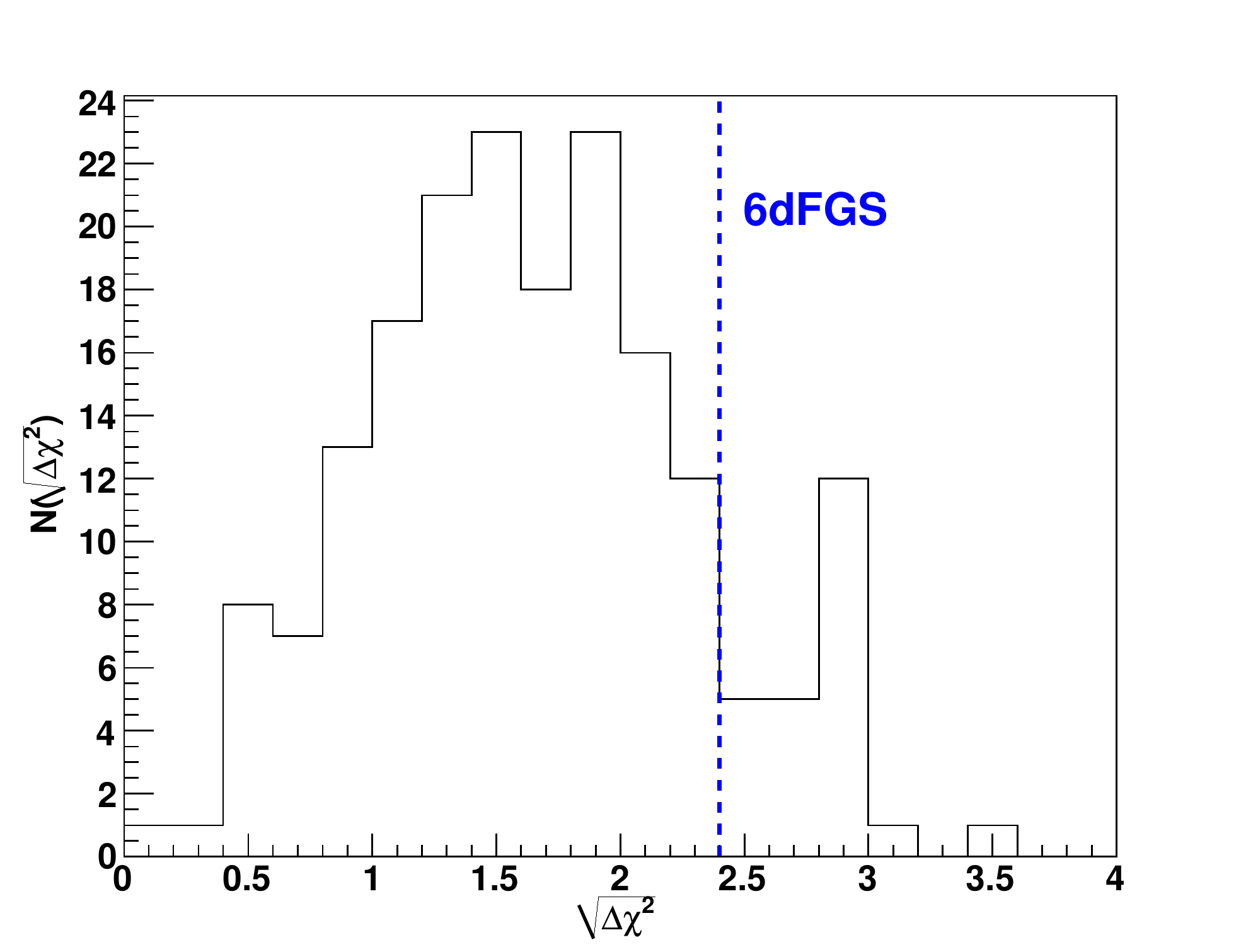,width=10cm}
\caption{The number of log-normal realisations found with a certain $\sqrt{\Delta\chi^2}$, where the $\Delta\chi^2$ is obtained by comparing a fit using a $\Lambda$CDM correlation function model with a no-baryon model. The blue line indicates the 6dFGS result.}
\label{fig:sigma}
\end{center}
\end{figure}

\begin{figure}[tb]
\begin{center}
\epsfig{file=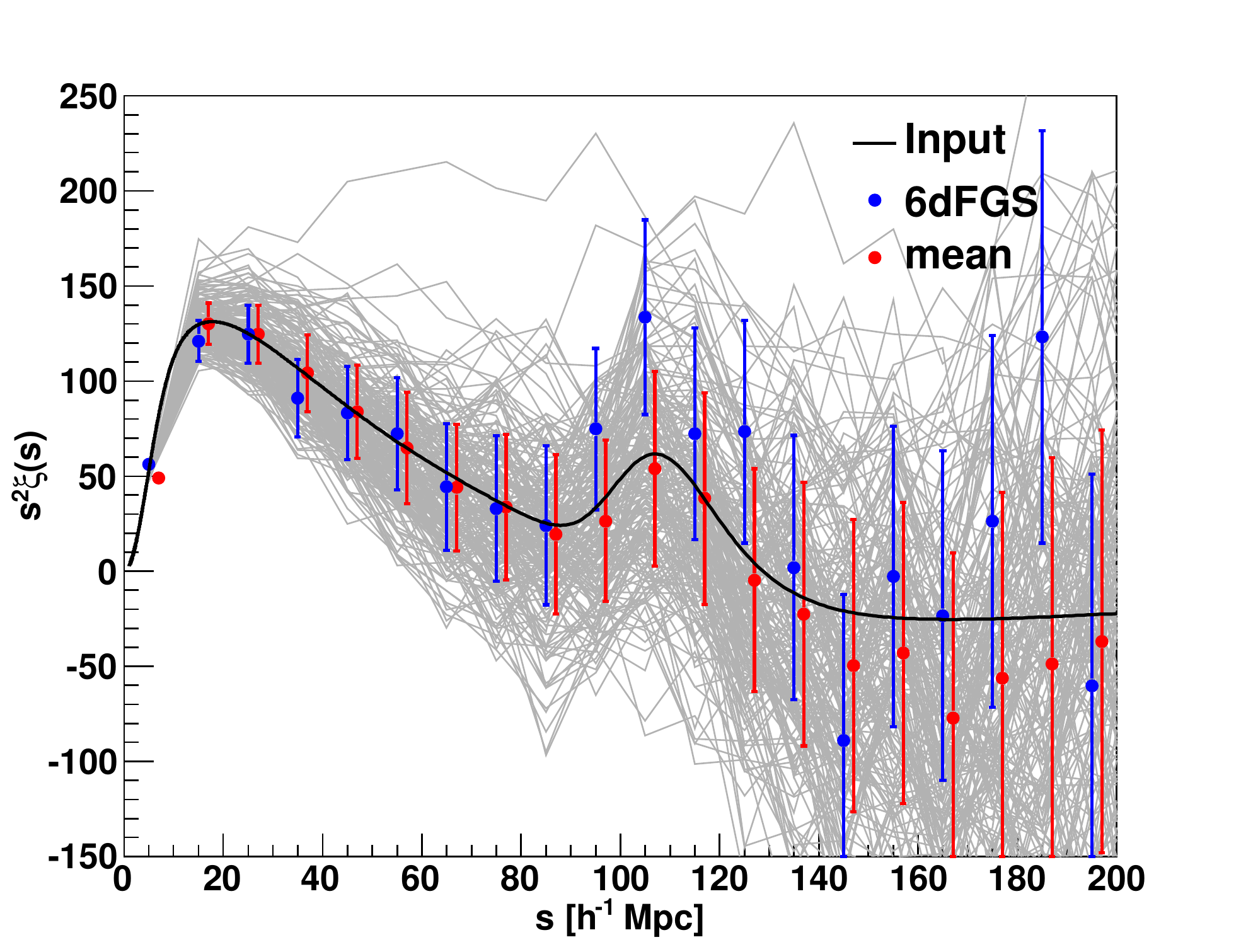,width=10cm}
\caption{The different log-normal realisations used to calculate the covariance matrix (shown in grey). The red points indicate the mean values, while the blue points show actual 6dFGS data (the data point at $5h^{-1}\;$Mpc is not included in the fit). The red data points are shifted by $2h^{-1}$\;Mpc to the right for clarity.}
\label{fig:log_6df}
\end{center}
\end{figure}

\begin{figure}[tb]
\begin{center}
\epsfig{file=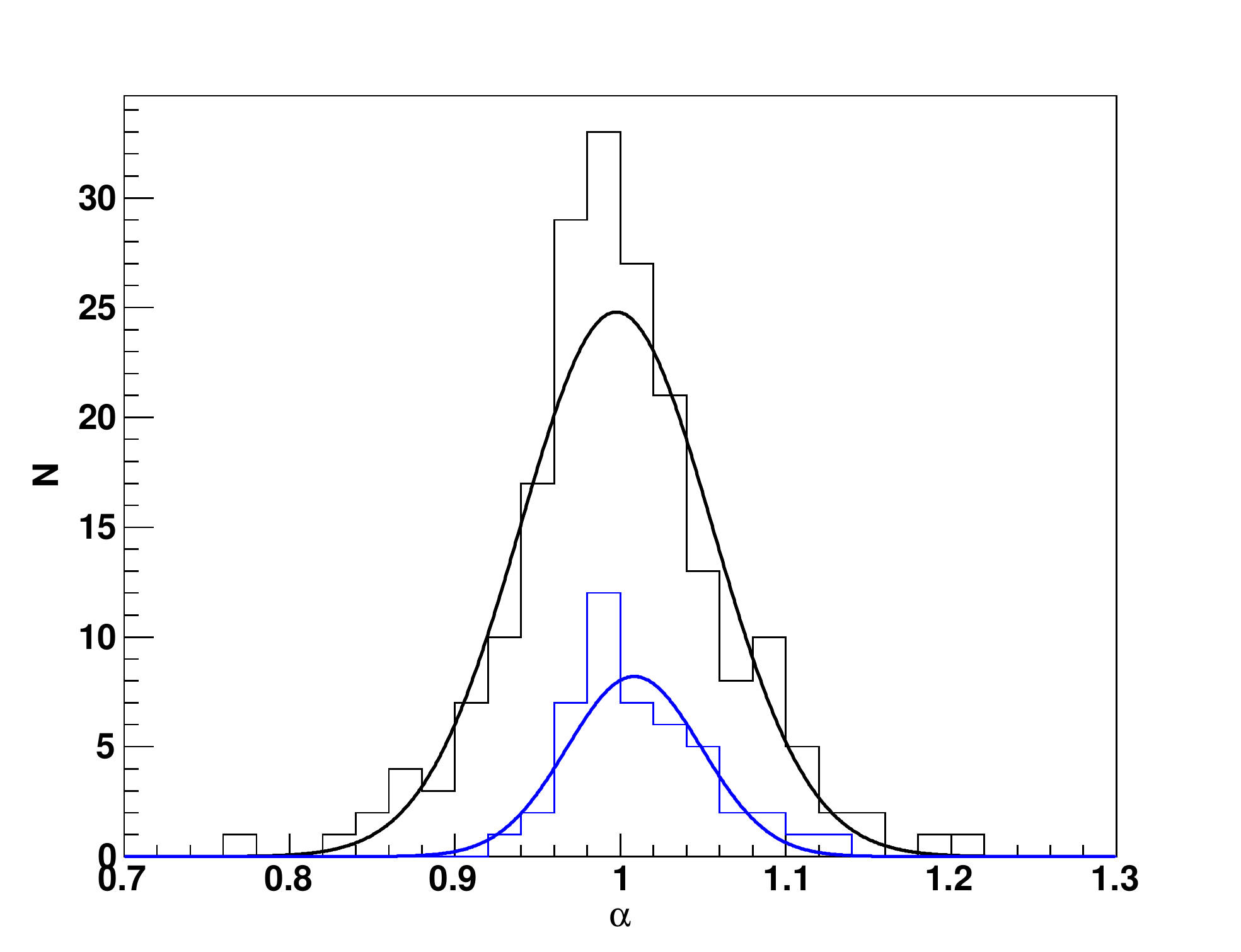,width=10cm}
\caption{This plot shows the distribution of the parameter $\alpha$ derived from the $200$ log-normal realisations (black). The distribution is well fit by a Gaussian with a mean of $\mu = 0.998\pm0.004$ and a width of $\sigma = 0.057\pm0.005$. In blue we show the same distribution selecting only the log-normal realisations with a strong BAO peak ($> 2\sigma$). The Gaussian distribution in this case gives a mean of $1.007\pm0.007$ and $\sigma=0.041\pm0.008$.}
\label{fig:alpha_distri}
\end{center}
\end{figure}

To test the significance of our detection of the BAO signature we follow~\citet{Eisenstein:2005su} and perform a fit with a fixed $\Omega_b = 0$, which corresponds to a pure CDM model without a BAO signature. The best fit has $\chi^2 = 21.4$ with $14$ degrees of freedom and is shown as the red dashed line in Figure~\ref{fig:bao}. The parameter values of this fit depend on the parameter priors, which we set to $0.7 < \alpha < 1.3$ and $0.1 < \Omega_mh^2 < 0.2$. Values of $\alpha$ much further away from $1$ are problematic since eq.~\ref{eq:alpha} is only valid for $\alpha$ close to $1$. Comparing the best pure CDM model with our previous fit, we estimate that the BAO signal is detected with a significance of $2.4\sigma$ (corresponding to $\Delta\chi^2 = 5.6$). As a more qualitative argument for the detection of the BAO signal we would like to refer again to Figure~\ref{fig:chi2} where the direction of the degeneracy clearly indicates the sensitivity to the BAO peak.
 
We can also use the log-normal realisations to determine how likely it is to find a BAO detection in a survey like 6dFGS. To do this, we produced $200$ log-normal mock catalogues and calculated the correlation function for each of them. We can now fit our correlation function model to these realisations. Furthermore, we fit a no-baryon model to the correlation function and calculate $\Delta\chi^2$, the distribution of which is shown in Figure~\ref{fig:sigma}. We find that $26\%$ of all realisations have {\sl at least} a $2\sigma$ BAO detection, and that  $12\%$ have a detection $> 2.4\sigma$. The log-normal realisations show a mean significance of the BAO detection of $1.7\pm0.7\sigma$, where the error describes the variance around the mean.

Figure~\ref{fig:log_6df} shows the 6dFGS data points together with all $200$ log-normal realisations (grey). The red data points indicate the mean for each bin and the black line is the input model derived as explained in Section~\ref{sec:log}. This comparison shows that the 6dFGS data contain a BAO peak slightly larger than expected in $\Lambda$CDM.

The amplitude of the acoustic feature relative to the overall normalisation of the galaxy correlation function is quite sensitive to the baryon fraction, $f_b = \Omega_b/\Omega_m$~\citep{Matsubara:2004fr}. A higher BAO peak could hence point towards a larger baryon fraction in the local Universe. However since the correlation function model seems to agree very well with the data (with a reduced $\chi^2$ of $1.12$) and is within the range spanned by our log-normal realisations, we can not claim any discrepancy with $\Lambda$CDM. Therefore, the most likely explanation for the excess correlation in the BAO peak is sample variance.

In Figure~\ref{fig:alpha_distri} we show the distribution of the parameter $\alpha$ obtained from the $200$ log-normal realisations. The distribution is well described by a Gaussian with $\chi^2/\text{d.o.f.} = 14.2/20$, where we employed Poisson errors for each bin. This confirms that $\alpha$ has Gaussian distributed errors in the approximation that the 6dFGS sample is well-described by log-normal realisations of an underlying $\Lambda$CDM power spectrum. This result increases our confidence that the application of Gaussian errors for the cosmological parameter fits is correct. The mean of the Gaussian distribution is at $0.998\pm0.004$ in agreement with unity, which shows, that we are able to recover the input model. The width of the distribution shows the mean expected error in $\alpha$ in a $\Lambda$CDM universe for a 6dFGS-like survey. We found $\sigma = 0.057\pm0.005$ which is in agreement with our error in $\alpha$ of $5.9\%$. Figure~\ref{fig:alpha_distri} also contains the distribution of $\alpha$, selecting only the log-normal realisations with a strong ($>2\sigma$) BAO peak (blue data). We included this selection to show, that a stronger BAO peak does not bias the estimate of $\alpha$ in any direction. The Gaussian fit gives $\chi^2/\text{d.o.f.} = 5/11$ with a mean of $1.007\pm0.007$. The distribution of $\alpha$ shows a smaller spread with $\sigma = 0.041\pm0.008$, about $2\sigma$ below our error on $\alpha$. This result shows, that a survey like 6dFGS is able to constrain $\alpha$ (and hence $D_V$ and $H_0$) to the precision we report in this paper.

\section{Future all sky surveys}
\label{sec:future}

\begin{figure}[tb]
\begin{center}
\epsfig{file=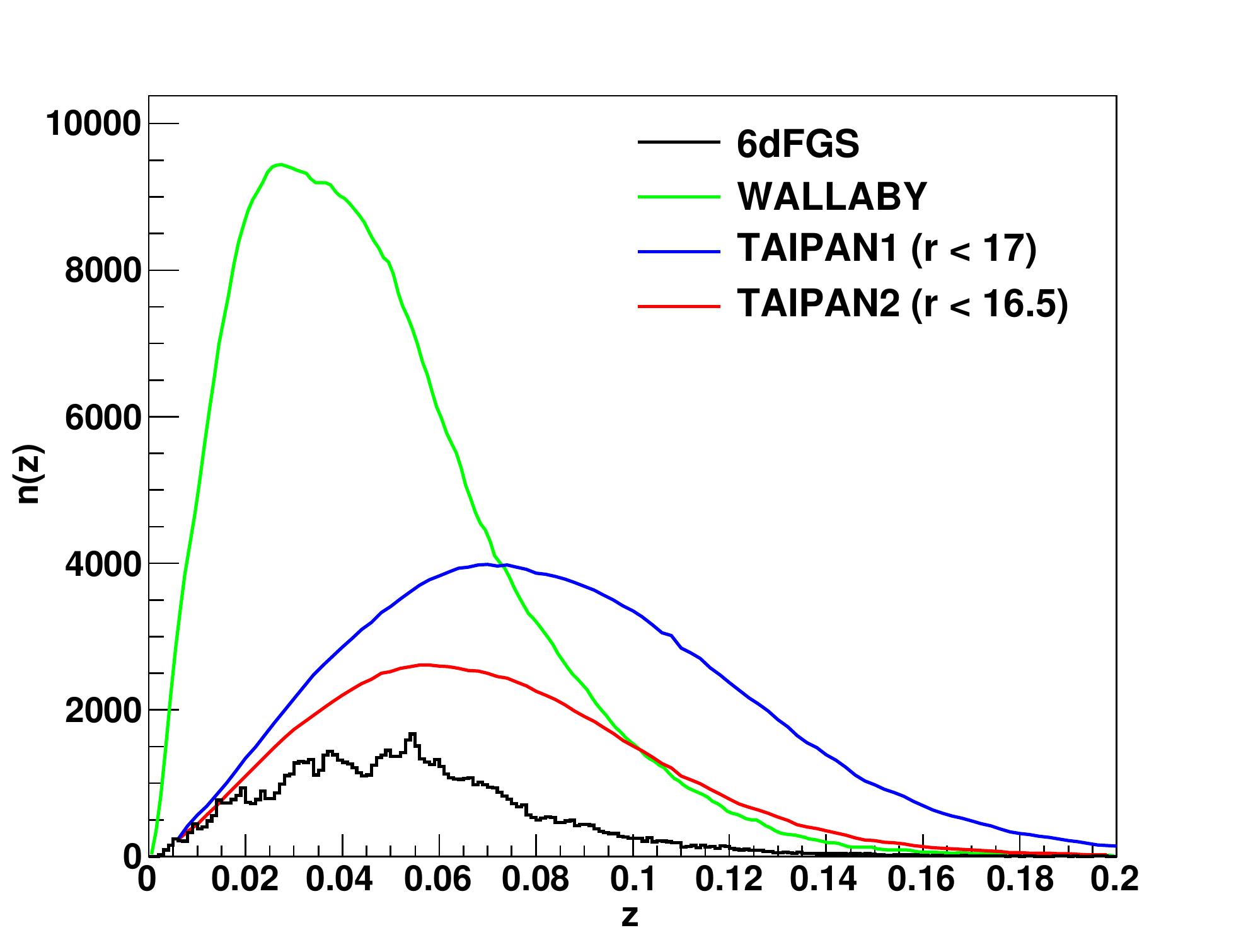,width=10cm}
\caption{Redshift distribution of 6dFGS, WALLABY and two different versions of the proposed TAIPAN survey.
See text for details.}
\label{fig:red_comp}
\end{center}
\end{figure}

\begin{figure}[tb]
\begin{center}
\epsfig{file=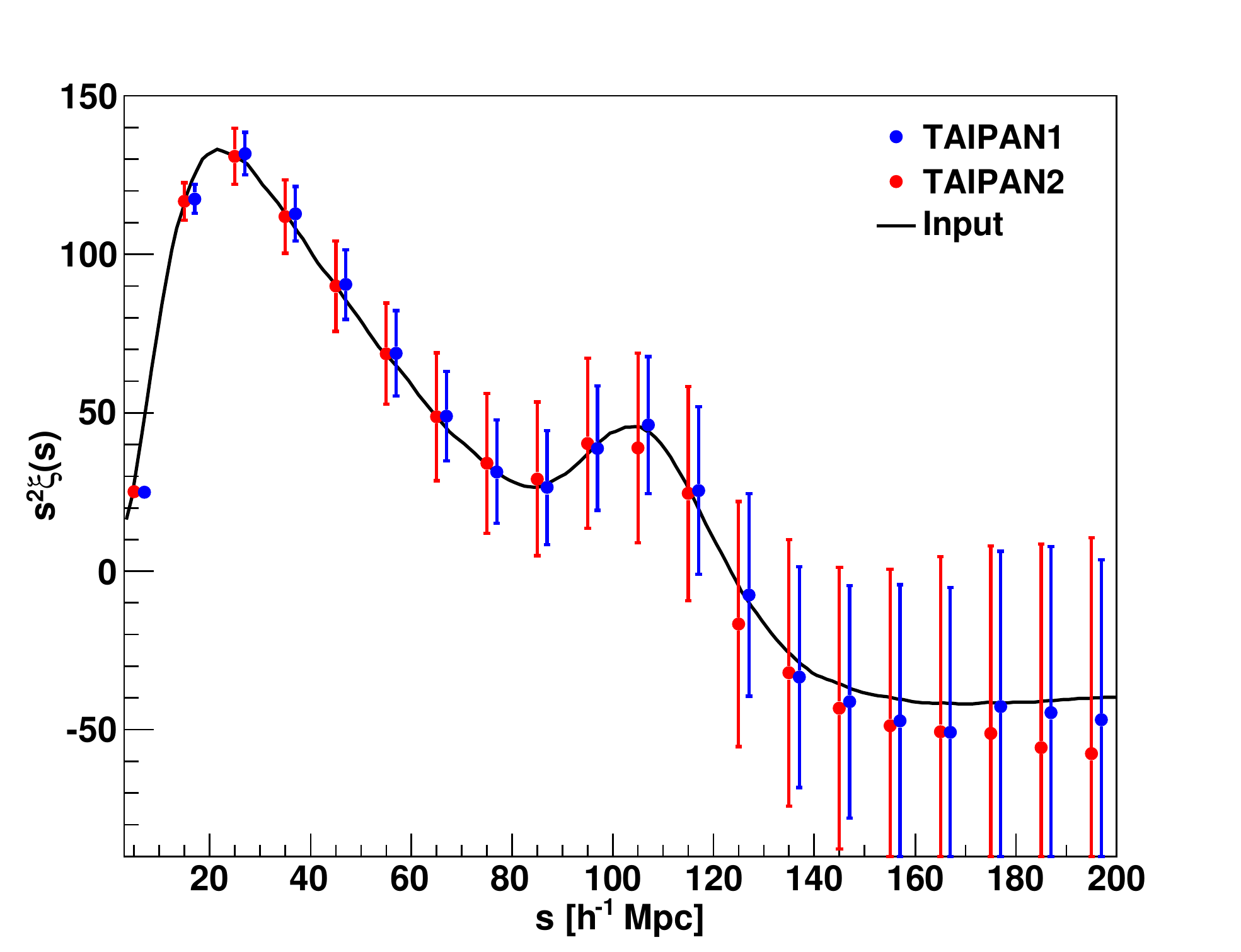,width=10cm}
\caption{Predictions for two versions of the proposed TAIPAN survey. Both predictions assume a $2\pi$\;steradian southern sky-coverage, excluding the Galactic plane (i.e. $|b| > 10^\circ$). TAIPAN1 contains $406\,000$ galaxies while TAIPAN2 contains $221\,000$, (see Figure~\ref{fig:red_comp}). The blue points are shifted by $2h^{-1}$\;Mpc to the right for clarity. The black line is the input model, which is a $\Lambda$CDM model with a bias of $1.6$, $\beta = 0.3$ and $k_* = 0.17h\;$Mpc$^{-1}$. For a large number of realisations, the difference between the input model and the mean (the data points) is only the convolution with the window function.}
\label{fig:log_taipan_both}
\end{center}
\end{figure}

\begin{figure}[tb]
\begin{center}
\epsfig{file=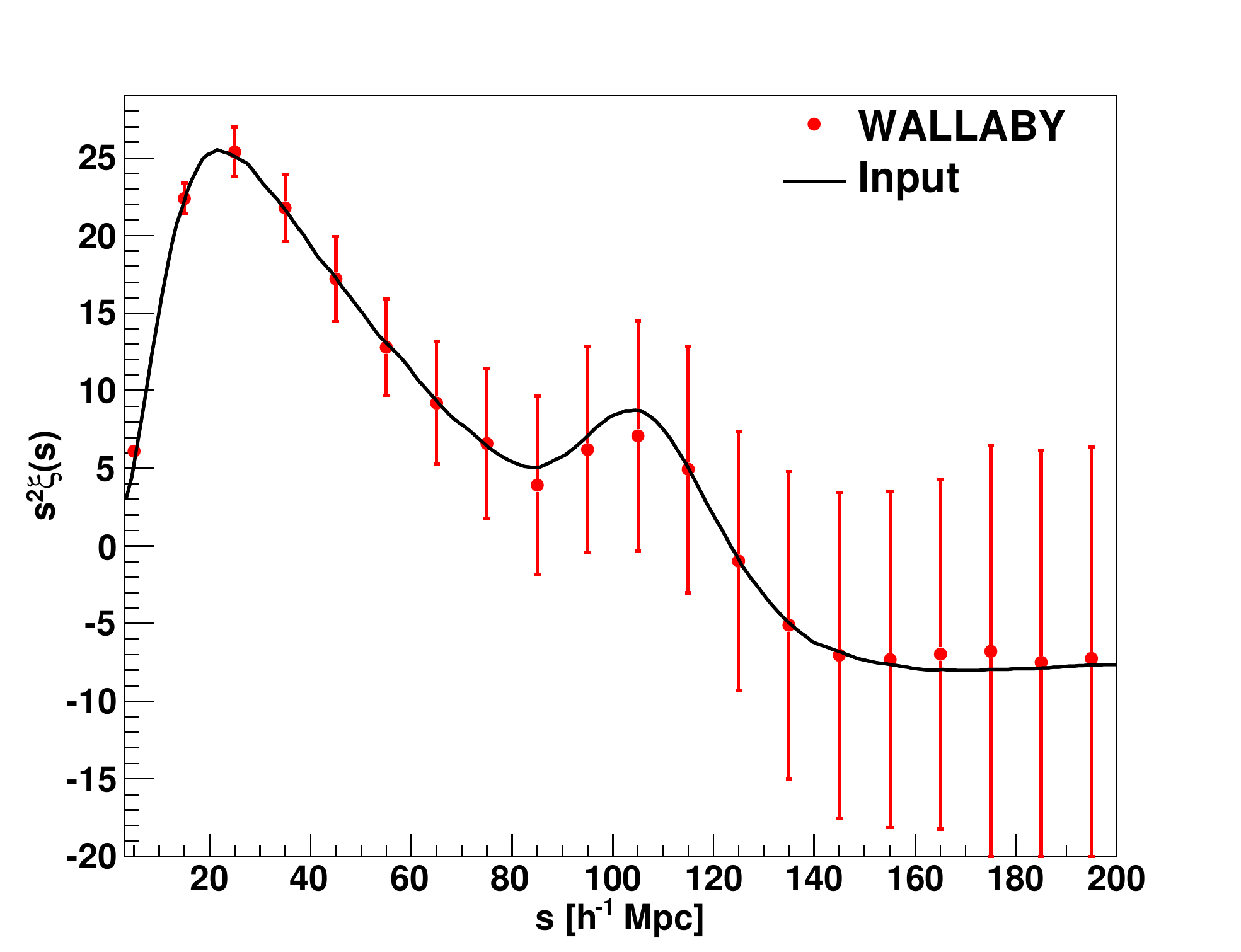,width=10cm}
\caption{Prediction for the WALLABY survey. We have assumed a $4\pi$\;steradian survey with $602\,000$ galaxies, $b = 0.7$, $\beta = 0.7$ and $k_* = 0.17h$\;Mpc$^{-1}$.}
\label{fig:log_wallaby}
\end{center}
\end{figure}

A major new wide-sky survey of the local Universe will be the Wide field ASKAP L-band Legacy All-sky Blind surveY (WALLABY)\footnote{http://www.atnf.csiro.au/research/WALLABY}. This is a blind H{\sc I} survey planned for the Australian SKA Pathfinder telescope (ASKAP), currently under construction at the Murchison Radio-astronomy Observatory (MRO) in Western Australia.

The survey will cover at least $75\%$ of the sky with the potential to cover $4\pi$ of sky if the Westerbork Radio Telescope delivers complementary northern coverage. Compared to 6dFGS, WALLABY will more than double the sky coverage including the Galactic plane. WALLABY will contain $\sim 500\,000$ to $600\,000$ galaxies with a mean redshift of around $0.04$, giving it around 4 times greater galaxy density compared to 6dFGS. In the calculations that follow, we assume for WALLABY a $4\pi$ survey without any exclusion around the Galactic plane. The effective volume in this case turns out to be $0.12h^{-3}$\;Gpc$^3$.

The TAIPAN survey\footnote{TAIPAN: Transforming Astronomical Imaging surveys through Polychromatic Analysis of Nebulae} proposed for the UK Schmidt Telescope at Siding Spring Observatory, will cover a comparable area of sky, and will extend 6dFGS in both depth and redshift ($z\simeq 0.08$).

The redshift distribution of both surveys is shown in Figure~\ref{fig:red_comp}, alongside 6dFGS. Since the TAIPAN survey is still in the early planning stage we consider two realisations: TAIPAN1 ($406\,000$ galaxies to a faint magnitude limit of $r=17$) and the shallower TAIPAN2 ($221\,000$ galaxies to $r=16.5$). We have adopted the same survey window as was used for 6dFGS, meaning that it covers the whole southern sky excluding a $10^{\circ}$ strip around the Galactic plane. The effective volumes of TAIPAN1 and TAIPAN2 are $0.23h^{-3}$\;Gpc$^3$ and $0.13h^{-3}$\;Gpc$^3$, respectively.

To predict the ability of  these surveys to measure the large scale correlation function we produced $100$ log-normal realisations for TAIPAN1 and WALLABY and $200$ log-normal realisations for TAIPAN2. Figures~\ref{fig:log_taipan_both} and \ref{fig:log_wallaby} show the results in each case. The data points are the mean of the different realisations, and the error bars are the diagonal of the covariance matrix. The black line represents the input model which is a $\Lambda$CDM prediction convolved with a Gaussian damping term using $k_* = 0.17h$\;Mpc$^{-1}$ (see eq.~\ref{eq:damp}). We used a bias parameter of $1.6$ for TAIPAN and following our fiducial model we get $\beta = 0.3$, resulting in $A = b^2(1 + 2\beta/3 + \beta^2/5) = 3.1$. For WALLABY we used a bias of $0.7$~\citep[based on the results found in the HIPASS survey; ][]{Basilakos:2007wq}. This results in $\beta = 0.7$ and $A = 0.76$. To calculate the correlation function we used $P_0 = 40\,000h^3\;$Mpc$^3$ for TAIPAN and $P_0 = 5\,000h^3\;$Mpc$^3$ for WALLABY.

The error bar for TAIPAN1 is smaller by roughly a factor of $1.7$ relative to 6dFGS, which is consistent with scaling by $\sqrt{V_{\rm eff}}$ and is comparable to the SDSS-LRG sample. We calculate the significance of the BAO detection for each log-normal realisation by performing fits to the correlation function using $\Lambda$CDM parameters and $\Omega_b = 0$, in exactly the same manner as the 6dFGS analysis described earlier. We find a $3.5\pm0.8\sigma$ significance for the BAO detection for TAIPAN1, $2.1\pm0.7\sigma$ for TAIPAN2 and $2.1\pm0.7\sigma$ for WALLABY, where the error again describes the variance around the mean. 

We then fit a correlation function model to the mean values of the log-normal realisations for each survey, using the covariance matrix derived from these log-normal realisations. We evaluated the correlation function of WALLABY, TAIPAN2 and TAIPAN1 at the effective redshifts of $0.1$, $0.12$ and $0.14$, respectively. With these in hand, we are able to derive distance constraints to respective precisions of $7\%$, $6\%$ and $3\%$. The predicted value for WALLABY is not significantly better than that from 6dFGS. This is due to the significance of the 6dFGS BAO peak in the data, allowing us to place tight constraints on the distance. As an alternative figure-of-merit, we derive the constraints on the Hubble constant. All surveys recover the input parameter of $H_0 = 70\;$km\;s$^{-1}$Mpc$^{-1}$, with absolute uncertainties of $3.7$, $3$ and $2.2\;$km\;s$^{-1}$Mpc$^{-1}$ for WALLABY, TAIPAN2 and TAIPAN1, respectively. Hence, TAIPAN1 is able to constrain the Hubble constant to $3\%$ precision. These constraints might improve when combined with Planck constraints on $\Omega_bh^2$ and $\Omega_mh^2$ which will be available when these surveys come along. 

Since there is significant overlap between the survey volume of 6dFGS, TAIPAN and WALLABY, it might be interesting to test whether the BAO analysis of the local Universe can make use of a multiple tracer analysis, as suggested recently by~\citet{ArnalteMur:2011xp}. These authors claim that by employing {\sl two} different tracers of the matter density field --
one with high bias to trace the central over-densities, and one with low bias to trace the small density fluctuations -- one 
can improve the detection and measurement of the BAO signal. \citet{ArnalteMur:2011xp} test this approach using the SDSS-LRG sample (with a very large bias) and the SDSS-main sample (with a low bias). Although the volume is limited by the amount of sample overlap, they detect the BAO peak at $4.1\sigma$. Likewise, we expect that the contrasting high bias of 6dFGS and TAIPAN, when used  in conjunction with the low bias of WALLABY, would furnish a combined sample that would be ideal for such an analysis. 

Neither TAIPAN nor WALLABY are designed as BAO surveys, with their primary goals relating to galaxy formation and
the local Universe. However, we have found that TAIPAN1 would be able to improve the measurement of the local Hubble constant by about $30\%$ compared to 6dFGS going to only slightly higher redshift. WALLABY could make some interesting contributions in the form of a multiple tracer analysis.

\section{Conclusion}
\label{sec:conclusion}

We have calculated the large-scale correlation function of the 6dF Galaxy Survey and detected a BAO peak with a significance of $2.4\sigma$. Although 6dFGS was never designed as a BAO survey, the peak is detectable because the survey
contains a large number of very bright, highly biased galaxies, within a sufficiently large effective volume of $0.08h^{-3}$\;Gpc$^3$. We draw the following conclusions from our work:

\begin{itemize}
\item The 6dFGS BAO detection confirms the finding by SDSS and 2dFGRS of a peak in the correlation function at around $105h^{-1}$\;Mpc, consistent with $\Lambda$CDM. This is important because 6dFGS is an independent sample, with a different target selection,  redshift distribution, and bias compared to previous studies. The 6dFGS BAO measurement is the lowest redshift BAO measurement ever made.
\item We do not see any excess correlation at large scales as seen in the SDSS-LRG sample. Our correlation function is consistent with a crossover to negative values at $140h^{-1}$\;Mpc, as expected from $\Lambda$CDM models.
\item We derive the distance to the effective redshift as $D_V(z_{\rm eff}) = 456\pm27\;$Mpc ($5.9\%$ precision). Alternatively, we can derive $r_s(z_d)/D_V(z_{\rm eff}) = 0.336\pm0.015$ ($4.5\%$ precision). All parameter constraints are summarised in Table~\ref{tab:para}.
\item Using a prior on $\Omega_mh^2$ from WMAP-7, we find $\Omega_m = 0.296\pm0.028$. Independent of WMAP-7, and taking into account curvature and the dark energy equation of state, we derive $\Omega_m = 0.287 + 0.039(1+w) + 0.039\Omega_k \pm 0.027$. This agrees very well with the first value, and shows the very small dependence on cosmology for parameter derivations from 6dFGS given its low redshift.
\item We are able to measure the Hubble constant, $H_0 = 67.0\pm3.2\;$km s$^{-1}$Mpc$^{-1}$, to $4.8\%$ precision, using only the standard ruler calibration by the CMB (in form of $\Omega_mh^2$ and $\Omega_bh^2$). Compared to previous BAO measurements, 6dFGS is almost completely independent of cosmological parameters (e.g. $\Omega_k$ and $w$), similar to Cepheid and low-$z$ supernovae methods. However, in contrast to these methods, the BAO derivation of the Hubble constant depends on very basic early Universe physics and avoids possible systematic errors coming from the build up of a distance ladder.
\item By combining the 6dFGS BAO measurement with those of  WMAP-7 and previous redshift samples \citet[from SDSS-DR7, SDDS-LRG and 2dFGRS; ][]{Percival:2009xn}, we can further improve the constraints on the dark energy equation of state, $w$, by breaking the $H_0-w$ degeneracy in the CMB data. Doing this, we find $w=-0.97\pm0.13$, which is an improvement of $24\%$ compared to previous combinations of BAO and WMAP-7 data.
\item We have made detailed predictions for two next-generation low redshift surveys,  WALLABY and TAIPAN. Using our 6dFGS result, we predict that both surveys will detect the BAO signal, and that WALLABY may be the first radio galaxy survey to do so. Furthermore, we predict that TAIPAN has the potential to constrain the Hubble constant to a precision of $3\%$ improving the 6dFGS measurement by $30\%$.
\end{itemize}

\cleardoublepage


\cleardoublepage

\renewcommand*\abstractname{\flushleft \Large Abstract\hfill}

\chapter{$\lowercase{z} \approx 0$ measurements of the growth rate and $\sigma_8$}
\label{ch:RSD}
\begin{center}
\emph{\textbf{\citeauthor{Beutler:2012px}}}\\
\emph{\textbf{MNRAS 423, 3430B (2012)}}
\end{center}

\begin{abstract}
We present a detailed analysis of redshift-space distortions in the two-point correlation function of the 6dF Galaxy Survey (6dFGS). The $K$-band selected sub-sample which we employ in this study contains $81\,971$ galaxies distributed over $17\,000\,$deg$^2$ with an effective redshift $z_{\rm eff} = 0.067$. By modelling the 2D galaxy correlation function, $\xi(r_p,\pi)$, we measure the parameter combination $f(z_{\rm eff})\sigma_8(z_{\rm eff}) = 0.423 \pm 0.055$, where $f \simeq \Omega_m^{\gamma}(z)$ is the growth rate of cosmic structure and $\sigma_8$ is the r.m.s. of matter fluctuations in $8h^{-1}\,$Mpc spheres.\\
Alternatively, by assuming standard gravity we can break the degeneracy between $\sigma_8$ and the galaxy bias parameter, $b$.  Combining our data with the Hubble constant prior from~\citet{Riess:2011yx}, we measure $\sigma_8 = 0.76 \pm 0.11$ and $\Omega_m = 0.250 \pm 0.022$, consistent with constraints from other galaxy surveys and the Cosmic Microwave Background data from WMAP7.\\
Combining our measurement of $f \sigma_8$ with WMAP7 allows us to test the cosmic growth history and the relationship between matter and gravity on cosmic scales by constraining the growth index of density fluctuations, $\gamma$. Using only 6dFGS and WMAP7 data we find $\gamma = 0.547 \pm 0.088$, consistent with the prediction of General Relativity. We note that because of the low effective redshift of 6dFGS our measurement of the growth rate is independent of the fiducial cosmological model (Alcock-Paczynski effect).  We also show that our conclusions are not sensitive to the model adopted for non-linear redshift-space distortions.\\
Using a Fisher matrix analysis we report predictions for constraints on $f\sigma_8$ for the WALLABY survey and the proposed TAIPAN survey. The WALLABY survey will be able to measure $f\sigma_8$ with a precision of $4-10\%$, depending on the modelling of non-linear structure formation. This is comparable to the predicted precision for the best redshift bins of the Baryon Oscillation Spectroscopic Survey (BOSS), demonstrating that low-redshift surveys have a significant role to play in future tests of dark energy and modified gravity.
\end{abstract}

\section{Introduction}

The distribution of matter in the cosmos depends on the gravitational interaction and the expansion history of the Universe. Assuming that galaxies trace the mass distribution, a measurement of galaxy clustering can be used to derive fundamental properties of the Universe. 

On large scales the movement of galaxies is dominated by the Hubble recession, while on small scales the gravitational field introduces so-called peculiar velocities. Individual galaxy redshifts combine both Hubble recession and peculiar velocities indistinguishably. However, these effects can be statistically distinguished in a large sample of galaxy redshifts. This is the purpose of this paper. The difference between the redshift-inferred distance and the true distance is known as redshift-space distortion. Redshift-space distortions effectively couple the density and velocity fields, complicating the models needed to accurately describe observed galaxy samples. On the other hand they permit measurements of the properties of the galaxy velocity field, which are difficult to access otherwise. In standard gravity we can use the amplitude of peculiar velocities to measure parameters that describe the matter content of the Universe such as $\Omega_m$ and $\sigma_8$. 

In this paper we report measurements of the parameter combination $f(z_{\rm eff})\sigma_8(z_{\rm eff})$, where $f = d\ln(D)/d\ln(a)$ is the growth rate of cosmic structure (in terms of the linear growth factor $D$ and cosmic scale factor $a$) and $\sigma_8$ is the r.m.s. of the matter fluctuations in spheres of $8h^{-1}\,$Mpc. The measurement of $f(z_{\rm eff})\sigma_8(z_{\rm eff})$ can be used to test theories of dark energy and modified gravity, since a stronger gravitational interaction causes a larger growth rate $f$. It is interesting to note in this context the fact that a different form of gravitational interaction on large scales could be responsible for the accelerating expansion of the Universe (e.g.~\citealt{Dvali:2000hr, Wang:2007ht}). Probes such as type Ia supernovae, the Cosmic Microwave Background (CMB) or Baryon Acoustic Oscillations, which have proven the existence of the current acceleration of the expansion of the Universe (e.g.~\citealt{Blake:2011en}), cannot distinguish between acceleration due to a dark energy component with negative pressure or due to a modification of General Relativity.  However, measurements of the growth of structure are able to distinguish between these models.

In order to measure $f(z_{\rm eff})\sigma_8(z_{\rm eff})$ we have to model the effect of redshift-space distortions on the correlation function. While on large scales linear theory can be used to model these effects, on smaller scales non-linear contributions complicate the process. 
Several new approaches have been suggested in recent years to extend linear theory. We will discuss some of these models and apply them to our dataset.

Redshift-space distortions have previously been analysed using both the correlation function and power spectrum using data from the 2dF Galaxy Redshift Survey (2dFGRS;~\citealt{Peacock:2001gs, Hawkins:2002sg, Cole:2005sx}) and the Sloan Digital Sky Survey (SDSS;~\citealt{Tegmark:2003ud,Zehavi:2004zn,Tegmark:2006az,Cabre:2008sz,Song:2010kq,Samushia:2011cs}). More recently it has become possible to do similar studies at higher redshift using the VVDS~\citep{Guzzo:2008ac}, WiggleZ~\citep{Blake:2011rj} and VLT VIMOS surveys~\citep{Bielby:2010ps}. 

Redshift-space distortion measurements are also sensitive to the overall amplitude of the clustering pattern of matter, commonly parameterised by $\sigma_8$~\citep{Lahav:2001sg}. This parameter is used to normalise the amplitude of clustering statistics such as the correlation function, $\xi \propto \sigma_8^2$. From the CMB we have a very accurate measurement of the matter fluctuations in the early universe (the scalar amplitude $A_s$) at the time of decoupling, $z_*$. In order to derive $\sigma_8(z{=}0)$ we have to extrapolate this measurement to redshift zero, involving assumptions about the expansion history of the Universe. The CMB constraint on $\sigma_8$ heavily depends on these assumptions. Hence there is a clear advantage in obtaining low-redshift measurements of this parameter. The 6dF Galaxy Survey, which we analyse in this study, is one of the largest galaxy redshift surveys available. Its very small effective redshift and wide areal coverage ($41\%$ of the sky) make it a powerful sample for the study of the local galaxy distribution. 

Galaxy surveys usually have to consider degeneracies between redshift-space distortions and the Alcock-Paczynski effect, which arises from the need to assume a cosmological model to transform redshifts into distances. At low redshift this effect is very small, meaning that our measurement is fairly independent of the choice of the fiducial cosmological model.

While at high redshift ($z > 1$) the matter density dominates both the expansion of the Universe and the growth of perturbations, at low redshift these two are partially decoupled, with dark energy mostly dominating the background expansion and the matter density dominating the growth of perturbations. As a result in $\Lambda$CDM and most proposed modified gravity models, low redshift measurements of the growth rate have a better constraining power than high redshift measurements. The measurement of the growth rate in 6dFGS therefore not only provides a new independent data point at very low redshift, but also promises to make a valuable contribution to tests of General Relativity on cosmic scales.

The outline of this paper is as follows: In section~\ref{sec:survey2} we introduce the 6dF Galaxy Survey. In section~\ref{sec:data} we describe the details of the correlation function estimate and introduce the 2D correlation function of 6dFGS. In section~\ref{sec:error} we derive the covariance matrix for the 2D correlation function. In section~\ref{sec:theory} we summarise the theory of redshift-space distortions, including extensions to the standard linear approach. We also discuss wide-angle effects and other systematics, such as the Alcock-Paczynski effect. In section~\ref{sec:fit} we fit the 2D correlation function to derive $g_{\theta}(z_{\rm eff})=f(z_{\rm eff})\sigma_8(z_{\rm eff})$ and $\sigma_8$. Cosmological implications are investigated in section~\ref{sec:impl}. In section~\ref{sec:future2} we make Fisher matrix predictions for two future low redshift galaxy surveys, WALLABY and the proposed TAIPAN survey. We conclude in section~\ref{sec:conc}.

Throughout the paper we use $r$ to denote real space separations and $s$ to denote separations in redshift-space. Our fiducial model assumes a flat universe with $\Omega^{\rm fid}_m = 0.27$, $w^{\rm fid} = -1$ and $\Omega_k^{\rm fid} = 0$. The Hubble constant is set to $H_0 = 100h\,$km s$^{-1}$Mpc$^{-1}$.

\section{The 6\lowercase{d}F Galaxy Survey}
\label{sec:survey2}

The 6dF Galaxy Survey~\citep[6dFGS;][]{Jones:2004zy,Jones:2006xy,Jones:2009yz} is a near-infrared selected ($JHK$) redshift survey of $125\,000$ galaxies across four-fifths of the southern sky, with secondary samples selected in $b_{\rm J}$ and $r_{\rm F}$. The $|b| < 10^\circ$ region around the Galactic Plane is avoided by the $JHK$ surveys to minimise Galactic extinction and foreground source confusion in the Plane (as is $|b| < 20^\circ$ for $b_{\rm J}$ and $r_{\rm F}$). The near-infrared photometric selection was based on total magnitudes from the Two-Micron All-Sky Survey -- Extended Source Catalog~\citep[2MASS XSC;][]{Jarrett:2000me}. The spectroscopic redshifts of 6dFGS were obtained with the Six-Degree Field (6dF) multi-object spectrograph of the UK Schmidt Telescope (UKST) between 2001 and 2006. The effective volume of 6dFGS is about the same as the 2dF Galaxy Redshift Survey~\citep[2dFGRS;][]{Colless:2001gk} and is a little under a third that of the Sloan Digital Sky Survey main spectroscopic sample at its Seventh Data Release~\citep[SDSS DR7;][]{Abazajian:2008wr}. A subset of early-type 6dFGS galaxies (approximately $10\,000$) have measured line-widths that will be used to derive Fundamental Plane distances and peculiar motions.

The 6dFGS $K$-selected sample used in this paper contains 81\,971 galaxies selected to a faint limit of $K = 12.75$. The 2MASS magnitudes are on the Vega system. The mean completeness of the 6dFGS is 92 percent and median redshift is $z = 0.05$. Completeness corrections are derived by normalising completeness-apparent magnitude functions so that, when integrated over all magnitudes, they equal the measured total completeness on a particular patch of sky. This procedure is outlined in the luminosity function evaluation of~\citet{Jones:2006xy} and also in Jones et al., (in prep). The original survey papers~\citep{Jones:2004zy,Jones:2009yz} describe in full detail the implementation of the survey and its associated online database.

\begin{figure}[tb]
\begin{center}
\epsfig{file=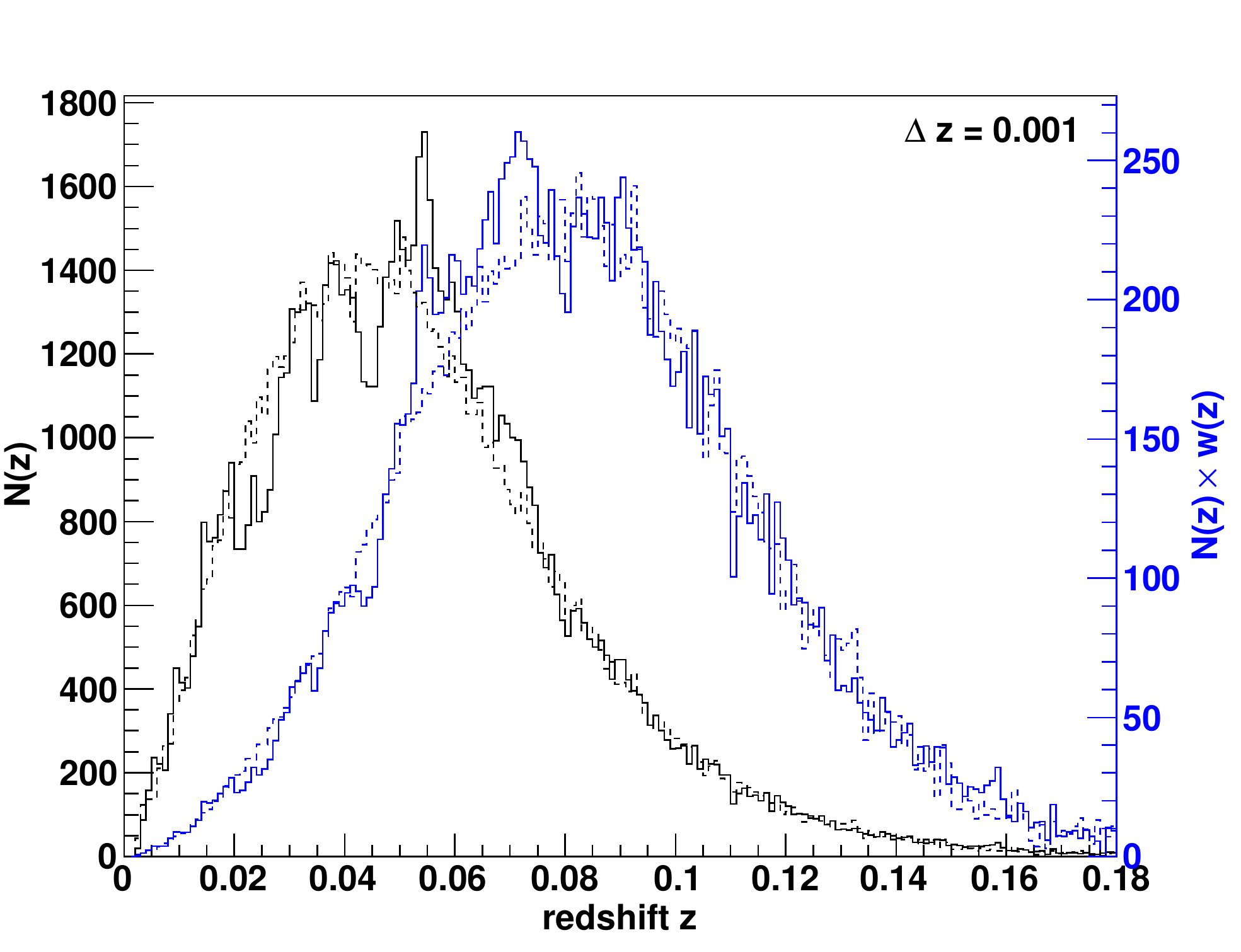,width=10cm}
\caption{The solid black line shows the 6dFGS redshift distribution, while the dashed black line shows one of the random mock catalogues containing the same number of galaxies. The blue solid and dashed lines show the distribution after weighting with $P_0 = 1600h^3\,$Mpc$^{-3}$ (see section~\ref{sec:dweight} for more details on the employed weighting scheme).}
\label{fig:wred2}
\end{center}
\end{figure}

The clustering in a galaxy survey is estimated relative to a random (unclustered) distribution which follows the same angular and redshift selection function as the galaxy sample itself. We base our random mock catalogue generation on the 6dFGS luminosity function, where we use random numbers to pick volume-weighted redshifts and luminosity function-weighted absolute magnitudes. We then test whether the redshift-magnitude combination falls within the 6dFGS $K$-band faint and bright apparent magnitude limits ($8.75 \leq K \leq 12.75$).

Figure~\ref{fig:wred2} shows the redshift distribution of the 6dFGS $K$-selected sample (black solid line) compared to a mock catalogue with the same number of galaxies (black dashed line).

\section{Correlation function measurement}
\label{sec:data}

We calculate the co-moving distances for each galaxy using the measured redshift
\begin{equation}
D_C = \frac{c}{H_0}\int^z_0\frac{dz'}{E(z')}
\end{equation} 
with
\begin{equation}
E(z) = \left[\Omega^{\rm fid}_m(1+z)^3 + \Omega^{\rm fid}_{\Lambda}\right]^{1/2},
\end{equation}
where we assume a flat universe with $\Omega_k^{\rm fid} = 0$ and $\Omega^{\rm fid}_{\Lambda} = 1 - \Omega^{\rm fid}_m$ and describe dark energy as a cosmological constant ($w^{\rm fid} = -1$). Given the low redshift of our dataset, these assumptions have a very small impact on our final results (see section~\ref{sec:AP}).

We define the positions of two galaxies as $\vec{s_1}$ and $\vec{s_2}$. The redshift-space separation is then given by $\vec{h} = \vec{s}_1 - \vec{s}_2$, while $\vec{s} = (\vec{s}_1 + \vec{s}_2)/2$ is the mean distance to the pair of galaxies. Now we can calculate the separation along the line-of-sight $\pi$ and the separation across the line-of-sight $r_p$
\begin{align}
\pi &= \frac{|\vec{s}\cdot\vec{h}|}{|\vec{s}|},\\
r_p &= \sqrt{|\vec{h}|^2 - \pi^2}.
\end{align}
The absolute separation is then given by $s = \sqrt{\pi^2 + r_p^2}$.

We measure the separation between all galaxy pairs in our survey and count the number of such pairs in each separation bin. We do this for the 6dFGS data catalogue, a random catalogue with the same selection function, and a combination of data-random pairs. We call the pair-separation distributions obtained from this analysis step $DD$, $RR$ and $DR$, respectively. In the analysis we used $30$ random catalogues with the same size as the real data catalogue and average $DR$ and $RR$. The redshift-space correlation function itself is then given by~\cite{Landy:1993yu}:
\begin{equation}
\xi'_{\rm data} = 1 + \frac{DD}{RR} \left(\frac{n_r}{n_d} \right)^2 - 2\frac{DR}{RR} \left(\frac{n_r}{n_d} \right),
\label{eq:LS3}
\end{equation}
where the ratio $n_r/n_d$ is given by
\begin{equation}
\frac{n_r}{n_d} = \frac{\sum^{N_r}_iw_i}{\sum^{N_d}_jw_j}
\end{equation}
and the sums go over all random ($N_r$) and data ($N_d$) galaxies. The galaxies are weighted by the inverse completeness $C_i$ of their area of the sky
\begin{equation}
w_i(z) = C_i.
\label{eq;compl}
\end{equation}
We will discuss further weighting techniques in the next section. 

There is a possible bias in the estimation of the correlation function due to the fact that we estimate both the mean density and the pair counts from the same survey. This leads to a non-zero difference between the true correlation function estimate of an ensemble of surveys and the ensemble average of $\xi(s)$ from each survey. This is commonly known as the integral constraint (e.g.~\citealt{Peebles:1980}), which can be calculated as (see e.g.~\citealt{Roche:2002vj})
\begin{equation}
ic = \frac{\sum \xi_{\rm model}RR}{\sum RR}
\end{equation}
and enters our correlation function estimate as
\begin{equation}
\xi_{\rm data} = \xi'_{\rm data} + ic,
\end{equation}
where $\xi'_{\rm data}$ is the redshift-space correlation function from eq.~\ref{eq:LS3} and $\xi_{\rm model}$ is the model for the correlation function. In 6dFGS $ic$ is typically around $6\times 10^{-4}$ and so has no significant impact on the final result.

\begin{figure}[tb]
\begin{center}
\epsfig{file=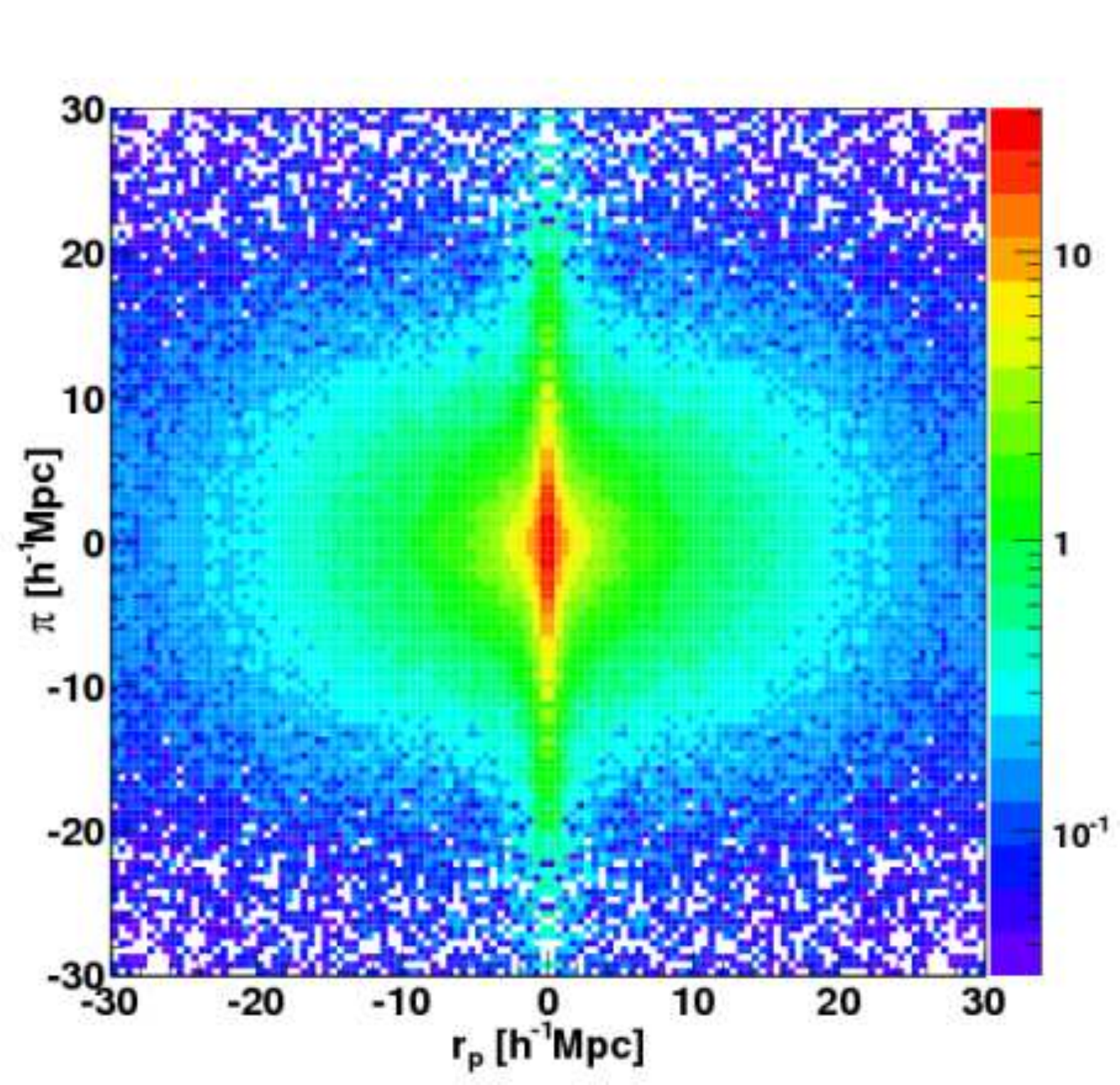,width=10cm}
\caption{The 2D correlation function of 6dFGS using a density weighting with $P_0 = 1600h^3\,$Mpc$^{-3}$. For reasons of presentation we binned the correlation function in $0.5h^{-1}\,$Mpc bins, while in the analysis we use larger bins of $2h^{-1}\,$Mpc. Both redshift-space distortion effects are visible: the ``finger-of-God'' effect at small angular separation $r_p$, and the anisotropic (non-circular) shape of the correlation function at large angular separations.}
\label{fig:kaiser}
\end{center}
\end{figure}

In Figure~\ref{fig:kaiser} we show the 2D correlation function calculated from the 6dFGS dataset. In this Figure we use bins of $0.5h^{-1}\,$Mpc, while for the analysis later on we use larger bins of $2h^{-1}\,$Mpc (see Figure~\ref{fig:kaiser_models}). The figure shows clearly the two effects of redshift-space distortions which we will discuss later in section~\ref{sec:theory}, the ``finger-of-God'' effect at small $r_p$, and the linear infall effect at larger $r_p$ which gives the correlation function a non-circular shape.

\subsection{Density weighting}
\label{sec:dweight}

In Fourier space the error  in measuring the amplitude of a mode of the linear power spectrum\footnote{As the correlation function and power spectrum are related by a Fourier transform, the following discussion also holds true for a correlation function measurement.} is given by
\begin{equation}
\sigma_{P(k)} = (b + f\mu^2)^2P(k) + \langle N \rangle,
\end{equation}
where $b$ is the linear bias, $f$ is the growth rate, $\mu$ is the cosine of the angle to the line of sight and $P(k)$ is the matter power spectrum. The first term on the right hand side of this equation represents the sample-variance error, while the second term ($\langle N \rangle$) represents the Poisson error.

\begin{figure}[p]
\centering
\vspace{-0.7cm}
\subfigure[]{
   \epsfig{file=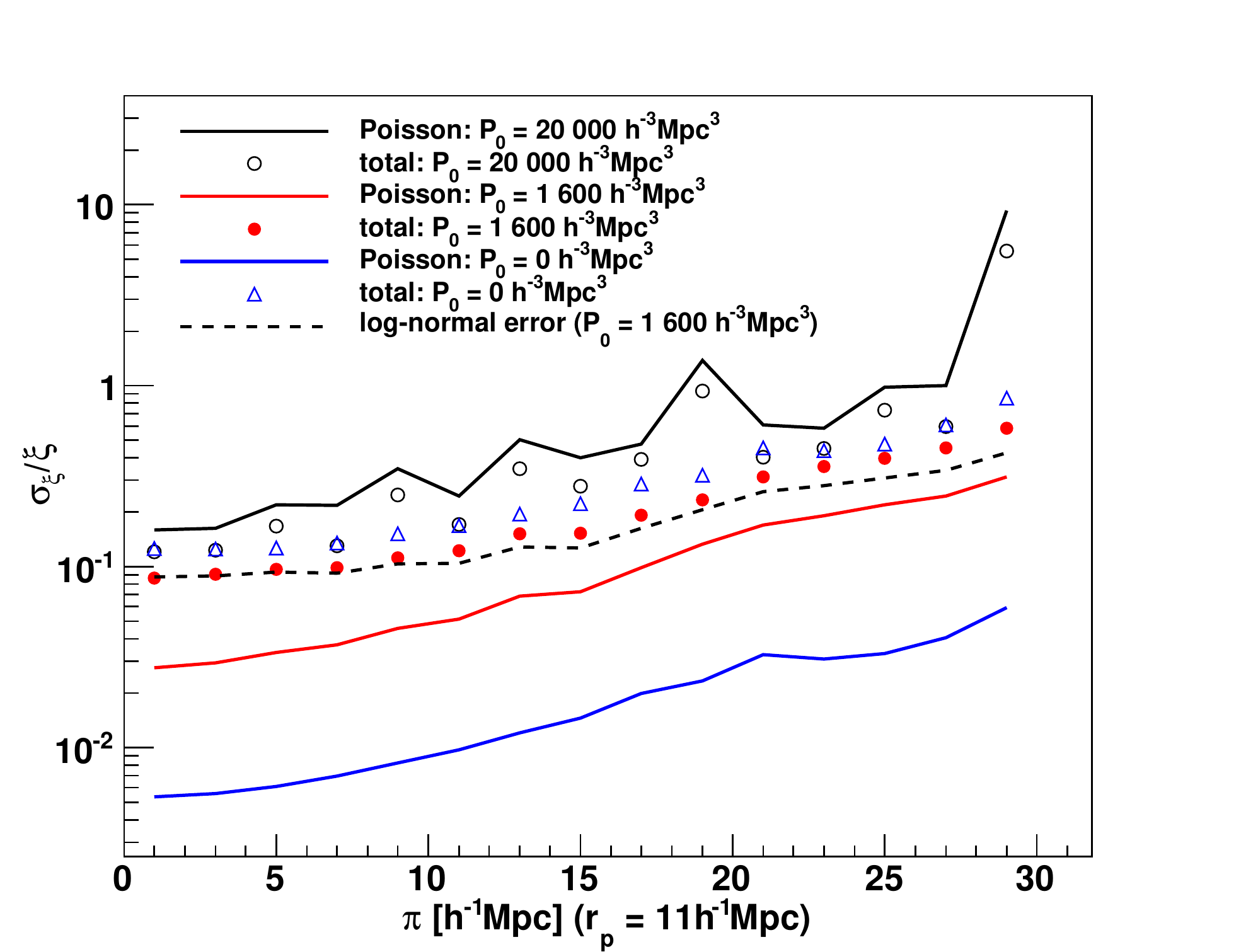,width=9.5cm}
   \label{fig:kaiser_error}
 }\\
 \vspace{-0.4cm}
 \subfigure[]{
   \epsfig{file=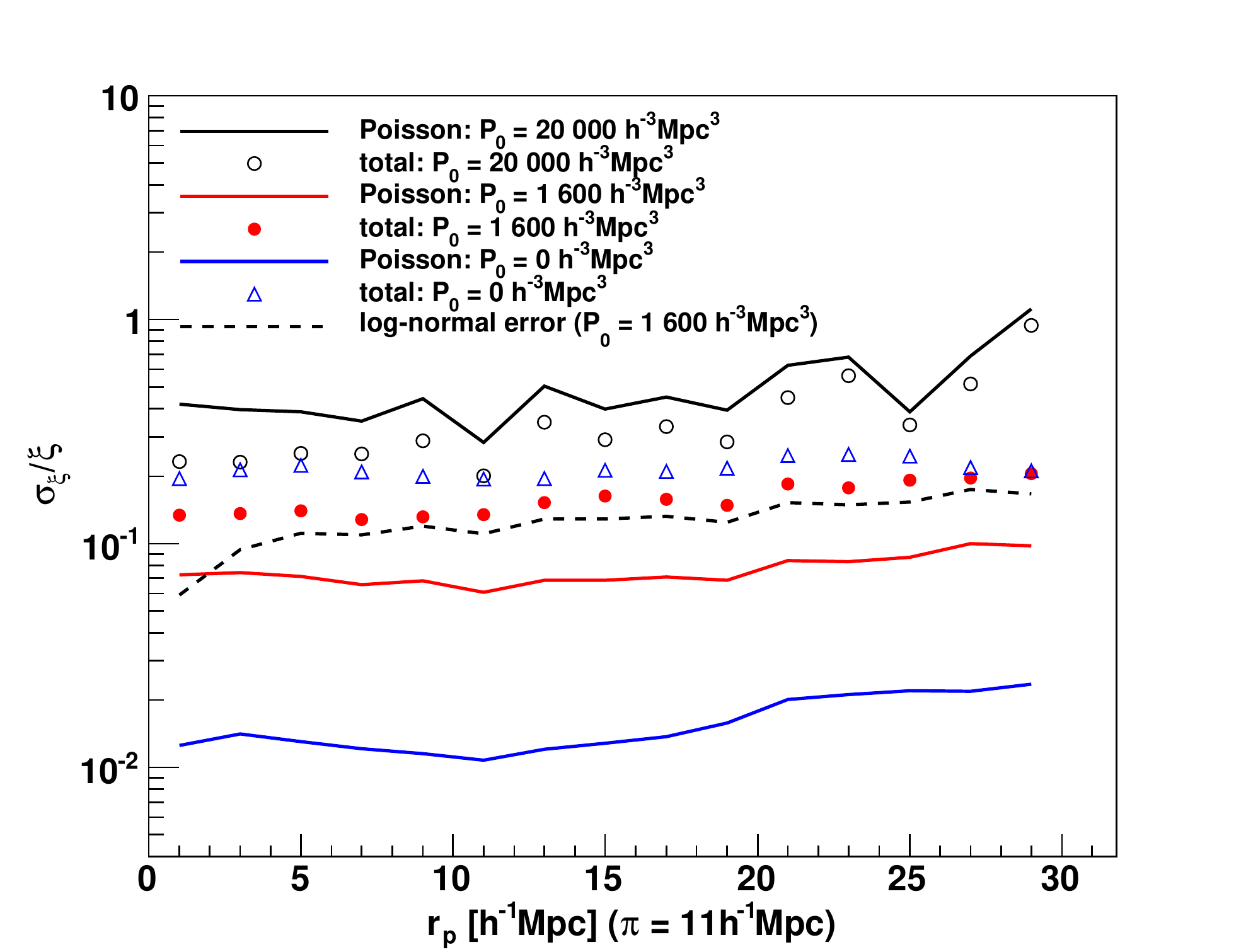,width=9.5cm}
   \label{fig:kaiser_error2}
 }
\label{fig:kaiser_errors}
\caption{(a) The relative error in the 2D correlation function as a function of line-of-sight separation $\pi$ at a fixed $r_p = 11h^{-1}\,$Mpc. Other regions of the 2D correlation function behave in a similar manner. The solid lines show the Poisson error for different values of $P_0$, while the data points show the total (Poisson + sample variance) error obtained as the diagonal of the covariance matrix derived using jack-knife re-sampling. The purpose of the weighting ($P_0$) is to minimise the total error, which is achieved for a value of $P_0 \approx 1600h^{-3}\,$Mpc$^{3}$. The weighting reduces the error by almost a factor of two on most scales. The dashed line shows the error derived from log-normal realisations using $P_0 = 1600h^{-3}\,$Mpc$^{3}$ and is in very good agreement with the jack-knife error.
(b) same as (a) for a fixed line-of-sight separation $\pi = 11h^{-1}\,$Mpc.}
\end{figure}

If the sample-variance error is dominant we can reduce the power spectrum error by employing a weighting scheme which depends upon the galaxy density $n(z)$, such as the one suggested by~\citet{Feldman:1993ky}
\begin{equation}
w_i(z) = \frac{1}{1 + n(z)P_0},
\label{eq:static}
\end{equation}
where $P_0$ describes the amplitude of the weighting. A stronger weighting (larger value of $P_0$) yields a smaller sample-variance error since it increases the survey volume by up-weighting sparsely sampled regions. However, such a weighting scheme also increases the Poisson error because it shifts the effective redshift to larger values with a smaller galaxy number density. This is illustrated in Figure~\ref{fig:kaiser_error} and~\ref{fig:kaiser_error2}. Such a weighting scheme is standard for large scale structure analyses.

In a magnitude-limited sample such as 6dFGS, up-weighting higher redshift galaxies also has the effect of shifting the galaxy bias to larger values. The sample-variance error is proportional to the clustering amplitude, and so a larger bias results in a larger error. However, the weighting will still ensure that the relative error of the power spectrum, $\sigma_{P(k)}/P(k)$, is minimised. The redshift-space distortion signal is inversely proportional to the galaxy bias, $\beta \simeq \Omega_m^{\gamma}(z)/b$. If weighting increases the bias $b$, it also reduces the signal we are trying to measure. We therefore must investigate whether the advantage of the weighting (the reduced relative error) outweighs the disadvantage (increasing galaxy bias).

The situation is very different for measuring a signal that is proportional to the clustering amplitude, such as the baryon acoustic peak. In this case the error and the signal are proportional to the bias, and so weighting will always be beneficial. We stress that an increasing bias with redshift is expected in almost all galaxy redshift surveys. Therefore redshift-space distortion studies should first test whether galaxy weighting improves the measurement. The 6dF Galaxy Survey is quite sensitive to the weighting scheme employed because it has a high galaxy density, making the sample-variance error by far the dominant source of error.

Finally, we have to consider the correlation between the bins in the measured power spectrum or correlation function. If the error is sample-variance dominated, the bins will show large correlation (especially in the correlation function), while in the case of Poisson-noise dominated errors, the correlation is much smaller. Weighting will always increase the Poisson noise and hence reduce the correlation between bins.

All of the density weighting effects discussed above need to be considered, before deciding which weighting is best suited to the specific analysis. We summarise as follows:
\begin{itemize}
\item If a galaxy sample is sample-variance limited, density weighting will reduce the (relative) error of the clustering measurement.
\item In most galaxy redshift surveys, the density weighting increases the galaxy bias, which reduces the redshift-space distortion signal, by flattening the clustering anisotropy. 
\item Galaxy weighting also reduces the correlation between bins in the power spectrum and correlation function.
\end{itemize}
Whether the signal-to-noise is actually improved by density weighting depends on the specific sample. For the 6dFGS redshift-space distortion analysis we found $P_0 \approx 1600h^{-3}\,$Mpc$^{3}$ leads to the most accurate constraint on the growth rate.

This discussion also indicates that low-biased galaxy samples have an advantage over a highly biased galaxy sample in measuring redshift-space distortions. We will discuss this point further in section~\ref{sec:future2}.

\section{Error estimate}
\label{sec:error}

In this section we will derive a covariance matrix for the 2D correlation function using jack-knife re-sampling. We also use log-normal realisations to test the jack-knife covariance matrix.

\subsection{Jack-knife re-sampling}

We divide the dataset into $N = 480$ subsets, selected in R.A. and Dec. Each re-sampling step excludes one subset before calculating the correlation function. The $N-1$ remaining subsets have a volume which is $(N-1)/N$ times the volume of the original data. The covariance matrix is then given by
\begin{equation}
C_{ij} = \frac{(N-1)}{N}\sum^N_{k=1}\left[\xi^k(s_i) - \overline{\xi}(s_i)\right]\left[\xi^k(s_j) - \overline{\xi}(s_j)\right],
\end{equation}
where $\xi^k(s_i)$ is the correlation function estimate at separation $s_i$ with the exclusion of subset $k$. The mean value is defined as
\begin{equation}
\overline{\xi}(s_i) = \frac{1}{N}\sum^N_{k=1}\xi^k(s_i).
\end{equation}
The case $i = j$ gives the error ignoring correlations between bins $\sigma_i^2 = C_{ii}$. 

\subsection{Log-normal realisations}

We can create a log-normal realisation~\citep{Coles:1991if,Cole:2005sx, Percival:2006gt, Blake:2011rj, Beutler:2011hx} of a galaxy survey by deriving a density field from a model power spectrum, $P(k)$, assuming Gaussian fluctuations. This density field is then Poisson sampled, taking into account the window function and the total number of galaxies. The assumption that the input power spectrum has Gaussian fluctuations can only be used if the fluctuations are much smaller than the mean density, otherwise the Gaussian model assigns a non-zero probability to regions of negative density. A log-normal random field, $LN(\vec{x})$, can avoid this unphysical behaviour. It is obtained from a Gaussian field $G(\vec{x})$ by
\begin{equation}
LN(\vec{x}) = \exp[G(\vec{x}) - \sigma_G^2/2] - 1,
\end{equation}
where $\sigma_G$ is the variance of the field. $LN(\vec{x})$ is  positive-definite but approaches $G(\vec{x})$ whenever the perturbations are small. As an input power spectrum for the log-normal field we use the linear model~\citep{Kaiser:1987qv}, given by
\begin{equation}
P_g(k,\mu) = b^2(1 + \beta\mu^2)^2P_{\rm \delta\delta}(k)
\end{equation}
with $b = 1.47$ and $\beta = 0.35$, which is consistent with the parameters we measure for 6dFGS (see section~\ref{sec:fit}). $P_{\rm \delta\delta}(k)$ is a linear density power spectrum in real space obtained from CAMB~\citep{Lewis:1999bs} and $P_g(k,\mu)$ is the galaxy power spectrum in redshift-space. Our method is explained in more detail in~\citet{Beutler:2011hx}, appendix A.

We produce $N = 1500$ such realisations and calculate the 2D correlation function for each of them, deriving a covariance matrix
\begin{equation}
\begin{split}
C_{ij} = \frac{1}{N-1}&\sum^N_{n=1}\left[\xi^i_n(r_p,\pi) - \overline{\xi^i}(r_p,\pi)\right]\cr
&\times\left[\xi^j_n(r_p,\pi) - \overline{\xi^j}(r_p,\pi)\right],
\end{split}
\end{equation}
where the mean value $\overline{\xi^i}(r_p,\pi)$ is defined as
\begin{equation}
\overline{\xi^i}(r_p,\pi) = \frac{1}{N}\sum^N_{n=1}\xi^i_n(r_p,\pi)
\end{equation}
and $\xi^i_n(r_p,\pi)$ is the 2D correlation function estimate of realisation $n$ at a specific separation $(r_p, \pi)$.

\subsection{Discussion: Error analysis} 

\begin{figure}[tb]
\begin{center}
   \epsfig{file=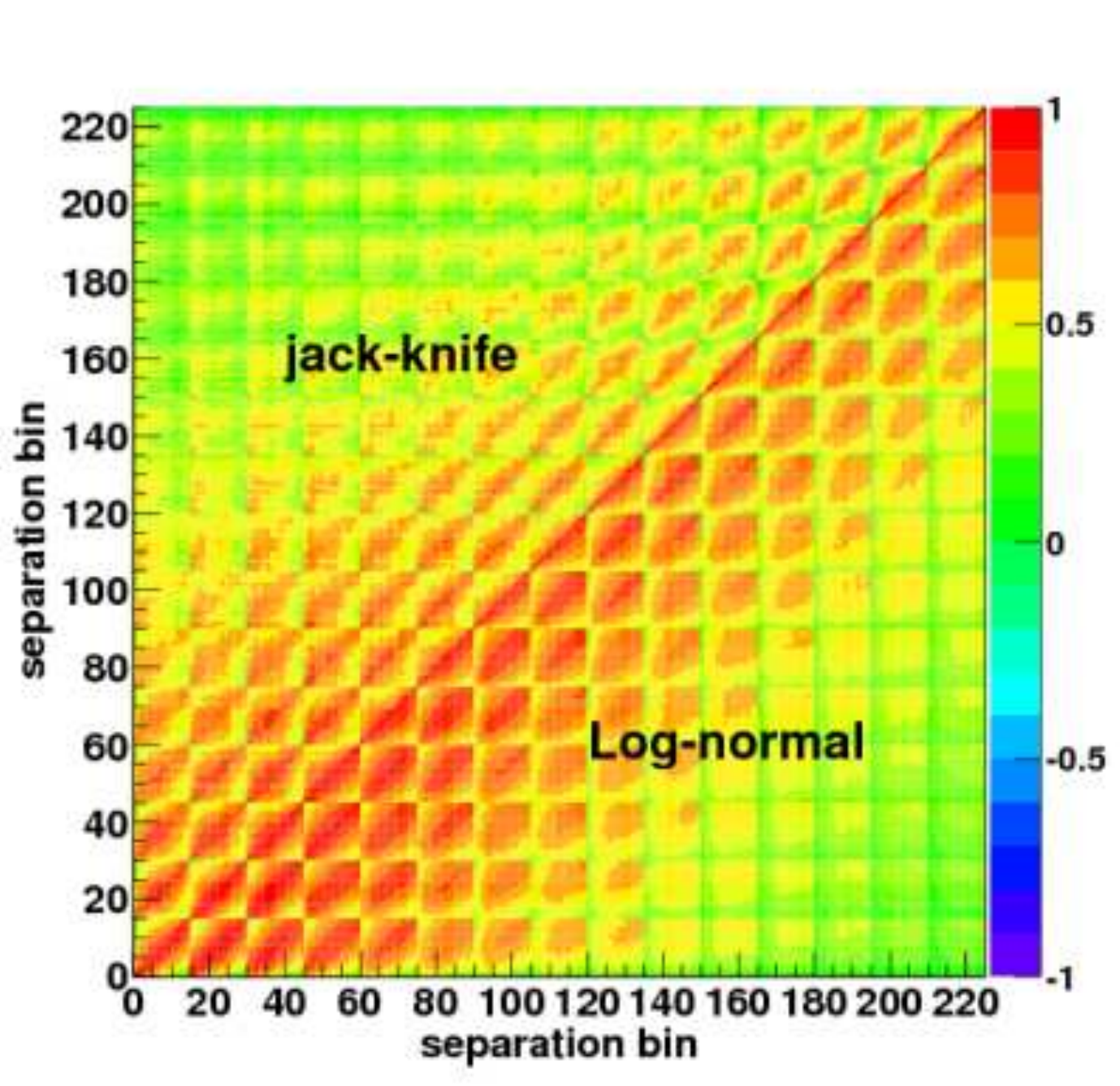,width=10cm}
\caption{The correlation matrix for the 2D correlation function $\xi(r_p,\pi)$ with a bin size of $2\times 2h^{-1}\,$Mpc. The upper-left corner shows the jack-knife estimate, while the lower-right corner shows the result of using $1500$ log-normal realisations. Since this plot shows the correlation of all $15\times 15$ bins it contains $225\times 225$ entries.}
   \label{fig:cov2D}
\end{center}
\end{figure}

\begin{figure}[tb]
\begin{center}
   \epsfig{file=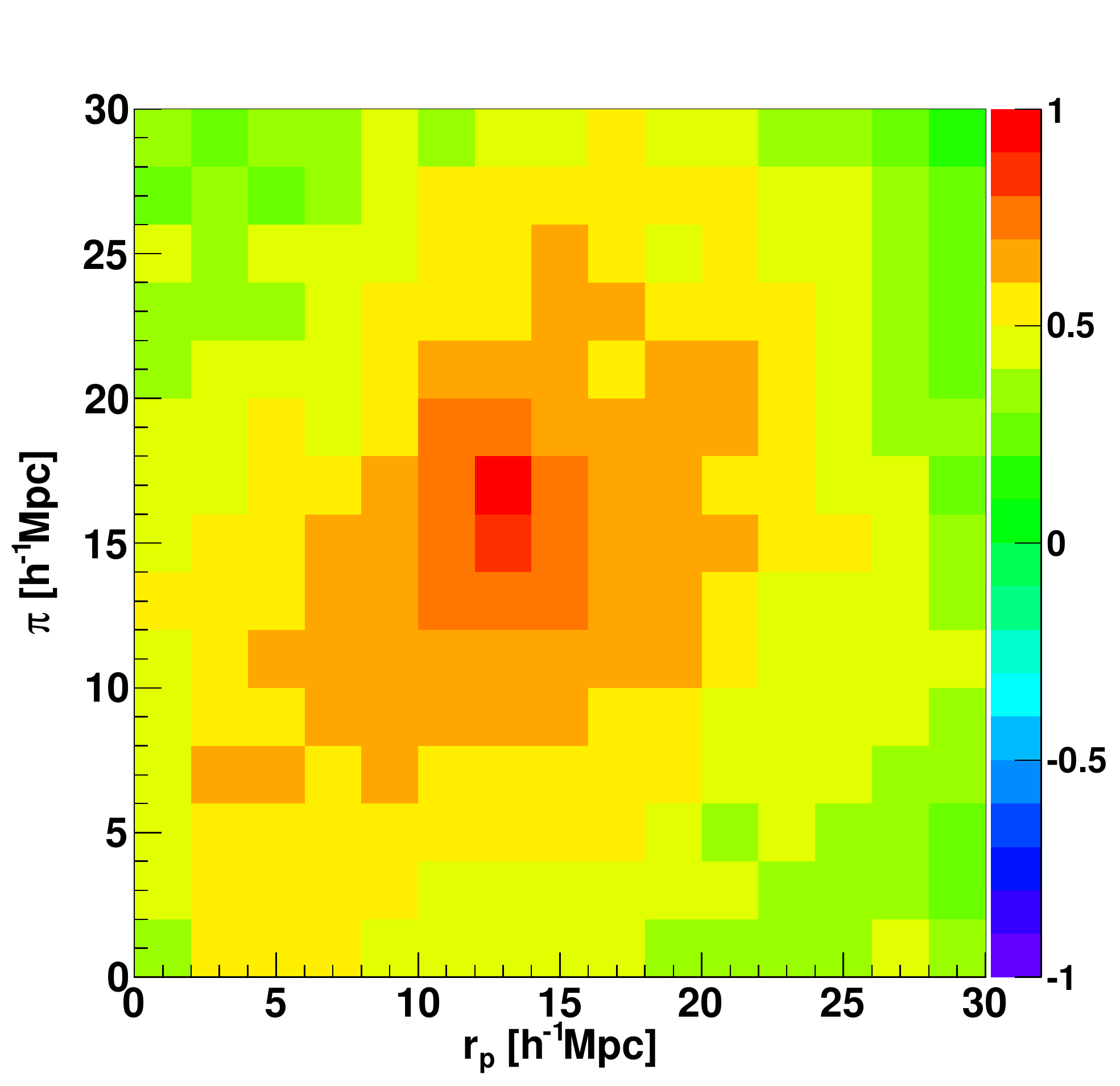,width=10cm}
\caption{This plot shows the correlation of bin $127$ ($r_p = 13h^{-1}\,$Mpc, $\pi = 17h^{-1}\,$Mpc) with all other bins in the 2D correlation function, derived using jack-knife re-sampling. It corresponds to row/column $127$ of the jack-knife correlation matrix which is shown in Figure~\ref{fig:cov2D} (upper-left corner).}
   \label{fig:cov2D_2}
\end{center}
\end{figure}

Figure~\ref{fig:cov2D} shows the correlation matrix
\begin{equation}
r_{ij} = \frac{C_{ij}}{\sqrt{C_{ii}C_{jj}}},
\end{equation}
derived from the two different covariance matrices, $C_{ij}$. We plot the jack-knife result in the upper-left corner and the log-normal result in the lower-right corner. Both the correlation and covariance matrix are symmetric. While the log-normal correlation matrix indicates somewhat more correlation between bins, overall the correlation matrices are in rough agreement. The diagonal errors for both the log-normal and jack-knife covariance matrices are plotted in Figures~\ref{fig:kaiser_error} and \ref{fig:kaiser_error2} and show very good agreement at large scales.

Every row/column in Figure~\ref{fig:cov2D} shows the correlation of one bin with all other bins in $\xi(r_p, \pi)$. Figure~\ref{fig:cov2D_2} shows an example of such a row/column obtained from jack-knife re-sampling as a $15\times 15$ matrix, in this case for bin $127$ ($r_p = 13h^{-1}\,$Mpc, $\pi = 17h^{-1}\,$Mpc).

Log-normal realisations do not account for non-linear mode coupling and are very model-dependent in the quasi-linear and non-linear regime. 
Since our analysis relies on fits to fairly small scales, we decided to use the jack-knife covariance matrix in our analysis. However, we find that none of the results reported in this paper depend significantly on which of the two covariance matrices is used.

\section{Modelling the 2D correlation function}
\label{sec:theory}

In this section we discuss the theory of redshift-space distortions, starting with the standard linear perturbation theory. We then describe different approaches for extending the linear model to include non-linear structure formation. We also discuss deviations from the plane-parallel approximation.

\subsection{Linear redshift-space distortions}

The position of a galaxy in real space $\vec{r} = (x,y,z)$ is mapped to the position in redshift-space, $\vec{s}$, via
\begin{equation}
\vec{s} = \vec{r} + \frac{v_z(\vec{r})}{aH(z)}\hat{z},
\end{equation}
where the unit vector $\hat{z}$ indicates the line-of-sight direction, and the quantity $v_z$ is the line-of-sight component of the velocity field, namely $v_z = \vec{v}\cdot\hat{z}$. The scale factor $a$ is defined as $1/(1+z)$ and $H(z)$ is the Hubble constant at redshift $z$. The second term in the equation above represents peculiar velocities caused by gravitational interaction. 
On small scales this elongates structures along the line-of-sight and leads to the so-called ``finger-of-God'' effect~\citep{Jackson:2008yv}. On large scales matter falls in towards over-dense regions and systematically influences our distance measurement, making the over-densities appear more over-dense. The latter effect can be described by linear theory, while the ``finger-of-God'' effect is a non-linear phenomenon.

The model of linear redshift-space distortions has been developed by~\citet{Kaiser:1987qv} and~\citet{Hamilton:1992} assuming a plane-parallel approximation. In Fourier space the linear model can be written as 
\begin{equation}
P_g(k,\mu) = b^2(1 + \beta\mu^2)^2P_{\delta\delta}(k),
\end{equation}
where $P_{\delta\delta}(k)$ is the matter density power spectrum and $P_g(k,\mu)$ is the galaxy density power spectrum. In this model, linear redshift-space distortions are quantified by the parameter $\beta$, which is defined as
\begin{equation}
\beta = \frac{1}{b}\frac{d\ln D(z)}{d\ln(a)} \simeq \frac{\Omega_m^{\gamma}(z)}{b},
\end{equation} 
where $b$ is the linear galaxy bias factor, $D(z)$ is the growth factor and $\gamma$ is the gravitational growth index, which takes the value $\gamma=0.55$ in $\Lambda$CDM~\citep{Linder:2005in}.
$\Omega_m(z)$ is the matter density at redshift $z$, and is defined as 
\begin{equation}
\Omega_m(z) = \frac{H_0^2}{H(z)^2}\Omega_m(z=0)(1+z)^{3}
\end{equation}
with
\begin{equation}
\frac{H^2_0}{H(z)^2} = \left[\Omega_m(1+z)^{3}  + \Omega_{\Lambda}\right]^{-1},
\end{equation}
where we again follow our fiducial model of $\Omega^{\rm fid}_k = 0$ and $w^{\rm fid} = -1$. Modifications of the gravitational force mainly affect $\gamma$, while changes in the expansion history of the Universe affect $\Omega_m(z)$.

\subsection{Parameterisation}

From now on we will formulate our equations in terms of $g_{\theta}(z) = f(z)\sigma_8(z)$ and $g_b(z) = b\sigma_8(z)$. We choose the parameter set [$g_{\theta}$, $g_b$] instead of [$\beta$, $\sigma_8$, $b$], because $\sigma_8$ and the linear bias, $b$, are degenerate and difficult to disentangle~\citep{White:2008jy,Song:2008qt, Song:2010kq}. Within this parameterisation the power spectrum is expressed as 
\begin{equation}
P_g(k) = b^2 P_{\delta\delta}(k) = g^2_{b}Q_{\delta\delta}(k),
\end{equation}
where $Q_{\delta\delta}(k)$ is the unnormalised matter density power spectrum (see eq.~\ref{eq:D3}).  In terms of our new parameters, we can write
\begin{equation}
\beta = \frac{g_{\theta}}{g_{b}}.
\end{equation}

\subsection{Extensions to linear theory}
\label{sec:model2}

For the remainder of this section we will discuss possible extensions of the linear model to include non-linear structure formation and wide-angle effects. The simplest model of non-linearities is the so-called streaming model~\citep{Peebles:1980,Hatton:1997xs}. Here, the redshift-space correlation function is just a convolution of the linear correlation function in redshift-space with a pairwise velocity probability density function, $F(v)$,
\begin{equation}
\xi_{\rm st}(r_p,\pi) = \int^{\infty}_{-\infty} \xi\left(r_p, \pi - \frac{v}{H(z_{\rm eff})a_{\rm eff}}\right)F(v)dv,
\label{eq:stream}
\end{equation}
where $a_{\rm eff} = 1/(1+z_{\rm eff})$ is the scale factor and $H(z_{\rm eff})$ is the Hubble constant at the effective redshift. We chose $F(v)$ to be an exponential distribution~\citep{Peacock:1996ci}
\begin{equation}
F(v) = \frac{1}{\sigma_p\sqrt{2}}\exp\left[\frac{-\sqrt{2}|v|}{\sigma_p}\right],
\label{eq:fv}
\end{equation}
which has been shown to successfully describe observations~\citep{Davis:1982gc,Fisher:1993ye,Marzke:1995yk,Landy:2002xg}. Although the pairwise velocity dispersion within a halo is expected to follow a Gaussian distribution instead of an exponential, galaxies populate halos of a wide range of masses and velocity dispersions, which combine to approximately form an exponential function~\citep{Sheth:1995is,Diaferio:1996de,Seto:1997mv}. The parameter $\sigma_p$ depends on galaxy type (e.g.~\citealt{Madgwick:2003bd}) and hence its use for cosmological constraints is limited. 

In recent years many improvements to the model discussed above have been suggested. We will initially discuss these models in Fourier space because the theoretical motivation is clearer and many expressions needed for these models simplify considerably. However, since our data is in real space we will also give real-space expressions later on. We start with the streaming model of eq.~\ref{eq:stream}, which in Fourier space is given by
\begin{equation}
P_g(k,\mu) = b^2(1 + \beta\mu^2)^2P_{\delta\delta}(k)\frac{1}{1 + k^2\mu^2\sigma_p^2/2}.
\label{eq:nl1}
\end{equation}
Hence in Fourier space, the convolution with an exponential function becomes a multiplication by a Lorentzian distribution.

The model above assumes that there is a perfect correlation between the velocity field and the density field, which is given by $P_{\delta\delta}(k) = P_{\delta\theta}(k) = P_{\theta\theta}(k)$ where $P_{\delta\delta}(k)$ is the matter density power spectrum as before. $P_{\theta\theta}(k) = \langle|\theta_k|^2\rangle$ is the velocity divergence power spectrum where $\theta =  \vec{\nabla} \cdot \vec{v}$ is the velocity divergence, and $P_{\delta\theta}(k)$ is the cross power spectrum. Non-linear effects will violate these assumptions, since the density power spectrum is expected to increase in amplitude at small scales because of non-linear effects, while the velocity field becomes randomised at small scales (e.g. within virialized galaxy clusters) and hence $P_{\theta\theta}(k)$ will decrease in amplitude (e.g.~\citealt{Carlson:2009it}). \citet{Scoccimarro:2004tg} suggested expressing the 2D power spectrum without the assumption of linear relations between the density field and velocity field, by
\begin{equation}
\begin{split}
P_g(k,\mu) &= F_q(k,\mu,\sigma_v)\cr
&\times\left[b^2P_{\delta\delta}(k) + 2\mu^2bfP_{\delta\theta}(k) + \mu^4f^2P_{\theta\theta}(k)\right]\cr
&= F_q(k,\mu,\sigma_v)\cr
&\times\left[g_b^2Q_{\delta\delta}(k) + 2\mu^2g_bg_{\theta}Q_{\delta\theta}(k) + \mu^4g_{\theta}^2Q_{\theta\theta}(k)\right],
\end{split}
\label{eq:nl4}
\end{equation}
where the different $Q_{xy}$ are defined as 
\begin{equation}
\begin{split}
Q_{\delta\delta}(k) &= P_{\delta\delta}(k)/\sigma_8(z_{\rm eff})^2,\\
Q_{\delta\theta}(k) &= P_{\delta\theta}(k)/\sigma_8(z_{\rm eff})^2,\\
Q_{\theta\theta}(k) &= P_{\theta\theta}(k)/\sigma_8(z_{\rm eff})^2.
\end{split}
\label{eq:D3}
\end{equation}
and the damping function $F_q(k,\mu,\sigma_v)$ is usually chosen to be a Gaussian of the form 
\begin{equation}
F_q(k,\mu,\sigma_v) = e^{-(k\mu\sigma_v)^2}.
\end{equation}
The parameter $\sigma_v$ quantifies the non-linear dispersion in the bulk motion of halos. It is different to the $\sigma_p$ parameter we introduced earlier that describes small-scale randomised motion (e.g. of galaxies within a halo). We can derive $\sigma_v$ from the velocity power spectrum as
\begin{equation}
\sigma_v^2(z) = \frac{g_{\theta}(z)^2}{6\pi^2}\int^{\infty}_{0}Q_{\theta\theta}(k)dk.
\label{eq:sigv}
\end{equation}
\citet{Jennings:2010uv} provide fitting formulae for $P_{\delta\theta}(k)$ and $P_{\theta\theta}(k)$ derived from N-body simulations. They find the following relation between the different power spectra
\begin{equation}
P_{xy}(k) = \frac{\alpha_0\sqrt{P_{\delta\delta}(k)} + \alpha_1P^2_{\delta\delta}(k)}{\alpha_2 + \alpha_3P_{\delta\delta}(k)},
\label{eq:nl5}
\end{equation}
where $P_{\delta\delta}(k)$ can be obtained from CAMB by including halofit~\citep{Smith:2002dz}. For the cross power spectrum $P_{xy}(k) = P_{\delta\theta}(k)$ we use the (updated) parameters ($\alpha_0$, $\alpha_1$, $\alpha_2$, $\alpha_3$) = ($-12\,483.8$, $2.55430$, $1\,381.29$, $2.54069$) and for $P_{xy}(k) = P_{\theta\theta}(k)$ we use ($-12\,480.5$, $1.52404$, $2\,165.87$, $1.79640$). The fitting formula reproduces $P_{\theta\theta}$ to better than $1\%$ for $k<0.4h\,$Mpc$^{-1}$, to $10\%$ for $0.4 < k < 0.7h\,$Mpc$^{-1}$, and to $15\%$ for $0.7 < k < 1h\,$Mpc$^{-1}$. It also reproduces $P_{\delta\theta}$ to less than $4\%$ over the whole range $k < 1 h\,$Mpc$^{-1}$ (Jennings, private communication). We cut off the integral when the Jennings formula predicts negative values, although because of the high precision of the fitting formula up to large $k$, such a cut-off will not affect our measurement.

We can express~eq.~\ref{eq:nl4} in real-space as
\begin{equation}
\begin{split}
\xi_{\rm Sc}(r_p,\pi) &= \left[g_b^2\xi_{0,\delta\delta}(r) + \frac{2}{3}g_bg_{\theta}\xi_{0,\delta\theta}(r) + \frac{1}{5}g^2_{\theta}\xi_{0,\theta\theta}(r)\right]\mathcal{P}_0(\mu)\cr
&+\left[\frac{4}{3}g_bg_{\theta}\xi_{2,\delta\theta}(r) + \frac{4}{7}g^2_{\theta}\xi_{2,\theta\theta}(r)\right]\mathcal{P}_2(\mu)\cr
&+ \frac{8}{35}g^2_{\theta}\xi_{4,\theta\theta}(r)\mathcal{P}_4(\mu),
\end{split}
\label{eq:song}
\end{equation}
where $\mathcal{P}_{\ell}(\mu)$ are the Legendre polynomials and the spherical harmonic moments $\xi_{\ell,xy}(r)$ are given by
\begin{equation}
\begin{split}
\xi_{\ell,xy}(r) &= \int^{\infty}_{0}\int^1_{-1}\frac{k^2dkd\mu}{(2\pi)^2}e^{-(k\mu\sigma_v)^2}\cr
&\times \cos(kr\mu)Q_{xy}(k)\mathcal{P}_{\ell}(\mu).
\end{split}
\label{eq:xilm1}
\end{equation}
Appendix~\ref{sec:ana} shows a partial analytic solution for the double integral above.

We therefore have two different models which we will apply to our data: The simple streaming model $\xi_{\rm st}(r_p,\pi)$ and the Scoccimarro models $\xi_{\rm Sc}(r_p,\pi)$. Note that $\xi_{\rm Sc}(r_p,\pi)$ does not include the parameter $\sigma_p$ and hence has one less free parameter than the streaming model. 

All equations above are based on the plane-parallel approximation. In the case of 6dFGS we also need to account for wide-angle effects. This means we will replace the equations above with more general descriptions, which we discuss in the next section.

\subsection{Wide-angle formalism}
\label{sec:wide}

So far we have assumed that the separation between galaxy pairs is much smaller than the distance of the galaxies from the observer. The 6dF Galaxy Survey has a maximum opening angle of $180^{\circ}$ and the (effective) redshift is fairly low at $z_{\rm eff} = 0.067$ (see section~\ref{sec:data}). We therefore include wide-angle correction terms. The wide-angle description of redshift-space distortions has been laid out in several papers~\citep{Szalay:1997cc,Szapudi:2004gh,Matsubara:2004fr,Papai:2008bd,Raccanelli:2010hk}. Here we will expand on this work by formulating the equations in terms of $g_{\theta}$ and $g_{b}$ and by distinguishing between the density-density, velocity-velocity and density-velocity contributions ($\xi_{\delta\delta}, \xi_{\theta\theta}$ and $\xi_{\delta\theta}$). The model will then correspond to the Scoccimarro model (eq.~\ref{eq:song}) in the last section.

The general redshift-space correlation function (ignoring the plane-parallel approximation) depends on $\phi$, $\theta$ and $s$. Here, $s$ is the separation between the galaxy pair, $\theta$ is the half opening angle, and $\phi$ is the angle of $s$ to the line-of-sight (see Figure~$1$ in~\citealt{Raccanelli:2010hk}). The angles $\phi$ and $\theta$ are not independent, but the relation between them is usually expressed through the two angles $\phi_1$ and $\phi_2$ given by $\phi = \frac{1}{2}(\phi_1 + \phi_2)$ and $\theta = \frac{1}{2}(\phi_1 - \phi_2)$. The total correlation function model, including $O(\theta^2)$ correction terms, is then given by~\citep{Papai:2008bd}
\begin{equation}
\begin{split}
\xi(\phi,\theta,s) &= a_{00} + 2a_{02}\cos(2\phi) + a_{22}\cos(2\phi) + b_{22}\sin^2(2\phi)\\
& +\Big[ - 4a_{02}\cos(2\phi) - 4a_{22} - 4b_{22} - 4a_{10}\cot^2(\phi)\\
& + 4a_{11}\cot^2(\phi) - 4a_{12}\cot^2(\phi)\cos(2\phi) + 4b_{11}\\
& - 8b_{12}\cos^2(\phi)\Big]\theta^2 + O(\theta^4).
\end{split} 
\label{eq:wide10}
\end{equation}
This equation reduces to the plane-parallel approximation if $\theta = 0$. The factors $a_{xy}$ and $b_{xy}$ in this equation are given by
\begin{equation}
\begin{split}
a_{00} &= g_b^2\xi_{0,\delta\delta}^2(r) + \frac{2g_bg_{\theta}}{3}\xi_{0,\delta\theta}^2(r) + \frac{2g_{\theta}^2}{15}\xi_{0,\theta\theta}^2(r)\\
& - \frac{g_bg_{\theta}}{3}\xi^2_{2,\delta\theta}(r) + \frac{2g_{\theta}^2}{21}\xi^2_{2,\theta\theta}(r) + \frac{3g_{\theta}^2}{140}\xi^2_{4,\theta\theta}(r)\\
a_{02} &= -\frac{g_bg_{\theta}}{2}\xi^2_{2,\delta\theta}(r) + \frac{3g_{\theta}^2}{14}\xi^2_{2,\theta\theta}(r) + \frac{g_{\theta}^2}{28}\xi^2_{4,\theta\theta}(r)\\
a_{22} &= \frac{g_{\theta}^2}{15}\xi^2_{0,\theta\theta}(r) - \frac{g_{\theta}^2}{21}\xi^2_{2,\theta\theta}(r) + \frac{19g_{\theta}^2}{140}\xi^2_{4,\theta\theta}(r)\\
b_{22} &= \frac{g_{\theta}^2}{15}\xi_{0,\theta\theta}^2(r) - \frac{g_{\theta}^2}{21}\xi^2_{2,\theta\theta}(r) - \frac{4g_{\theta}^2}{35}\xi^2_{4,\theta\theta}(r)\\
a_{10} &= \left[2g_bg_{\theta}\xi^1_{1,\delta\theta}(r) + \frac{4g_{\theta}^2}{5}\xi^1_{1,\theta\theta}(r)\right]\frac{1}{r} - \frac{g_{\theta}^2}{5r}\xi^1_{3,\theta\theta}(r)\\
a_{11} &= \frac{4g_{\theta}^2}{3r^2}\left[\xi^0_{0,\theta\theta}(r) - 2\xi^0_{2,\theta\theta}(r)\right]\\
a_{12} &= \frac{g_{\theta}^2}{5r}\left[2\xi^1_{1,\theta\theta}(r) -3\xi^1_{3,\theta\theta}(r)\right]\\
b_{11} &= \frac{4g_{\theta}^2}{3r^2}\left[\xi^0_{0,\theta\theta}(r) + \xi^0_{2,\theta\theta}(r)\right]\\
b_{12} &= \frac{2g_{\theta}^2}{5r}\left[\xi^1_{1,\theta\theta}(r) + \xi^1_{3,\theta\theta}(r)\right] ,
\end{split}
\label{eq:co}
\end{equation}
where the spherical harmonic moments, $\xi^m_{\ell,xy}(r)$, are
\begin{equation}
\begin{split}
\xi^m_{\ell,xy}(r) &= \int^{\infty}_{0}\int^1_{-1}\frac{k^mdkd\mu}{(2\pi)^2}e^{-(k\mu\sigma_v)^2}\cr
&\times \cos(kr\mu)Q_{xy}(k)\mathcal{P}_{\ell}(\mu),
\end{split}
\label{eq:xilm2}
\end{equation}
with $Q_{\delta\delta}(k)$, $Q_{\delta\theta}(k)$ and $Q_{\theta\theta}(k)$ as defined in equation~\ref{eq:D3} (see appendix~\ref{sec:ana} for an analytic solution).

In order to obtain a model for the 2D correlation function (including wide-angle effects), we can simply integrate eq.~\ref{eq:wide10} over $\theta$. In case of the plane-parallel approximation this would correspond to eq.~\ref{eq:song}. To reduce the equation to the simple streaming model we have to set $\sigma_v = 0$, $Q_{\delta\delta}(k) = Q_{\delta\theta}(k) = Q_{\theta\theta}(k)$ and convolve with $F(v)$.

\citet{Samushia:2011cs} studied wide angle effects in the SDSS-LRG sample and found that such effects are very small. The lower redshift and larger sky coverage of the 6dF Galaxy Survey mean that wide-angle effects are certainly larger in our data set than SDSS-LRG. However, we found no significant impact of such effects on our results. This is mainly because our analysis includes only rather small scales ($\pi < 30h^{-1}\,$Mpc, $r_p < 30h^{-1}\,$Mpc, see section~\ref{sec:fit}). This agrees with the findings of~\citet{Beutler:2011hx}.

We stress that the correction terms discussed above only capture first-order effects and so we additionally restrict our analysis to $\theta < 50^{\circ}$.

Many authors have found, using N-body simulations, that both the simple streaming and Scoccimarro models do not capture all non-linear and quasi-linear effects present in redshift surveys (e.g.~\citealt{Jennings:2010ne,Kwan:2011hr,Torre:2012} and references therein). For example the linear bias model seems to be too simplistic at small scales. In this study we restrict ourselves to the models discussed above but would like to point out that many other models have been suggested~\citep{Matsubara:2007wj,Matsubara:2008wx,Taruya:2010mx,Reid:2011ar,Seljak:2011tx} which could be compared to our data in future work.

\subsection{Systematics and the Alcock-Paczynski effect}
\label{sec:AP}

A dataset such as 6dFGS, containing galaxies with a high linear bias factor, may be prone to scale-dependent galaxy bias on small scales. We used the GiggleZ simulations~(Poole et al., in preparation) to derive the form of the scale-dependent bias. We find a $1.7\%$ correction to a linear bias at scales of $s = 10h^{-1}\,$Mpc, and note that this correction has a small but non-negligible impact on our results, depending on the smallest scales included in the fit.

A further systematic uncertainty comes from the so-called Alcock-Paczynski (AP) effect~\citep{Alcock:1979mp}. By assuming a cosmological model for the conversion of redshifts and angles to distances (see section~\ref{sec:data}) we could introduce an additional anisotropic signal in the correlation function which depends on the angle to the line-of-sight, if the true cosmology differs from our fiducial cosmological model. These distortions are degenerate with the linear redshift-space distortion signal and hence both effects need to be modelled to avoid systematic bias~\citep{Ballinger:1996cd,Simpson:2009zj}. Using the two scaling factors 
\begin{align}
f_{\parallel} &= \frac{H^{\rm fid}(z)}{H(z)}\\
f_{\perp} &= \frac{D_A(z)}{D^{\rm fid}_A(z)},
\end{align}
we can account for the AP effect by rescaling the corresponding axis
\begin{align}
\pi' &= \frac{\pi}{f_{\parallel}}\\
r_p' &= \frac{r_p}{f_{\perp}}.
\end{align}
While this is a matter of concern for high-redshift surveys, the 6dF Galaxy Survey is very nearly independent of the Alcock-Paczynski distortions because the distances $\pi$ and $r_p$ are almost independent of the fiducial cosmological model when expressed in units of $h^{-1}$Mpc. For example the value of $f_{\perp}$ at $z = 0.067$ changes by only ($0.3\%$, $1.5\%$) for $10\%$ changes in ($\Omega_m$, $w$). The corresponding values for $f_{\parallel}$ are ($0.7\%$, $1.6\%$). The relative tangential/ radial distortion depends on the value of $(1+z)D_A(z)H(z)/c$ relative to the fiducial model:
\begin{equation}
\frac{D_A(z)H(z)}{D_A^{\rm fid}(z)H^{\rm fid}(z)} = \frac{f_{\perp}}{f_{\parallel}}.
\end{equation}
This parameter combination changes by only $0.3$ ($0.1$)\% for $10\%$ changes in $\Omega_m$ ($w$).

Finally, it has been shown that $\gamma$ has a degeneracy with the equation of state parameter for dark energy, $w$~\citep{Simpson:2009zj}. High redshift measurements usually assume $w=-1$, which could introduce a bias in the measured value of $\gamma$ if dark energy is not exactly a cosmological constant. Again 6dFGS is robust against such effects.

\section{Fitting the 2D correlation function}
\label{sec:fit}

\begin{table}[tb]
\begin{center}
\caption{Cosmological parameters derived from the 6dFGS 2D correlation function. The effective redshift is $z_{\rm eff} = 0.067$. The last column indicates the priors/assumptions which go into each individual parameter measurement. The prior on the Hubble constant comes from~{\protect \citet{Riess:2011yx}} and the WMAP7 prior from~{\protect \citet{Komatsu:2010fb}}. The asterisks denote parameters which are derived from fitting parameters.}
\vspace{0.4cm}
\begin{tabular}{lll}
\hline
\multicolumn{3}{c}{Summary of parameter measurements from 6dFGS}\\
\hline
$g_{\theta}(z_{\rm eff})$ & $0.423\pm0.055$ & \\
$g_b(z_{\rm eff})$ & $1.134\pm0.073$ & \\
$\beta^*$ & $0.373\pm 0.054$ & \\
\hline
$\sigma_8$ & $0.76\pm 0.11$ & $\left[H_0=73.8\pm2.4, \gamma = 0.55\right]$\\
$\Omega_m$ & $0.250\pm0.022$ & $\left[H_0=73.8\pm2.4, \gamma = 0.55\right]$\\
b & $1.48\pm0.27$ & $\left[H_0=73.8\pm2.4, \gamma = 0.55\right]$\\
$f^*(z_{\rm eff})$ & $0.58\pm0.11$ & $\left[g_{\theta}+\sigma_8 \text{ from 6dFGS}\right]$\\
\hline
$\gamma$ & $0.547\pm 0.088$ & $\left[ \rm WMAP7\right]$\\
$\Omega_m$ & $0.271\pm0.027$ & $\left[ \rm WMAP7\right]$\\
\hline
\end{tabular}
\label{tab:para2}
\end{center}
\end{table}

We now fit the two models for the 2D correlation function we developed earlier ($\xi_{\rm st}(r_p,\pi)$ and $\xi_{\rm Sc}(r_p,\pi)$) to our data. For the final result we use the $\xi_{\rm Sc}(r_p,\pi)$ model since it gives similar measurements to the streaming model with one less free parameter.

Other studies (e.g.~\citealt{Samushia:2011cs}) prefer to analyse the correlation function moments $\xi_0, \xi_2$ (and if possible $\xi_4$), which carry the same information as the 2D correlation function. The correlation function moments have the advantage that the number of bins grows linearly with the highest scales analysed, while for $\xi(r_p,\pi)$ the number of bins grows quadratically. This makes it easier to get reliable covariance matrices. \citet{Samushia:2011cs} also show that the measurement errors of $\xi_{\ell}$ are more Gaussian. However, the correlation function moments are integrals over $\mu$ and hence carry information from all directions, including $\mu = 0$. Finger-of-God distortions can influence the correlation function moments up to large scales ($20$ - $30h^{-1}\,$Mpc), while in the 2D correlation function they can be excluded via a cut in $r_p$. In 6dFGS we have found that these non-linear effects have a strong impact on the correlation function moments up to $30h^{-1}\,$Mpc. We have therefore decided to focus on the 2D correlation function instead of the correlation function moments.

\subsection{Derivation of the growth rate, $g_{\theta} = f\sigma_8$}

\begin{figure}[tb]
\begin{center}
   \epsfig{file=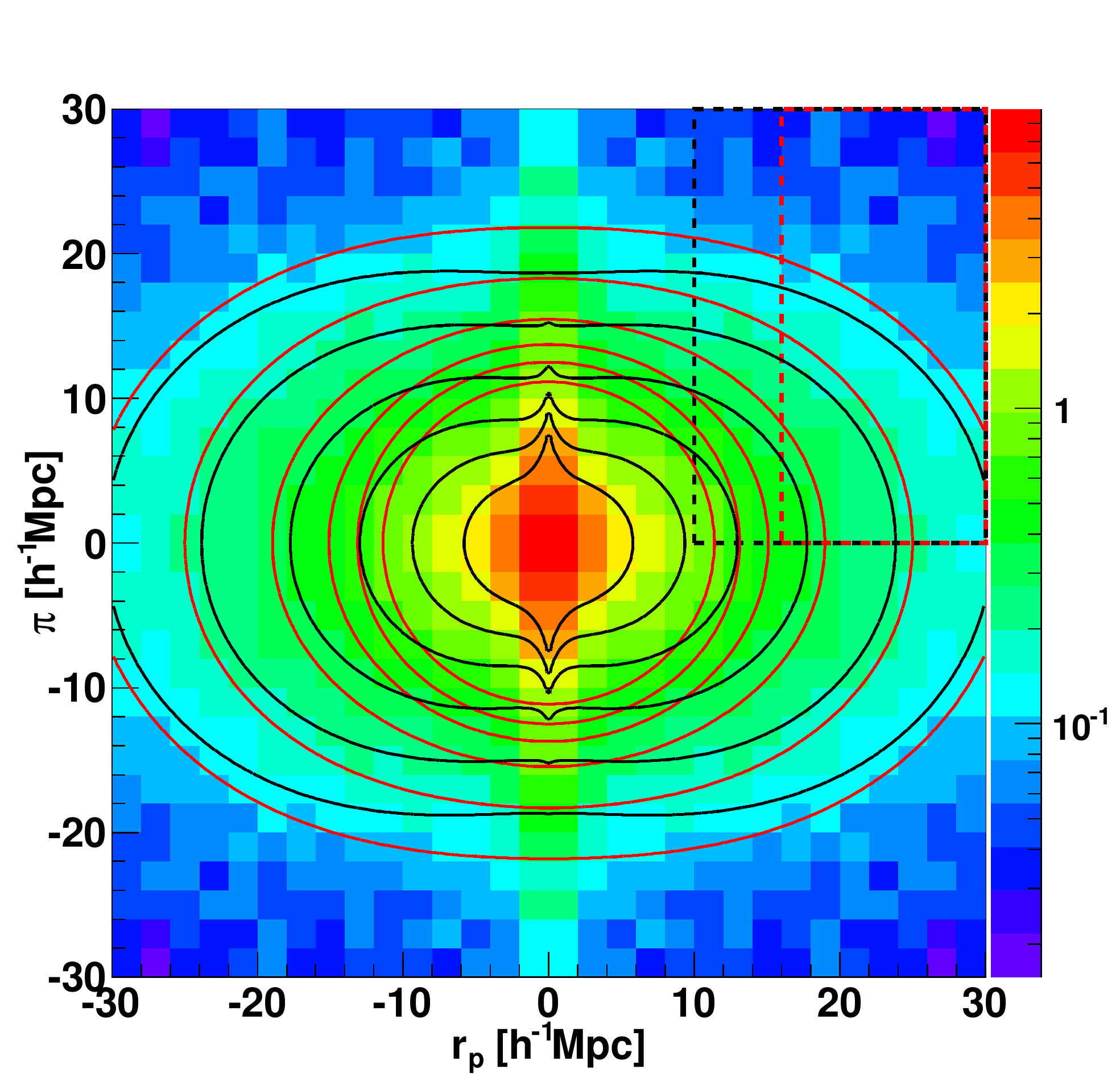,width=9cm}
\caption{The 2D correlation function in $2h^{-1}\,$Mpc bins. The fitting area is indicated by the dashed lines, where black corresponds to the streaming model, ($\xi_{\rm st}(r_p,\pi)$) and red corresponds to the Scoccimarro model ($\xi_{\rm Sc}(r_p,\pi)$). The black and red contours show the best fitting models for $\xi_{\rm st}(r_p,\pi)$ (black) and $\xi_{\rm Sc}(r_p,\pi)$ (red). The deviations seen in the two contours at large scales are well within the error bars of the two models, which can be seen in Figure~\ref{fig:chi2_1}. At small scales ($< 14h^{-1}\,$Mpc) the Scoccimarro model predicts much more clustering, while in the real data this clustering is smeared out along the line of sight because of the finger-of-God effect.}
   \label{fig:kaiser_models}
\end{center}
\end{figure}

\begin{figure}[tb]
\begin{center}
   \epsfig{file=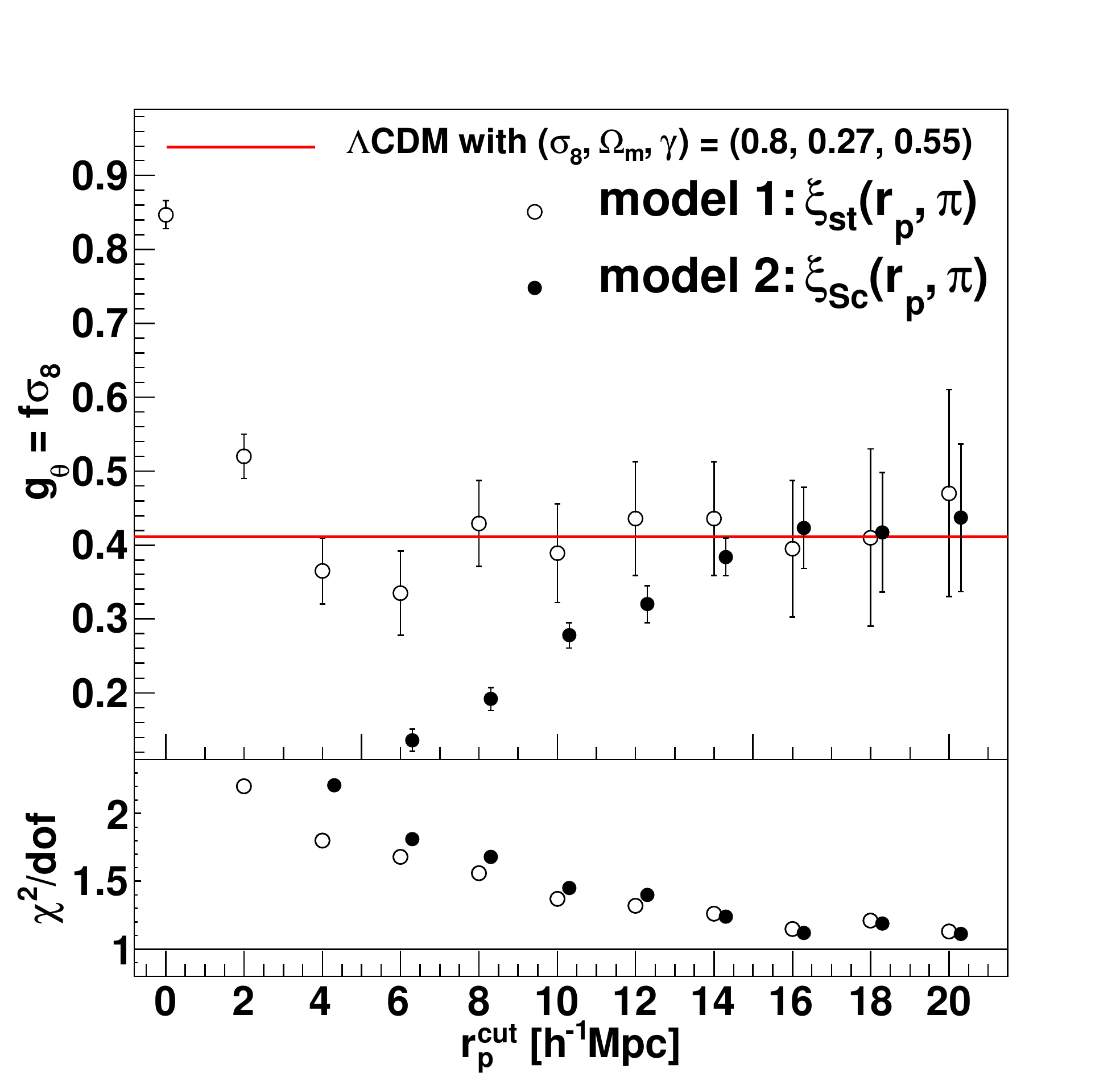,width=9cm}
\caption{The value of $g_{\theta}(z_{\rm eff}) = f(z_{\rm eff})\sigma_8(z_{\rm eff})$ as a function of the cut-off scale $r_p^{\rm cut}$, obtained by fitting the 6dFGS 2D correlation function with two different models (as described in section~\ref{sec:model2} and~\ref{sec:wide}). At large scales the two models converge to similar values, while on small scales the models deviate from each other because of the different descriptions of non-linear evolution. For the final parameter measurements in Table~{\protect \ref{tab:para2}} we chose model 2, $\xi_{\rm Sc}(r_p,\pi)$, with a conservative cut-off scale of $r_p^{\rm cut} = 16h^{-1}\,$Mpc. In the lower panel we plot the reduced $\chi^2$ as an indicator of the quality of the fit.}
   \label{fig:can}
\end{center}
\end{figure}

In Figure~\ref{fig:kaiser_models} we show the 6dFGS 2D correlation function. For our analysis we bin the data in $2\times 2h^{-1}\,$Mpc bins from $0$ to $30h^{-1}\,$Mpc in $r_p$ and $\pi$. Including larger scales does not add further information. At small $r_p$, the finger-of-God effect becomes dominant and we expect any linear model to fail. Since our description of non-linearities, in both of our models, is limited in its capability to capture all non-linear effects, it is necessary to include a cut-off scale $r^{\rm cut}_p$ marking a lower limit of the fitting range in $r_p$. 

Figure~\ref{fig:can} shows the measured value of $g_{\theta}$ as a function of the cut-off scale $r_p^{\rm cut}$ for our two different models. Above $r_p^{\rm cut} \approx 8h^{-1}\,$Mpc the streaming model, $\xi_{\rm st}(r_p,\pi)$, approaches a constant value of $g_{\theta}$. Our second model, $\xi_{\rm Sc}(r_p,\pi)$, contains a systematic error up to much larger scales, before it comes into agreement with the streaming model at about $r_p^{\rm cut} = 16h^{-1}\,$Mpc. This is expected since this model does not include a description of effects in the non-linear regime. For the final constraints we choose $r_p^{\rm cut} = 10h^{-1}\,$Mpc for the streaming model and $r_p^{\rm cut} = 16h^{-1}\,$Mpc for the Scoccimarro model. We also note that since the Scoccimarro model is based on only two free parameters ($g_{\theta}$ and $g_{b}$), the error is generally smaller compared to the streaming model, which has three free parameters ($g_{\theta}$, $g_b$ and $\sigma_p$). Other studies fit for the parameter $\sigma_v$ (e.g.~\citealt{Torre:2012}) in the Scoccimarro model, but we derive it using eq.~\ref{eq:sigv}.

For $\xi_{\rm Sc}(r_p,\pi)$ we use the fitting range $0 < \pi < 30h^{-1}\,$Mpc and $16 < r_p < 30h^{-1}\,$Mpc, which results in a total of $105$ bins. The best-fitting results are $g_{\theta} = 0.423\pm 0.055$ and $g_b = 1.134\pm 0.073$, where the errors for each parameter are derived by marginalising over all other parameters. The $\chi^2$ of this fit is $115$ with $103$ degrees of freedom (d.o.f.), indicating a good fit to the data. 

\begin{figure}[tb]
\begin{center}
   \epsfig{file=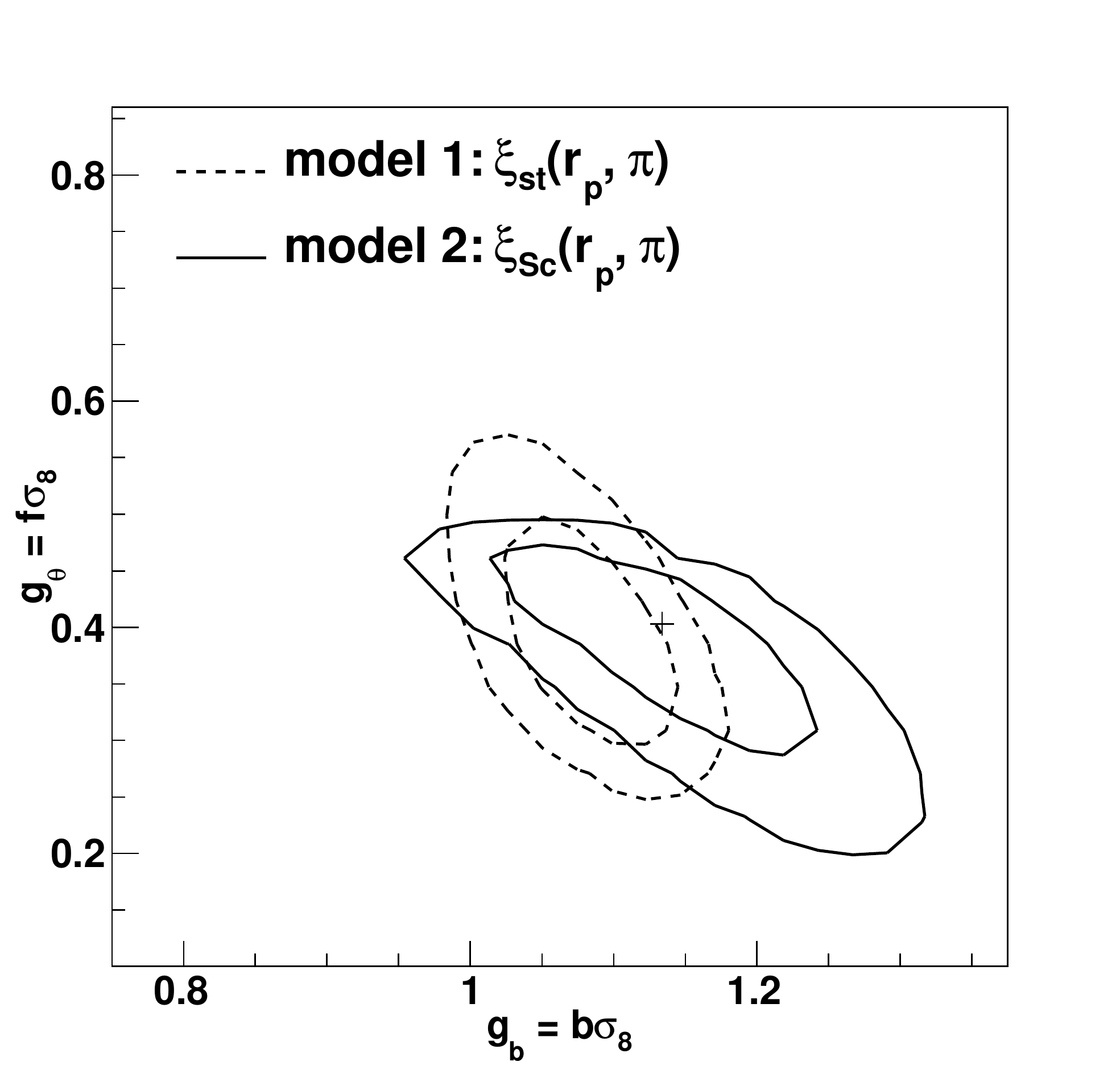,width=10cm}
\caption{Likelihood distribution of $g_{\theta}$ and $g_b$ derived from the fit to the 2D correlation function. The solid black contours show model $\xi_{\rm Sc}(r_p,\pi)$, while the dashed contours show the streaming model (see section~\ref{sec:model2} and~\ref{sec:wide} for details of the modelling). The fitting range is $0 < \pi < 30h^{-1}\,$Mpc and $10 < r_p < 30h^{-1}\,$Mpc for $\xi_{\rm st}(r_p,\pi)$ and $0 < \pi < 30h^{-1}\,$Mpc and $16 < r_p < 30h^{-1}\,$Mpc for $\xi_{\rm Sc}(r_p,\pi)$. The black cross indicates the best-fitting value for the solid black contours.}
   \label{fig:chi2_1}
\end{center}
\end{figure}

For $\xi_{\rm st}(r_p,\pi)$, we have the fitting range $0 < \pi < 30h^{-1}\,$Mpc and $10 < r_p < 30h^{-1}\,$Mpc, which results in a total of $150$ bins. The best fitting parameters are $g_{\theta} = 0.389\pm 0.067$, $g_b = 1.084\pm 0.036$ and $\sigma_p = 198\pm81\,$km/s. The reduced $\chi^2$ of this fit is given by $\chi^2/\rm d.o.f. = 202/147 = 1.37$. We compare the constraints on $g_{\theta}$ and $g_b$ from both models in Figure~\ref{fig:chi2_1}.

In the Scoccimarro model we could use the parameter $\sigma_v \propto g_{\theta}$ instead of $g_{\theta}$ to test cosmology, as suggested by~\citet{Song:2010kq}. Our best fit gave $\sigma_v = 2.59\pm0.34h^{-1}\,$Mpc. However, this parameter depends on an additional integral over the velocity power spectrum, which adds a theoretical uncertainty. We therefore prefer to use $g_{\theta}$ in the following discussions. 

We can also express our results in terms of $\beta$ which is given by $\beta = g_{\theta}/g_b = 0.373\pm 0.054$. We summarise all measured and derived parameters in Table~\ref{tab:para2}.

\subsection{Derivation of $\sigma_8$ and $\Omega_m$}
\label{sec:sig8}

\begin{figure}[tb]
\begin{center}
\epsfig{file=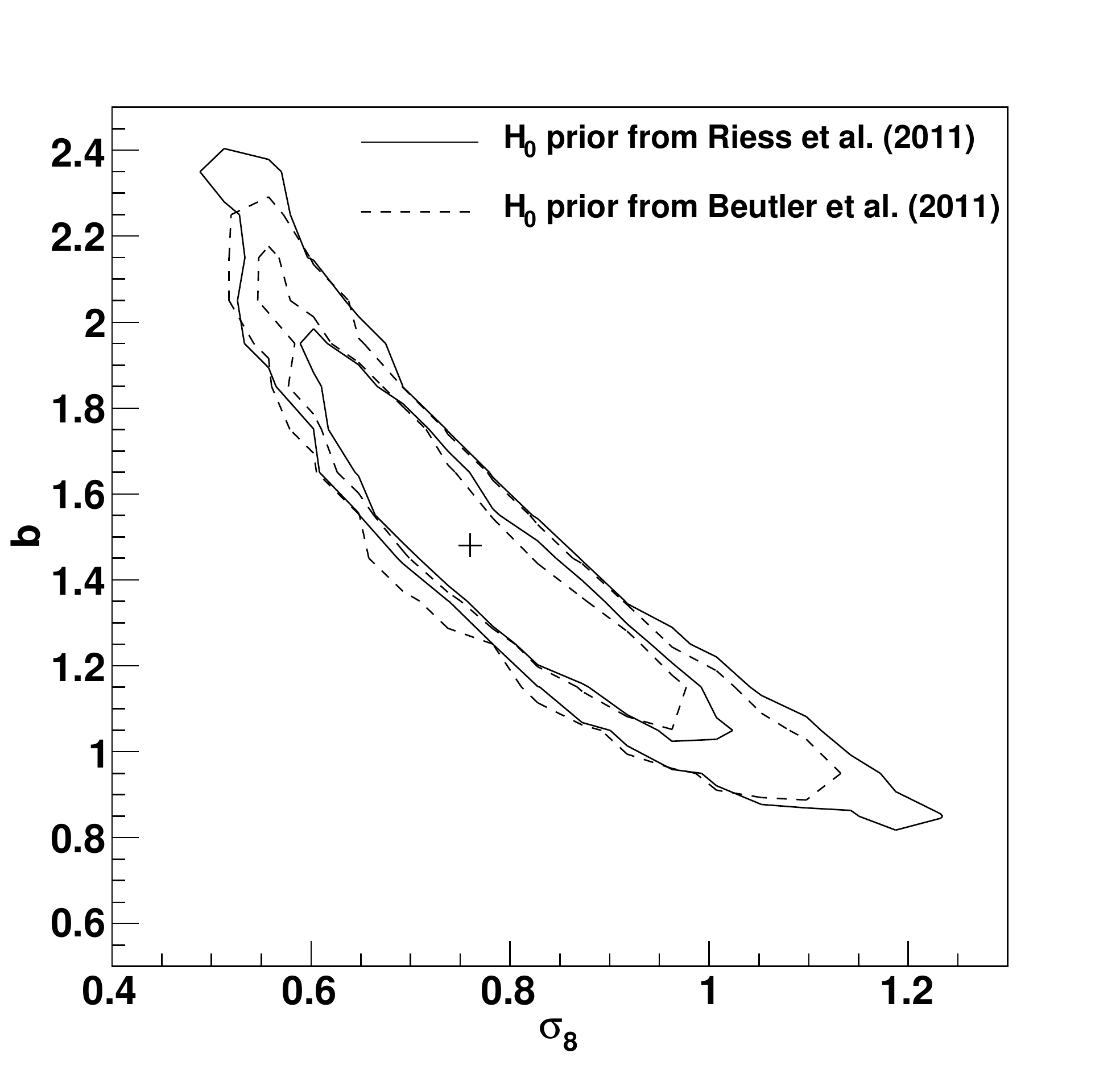,width=9cm}
\caption{This plot shows the likelihood distribution of the galaxy bias $b$ and $\sigma_8$, which we obtained by fitting the 6dFGS 2D correlation function assuming $\gamma = 0.55$. The solid black line shows the result using a prior on the Hubble constant of $H_0 = 73.8\pm2.4\,$km\,s$^{-1}$\,Mpc$^{-1}$ from~\citet{Riess:2011yx}, while the dashed black line uses a prior of $H_0 = 67.0\pm 3.2\,$km\,s$^{-1}$\,Mpc$^{-1}$ from~\citet{Beutler:2011hx}. Although the detection of redshift-space distortions can partially break the degeneracy between $b$ and $\sigma_8$ which exists in the 1D correlation function, there is still a significant residual degeneracy. The black cross marks the maximum likelihood value for the solid black lines.}
\label{fig:chi2_sigma8_b}
\end{center}
\end{figure}

In this section we use redshift-space distortions to directly measure $\sigma_8$. The angular dependence of the redshift-space distortion signal in the 2D correlation function allows us to measure $\beta$, which quantifies the amplitude of redshift-space distortions. Together with $\Omega_m(z)$ and $\gamma = 0.55$, this constrains the linear bias $b$ through the equation
\begin{equation}
b \simeq \frac{\Omega_m^{\gamma}(z)}{\beta}.
\end{equation}
Knowing $b$ we can use the absolute amplitude of the correlation function, $\left[b\sigma_8(z)\right]^2$, to constrain $\sigma_8(z{=}0) = [D(z{=}0)/D(z_{\rm eff})] \times \sigma_8(z_{\rm eff})$. 

For computational reasons we use our first model, $\xi_{\rm st}(r_p,\pi)$, in this sub-section and fit the five parameters $\sigma_8$, $\Omega_m$, $b$, $H_0$ and $\sigma_p$ using an MCMC approach. Since the shape of the correlation function is only sensitive to $\Gamma = \Omega_mh$, we cannot constrain $\Omega_m$ and $H_0$ at the same time. For the final results we include a prior on the Hubble constant ($H_0 = 73.8\pm2.4\,$km\,s$^{-1}$\,Mpc$^{-1}$,~\citealt{Riess:2011yx}, from now on referred to as HST prior) and marginalise over it. We use the same binning and fitting ranges as in the previous section.

The best-fitting model results in $\chi^2/\rm d.o.f = 1.35$. We find $\sigma_8 = 0.76\pm0.11$, $\Omega_m = 0.250\pm 0.022$, $b = 1.48\pm0.27$ and $\sigma_p = 174\pm 73\,$km/s. The remaining degeneracy between the bias $b$ and $\sigma_8$ is illustrated in Figure~\ref{fig:chi2_sigma8_b}. We include all these results in Table~\ref{tab:para2}.

\begin{figure}[tb]
\begin{center}
\epsfig{file=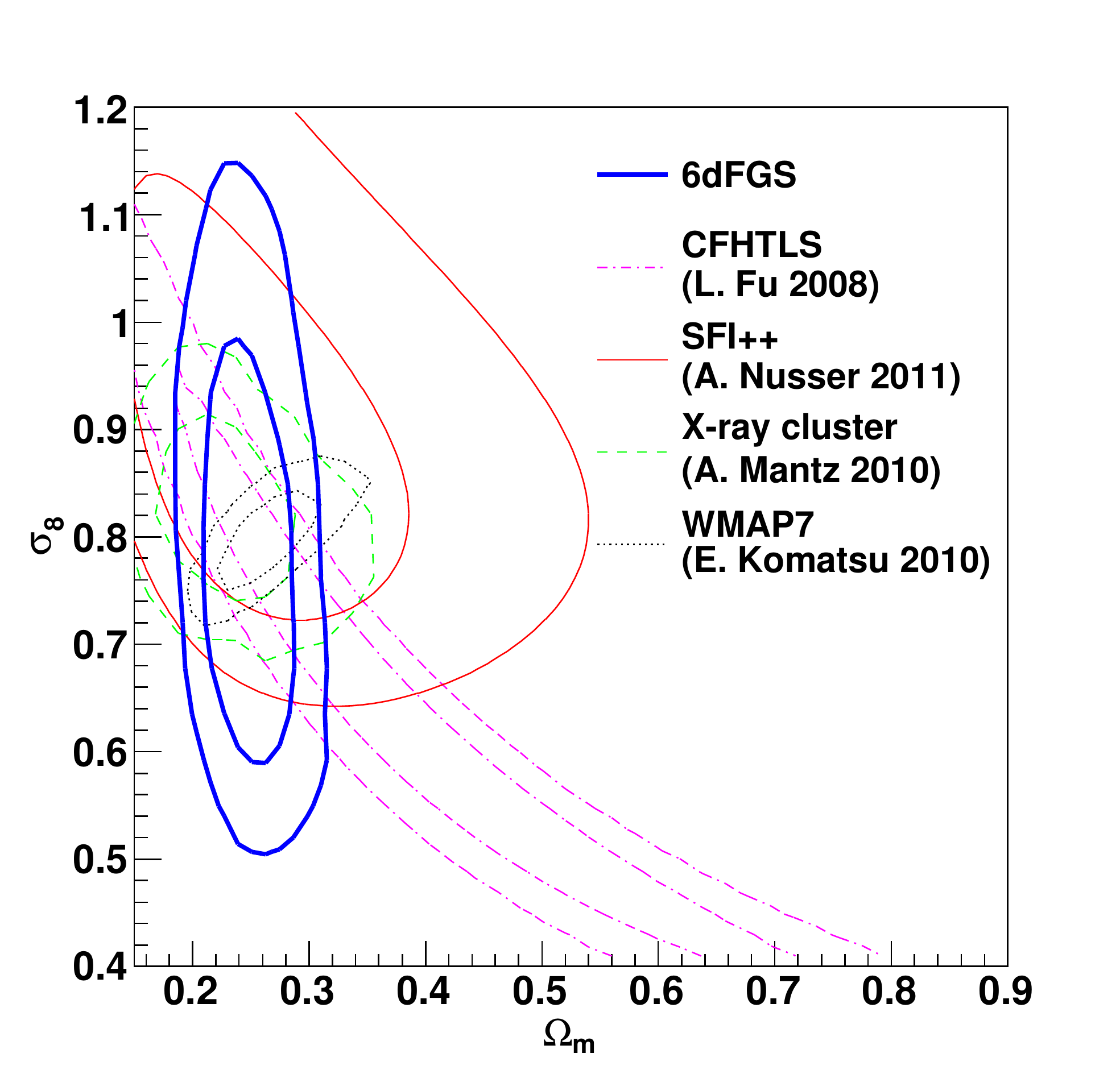,width=10cm}
\caption{This plot shows the likelihood distribution in $\sigma_8$ and $\Omega_m$ for different cosmological probes. The solid blue contours show the 6dFGS result, the magenta dotted dashed contours shows the currently best result of weak lensing from the CFHT wide synoptic Legacy Survey~\citep{Fu:2007qq}, the red solid contours show the result of the SFI++ peculiar velocity survey~\citep{Nusser:2011tu}, the green dashed contours show the result of~\citet{Mantz:2009fw} using cluster abundances and the black dotted contours are from WMAP7~\citep{Komatsu:2010fb}.}
\label{fig:chi2_om_sigma8}
\end{center}
\end{figure}

\begin{figure}[tb]
\begin{center}
\epsfig{file=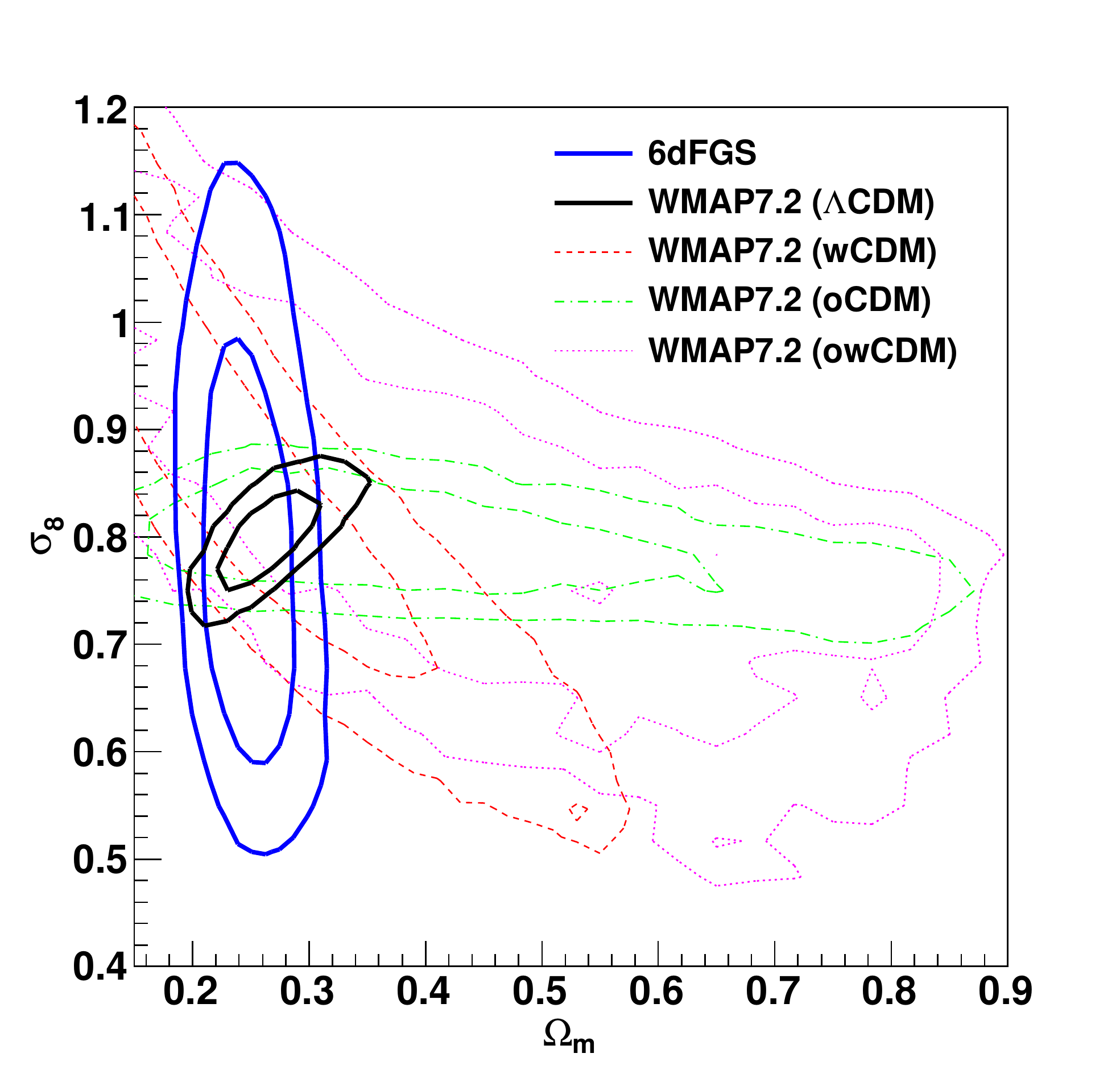,width=10cm}
\caption{Likelihood distribution of $\sigma_8-\Omega_m$ from WMAP7 for a $\Lambda$CDM model (solid black line), $w$CDM model (dashed red line), oCDM model (dotted dashed green line) and o$w$CDM model (dotted magenta line). In blue we show the 6dFGS result. Both parameters are defined at redshift zero and WMAP constraints on these parameters depend on assumptions about the expansion history of the Universe. We use CosmoMC~\citep{Lewis:2002ah}, together with the WMAP7.2~\citep{Komatsu:2010fb} dataset to produce these likelihood distributions.}
\label{fig:WMAP_om_sigma8}
\end{center}
\end{figure}

Figure~\ref{fig:chi2_om_sigma8} compares the 6dFGS $\Omega_m-\sigma_8$ probability distribution to measurements from several other datasets: The CFHT wide synoptic Legacy Survey (CFHTLS)~\citep{Fu:2007qq}, the SFI++ peculiar velocity survey~\citep{Nusser:2011tu}, cluster abundance from X-ray surveys~\citep{Mantz:2009fw} and WMAP7~\citep{Komatsu:2010fb}. Many of the experiments shown in this figure have systematic modelling uncertainties when extrapolating to $z = 0$ arising from assumptions about the expansion history of the Universe. Only 6dFGS and SFI++ are at sufficiently low redshift to be independent of such effects. To illustrate the impact of these effects we plot in Figure~\ref{fig:WMAP_om_sigma8} the probability distribution $\Omega_m-\sigma_8$ from WMAP7 for different cosmological models. The CMB measures the scalar amplitude $A_s$, which needs to be extrapolated from redshift $z_* \approx 1100$ to redshift zero to obtain $\sigma_8$. Every parameter that influences the expansion history of the Universe in this period affects the value of $\sigma_8$ derived from the CMB alone. The relation between $A_s$ and $\sigma_8$ is given by (e.g.~\citealt{Takada:2005si})
\begin{align}
\sigma_8^2(z) &= A_s\left(\frac{2c^2}{5\Omega_m H_0^2}\right)^2 \int^{\infty}_{0}k^3dkD^2(k,z)T^2(k)\left(\frac{k}{k_{*}}\right)^{n_s-1}\cr
&\times \left[\frac{3\sin(kR)}{(kR)^3} - \frac{3\cos(kR)}{(kR)^2}\right]^2,
\label{eq:sig8z}
\end{align}
where $R = 8h^{-1}\,$Mpc, $k_* = 0.02\,$Mpc$^{-1}$ and $A_s = (2.21\pm0.09) \times 10^{-9}$~\citep{Komatsu:2008hk}. $D(k,z)$ is the growth factor at redshift $z$ and $T(k)$ is the transfer function. If higher-order cosmological parameters such as the dark energy equation of state parameter $w$ are marginalised over, then the measurements of $\Omega_m$ and $\sigma_8$ weaken considerably (see Figure~\ref{fig:WMAP_om_sigma8}).

We now assess the influence of the $H_0$ prior. We replace the result of~\citet{Riess:2011yx} with a measurement derived from the 6dFGS dataset using Baryon Acoustic Oscillations~\citep{Beutler:2011hx}. The prior from this study is lower than the former value and is given by $H_0 = 67.0\pm 3.2\,$km\,s$^{-1}$\,Mpc$^{-1}$. Using the 6dFGS value of $H_0$ results in $\sigma_8 = 0.75\pm0.13$, $\Omega_m = 0.279\pm 0.028$, $b = 1.52\pm0.29$ and $\sigma_p = 174\pm 106\,$km/s. The quality of the fit is $\chi^2/\rm d.o.f. = 1.35$, very similar to the value obtained with the HST prior. Comparing the two results shows that a different prior in $H_0$ shifts the constraint in $\sigma_8$ and $b$ along the degeneracy shown in Figure~\ref{fig:chi2_sigma8_b}. However, we note that the 6dFGS measurement of $H_0$ is derived from the same dataset as our present study and hence could be correlated with our measured growth rate. We use these results only for comparison, and include the values obtained using the HST prior in Table~\ref{tab:para2} as our final results of this section.

Alternative methods for deriving $\sigma_8$ or $f\sigma_8$ at low redshift are provided by peculiar velocity surveys (e.g.~\citealt{Gordon:2007zw,Abate:2009kd,Nusser:2011tu,Turnbull:2011ty,Davis:2010sw,Hudson:2012gt}). 6dFGS will soon provide its own peculiar velocity survey of around $10\,000$ galaxies. Velocity surveys have the advantage of tracing the matter density field directly, without the complication of a galaxy bias. However, they are much harder to obtain and current velocity surveys are $1$ - $2$ orders of magnitude smaller than galaxy redshift surveys. 

\section{Cosmological implications}
\label{sec:impl}

\begin{figure}[tb]
\begin{center}
\epsfig{file=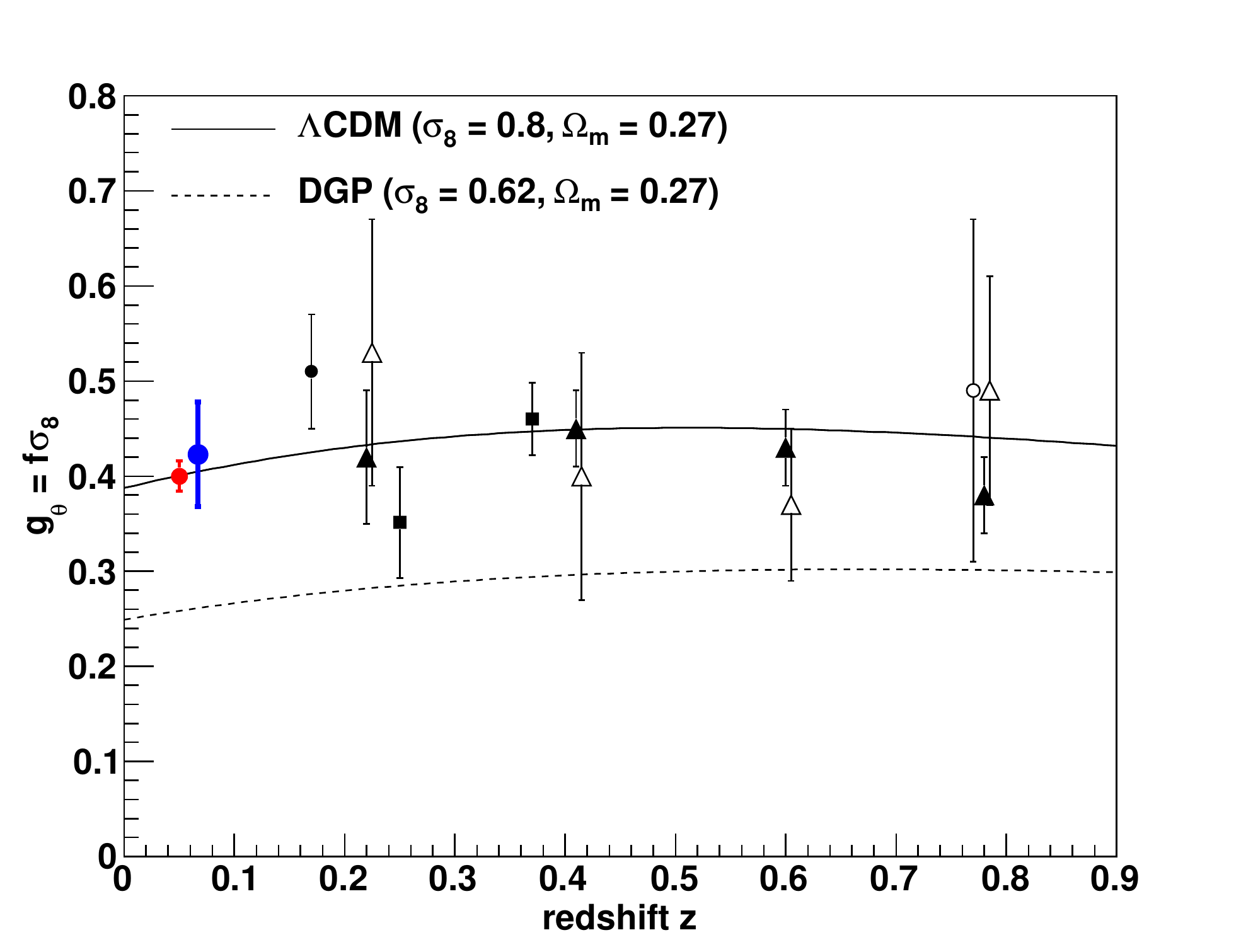,width=10cm}
\caption{Comparison of measurements of the growth of structure using galaxy surveys at different redshifts. The different data points belong to 6dFGS (solid blue circle, this paper), 2dFGRS (solid black circle;~\citealt{Hawkins:2002sg}), SDSS (solid black boxes;~\citealt{Samushia:2011cs}), WiggleZ (solid black triangles;~\citealt{Blake:2011rj}) and VVDS (empty circle;~\citealt{Guzzo:2008ac}). We also included a WALLABY forecast with a $4\%$ error-bar in red (see section~\ref{sec:future2}). For the WiggleZ survey we also include the data points from~\citet{Blake:2011ep} (empty triangles, shifted by $\Delta z=0.005$ to the right for visibility), where the Alcock-Paczynski effect has been taken into account. We plot a $\Lambda$CDM model as well as a DGP model for comparison.}
\label{fig:growth}
\end{center}
\end{figure}

In this section we test General Relativity by measuring the growth index $\gamma$. We would like to stress that the $\gamma$-parameterisation of modified gravity has its limitations and other more general parameterisations have been proposed (see e.g.~\citealt{Silvestri:2009hh,Bean:2010zq,Daniel:2010yt,Clifton:2011jh,Hojjati:2011ix, Baker:2011jy}). However, this is going beyond the scope of this paper.

We combine our result for $g_{\theta}(z_{\rm eff})$ with the latest results from WMAP7~\citep{Komatsu:2010fb}, where we use the WMAP7.2 dataset provided on the NASA webpage\footnote{\url{http://lambda.gsfc.nasa.gov/product/map/dr4/likelihood_get.cfm}}. While $\Lambda$CDM predicts $\gamma \approx 0.55$, alternative theories of gravity deviate from this value. One example of such an alternative model is the DGP braneworld model of~\citet{Dvali:2000hr}, in which our observable Universe is considered to be a brane embedded in a higher dimensional bulk space-time and the leakage of gravity force propagating into the bulk can lead to the current accelerated expansion of the Universe. Because of the missing dark energy component, this model predicts a larger growth index of $\gamma \approx 0.69$~\citep{Linder:2005in}.

In Figure~\ref{fig:growth} we compare measurements of the growth of structure $g_{\theta}$ from different galaxy redshift surveys. For the WiggleZ survey we include data points which assume a correct fiducial cosmology (solid black triangles) as well as data points which account for the Alcock-Paczynski effect (empty black triangles). The degeneracy between the Alcock-Paczynski effect and the linear redshift-space distortion signal increases the error by about a factor of two. As we showed in section~\ref{sec:AP}, the Alcock-Paczynski effect is very small in 6dFGS, which therefore yields a direct measurement of the redshift-space distortion signal.

All data points seem to be in good agreement with the $\Lambda$CDM model (black solid line), while the DGP model generally predicts smaller values of $g_{\theta}$. The value of $\sigma_8$ for the two different models has been derived from the CMB scalar amplitude $A_s$~\citep{Komatsu:2010fb}, where we use the corresponding Friedmann equation to calculate $\sigma_8$.

The analysis method we apply in this section is summarised in the following four points:
\begin{enumerate}
\item We produce a Monte Carlo Markov Chain (MCMC) with CosmoMC~\citep{Lewis:2002ah} for a $\Lambda$CDM universe by fitting the WMAP7 dataset. The CMB depends on dark energy through the distance of last scattering and the late-time Integrated Sachs-Wolfe (ISW) effect. We avoid the contributions of the ISW effect by limiting the WMAP7 dataset to multipole moments $\ell > 100$.
\item Now we importance-sample the CosmoMC chain by randomly choosing a value of $\gamma$ in the range $0 \leq \gamma \leq 1$ for each chain element. Since the value of $\sigma_8(z_{\rm eff})$ depends on $\gamma$ we have to recalculate this value for each chain element. First we derive the growth factor
\begin{equation}
D(a_{\rm eff}) = \exp\left[ -\int_{a_{\rm eff}}^1da'\,f(a')/a'\right],
\end{equation}
where $a_{\rm eff}$ is the scale factor at the effective redshift $a_{\rm eff} = 1/(1 + z_{\rm eff})$. In order to derive $\sigma_{8,\gamma}(z_{\rm eff})$ we have to extrapolate from the matter dominated region to the effective redshift,
\begin{equation}
\sigma_{8,\gamma}(z_{\rm eff}) = \frac{D_{\gamma}(z_{\rm eff})}{D(z_{hi})}\sigma_8(z_{hi}),
\end{equation} 
where we use $\sigma_8(z_{hi})$ from eq.~\ref{eq:sig8z} and $z_{hi} = 50$, well in the matter-dominated regime.
\item We now calculate the growth rate using $f_{\gamma}(z_{\rm eff}) \simeq \Omega_m^{\gamma}(z_{\rm eff})$ and construct $g_{\theta,\gamma}(z_{\rm eff}) = f_{\gamma}(z_{\rm eff})\sigma_{8,\gamma}(z_{\rm eff})$.
\item Finally we compare the model with $g_{\theta} = 0.423\pm 0.055$ from Table~\ref{tab:para2} and combine the likelihood from this comparison with the WMAP7 likelihood. 
\end{enumerate}

\begin{figure}[tb]
\begin{center}
\epsfig{file=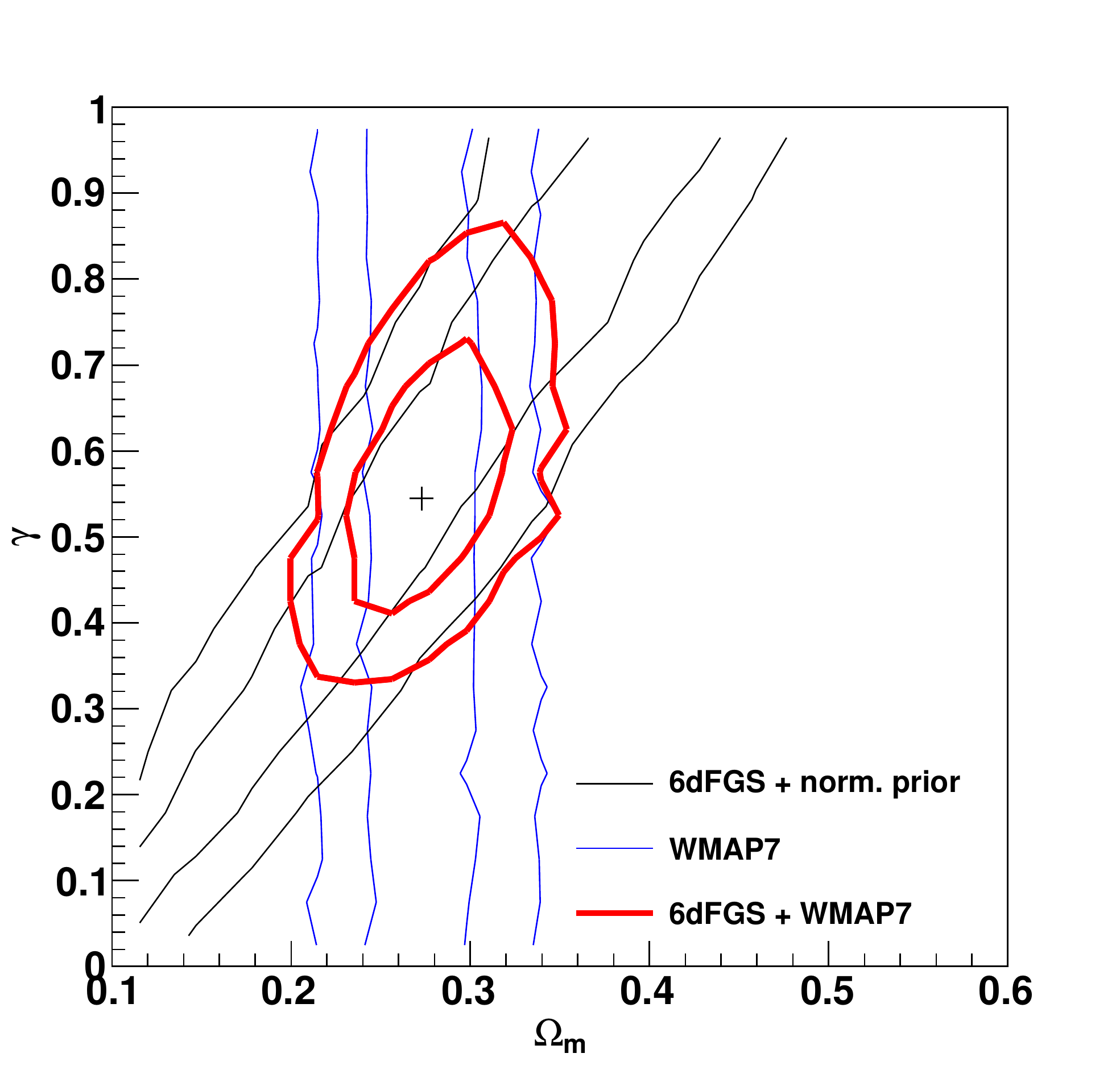,width=10cm}
\caption{Likelihood distribution of $\gamma-\Omega_m$ for our fit to $g_{\theta} = 0.423\pm0.055$ from 6dFGS and WMAP7~\citep{Komatsu:2010fb}. The 6dFGS contours (black) include a normalisation prior from WMAP7. Marginalising over the different parameters in the MCMC chain gives $\Omega_m = 0.271\pm 0.027$ and $\gamma = 0.547\pm 0.088$. The low redshift of 6dFGS makes this measurement particularly sensitive to $\gamma$ and independent of systematic effects like the Alcock-Paczynski distortion.}
\label{fig:gamma_om}
\end{center}
\end{figure}

The result is shown in Figure~\ref{fig:gamma_om}. Marginalising over the remaining parameters we get $\gamma = 0.547\pm 0.088$ and $\Omega_m = 0.271\pm 0.027$, which is in agreement with the prediction of a $\Lambda$CDM universe ($\gamma \approx 0.55$).  Our analysis depends only on the growth rate measured in 6dFGS and WMAP7. This makes our measurement of $\gamma$ independent of systematic effects like the Alcock-Paczynski distortion which is a matter of concern for galaxy redshift surveys at higher redshift. 

\section{Future low-redshift galaxy surveys: WALLABY and TAIPAN}
\label{sec:future2}

In this section we make predictions for the accuracy of $f\sigma_8$ measurements from future low-redshift galaxy surveys using a Fisher matrix analysis based on~\citet{White:2008jy}

The Wide-field ASKAP L-band Legacy All-sky Blind surveY (WALLABY)\footnote{http://www.atnf.csiro.au/research/WALLABY} is an HI survey planned for the Australian SKA Pathfinder telescope (ASKAP), currently under construction at the Murchison Radio-astronomy Observatory (MRO) in Western Australia. The survey will cover $75\%$ of the sky and a proposal exists to fill up the remaining $25\%$ using the Westerbork Radio Telescope. In this analysis we follow the survey parameters employed by~\citet{Beutler:2011hx} (see also Duffy et al., in preparation), where for WALLABY we assume a $4\pi$ survey containing $600\,000$ galaxies at a mean redshift of $z = 0 .04$. The linear bias of a typical WALLABY galaxy is $0.7$~\citep{Basilakos:2007wq} and the volume of the survey is $0.12h^{-3}$Gpc$^3$.

The TAIPAN survey\footnote{TAIPAN: Transforming Astronomical Imaging surveys through Polychromatic Analysis of Nebulae} proposed for the UK Schmidt telescope at Siding Spring Observatory in New South Wales will cover a similar sky area as 6dFGS but will extend to a larger redshift such that $\overline{z} = 0.08$. For our TAIPAN forecast we assumed the same sky-coverage as 6dFGS ($f_{\rm sky} = 0.41$), a bias of $b = 1.4$, a total of $400\,000$ galaxies and a volume of $0.23h^{-3}$\,Gpc$^3$.

First we test the Fisher matrix prediction for $f\sigma_8$ in the case of 6dFGS. We assume a survey volume of $0.08h^{-3}\,$Gpc$^3$, with $81\,971$ galaxies. Using $k_{\rm max} = 0.1h\,$Mpc$^{-1}$ we forecast a measurement of $f\sigma_8$ of $23\%$, while using $k_{\rm max} = 0.2h\,$Mpc$^{-1}$ produces a $8.3\%$ error. The actual error in $f\sigma_8$ we found in this paper is $13\%$, somewhere between these two values. For WALLABY and TAIPAN we will report constraints for both $k_{\rm max} = 0.1h\,$Mpc$^{-1}$ and $k_{\rm max} = 0.2h\,$Mpc$^{-1}$.

With the specifications given above and using $k_{\rm max} = 0.1\; (0.2)h\,$Mpc$^{-1}$, the Fisher matrix forecast for WALLABY is a measurement of $f\sigma_8$ with $10.5\; (3.9)\%$ error. We included the WALLABY forecast with a $4\%$ error-bar in Figure~\ref{fig:growth}. The model TAIPAN survey produces forecast errors of $13.2\;(4.9)\%$, improving the results from 6dFGS by almost a factor of two. Although TAIPAN maps a larger volume of the Universe compared to WALLABY, it does not produce a better measurement of $f\sigma_8$. WALLABY has a smaller galaxy bias, which increases the redshift-space distortion signal by a factor of two compared to TAIPAN. This will also be very useful for breaking the degeneracy between bias and $\sigma_8$ using the technique of section~\ref{sec:sig8}.

The WALLABY survey will target galaxies rich in HI gas. Such galaxies will mostly populate under-dense regions of the Universe, because in groups and clusters galaxies are stripped of their gas by interactions with other galaxies and the intra-group and intra-cluster medium.  This is the reason that HI-selected galaxies possess a low bias ($\sim 0.7$).  However, this fact also implies that WALLABY galaxies sample the density field in a manner that avoids high-density regions. These high-density regions are an important source of non-linear redshift-space distortions (``finger-of-God'' effect). We can hence suppose that non-linear effects will be smaller in amplitude in an HI survey compared to highly-biased surveys such as 6dFGS (see e.g.~\citealt{Simpson:2011vn} or Figure 4 in~\citealt{Reid:2011ar}). This should allow the inclusion of much smaller scales in the analysis, producing more accurate measurements. A more detailed analysis using WALLABY mock catalogues is in preparation.

\citet{McDonald:2008sh} have suggested that multiple tracers within the same cosmic volume can be used to reduce the sampling variance and improve cosmological parameter constraints (see also \citealt{Seljak:2008xr,Slosar:2009aa,Bernstein:2011ju}). Using the ratio of the perturbation amplitudes of two surveys with different bias factors gives
\begin{equation}
\frac{b_1 + f\mu^2}{b_2 + f\mu^2} = \frac{\alpha b_2 + f\mu^2}{b_2 + f\mu^2},
\end{equation} 
where $\alpha = b_1/b_2$. The angular dependence of this expression allows one to extract $f$ without any dependence on the density field (see Figure 1 in \citealt{Bernstein:2011ju}). The density field is the source of the sampling variance error, since it will change, depending on the patch of the sky which is observed. Using the ratio of two tracers, the precision with which the growth rate $f$ can be determined is (in principle) only limited by the shot noise, and not by the sampling variance.

The three surveys discussed above, 6dFGS, WALLABY and TAIPAN, have a large overlapping volume which allows the potential application of this method. The technique works best for densely-sampled surveys with very different bias factors, $b_1$ and $b_2$. While TAIPAN and 6dFGS have very similar bias, the bias of WALLABY will be much smaller.

We assume an overlap volume of $0.41\times 0.12h^{-3}\,$Gpc$^3 = 0.049h^{-3}\,$Gpc$^3$, where we multiply the sky coverage of 6dFGS and TAIPAN with the effective volume of WALLABY. For the different surveys we use the parameters as stated above. We forecast Fisher matrix constraints on $f\sigma_8$ of  $10.3\;(5)\%$ using $k_{\rm max} = 0.1\;(0.2)h\,$Mpc$^{-1}$. Because the overlap volume is only $\approx 1/3$ of the WALLABY volume, this result does not improve the measurement arising from WALLABY alone, especially considering that WALLABY itself may be able to include modes up to large $k_{\rm max}$ in the fitting process.

Our results show that future surveys such as WALLABY and TAIPAN will provide an accurate measurement of $f\sigma_8$ at low redshift, and will be able to complement future high-redshift surveys such as BOSS, which will have a similar accuracy for several data points over the higher redshift range $0.2$ - $0.6$~\citep{Song:2008qt,White:2008jy,Reid:2011ar}.

\section{Conclusion}
\label{sec:conc}

In this paper we have measured the 2D correlation function of the 6dF Galaxy Survey. We derived a covariance matrix using jack-knife resampling as well as log-normal realisations and showed that both techniques give comparable results. We have modelled the 2D correlation function with a simple streaming model and a more advanced approach suggested by~\citet{Scoccimarro:2004tg} combined with the N-body calibrated results from~\citet{Jennings:2010uv}. We formulated these models in real-space including wide-angle corrections. For the final results on $f\sigma_8$ we chose the model by~\citet{Scoccimarro:2004tg}, although we found that both models gave consistent results at sufficiently large scales. 

We analysed the measurement in two different ways. First we fitted for the two parameters $g_{\theta}(z_{\rm eff}) = f(z_{\rm eff})\sigma_8(z_{\rm eff})$ and $g_b(z_{\rm eff}) = b\sigma_8(z_{\rm eff})$, where these constraints depend only on the 6dFGS data. Our second analysis method assumes a growth index from standard gravity ($\gamma \approx 0.55$) and fits for $\sigma_8$, $b$, $\Omega_m$, $H_0$ and $\sigma_p$, where we combine the 6dFGS measurement with a prior in the Hubble constant. All parameter measurements are summarised in Table~\ref{tab:para2}. We can summarise the results as follows:
\begin{itemize}
\item Our first analysis method found $g_{\theta}(z_{\rm eff}) = f(z_{\rm eff})\sigma_8(z_{\rm eff}) = 0.423\pm0.055$ and $g_b(z_{\rm eff}) = b\sigma_8(z_{\rm eff}) = 1.134\pm0.073$, at an effective redshift of $z_{\rm eff} = 0.067$. The 6dFGS measurement of $g_{\theta}$, unlike high-redshift measurements, does not depend on assumptions about the expansion history of the Universe and the Alcock-Paczynski distortion.
\item In our second analysis method we used the angle dependence of redshift-space distortions in the 2D correlation function to break the degeneracy between the galaxy bias $b$ and the normalisation of the matter clustering statistic $\sigma_8$, assuming standard gravity. We found $\sigma_8 = 0.76\pm0.11$, $\Omega_m = 0.250\pm 0.022$, $b = 1.48\pm0.27$ and $\sigma_p = 174\pm73$km/s. This result uses a prior on $H_0$ from~\citet{Riess:2011yx}.
\item Combining our measurement of $g_{\theta}(z_{\rm eff})$ with WMAP7~\citep{Komatsu:2010fb} allows us to measure the growth index $\gamma$, directly testing General Relativity. We found $\gamma = 0.547\pm 0.088$ and $\Omega_m = 0.271\pm 0.027$, in agreement with the predictions of General Relativity ($\gamma \approx 0.55$). The 6dFGS measurement of this parameter is independent of possible degeneracies of $\gamma$ with other parameters which affect the correlation function at high redshift, such as the dark energy equation of state parameter $w$.
\item We used a Fisher matrix analysis to forecast the constraints on $f\sigma_8$ that would be obtained from two future low-redshift galaxy surveys, WALLABY and TAIPAN. We found that WALLABY will be able to measure $f\sigma_8$ to a forecast accuracy of $10.5\%$ for $k_{\rm max} = 0.1h\,$Mpc$^{-1}$ and $3.9 \%$ for $k_{\rm max}=0.2h\,$Mpc$^{-1}$. A combination of 6dFGS, TAIPAN and WALLABY, using the multiple-tracer method proposed by \citet{McDonald:2008sh}, will be able to constrain $f\sigma_8$ to $5$ - $10.3\%$. These measurements would complement future large-volume surveys such as BOSS, which will measure the growth rate at much higher redshift ($z > 0.2$), and contribute to future precision tests of General Relativity on cosmic scales.
\end{itemize}

\cleardoublepage

\chapter{Dependence of halo occupation on stellar mass}
\label{ch:HOD}
\begin{center}
\emph{\textbf{\citeauthor{Beutler:2012yr}}}\\
\emph{\textbf{MNRAS 429, 3604B (2013)}}
\end{center}

\begin{abstract}
In this paper we study the stellar-mass dependence of galaxy clustering in the 6dF Galaxy Survey. The near-infrared selection of 6dFGS allows more reliable stellar mass estimates compared to optical bands used in other galaxy surveys. Using the Halo Occupation Distribution (HOD) model, we investigate the trend of dark matter halo mass and satellite fraction with stellar mass by measuring the projected correlation function, $w_p(r_p)$. We find that the typical halo mass ($M_1$) as well as the satellite power law index ($\alpha$) increase with stellar mass. This indicates, (1) that galaxies with higher stellar mass sit in more massive dark matter halos and (2) that these more massive dark matter halos accumulate satellites faster with growing mass compared to halos occupied by low stellar mass galaxies. 
Furthermore we find a relation between $M_1$ and the minimum dark matter halo mass ($M_{\rm min}$) of $M_1 \approx 22\,M_{\rm min}$, in agreement with similar findings for SDSS galaxies. The satellite fraction of 6dFGS galaxies declines with increasing stellar mass from $21\%$ at $M_{\rm stellar} = 2.6\times10^{10}h^{-2}\,M_{\odot}$ to $12\%$ at $M_{\rm stellar} = 5.4\times10^{10}h^{-2}\,M_{\odot}$ indicating that high stellar mass galaxies are more likely to be central galaxies. We compare our results to two different semi-analytic models derived from the Millennium Simulation, finding some disagreement. Our results can be used for placing new constraints on semi-analytic models in the future, particularly the behaviour of luminous red satellites. 
Finally we compare our results to studies of halo occupation using galaxy-galaxy weak lensing. We find good overall agreement, representing a valuable crosscheck for these two different tools of studying the matter distribution in the Universe.
\end{abstract}

\section{Introduction}
\label{sec:intro}
 
The first statistical studies of galaxy clustering~\citep{Totsuji:1969,Peebles:1973,Hauser:1973,Hauser:1974,Peebles:1974} found that the galaxy correlation function behaves like a power law, which is difficult to explain from first principles~\citep{Berlind:2001xk}. More recent studies, however, found deviations from a power law. For example~\citet{Zehavi:2004zn} showed that the projected correlation function $w_p(r_p)$ of SDSS galaxies exhibits a statistically significant departure from a power law. They also showed that a 3-parameter Halo Occupation Distribution (HOD) model (e.g.,~\citealt{Jing:1997nb, Ma:2000ik, Peacock:2000qk, Seljak:2000gq, Scoccimarro:2001, Berlind:2001xk, Cooray:2002dia}) together with a $\Lambda$CDM background cosmology, can account for this departure, reproducing the observed $w_p(r_p)$. 

Within the halo model the transition from the 1-halo term to the 2-halo term causes a ''dip'' in the correlation function at around $1-3h^{-1}\,$Mpc, corresponding to the exponential cutoff in the halo mass function. In case of a smooth transition between the one- and two-halo terms, this can mimic a power-law correlation function. Studies with Luminous Red Galaxies (LRGs) found that the deviation from a power-law is larger for highly clustered bright galaxies~\citep{Zehavi:2004zn, Zehavi:2004ii, Blake:2007xp, Zheng:2008np, Zehavi:2010bh}, and at high redshift~\citep{Conroy:2005aq}, which agrees with theoretical predictions~\citep{Watson:2011cz}.

While galaxy clustering is difficult to predict, dark matter clustering is dominated by gravity and can be predicted for a given cosmology using N-body simulations. Using models for how galaxies populate dark matter halos, which are usually motivated by N-body simulations, we can directly link galaxy clustering and matter clustering. This can be modelled in terms of the probability distribution $p(N|M)$ that a halo of virial mass $M$ contains $N$ galaxies of a given type. 
On strongly non-linear scales the dark matter distribution is given by the actual density distribution of the virialized halos, while on large and close to linear scales the dark matter distribution can be predicted from linear perturbation theory.

HOD modelling has been applied to galaxy clustering data from the 2-degree Field Galaxy Redshift Survey (2dFGRS)~\citep{Porciani:2004vi, Tinker:2006sk} and the Sloan Digital Sky Survey (SDSS)~\citep{vandenBosch:2002zn, Magliocchetti:2003ee, Zehavi:2004zn, Zehavi:2004ii, Tinker:2004gf, Yang:2004qi, Yang:2007pg, Zehavi:2010bh}. More recently it also became possible to model the clustering of high-z galaxies using VVDS~\citep{Abbas:2010hr}, Bo\"{o}tes~\citep{Brown:2008eb}, DEEP2~\citep{Coil:2005ku} and Lyman-break galaxies at high redshift in the GOODS survey~\citep{Lee:2005jha}. Such studies revealed, that the minimum mass, $M_{\rm min}$ for a halo to host a central galaxy  more luminous than some threshold, $L$ is proportional to $L$ at low luminosities, but steepens above $L_*$. Massive halos have red central galaxies with predominantly red satellites, while the fraction of blue central galaxies increases with decreasing host halo mass. Furthermore~\citet{Zehavi:2004ii} found that there is a scaling relation between the minimum mass of the host halos, $M_{\rm min}$ and the mass scale, $M_1$ of halos that on average host one satellite galaxy in addition to the central galaxy, $M_1 \approx 23\,M_{\rm min}$. Using a different HOD parameterization, \citet{Zheng:2007zg} found the relation to be $M_1 \approx 18\,M_{\rm min}$, very similar to~\citet{Zehavi:2010bh} who found $M_1 \approx 17\,M_{\rm min}$. 

The 6dF Galaxy Survey is one of the biggest galaxy surveys available today with a sky coverage of $42\%$ and an average redshift of $\overline{z} = 0.05$. The survey includes about $125\,000$ redshifts selected in the $J,H,K,b_J,r_F$-bands~\citep{Jones:2004zy,Jones:2005ya,Jones:2009yz}. The near infrared selection, the high completeness and the wide sky coverage make 6dFGS one of the best  surveys in the local Universe to study galaxy formation. This dataset has been used to study the large scale galaxy clustering to measure the Hubble constant using Baryon Acoustic Oscillations~\citep{Beutler:2011hx}, as well as the growth of structure at low redshift~\citep{Beutler:2012px}. While these previous studies used the $K$-band selected sample, in this analysis we use the $J$-band. The $J$-band allows the most reliable stellar mass estimate of the five bands available in 6dFGS, because of its lower background noise. Together with the $b_J-r_F$ colour we can derive stellar masses using the technique of~\citet{Bell:2000jt}, which leads to a dataset of $76\,833$ galaxies in total. The photometric near-infrared selection from 2MASS makes the stellar mass estimates in 6dFGS more reliable than stellar mass estimates in other large galaxy surveys which rely on optical bands~\citep{Drory:2004ib,Kannappan:2007ys,Longhetti:2008gv,Grillo:2007kg,Gallazzi:2009aj}.

Numerical N-body simulations are usually restricted to dark matter only. To understand galaxy formation, baryonic effects such as feedback and gas cooling, have to be included. Such simulations face severe theoretical and numerical challenges. Semi-analytic models build upon pre-calculated dark matter merger trees from cosmological simulations and include simplified, physically and observationally motivated, analytic recipes for different baryonic effects. Semi-analytic models have been shown to successfully reproduce observed statistical properties of galaxies over a large range of galaxy masses and redshifts (e.g.~\citealt{Croton:2005fe,Bower:2005vb,De Lucia:2006vua,Bertone:2007sj,Font:2008pc,Guo:2009fn}) and allow a level of understanding unavailable in N-body simulations. The underlying models are necessarily simplified and often use a large number of free parameters to fit different observations simultaneously. In this paper, we derive 6dFGS mock surveys from semi-analytic models based on the Millennium Simulation~\citep{Springel:2005nw} and compare the properties of these surveys with measurements in 6dFGS. Our results can be used to improve upon these semi-analytic models and further our understanding of baryonic feedback processes on galaxy clustering.

\citet{Mandelbaum:2005nx} studied halo occupation as a function of stellar mass and galaxy type using galaxy-galaxy weak lensing in SDSS. They found that for a given stellar mass, the halo mass is independent of morphology below $M_{\rm stellar} = 10^{11}\,M_{\odot}$, indicating that stellar mass is a good proxy for halo mass at that range. We compare our results with~\citet{Mandelbaum:2005nx} which represents a valuable crosscheck of the HOD analysis using two very different techniques, weak lensing and galaxy clustering. 

This paper is organised as follows: First we introduce the 6dF Galaxy Survey in section~\ref{sec:survey} together with the technique to derive the stellar masses. We also explain how we derive our four volume-limited sub-samples in stellar mass and redshift, which are then used for the further analysis. In section~\ref{sec:analysis} we calculate the projected correlation function, $w_p(r_p)$, for each sub-sample and use jack-knife re-sampling to derive the covariance matrices. In section~\ref{sec:pl} we fit power laws to the projected correlation functions of the four sub-samples. In section~\ref{sec:hod} we introduce the HOD framework and in section~\ref{sec:hodfits} we apply the HOD model to the data. In section~\ref{sec:sam} we derive 6dFGS mock samples from two different semi-analytic models, which we then compare to our results in section~\ref{sec:results} together with a general discussion of our findings. We conclude in section~\ref{sec:con}.

Throughout the paper we use $r$ to denote real space separations and $s$ to denote separations in redshift space. Our fiducial model to convert redshifts into distances is a flat universe with $\Omega_m = 0.27$, $w = -1$ and $\Omega_k = 0$. The Hubble constant is set to $H_0 = 100h\,$km\,s$^{-1}$Mpc$^{-1}$ which sets the unit of stellar masses to $h^{-2}M_{\odot}$, while most other masses are given in $h^{-1}M_{\odot}$. The HOD model uses cosmological parameters following WMAP7~\citep{Komatsu:2010fb}.

\section{The 6dF Galaxy survey}
\label{sec:survey}

The 6dF Galaxy Survey~\citep[6dFGS;][]{Jones:2004zy,Jones:2005ya,Jones:2009yz} is a near-infrared selected ($J,H,K$) redshift survey covering $17\,000\,$deg$^2$ of the southern sky. The $J$, $H$ and $K$ surveys avoids a $\pm10^\circ$ region around the Galactic Plane to minimise Galactic extinction and foreground source confusion in the Plane. The near-infrared photometric selection was based on total magnitudes from the Two-Micron All-Sky Survey Extended Source Catalog~\citep[2MASS XSC;][]{Jarrett:2000me}. The spectroscopic redshifts of 6dFGS were obtained with the Six-Degree Field (6dF) multi-object spectrograph of the UK Schmidt Telescope (UKST) between 2001 and 2006.

The 6dFGS $J$-selected sample used in this paper contains (after completeness cuts) $76\,833$ galaxies selected with $9.8 \leq J \leq 13.75$. We chose the $J$-band because it has the highest signal-to-noise of the three 2MASS bands. While there is slightly less extinction in the $K$-band compared to the $J$-band, for practical purposes, the $J$-band has better S/N because the night sky background glow is much less in the $J$- than in the $K$-band. The near infrared selection makes 6dFGS very reliable for stellar mass estimates.

The mean completeness of 6dFGS is $92$ percent and the median redshift is $z = 0.05$. Completeness corrections are derived by normalising completeness-apparent magnitude functions so that, when integrated over all magnitudes, they equal the measured total completeness on a particular patch of sky. This procedure is outlined in the luminosity function evaluation of~\citet{Jones:2006xy} and also in Jones et al., (in prep). The original survey papers~\citep{Jones:2004zy,Jones:2005ya,Jones:2009yz} describe in full detail the implementation of the survey and its associated online database.

The clustering in a galaxy survey is estimated relative to a random (unclustered) distribution which follows the same angular and redshift selection function as the galaxy sample itself. We base our random mock catalogue generation on the 6dFGS luminosity function~\citep{Jones:2006xy}, where we use random numbers to pick volume-weighted redshifts and luminosity function-weighted absolute magnitudes. We then test whether the redshift-magnitude combination falls within the 6dFGS $J$-band faint and bright apparent magnitude limits ($9.8 \leq J \leq 13.75$). We assigned a $b_J$-$r_F$ colour to each random galaxy using the redshift- $b_J$-$r_F$ colour relation measured in the data and used these to derive stellar masses for the random galaxies using the same technique as for the actual galaxies (see section~\ref{sec:stellar}).

\subsection{Stellar mass estimate and volume-limited sub-samples}
\label{sec:stellar}

To calculate the stellar mass for our dataset we use the stellar population synthesis results from~\citet{Bruzual A.:1993is} together with a scaled Salpeter initial mass function (IMF) as reported in~\citet{Bell:2000jt}
\begin{equation}
\begin{split}
\log_{10}(M_{\rm stellar}/L_J) &= -0.57C_{b_J-r_F} + 0.48\\
\log_{10}(L_J) &= (M_J^{\rm sun} - M_J)/2.5\\
\log_{10}(M_{\rm stellar}) &= \log_{10}(M_{\rm stellar}/L_J) + \log_{10}(L_J),
\label{eq:stellar}
\end{split}
\end{equation}
with the 2MASS $b_J$-$r_F$ colours, $C_{b_J-r_F}$ the $J$-band absolute magnitude, $M_J$ and the $J$-band absolute magnitude of the sun, $M_J^{\rm sun} = 3.70$~\citep{Worthey:1994iw}. The biggest uncertainty in stellar mass estimates of this type is the choice of the IMF. Assuming no trend in IMF with galaxy type, the range of IMFs presented in the literature cause uncertainties in the absolute normalisation of the stellar $M/L$-ratio of a factor of $2$ in the near-infrared~\citep{Bell:2000jt}. The 6dFGS stellar mass function as well as a comparison of different stellar mass estimates is currently in preparation (Jones et al. in prep).

\begin{figure}[tb]
\begin{center}
\epsfig{file=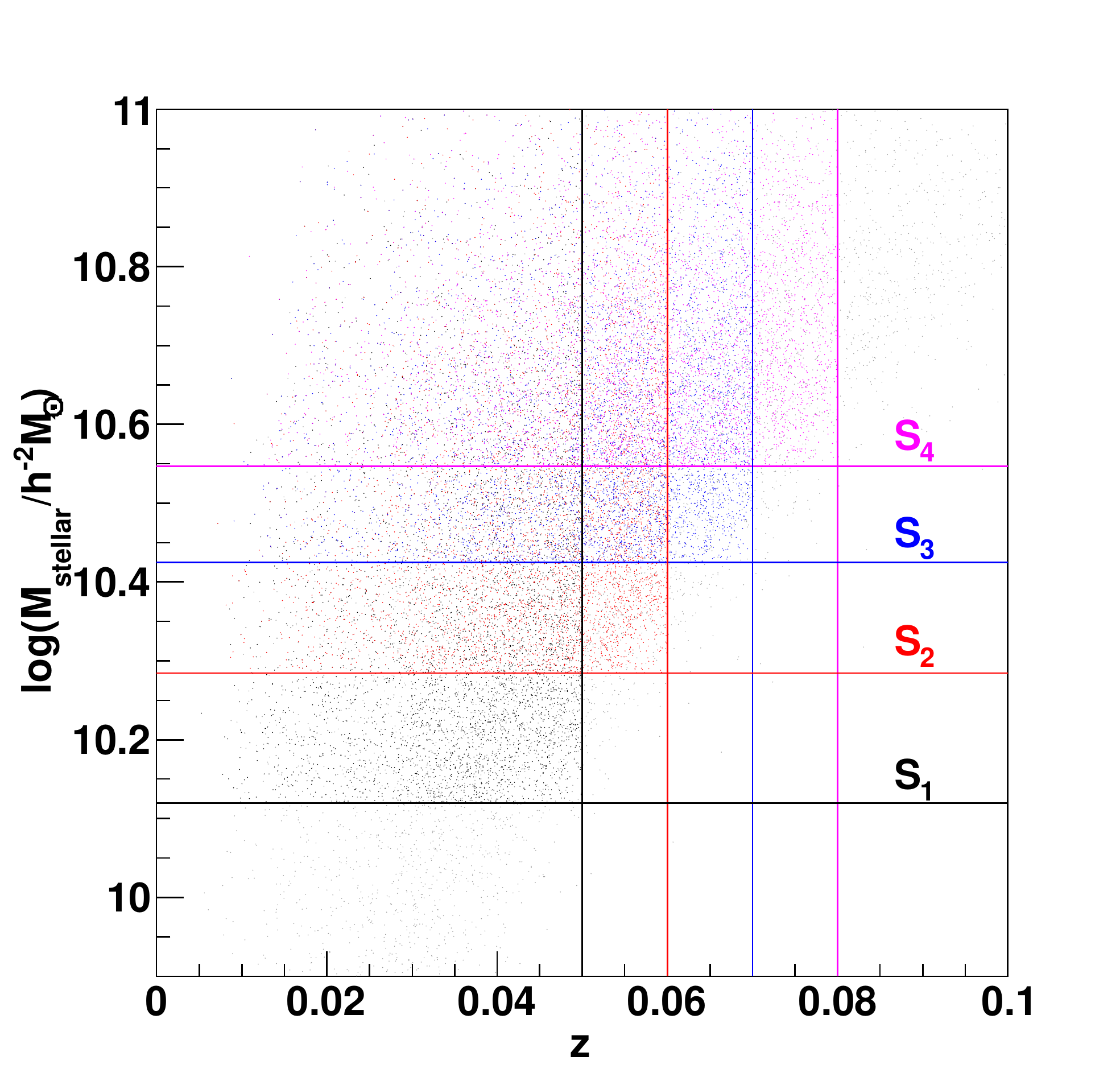,width=10cm}
\caption{The distribution of 6dFGS galaxies in log stellar mass and redshift. The redshift and stellar mass cuts imposed to create the four volume-limited samples ($S_1$ to $S_4$ see Table~\ref{tab:vls}) are shown by the coloured lines. All galaxies in the upper left quadrant created by the two correspondingly coloured lines are included in the volume-limited sub-samples. The plot shows a randomly chosen set of $20\%$ of all galaxies.}
\label{fig:z_vs_logSM2}
\end{center}
\end{figure}

\begin{figure}[tb]
\begin{center}
\epsfig{file=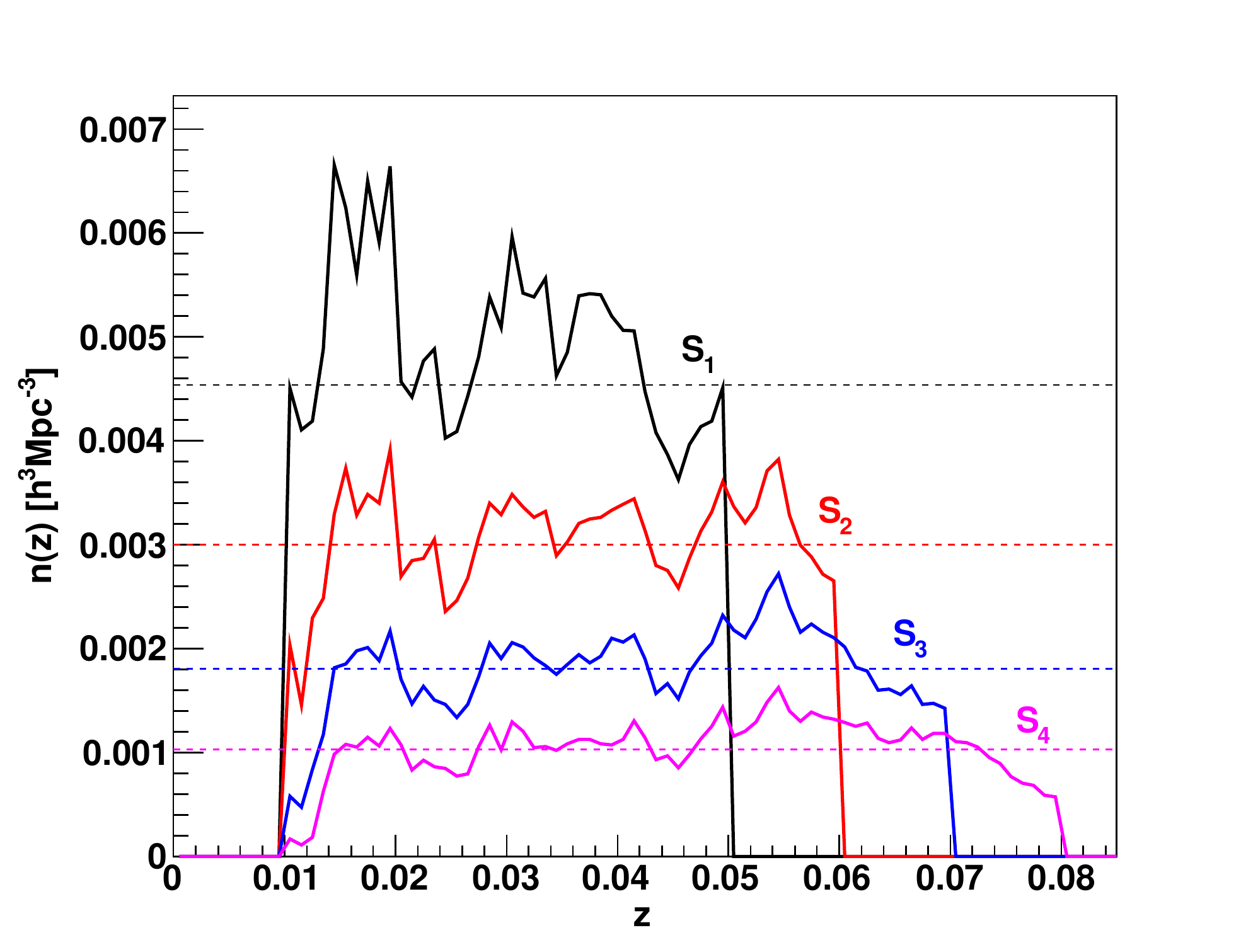,width=10cm}
\caption{Galaxy density as a function of redshift for the four different volume-limited sub-samples $S_1$ - $S_4$. All samples follow an approximately constant number density indicated by the dashed lines and listed in the last column of Table~\ref{tab:vls}.}
\label{fig:nz}
\end{center}
\end{figure}

We create four volume-limited sub-samples in redshift and stellar mass. This is done by choosing an upper limit in redshift ($z_{\rm max}$) and then maximising the number of galaxies by choosing a lower limit in stellar mass ($M^{\rm min}_{\rm stellar}$), meaning that every galaxy above that stellar mass will be detected in 6dFGS, if its redshift is below the redshift limit (see Figure~\ref{fig:z_vs_logSM2}). Because of the distribution in $b_J$-$r_F$ colour, a clear cut in absolute magnitude does not correspond to a clear cut in stellar mass and therefore our samples are not perfectly volume-limited. We use the absolute magnitude limit which corresponds to a chosen redshift limit and derive a stellar mass limit, using the $b_J$-$r_F$ colour corresponding to the $50\%$ height of the $b_J$-$r_F$ distribution ($b_J$-$r_F$($50\%$) $= 1.28$). We create four sub-samples ($S_1$-$S_4$) with upper redshift cuts at $z_{\rm max} = 0.05$, $0.06$, $0.07$ and $0.08$ and with the mean log stellar masses ranging from $\log_{10}(M_{\rm stellar}/h^{-2}\,M_{\odot}) = 10.41$ to $10.73$. We also include a low redshift cut-off at $z_{\rm min} = 0.01$. Since the stellar mass distributions for these samples overlap, especially for the higher stellar mass samples, the results are correlated to some extent (see Figure~\ref{fig:z_vs_logSM2}). All our sub-samples are summarised in Table~\ref{tab:vls}. Figure~\ref{fig:nz} shows the galaxy density as a function of redshift for the four sub-samples. The roughly constant number density with redshift shows that our sub-samples are close to volume-limited.

\begin{sidewaystable}
\begin{center}
\caption{Summary of the different volume-limited sub-samples, $S_1$ - $S_4$, used in this analysis. The effective redshift, $z_{\rm eff}$ and effective stellar mass are calculated as the mean of all galaxy pairs contributing to correlation function bins between $0.1$ and $40h^{-1}\,$Mpc and the error is the standard deviation from the mean. The galaxy density $n_g$ is calculated as the number of galaxies $N$ divided by the co-moving sample volume, which is calculated from the maximum redshift $z_{\rm max}$ and eq.~\ref{eq:codist}.}
\vspace{0.4cm}
\begin{tabular}{ccccccc}
\hline
sample & $z_{\rm max}$ & $\log_{10}\left(\frac{M^{\rm min}_{\rm stellar}}{h^{-2}M_{\odot}}\right)$ & $\langle z\rangle$ & $\log_{10}\left(\frac{\langle M_{\rm stellar}\rangle}{h^{-2}M_{\odot}}\right)$ & N & $n_g$ $[h^{3}\text{Mpc}^{-3}]$\\
\hline
$S_1$ & $0.05$ & $10.12$ & $0.0369\pm0.0010$ & $10.410\pm0.010$ & $24\,644$ & $4.536\times 10^{-3}$\\
$S_2$ & $0.06$ & $10.28$ & $0.0453\pm0.0011$ & $10.532\pm0.010$ & $27\,999$ & $3.001\times 10^{-3}$\\
$S_3$ & $0.07$ & $10.42$ & $0.0521\pm0.0012$ & $10.635\pm0.018$ & $26\,584$ & $1.806\times 10^{-3}$\\
$S_4$ & $0.08$ & $10.55$ & $0.0585\pm0.0020$ & $10.730\pm0.013$ & $22\,497$ & $1.030\times 10^{-3}$\\
\hline
\hline
\end{tabular}
\label{tab:vls}
\end{center}
\end{sidewaystable}

\section{Data analysis}
\label{sec:analysis}

In order to analyse the galaxy clustering in all three dimensions, we calculate the co-moving distances from us to each galaxy,
\begin{equation}
D_C = \frac{c}{H_0}\int^z_0\frac{dz'}{E(z')},
\label{eq:codist}
\end{equation} 
where $z$ is the measured redshift, and  
\begin{equation}
E(z) = \left[\Omega^{\rm fid}_m(1+z)^3 + \Omega^{\rm fid}_{\Lambda}\right]^{1/2}.
\end{equation}
We assume a flat universe with $\Omega_k^{\rm fid} = 0$ and $\Omega^{\rm fid}_{\Lambda} = 1 - \Omega^{\rm fid}_m$ and treat dark energy as a cosmological constant ($w^{\rm fid} = -1$). Given the low redshift of our dataset, these assumptions have no impact on the final results.

We measure the separation between galaxies in our survey along the line of sight ($\pi$) and perpendicular to the line of sight ($r_p$) and count the number of galaxy pairs on this two-dimensional grid. We do this for the 6dFGS data catalogue, a random catalogue with the same selection function, and a combination of data-random pairs. We call the pair-separation distributions obtained from this analysis step $DD(r_p,\pi)$, $RR(r_p,\pi)$ and $DR(r_p,\pi)$, respectively. In the analysis we used $30$ random catalogues with the same size as the real data catalogue and average $DR(r_p,\pi)$ and $RR(r_p,\pi)$. The random mocks are sampled from the 6dFGS luminosity function (\citet{Jones:2006xy} and also Jones et al, in prep.), and hence they contain the same evolution of luminosity with redshift that we see in 6dFGS itself. The redshift-space correlation function is then given by the~\citet{Landy:1993yu} estimator:
\begin{equation}
\xi(r_p,\pi) = 1 + \frac{DD(r_p,\pi)}{RR(r_p,\pi)} \left(\frac{n_r}{n_d} \right)^2 - 2\frac{DR(r_p,\pi)}{RR(r_p,\pi)} \left(\frac{n_r}{n_d} \right),
\label{eq:LS2}
\end{equation}
where the ratio $n_r/n_d$ is given by
\begin{equation}
\frac{n_r}{n_d} = \frac{\sum^{N_r}_iw_i}{\sum^{N_d}_jw_j}
\end{equation}
and the sums go over all random ($N_r$) and data ($N_d$) galaxies. Here we employ a completeness weighting, $w_i$, where we weight each galaxy by the inverse sky- and magnitude completeness at its area of the sky~(Jones et al., in prep.). 

\begin{figure*}
\begin{center}
\epsfig{file=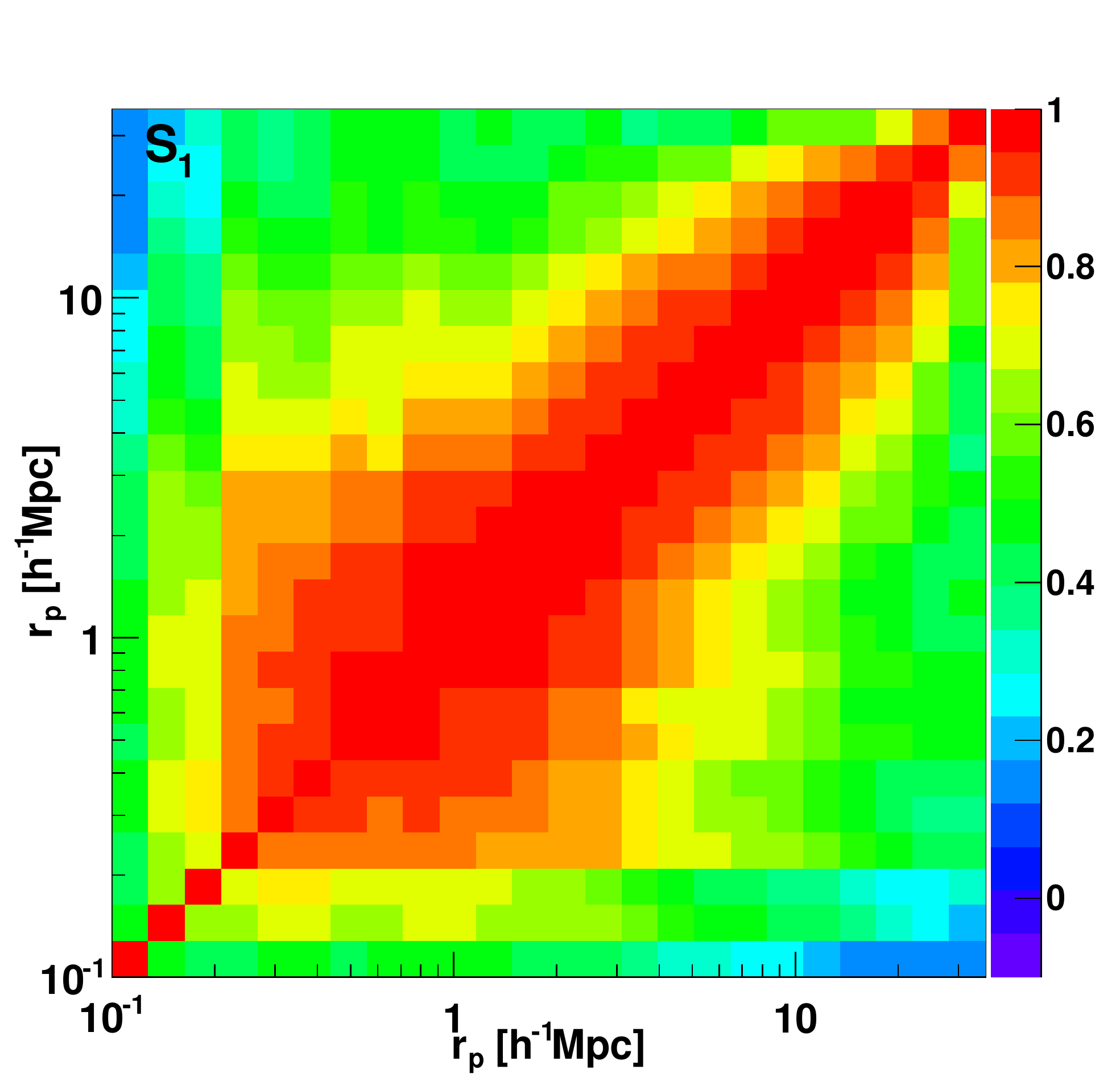,width=7.3cm}
\epsfig{file=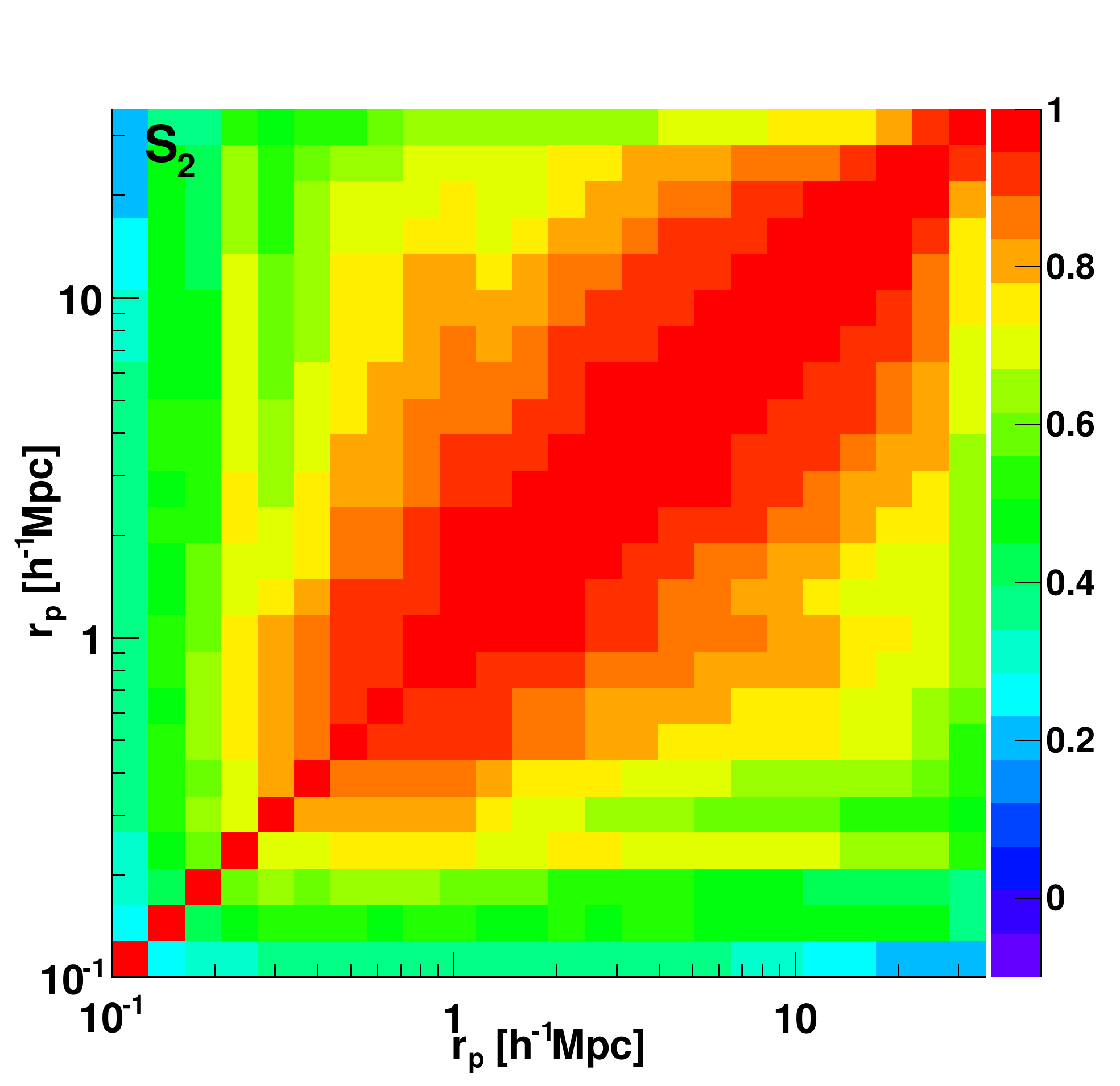,width=7.3cm}\\
\epsfig{file=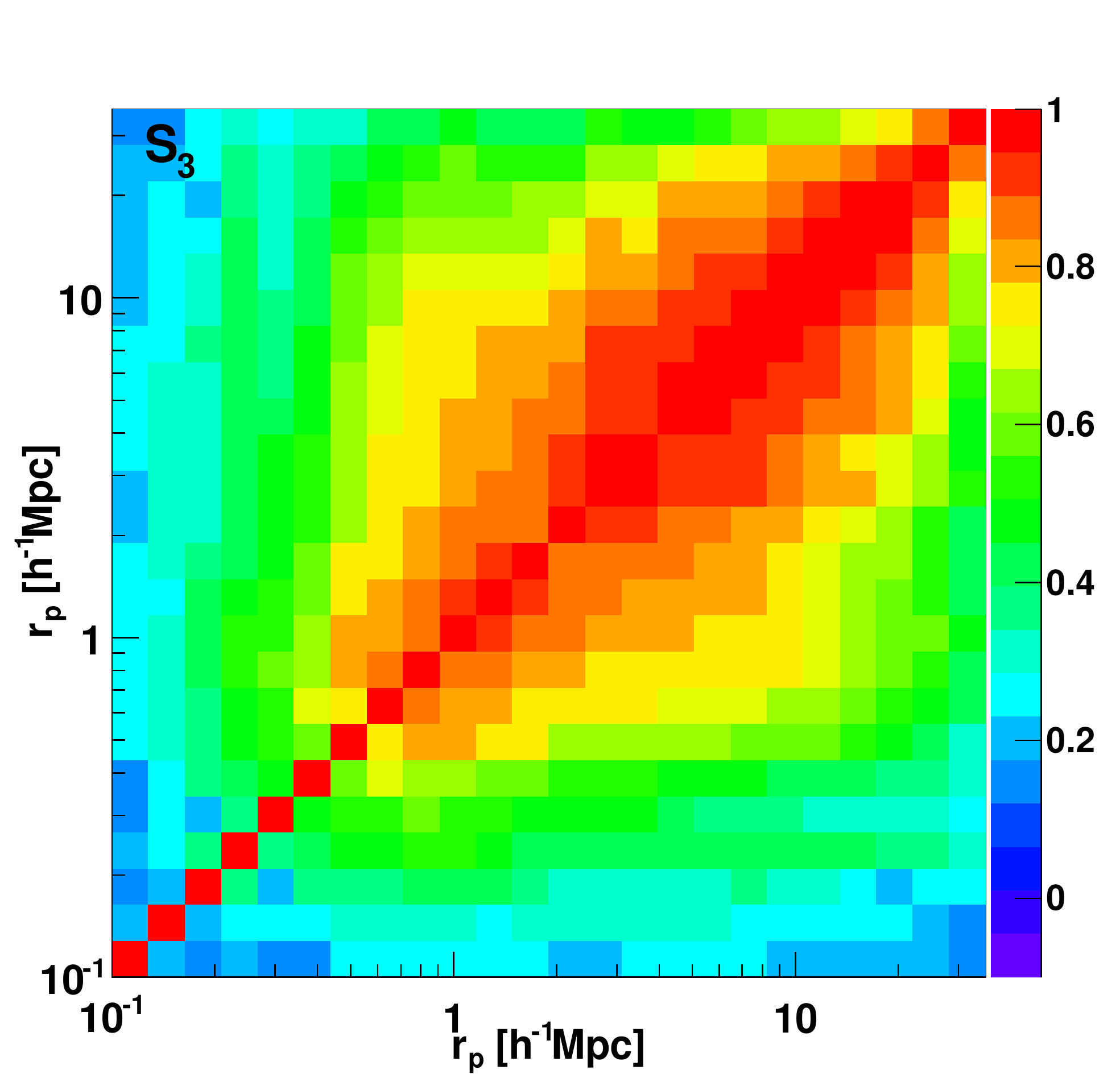,width=7.3cm}
\epsfig{file=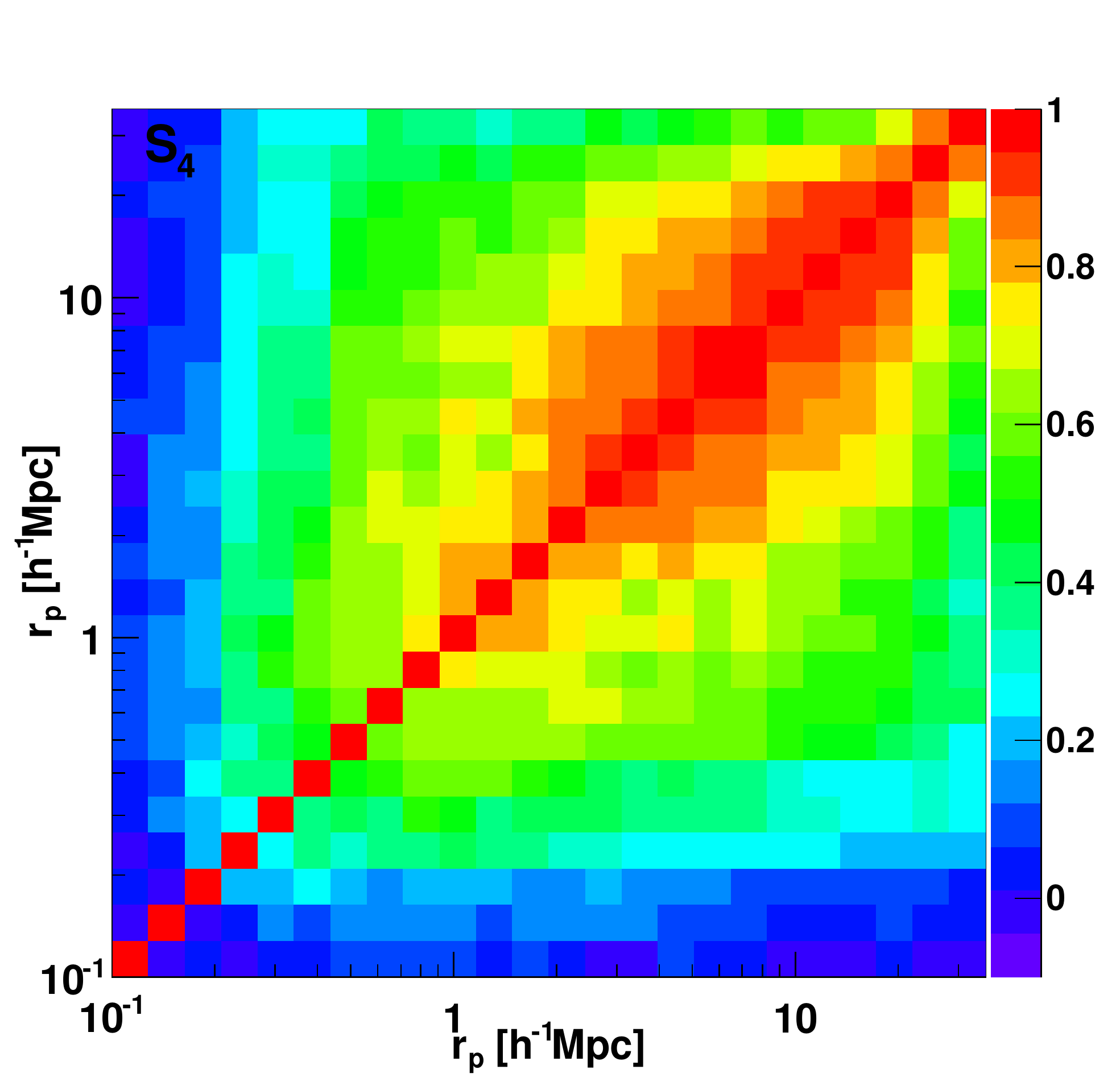,width=7.3cm}
\caption{The correlation matrix for the projected correlation functions, $w_p(r_p)$ derived from the covariance matrices for the four volume-limited 6dFGS sub-sample ($S_1$ - $S_4$) used in this analysis. The colour indicates the correlation with blue indicating no correlation and red indicating high correlation.}
\label{fig:matrix}
\end{center}
\end{figure*}

From the two-dimensional correlation function, $\xi(r_p,\pi)$, we calculate the projected correlation function
\begin{equation}
w_p(r_p) = 2\int^{\infty}_0d\pi\, \xi(r_p,\pi),
\label{eq:wp}
\end{equation}
where we bin $\xi(r_p,\pi)$ in $30$ logarithmic bins from $0.1$ to $100h^{-1}\,$Mpc in $r_p$ and $\pi$. The upper integration limit in eq.~\ref{eq:wp} was chosen to be $\pi^{\rm max} = 50h^{-1}$Mpc for all sub-samples. 

To derive a covariance matrix for the projected correlation function we use the method of jack-knife re-sampling our galaxy samples. First we divide the dataset into $N = 400$ subsets, selected in R.A. and Dec. Each re-sampling step excludes one subset before calculating the correlation function. The covariance matrix is then given by
\begin{equation}
C_{ij} = \frac{(N-1)}{N}\sum^N_{k=1}\left[w^k(r_p^i) - \overline{w}(r_p^i)\right]\left[w^k(r_p^j) - \overline{w}(r_p^j)\right],
\end{equation}
where $w^k(r_p^i)$ is the projected correlation function estimate at separation $r_p^i$ with the exclusion of subset $k$. The mean value is defined as
\begin{equation}
\overline{w}(r_p^i) = \frac{1}{N}\sum^N_{k=1}w^k(r_p^i).
\end{equation}
In Figure~\ref{fig:matrix} we show the four correlation matrices for the different volume-limited sub-samples, which is defined as 
\begin{equation}
r_{ij} = \frac{C_{ij}}{\sqrt{C_{ii} C_{jj}}}
\end{equation}
with $C_{ij}$ being the covariance matrix. The highly correlated part above $r_p \approx 0.1$ - $1h^{-1}\,$Mpc is caused by a cosmic variance dominated error, while on small scales the Poisson dominated error leads to almost uncorrelated bins. With increasing redshift, the number density decreases, which shifts the transition from the Poisson noise dominated part to the sample variance limited part to larger scales.

We also note that wide-angle effects can be neglected in this analysis, since we are interested in small-scale clustering (see~\citet{Beutler:2011hx} and~\citet{Beutler:2012px} for a detailed investigation of wide-angle effects in 6dFGS). 

\subsection{Fibre proximity limitations}
\label{sec:fibre}

The design of the 6dF instrument does not allow fibres to be placed closer than $5.7\,$arcmin~\citep{Jones:2004zy}, which corresponds to a distance of $r_p \approx 0.3h^{-1}\,$Mpc at redshift $z = 0.07$. This limitation is relaxed in 6dFGS where about $70\%$ of the survey area has been observed multiple times. However, for the remaining $30\%$ we have to expect to miss galaxy pairs with a separation smaller than $5.7\,$arcmin.

\begin{figure}
\begin{center}
\epsfig{file=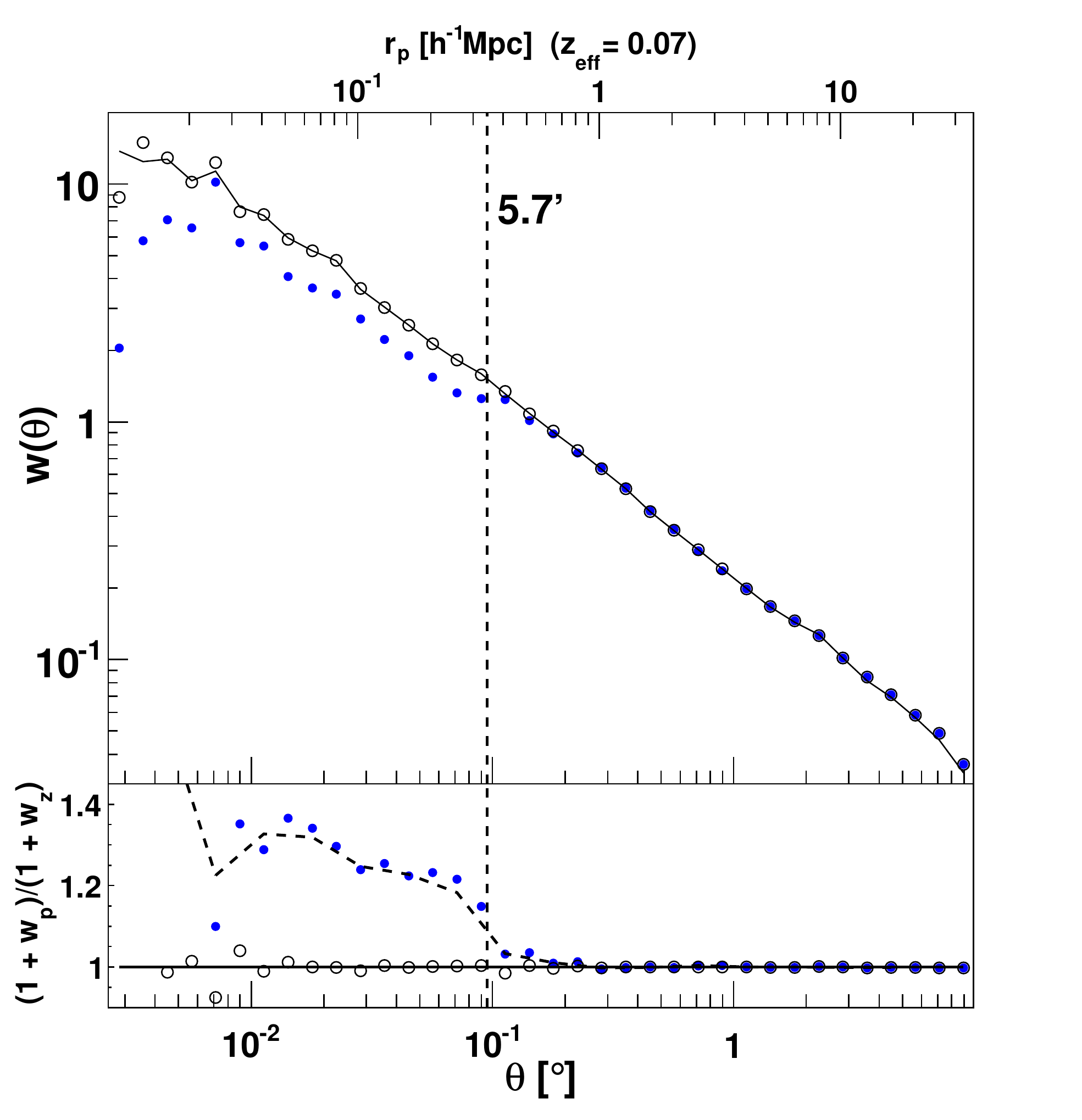,width=8cm}
\caption{The angular correlation function of 6dFGS (blue data points) as well as the target catalogue (black solid line). The deviation between the two angular correlation functions is caused by fibre proximity limitations. The lower panel shows the ratio of the two correlation functions, which can be used as a weight to correct for the fibre proximity limitations (see section~\ref{sec:fibre}). The open data points show the corrected 6dFGS angular correlation function which is in very good agreement with the target catalogue.}
\label{fig:fibre}
\end{center}
\end{figure}

Figure~\ref{fig:fibre} shows the angular correlation function for the 6dFGS redshift catalogue ($w_z$, blue data points) and the target catalogue ($w_p$, solid black line). While the target catalogue contains $104\,785$ galaxies, the redshift catalogue of galaxies which have a $J$-band, $b_J$-band and $r_F$-band magnitude contains $76\,833$. The dashed line indicates the angular scale of the fibres. The two angular correlation functions agree on scales $\theta > 0.1^\circ$, with the redshift catalogue falling below the target catalogue at lower scales.

In the lower panel of Figure~\ref{fig:fibre} we show the ratio $(1+w_p)/(1+w_z)$. The dashed line shows a spline fit to the blue data points, which than can be used to up-weight galaxy pairs with small angular separations and correct for the fibre proximity effect~\citep{Hawkins:2002sg}. This weighting is additional to the completeness weighting $w_i$ we introduced earlier. While the completeness weighting is applied to single data- as well as random galaxies, the fibre proximity weighting is only applied to data galaxy pairs. Applying this weight to the 6dFGS redshift catalogue results in the open data points in Figure~\ref{fig:fibre} which now agree very well with the target catalogue.

\section{Power law fits}
\label{sec:pl}

The projected correlation function can be related to the real-space correlation function $\xi(r)$, using~\citep{Davis:1982gc}
\begin{equation}
\begin{split}
w_p(r_p) &= 2\int^{\infty}_0dy\;\xi\left[(r_p^2 + y^2)^{1/2}\right]\\
&= 2\int^{\infty}_{r_p}r\,dr\,\xi(r)(r^2-r_p^2)^{-1/2}.
\end{split}
\label{eq:pro3}
\end{equation}
If the correlation function is assumed to follow a power law, $\xi(r) = (r/r_0)^{\gamma}$, with the clustering amplitude $r_0$ and the power law index $\gamma$, this can be written as
\begin{equation}
w_p(r_p) = r_p\left(\frac{r_p}{r_0}\right)^{-\gamma}\Gamma\left(\frac{1}{2}\right)\Gamma\left(\frac{\gamma-1}{2}\right)/\Gamma\left(\frac{\gamma}{2}\right),
\label{eq:pl}
\end{equation}
with $\Gamma$ being the Gamma-function. Using this equation we can infer the best-fit power law for $\xi(r)$ from $w_p(r_p)$. 

\begin{table}
\begin{center}
\caption{Best fitting parameters, $r_0$ and $\gamma$ for power law fits to the projected correlation functions $w_p(r_p)$ of the four volume-limited 6dFGS sub-samples ($S_1$ - $S_4$). The fitting range in all cases is $0.1 < r_p < 40\,h^{-1}$Mpc with $24$ bins and $2$ free parameters. The last column shows the reduced $\chi^2$ indicating the goodness of the fit.}
\vspace{0.4cm}
\begin{tabular}{cccc}
\hline
sample & $r_0$ [$h^{-1}$Mpc] & $\gamma$ & $\chi^2/\rm d.o.f.$\\
\hline
$S_1$ & $5.14\pm0.23$ & $1.849\pm0.025$ & $23.5/(24-2) = 1.07$\\
$S_2$ & $5.76\pm0.17$ & $1.826\pm0.019$ & $19.5/(24-2) = 0.89$\\
$S_3$ & $5.76\pm0.16$ & $1.847\pm0.019$ & $32.6/(24-2) = 1.48$\\
$S_4$ & $6.21\pm0.17$ & $1.846\pm0.019$ & $35.8/(24-2) = 1.63$\\
\hline
\hline
\end{tabular}
\label{tab:pls}
\end{center}
\end{table}

\begin{figure}[p]
\begin{center}
\epsfig{file=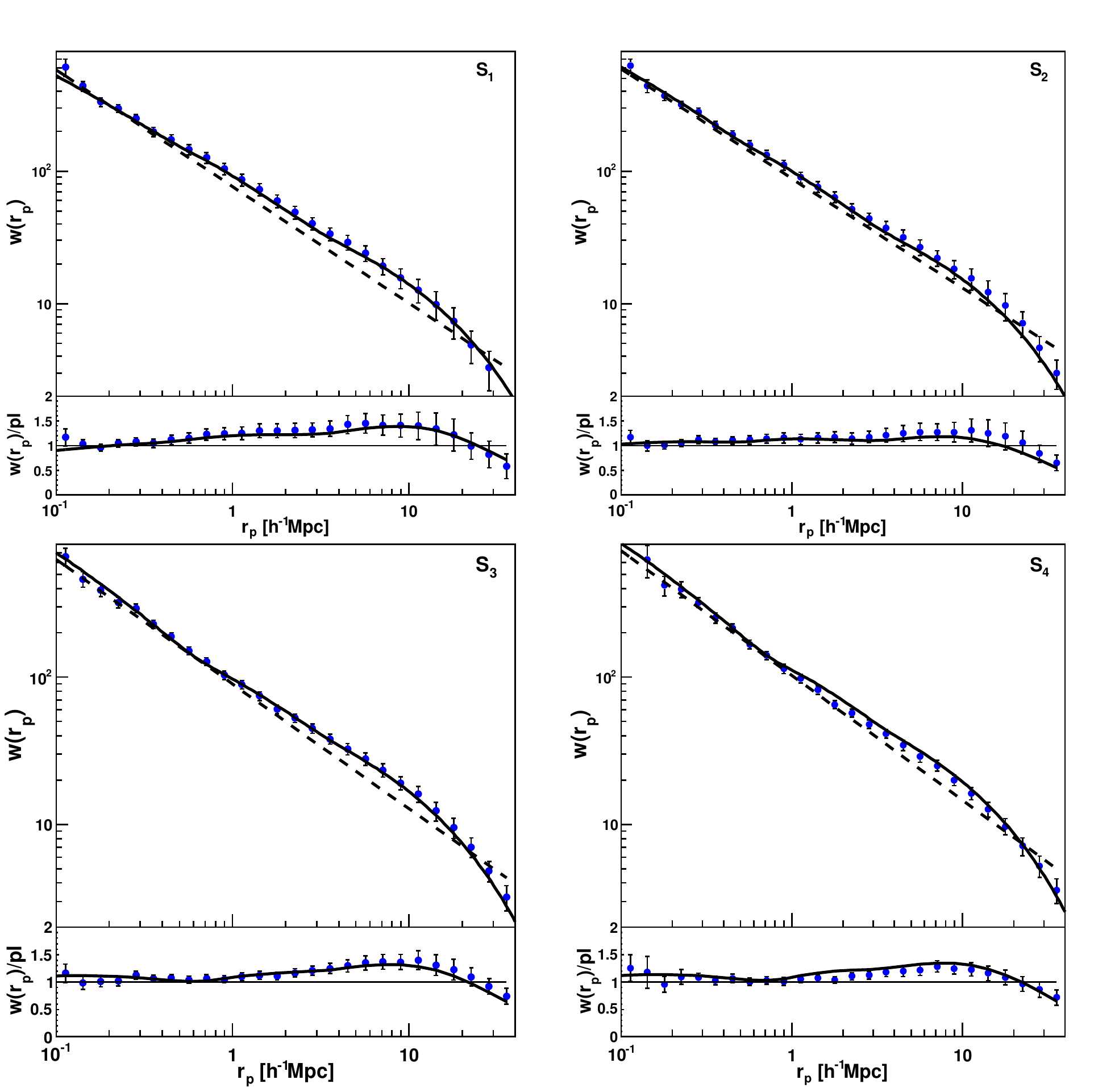,width=16cm}
\caption{The projected correlation function for the four different volume-limited 6dFGS sub-samples ($S_1$ - $S_4$). The dashed black lines show the best fitting power laws (see Table~\ref{tab:pls}) while the solid black lines show the best fitting HOD models, derived by fitting the data between $0.1 \leq r_p \leq 40h^{-1}\,$Mpc. The lower panel shows the data and HOD models divided by the best fitting power laws.}
\label{fig:xi_hod_all}
\end{center}
\end{figure}

Table~\ref{tab:pls} summarises the results of the power law fits to the four 6dFGS sub-samples. The best fitting power laws are also included in Figure~\ref{fig:xi_hod_all} together with the four projected correlation functions. We set the fitting range to be $0.1 < r_p < 40\,h^{-1}$Mpc which includes $24$ bins. Although 6dFGS has very good statistics at scales smaller than $0.1h^{-1}\,$Mpc we do not use them for our fits, since at such scales the fibre proximity correction becomes more than $30\%$ (see section~\ref{sec:fibre} and Figure~\ref{fig:fibre}). 

The value of the clustering amplitude, $r_0$, increases from $5.14$ to $6.21h^{-1}\,$Mpc with increasing stellar mass, while there doesn't seem to be a clear trend in $\gamma$, which varies around the value of $\gamma = 1.84$. The reduced $\chi^2$ in the last column of Table~\ref{tab:pls} indicates a good fit to the data for the first two sub-samples but grows to $\chi^2/\rm d.o.f. = 1.48$ and $1.63$ for the high stellar mass sub-samples, indicating deviations from a power law. The lower panels of Figure~\ref{fig:xi_hod_all} show the different projected correlation functions divided by the best fitting power law (blue data points). Here we can see that the deviations show systematic patterns. Such effects could be related to the strong correlations between bins as seen in Figure~\ref{fig:matrix}. Nevertheless, these patterns can be addressed with a full HOD analysis, which we will pursue in the next section. We will compare the result of our power law fits with other studies in a more detailed discussion in section~\ref{sec:results}.

We remark at this point, that there is no satisfying theoretical model that predicts a power law behaviour of the correlation function, and hence the motivation of such a fit can only be empirical. While alternative approaches (e.g. the HOD model) rely on assumptions about the clustering behaviour of galaxies and dark matter, they are physically motivated and hence allow us to learn more about galaxy clustering than a pure empirical power law fit.

\section{Theory: Halo Occupation Distribution}
\label{sec:hod}

The Halo Occupation Distribution (HOD) model describes the relation between galaxies and mass in terms of the probability distribution $p(N|M)$ that a halo of virial mass $M$ contains $N$ galaxies of a given type. Knowing how galaxies populate dark matter halos, we can use a dark matter correlation function and infer the galaxy correlation function. We use CAMB~\citep{Lewis:2002ah} to derive a model matter power spectrum which we turn into a correlation function using a Hankel transform
\begin{equation}
\xi(r) = \frac{1}{2\pi^2}\int^{\infty}_0dkP(k)k^2\frac{\sin(kr)}{kr}.
\end{equation}
The underlying cosmological model is fixed to ($\Omega_bh^2$, $\Omega_ch^2$, $n_s$, $\sigma_8$) $=$ ($0.02227$, $0.1116$, $0.966$, $0.8$) as reported in~\citet{Komatsu:2010fb} (we set $\sigma_8=0.9$ for one special case). We have to be aware of the fact that the fitted HOD parameters depend somewhat on the assumed values of $\Omega_m$ and $\sigma_8$ and hence the absolute values of the HOD parameters could be biased, if the assumed cosmology is wrong. However in this study we focus on the relative HOD parameters for different stellar mass selected sub-samples, which is fairly robust against such uncertainties.

Here we employ an analytic HOD methodology that is similar to that of~\citet{Zheng:2007zg,Blake:2007xp} and~\citet{Zehavi:2010bh}. We utilise analytic approximations for the halo mass function~\citep{Tinker:2008ff}, the biased clustering of halos~\citep{Tinker:2004gf}, the profile of dark matter within halos~\citep{Navarro:1996gj}, the concentration of dark matter halos as a function of mass~\citep{Duffy:2008pz} and the nonlinear dark matter power spectrum~\citep{Smith:2002dz}. In the next section we will discuss the basic inputs for our HOD model before discussing the HOD formalism itself.

\subsection{Halo mass function and the halo-mass relation}

The spatial density of halos as a function of their mass $M$ is specified by the halo mass function  
\begin{equation}
\frac{dn}{d\ln(M)} = f(\sigma)\rho_m\left|\frac{d\ln(\sigma)}{dM}\right|,
\label{eq:mass1}
\end{equation}
where $\sigma$ is the variance of the linear density field, when it is smoothed with a top-hat filter of mass scale $M=\frac{4}{3}\pi R^3\rho_m$ with the matter density $\rho_m = \Omega_m\rho_c$ and $\rho_c = 3H_0^2/8\pi G$ being the critical density of the Universe. $G$ is the gravitational constant.

In linear theory we can calculate $\sigma$ by
\begin{equation}
\sigma^2(M,z) = \frac{D(z)^2}{2\pi^2}\int^{\infty}_0 P(k)\hat{W}^2_M(k)k^2dk,
\end{equation}
where $P(k)$ is a linear model power spectrum at redshift $z = 0$, $D(z)$ is the linear growth factor normalised to unity at $z = 0$ and $\hat{W}_M(k)$ is the Fourier transform of the real-space top-hat window function at radius R
\begin{equation}
\hat{W}_M(k) = \frac{3}{(kR)^3}\left[\sin(kR) - kR\cos(kR)\right].
\end{equation}
To a first approximation, the function $f(\sigma)$ is universal, and the effects of power spectrum shape, redshift evolution, and background cosmological model (e.g. $\Omega_m$ and $\Omega_{\Lambda}$) enter only through determining $|d\ln\sigma/dM|$ and $\rho_m$. Using N-body simulations~\citet{Tinker:2008ff} found 
\begin{equation}
f(\sigma) = A\left[\left(\frac{\sigma}{b}\right)^{-a} + 1\right]e^{-c/\sigma^2},
\label{eq:mass2}
\end{equation}
with the best fitting values $A = 0.1086$, $a = 1.47$, $b = 2.57$, $c = 1.19$ for $z = 0$ halos, defined to be spherical regions centred on density peaks enclosing a mean interior over-density of $200$ times the cosmic mean density $\rho_m$. 

For a given peak height $\nu = \delta_c/\sigma(M,z)$ (with $\delta_c = 1.686$ being the linear over-density at which a spherical perturbation collapses), we can also calculate the halo bias $b(M)$, where we use the fitting function proposed by~\citet{Tinker:2010my}
\begin{equation}
b(M) = 1-A\frac{\nu^a}{\nu^a+\delta_c^a}+B\nu^b+C\nu^c
\label{eq:bias}
\end{equation}
with $A = 1.0000597$, $a = 0.13245$, $B = 0.183$, $b=1.5$, $C = 0.26523$ and $c = 2.4$. 

\subsection{Halo density profile}

N-body simulations provide many predictions about structure formation, one of which is the principal prediction that dark matter halos should have a cuspy density profile, on average well represented by the two-parameter NFW profile~\citep{Navarro:1996gj}
\begin{equation}
\rho(r, M) = \dfrac{\rho_s}{(r/r_{s})(1+r/r_{s})^2},
\label{eq:NFW}
\end{equation}
where $r_s$ is the characteristic scale radius and $\rho_s$ is given by
\begin{equation}
\rho_s = \frac{M}{4\pi r^3_s\left[\ln(1+c)-c/(1+c)\right]}
\label{eq:rhos}
\end{equation}
with the concentration parameter $c = c(M,z)$.
The profile is truncated at the virial radius, which for a given halo mass $M$ is defined by
\begin{equation}
r_{\rm vir} = \left(\frac{3M}{4\pi\Delta\overline{\rho}}\right)^{1/3},
\label{eq:rvir}
\end{equation}
with $\Delta = 200$.

The relation between the concentration parameter, the core radius $r_s$ and the virial radius $r_{\rm vir}$ is $c = r_{\rm vir}/r_s$. In this analysis we use the halo concentration mass relation from~\citet{Duffy:2008pz}
\begin{equation}
c(M,z) = 6.71\left(\frac{M}{M_{\text{pivot}}}\right)^{-0.091}(1+z)^{-0.44}
\label{eq:duffy}
\end{equation}
with $M_{\text{pivot}} = 2\times10^{12}h^{-1}M_{\odot}$. We tried replacing eq.~\ref{eq:duffy} with the form suggested by~\citet{Bullock:1999he}, and found that the best-fitting HOD parameters changed by much less than the statistical errors.

\subsection{HOD framework and formalism}

In the HOD parametrisation it is common to separate the clustering contributions from the most massive galaxies, which are assumed to sit in the halo centre, from satellite galaxies. This picture of how galaxies populate halos is supported by hydrodynamic simulations (e.g.~\citealt{Berlind:2002rn, Simha:2008hd}) and semi-analytic models (e.g.~\citealt{White:1991mr,Kauffmann:1993gv,Croton:2005fe,Bower:2005vb}). The mean central and satellite number density of galaxies that populate dark matter halos of mass $M$ is~\citep{Zheng:2004id}
\begin{equation}
\begin{split}
\langle N_{c}(M)\rangle &= \begin{cases} 0 & \text{ if }\;M < M_{\rm min}\cr
								1 & \text{ if }\;M \geq M_{\rm min}\end{cases},\\
\langle N_{s}(M)\rangle &= \left(\frac{M}{M_1}\right)^{\alpha},
\label{eq:hod}
\end{split}
\end{equation}
where $M_{\rm min}$ is the minimum dark matter halo mass which can host a central galaxy, $M_1$ corresponds to the mass of halos that contain, on average, one additional satellite galaxy ($\langle N_s(M_1)\rangle = 1$) and $\alpha$ sets the rate at which halos accumulate satellites when growing in mass. In very massive halos the number of satellites is proportional to halo mass $M$ to the power of $\alpha$. The total HOD number is given by
\begin{equation}
\langle N_{t}(M)\rangle = \langle N_{c}(M)\rangle \left[1 + \langle N_{s}(M)\rangle \right],
\end{equation}
so that a dark matter halo can only host a satellite galaxy if it contains already a central galaxy. In our model we assume a step like transition from $\langle N_{c}(M)\rangle = 0$ to $\langle N_{c}(M)\rangle = 1$. In reality this is more likely to be a gradual transition with a certain width $\sigma_{\rm log M}$~\citep{More:2008za}. To account for this, other studies (e.g.~\citealt{Zehavi:2010bh}) modify the HOD parametrisation of the central halo term to
\begin{equation}
\langle N_{c}(M)\rangle = \frac{1}{2}\left[1 + \text{erf}\left(\frac{\log_{10}(M) - \log_{10}(M_{\rm min})}{\sigma_{\rm log M}}\right)\right],
\label{eq:Nc2}
\end{equation}
which turns into eq.~\ref{eq:hod} for the case $\sigma_{\rm log M} = 0$. For our data we found that the reduction in $\chi^2$ obtained from fits including $\sigma_{\rm log M}$ as an additional parameter is not big enough to justify this parameterisation. For our largest sub-sample (S2) we found $\Delta\chi^2 = -0.78$, while a new parameter would be justified when $\Delta\chi^2 < -2$ following the Akaike information criterion~\citep{Akaike:1974}. 

There are also higher parameter models for the satellite fraction such as
\begin{equation}
\langle N_{s}(M)\rangle = \left(\frac{M-M_0}{M_1}\right)^{\alpha},
\end{equation}
where $M_0$ is the minimum halo mass at which satellites can exist. Again we tested this model and found that the reduction in $\chi^2$ does not justify this additional parameter. Hence we chose the two parameter model ($M_1$, $\alpha$) of eq~\ref{eq:hod}\footnote{$M_{\rm min}$ is fixed by the number density as will be explained in section~\ref{sec:twoh}}. 

Within the halo model we can account separately for the clustering amplitude of galaxies which sit in the same dark matter halo (one halo term) and galaxies which sit in different dark matter halos (two halo term). At small scales the clustering will be dominated by the one halo term and at large scales it will be dominated by the two halo term. For the correlation function this can be written as
\begin{equation}
\xi(r) = \xi_{1h}(r) + \xi_{2h}(r).
\end{equation}
where $\xi_{1h}(r)$ and $\xi_{2h}(r)$ represent the one halo and two halo terms respectively.

\subsection{The 1-halo term, $\xi_{1h}(r)$}

We separate the 1-halo term into contributions from central-satellite galaxy pairs and satellite-satellite galaxy pairs. The central-satellite contribution is given by
\begin{equation}
\xi_{1h}^{c-s}(r) = \frac{2}{n_g^2}\int^{\infty}_{M_{\rm vir}(r)} dM \frac{dn(M)}{dM}N_{c}(M)N_{s}(M)\frac{\rho(r,M)}{M},
\end{equation}
where $n_g$ is the galaxy number density and $\rho(r,M)$ is the halo density profile~(see eq.~\ref{eq:NFW}). The lower limit for the integral is the virial mass $M_{\rm vir}(r)$ corresponding to the virial separation $r_{\rm vir}$ (see eq.~\ref{eq:rvir}).\\
The satellite-satellite contribution is usually given by a convolution of the halo density profile with the halo mass function. Here we calculate this term in k-space since a convolution then turns into a simple multiplication
\begin{equation}
P_{1h}^{s-s}(k) = \frac{1}{n^2_g}\int^{\infty}_0dM \frac{dn(M)}{dM}N_{c}(M)N_{s}^2(M)\lambda(k,M)^2,
\label{eq:ssterm}
\end{equation} 
where $\lambda(k,M)$ is the normalised Fourier transform of the halo density profile $\rho(r,M)$ which is given by
\begin{equation}
\lambda(k,M) = \frac{4\pi}{M}\int_0^{\infty}dr\rho(r,M)r^2\frac{\sin(kr)}{kr}.
\end{equation}
The real-space expression of eq.~(\ref{eq:ssterm}) is
\begin{equation}
\xi_{1h}^{s-s}(r) = \frac{1}{2\pi^2}\int^{\infty}_0dk\,P_{1h}^{s-s}(k)k^2\frac{\sin(kr)}{kr},
\end{equation}
which can then be combined with the central-satellite contribution to obtain  the 1-halo term
\begin{equation}
\xi_{1h}(r) = \xi_{1h}^{c-s}(r) + \xi_{1h}^{s-s}(r).
\end{equation}

\subsection{The 2-halo term, $\xi_{2h}(r)$}
\label{sec:twoh}

The 2-halo term, $\xi_{2h}(r)$, can be calculated from the dark matter correlation function since on sufficiently large scales the galaxy and matter correlation function are related by a constant bias parameter. We calculate the two halo term in Fourier space as
\begin{equation}
\begin{split}
P_{2h}(k,r) = &P_m(k)\times\\
              &\left[\int^{M_{\rm lim}(r)}_0 dM\frac{dn(M)}{dM}b_h(M,r)\frac{N_{t}(M)}{n'_g(r)}\lambda(k,M)\right]^2,
\end{split}
\label{eq:term2}
\end{equation}
where $P_m(k)$ is the non-linear model power spectrum from CAMB including halofit~\citep{Smith:2002dz}, $M_{\rm lim}(r)$ is the halo mass limit for which we can find galaxy pairs with a separation larger than $r_{\rm vir}$ and $b_h(M,r)$ is the scale dependent halo bias at separation $r$ for which we assume the following model~\citep{Tinker:2004gf}
\begin{equation}
b_h^2(M,r) = b^2(M)\frac{\left[1+1.17\xi_m(r)\right]^{1.49}}{\left[1+0.69\xi_m(r)\right]^{2.09}},
\end{equation}
where $\xi_m(r)$ is the non-linear matter correlation function with $b(M)$ given by eq.~\ref{eq:bias}.

We define the restricted galaxy number density $n'_g(r)$ at separation $r$  
\begin{equation}
n'_g(r) = \int^{M_{\rm lim}(r)}_0 dM\frac{dn(M)}{dM}N_{t}(M).
\label{eq:ndash}
\end{equation}
For smaller separations $r$, only very small mass halos contribute to the 2-halo term because high mass halos have a virial radius $r_{\rm vir}$ which is larger than the galaxy separation, and hence the galaxy sits in the same halo. This mass limit is already given by $M_{\rm lim}(r)$ and can be derived using the $``n'_g$-matched$"$ approximation described by~\citet{Tinker:2005na}. First we calculate the restricted number density using 
\begin{equation}
\begin{split}
n'^2_g(r) = &\int^{\infty}_0dM_1\frac{dn(M_1)}{dM_1}N_{t}(M_1)\times \\
&\int^{\infty}_0dM_2\frac{dn(M_2)}{dM_2}N_{t}(M_2)P(r,M_1,M_2),
\end{split}
\end{equation}
where $p(r,M_1,M_2)$ quantifies the probability of non-overlapping halos of masses $M_1$ and $M_2$ with separation $r$. Defining $x = r/(r_{\rm vir,1} + r_{\rm vir,2})$, where $r_{\rm vir,1}$ and $r_{\rm vir,2}$ are the virial radii corresponding to masses $M_1$ and $M_2$ (see eq.~\ref{eq:rvir}). Using $y = (x-0.8)/0.29$, $p(y)$ is then given by 
\begin{equation}
p(y) = \begin{cases}1 &\text{ if }\; y > 1\\
0 &\text{ if }\; y < 0\\
3y^2-2y^3 &\text{ otherwise}.
\end{cases} 
\end{equation}
Given the value of $n'_g(r)$, we increase $M_{\rm lim}(r)$ in equation~\ref{eq:ndash} to produce a number density which matches the measured $n_g$. This value is then used in equation~\ref{eq:term2}.

We obtain the correlation function for the 2-halo term with 
\begin{equation}
\xi'_{2h}(r) = \frac{1}{2\pi^2}\int^{\infty}_0dk\,P_{2h}(k,r)k^2\frac{\sin(kr)}{kr}.
\end{equation}
Now we have to correct this term from the restricted galaxy density to the entire galaxy population which leads us to the final two halo term
\begin{equation}
1+\xi_{2h}(r) = \left(\frac{n'_g(r)}{n_g}\right)^2\left[1+\xi'_{2h}(r)\right].
\end{equation}
So far our HOD model has three free parameters: $M_{\rm min}$, $M_1$ and $\alpha$. However we use the galaxy number density to fix $M_{\rm min}$ for each set of $M_1$ and $\alpha$ via
\begin{align}
n_g &= \int^{\infty}_0 dM\frac{dn(M)}{dM}N_{t}(M)\\
&= \int^{\infty}_{M_{\rm min}} dM\frac{dn(M)}{dM}N_{t}(M).
\end{align}
We can now calculate the real-space correlation function, $\xi(r) = \left[1 + \xi_{1h}(r)\right] + \xi_{2h}(r)$ for any set of $M_1$ and $\alpha$, which we then turn into the projected correlation function, $w_p(r_p)$, using eq.~\ref{eq:pro3}.

\subsection{Derived quantities}
\label{sec:derived}

Since the HOD model is directly based on a description of dark matter clustering and its relation to galaxy clustering, an HOD model can tell us much more about a galaxy population, than just the two parameters $M_1$ and $\alpha$. 

For $\lambda(k,M) \rightarrow 1$ (which corresponds to large separations $r$ in real-space) the two halo term simplifies to
\begin{equation}
P_{2h}(k,r) \approx b^2_{\rm eff}P_m(k),
\end{equation}
where the effective bias is the galaxy number weighted halo bias factor~\citep{Tinker:2004gf}
\begin{equation}
b_{\rm eff} = \frac{1}{n_g}\int_0^{\infty}dM\frac{dn(M)}{dM}b_h(M)N_{t}(M).
\label{eq:beff}
\end{equation}
We can also ask what is the average group dark matter halo mass for a specific set of galaxies (often called host-halo mass). Such a quantity can be obtained as
\begin{equation}
M_{\rm eff} = \frac{1}{n_g}\int^{\infty}_0 dM\frac{dn(M)}{dM}MN_{t}(M),
\label{eq:Meff}
\end{equation} 
which represents a weighted sum over the halo mass function with the HOD number as a weight. 

The averaged ratio of satellite galaxies to the total number of galaxies is given by 
\begin{equation}
f_{s} = \frac{\int^{\infty}_0 dM \frac{dn(M)}{dM} N_{s}(M)}{\int^{\infty}_0 dM \frac{dn(M)}{dM} N_{c}(M)\left[1 + N_{s}(M)\right]}
\end{equation}
and the central galaxy ratio is given by $f_c = 1-f_s$.

\section{HOD parameter fits}
\label{sec:hodfits}

\begin{sidewaystable}
\begin{scriptsize}
\begin{center}
\caption{Summary of the best fitting parameters for HOD fits to the four 6dFGS sub-samples ($S_1$ - $S_4$) with the fitting range $0.1 < r_p < 40h^{-1}\,$Mpc. The last column shows the reduced $\chi^2$ derived from a fit to $24$ bins with $2$ free parameters and indicates good fits for all sub-samples. Error bars on the HOD parameters correspond to $1\sigma$, derived from the marginalised distributions. The sample $S_2'$ uses a different fitting range of $0.1 < r_p < 20h^{-1}\,$Mpc for the largest of our sub-samples, $S_2$. Furthermore we include a special fit to $S_2$ labeled $S_2^{\sigma_8=0.9}$, where we change our standard assumption of $\sigma_8=0.8$ to $\sigma_8=0.9$. The satellite fraction $f_s$, the effective dark matter halo mass $M_{\rm eff}$ and the effective galaxy bias $b_{\rm eff}$ are derived parameters (see section~\ref{sec:derived}). The bias depends on our initial assumption of $\sigma_8$ and hence this parameters should be treated as $b_{\rm eff}\times (\sigma_8/0.8)$.}
\vspace{0.4cm}
\begin{tabular}{cccccccc}
\hline
sample & $\log_{10}\left(\frac{M_1}{h^{-1}M_{\odot}}\right)$ & $\alpha$ & $\log_{10}\left(\frac{M_{\rm min}}{h^{-1}M_{\odot}}\right)$ & $f_{s}$ & $\log_{10}\left(\frac{M_{\rm eff}}{h^{-1}M_{\odot}}\right)$ & $b_{\rm eff}$ & $\chi^2/\rm d.o.f.$\\
\hline
$S_1$ & $13.396\pm0.017$ & $1.214\pm0.031$ & $12.0478\pm0.0049$ & $0.2106\pm0.0078$ & $13.501\pm0.015$ & $1.2704\pm0.0087$ & $13.5/(24-2) = 0.61$\\
$S_2$ & $13.568\pm0.013$ & $1.270\pm0.027$ & $12.2293\pm0.0043$ & $0.1879\pm0.0062$ & $13.532\pm0.012$ & $1.3443\pm0.0063$ & $16.6/(24-2) = 0.75$\\
$S_3$ & $13.788\pm0.012$ & $1.280\pm0.029$ & $12.4440\pm0.0039$ & $0.1578\pm0.0061$ & $13.546\pm0.012$ & $1.4002\pm0.0062$ & $21.4/(24-2) = 0.97$\\
$S_4$ & $14.022\pm0.011$ & $1.396\pm0.033$ & $12.6753\pm0.0032$ & $0.1243\pm0.0058$ & $13.617\pm0.012$ & $1.5170\pm0.0077$ & $24.4/(24-2) = 1.11$\\
\hline
$S_2^{\sigma_8=0.9}$ & $13.605\pm0.013$ & $1.193\pm0.023$ & $12.2382\pm0.0048$ & $0.2034\pm0.0077$ & $13.649\pm0.014$ & $1.2151\pm0.0065$ & $19.6/(24-2) = 0.89$\\
$S_2'$ & $13.568\pm0.013$ & $1.282\pm0.027$ & $12.2291\pm0.0042$ & $0.1890\pm0.0074$ & $13.542\pm0.010$ & $1.3492\pm0.0065$ & $19.1/(21-2) = 1.01$\\
\hline
\hline
\end{tabular}
\label{tab:results}
\end{center}
\end{scriptsize}
\end{sidewaystable}

To compare the HOD models to the data we use the same fitting range as for the power law fits earlier ($0.1 < r_p < 40\,h^{-1}\,$Mpc). The free parameters of the fit are $M_1$ and $\alpha$ as described in the last section and hence we have the same number of free parameters as for the power law fits.

\begin{SCfigure}
\epsfig{file=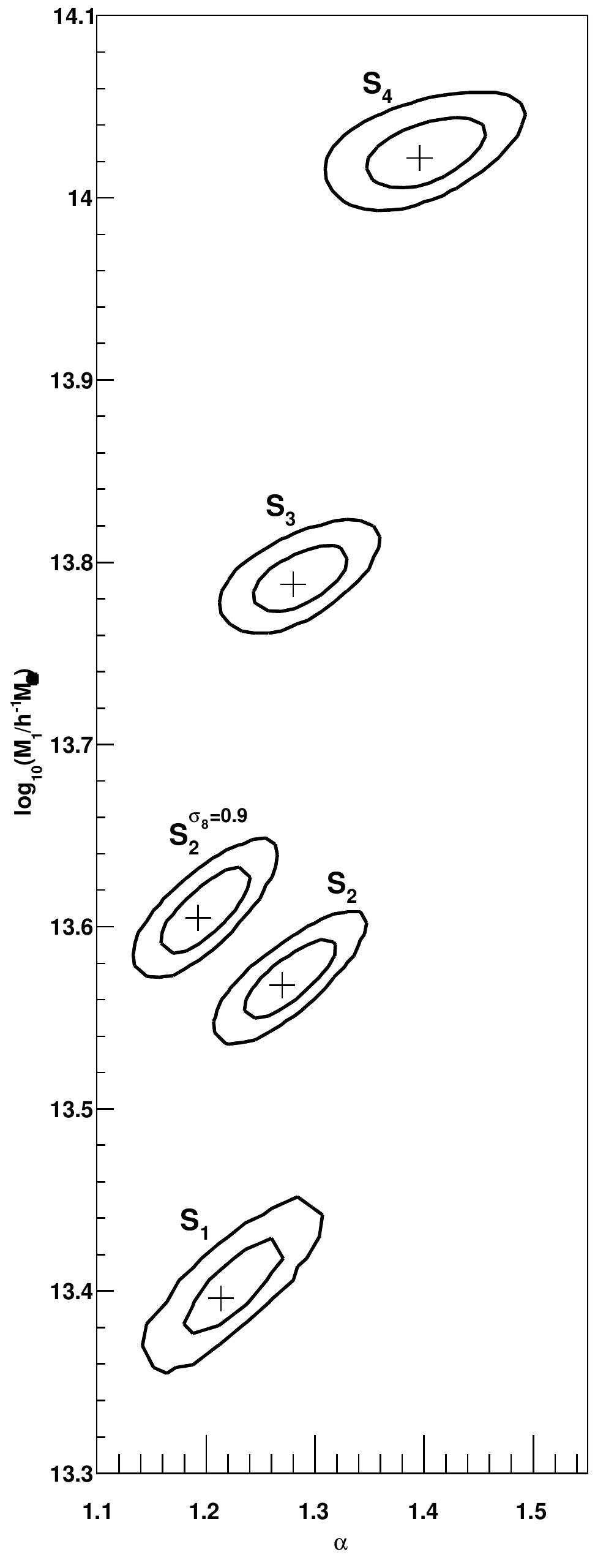,width=8cm}
\caption{Two dimensional probability distribution for $\log_{10}(M_1)$ and $\alpha$ for the HOD fits to the four 6dFGS volume-limited sub-sample ($S_1$ - $S_4$). We include two fits for the largest sample $S_2$, where for $S_2^{\sigma_8=0.9}$ we change our standard assumption of $\sigma_8=0.8$ to $\sigma_8=0.9$. The parameters derived from the fits are summarised in Table~\ref{tab:results}. The best fitting values are marker with black crosses.}
\label{fig:chi2_alpha_M0}
\end{SCfigure}

All the fitting results are summarised in Table~\ref{tab:results} and Figure~\ref{fig:xi_hod_all}. The reduced $\chi^2$ in the last column of Table~\ref{tab:results} indicates a good fit to the data in all cases and the ratio of data to best fitting power law in the lower panels of Figure~\ref{fig:xi_hod_all} shows that the HOD model reproduces the double peak structure present in the data. We also note that the reduced $\chi^2$ in case of the HOD fits is uniformly lower than for the power law fits indicating a better fit to the data for all sub-samples. 

Figure~\ref{fig:chi2_alpha_M0} shows the 2D probability distributions in $M_1$ and $\alpha$ for the four different volume-limited sub-samples. There is a strong trend of increasing $M_1$ with stellar mass and a weaker but still significant trend of increasing $\alpha$ with increasing stellar mass. This indicates that dark matter halos that host a central galaxy with higher stellar mass have their first satellite on average at a larger dark matter halo mass. However the number of satellites increases more steeply with halo mass for samples with higher stellar mass. This trend indicates that halos of higher mass have greater relative efficiency at producing multiple satellites. Similar trends were found in sub-samples of SDSS galaxies by~\citealt{Zehavi:2004ii}. As a test for the sensitivity of the fitting results to the upper fitting limit we performed a fit to sub-sample $S_2$ with the fitting range $0.1 < r_p < 20h^{-1}\,$Mpc. All parameters agree within $1\sigma$ with the fit to the larger fitting range. We called this fit $S_2'$ and included it in Table~\ref{tab:results}.

\begin{figure}[tb]
\begin{center}
\epsfig{file=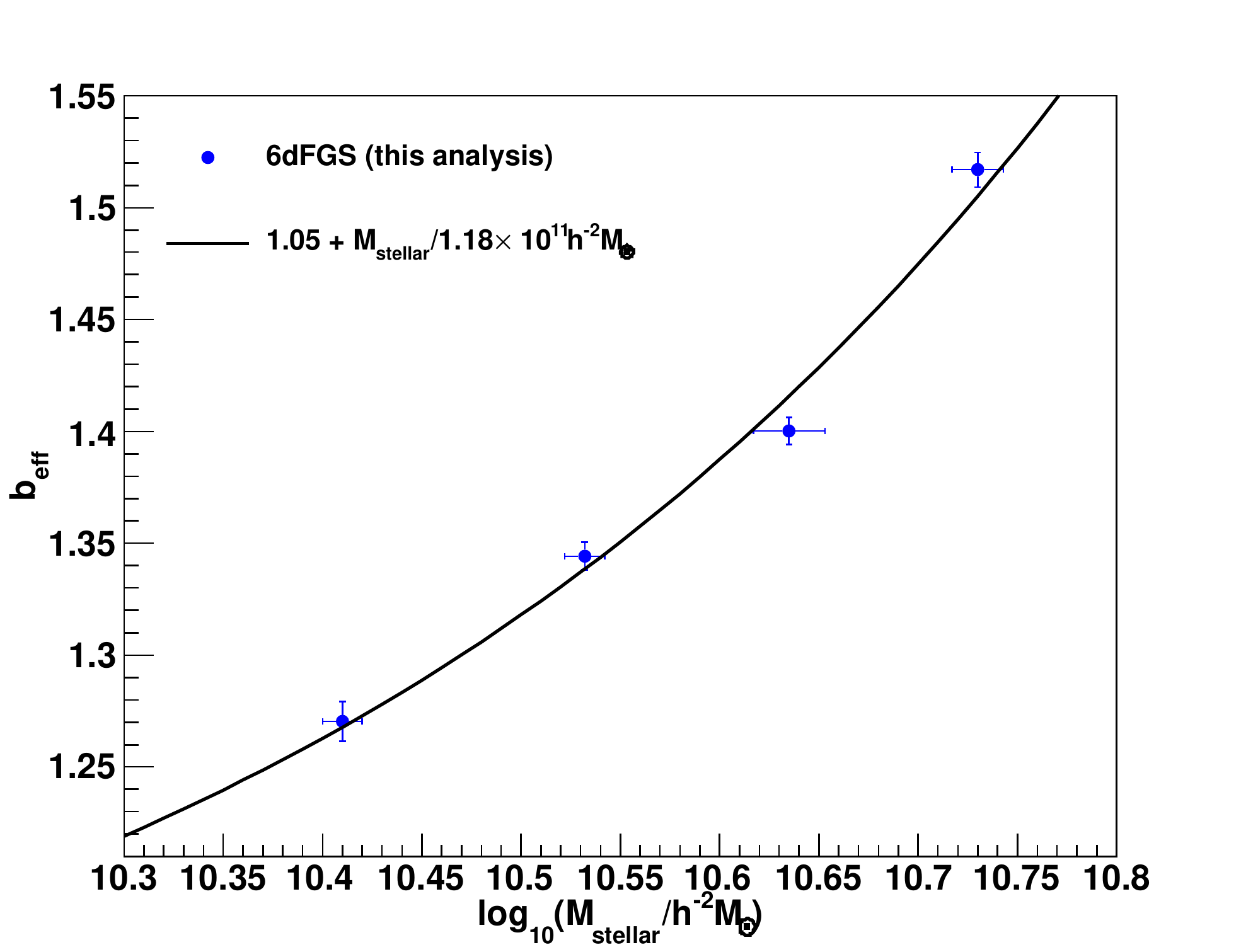,width=10cm}
\caption{The effective galaxy bias (see eq.~\ref{eq:beff}) as a function of log stellar mass for 6dFGS (blue data points). The increase in galaxy bias with stellar mass can be described by eq.~\ref{eq:bias2} which corresponds to the black line.}
\label{fig:beff}
\end{center}
\end{figure}

Table~\ref{tab:results} also includes derived parameters like the minimum dark matter halo mass, $M_{\rm min}$, the satellite fraction, $f_s$, the effective dark matter halo mass, $M_{\rm eff}$ and the effective galaxy bias, $b_{\rm eff}$. The errors on these parameters are calculated as the $68\%$-confidence level of their 1D probability distribution. 
The effective dark matter halo mass seems to be almost constant for all samples while the minimum dark matter halo mass $M_{\rm min}$ increases with stellar mass. The satellite fraction $f_{\rm s}$ decreases with increasing stellar mass, indicating that galaxies with high stellar mass have a higher probability to be central galaxies. The increasing effective galaxy bias indicates that galaxies with higher stellar mass are more strongly clustered and hence reside in high density regions of the Universe. The increasing effective galaxy bias is well described by a power law (in log stellar mass) of the form
\begin{equation}
b(M_{\rm stellar}) = \left(1.05 + M_{\rm stellar}/M_*\right)\times(0.8/\sigma_8)
\label{eq:bias2}
\end{equation} 
with $M_* = 1.18\times 10^{11}h^{-2}M_{\odot}$. We compare this function to the measurements in Figure~\ref{fig:beff}. The absolute stellar masses are subject to significant uncertainties and different methods to derive stellar masses can come to very different conclusions. However, most stellar mass estimates are related by a simple constant offset, and relation~\ref{eq:bias2} can be scaled accordingly. The equation above is only valid for the stellar mass range probed in this analysis ($M_{\rm stellar} = 2.6 - 5.4\times10^{10}h^{-2}\,M_{\odot}$) since the underlying dynamics are most likely not captured in eq.~\ref{eq:bias2}.

We will discuss the implications of all these results in the next sections. First we will derive 6dFGS mock samples using different semi-analytic models. We will then compare the predictions from these semi-analytic models to our data followed by a comparison to other studies.

\section{semi-analytic mock catalogues}
\label{sec:sam}

To compare our result with theory, we derive 6dFGS mock catalogues from two different semi-analytical models~\citep{Croton:2005fe,Bower:2005vb}, both based on the Millennium Simulation~\citep{Springel:2005nw} publicly available through the Millennium Simulation database\footnote{http://gavo.mpa-garching.mpg.de/MyMillennium/\\http://galaxy-catalogue.dur.ac.uk:8080/Millennium/}. Semi-analytic models are based on an underlying N-body simulation together with theoretically and observationally motivated descriptions of gas cooling, star formation and feedback processes.

The Millennium Simulation is a dark matter only N-body simulation which traces the hierarchical evolution of $2160^3$ particles in a periodic box of $500^3 h^{-3}\,$Mpc$^3$ from redshift $z = 127$ to $z = 0$. The underlying cosmological model follows WMAP1 cosmology~\citep{Spergel:2003cb} given by a matter density of $\Omega_m = \Omega_{dm} + \Omega_b = 0.25$, a cosmological constant of $\Omega_{\Lambda} = 0.75$, a Hubble constant of $H_0=75\,$km/s/Mpc, a spectral index of $n_s = 1$ and a r.m.s. of matter fluctuations in $8h^{-1}\,$Mpc spheres of $\sigma_8 = 0.9$. 
The individual particle mass of the simulation is $8.6\times 10^8 h^{-1}M_{\odot}$ and halos and sub-halos are identified from the spatial distribution of dark matter particles using a standard friends-of-friends algorithm and the SUBFIND algorithm~\citep{Springel:2000qu}. All sub-halos are then linked together to construct the halo merger trees which represent the basic input of the semi-analytic models.

Here we are using the $z=0$ output of the Millennium Simulation. To ensure that the stellar masses are calculated in a consistent and comparable way, we follow the following procedure to derive the 6dFGS mock catalogues:
\begin{enumerate}
\item We apply the 6dFGS $K$-band apparent magnitude limits of $8.85 \leq K \leq 12.75$ to the full $500h^{-3}\,$Mpc$^3$ simulation box. We have to use the $K$-band instead of the $J$-band, which is actually used in this analysis, because none of the semi-analytic models provide $J$-band magnitudes. However, the 6dFGS $J$-band and $K$-band samples have significant overlap, meaning that almost all galaxies which have a $K$-band magnitude also have a $J$-band magnitude. We also account for sky- and magnitude incompleteness.
\item We re-calculate stellar masses for each galaxy using the technique described in section~\ref{sec:stellar} but with the corresponding $K$-band relations~\citep{Bell:2000jt}, instead of $J$-band. The re-calculation of the stellar masses ensures that potential disagreement with our measurement is not caused by a different technique of deriving stellar masses or a different assumptions about the IMF. 
\item We apply the same redshift and stellar mass limits to the semi-analytic catalogues, which we used to produce the four volume-limited samples in the 6dFGS dataset (see Table~\ref{tab:vls}).
\end{enumerate}
We found that all mock 6dFGS catalogues derived from these semi-analytic models contain fewer galaxies than the data sample (by about $40\%$). This could be related to the slightly different cosmology used in these simulations, which should have its largest impact on large clusters, which are sampled in 6dFGS. We are not attempting to correct for such differences in the cosmological model. The aim of this part of our analysis is to test the current predictive power of semi-analytic models.

Semi-analytic models are often grouped into $``$Durham models$"$ and $``$Munich models$"$. The~\citet{Bower:2005vb} model belongs in the group of $"$Durham models$"$ while the~\citet{Croton:2005fe} model belongs in the group of $"$Munich models$"$.

\subsection{Durham models}

In the Durham models, merger trees are produced following~\citet{Helly:2003} which are independent of those generated by~\citet{Springel:2005nw}. When the satellite galaxy falls below a certain distance to the central galaxy given by $R_{\rm merge} = r_{c} + r_s$, where $r_c$ and $r_s$ are the half mass radii of the central and satellite galaxy, respectively, the satellite and central galaxy are treated as one. The largest of the galaxies contained within this new combined dark matter halo is assumed to be the central galaxy~\citep{Benson:2001au}, whilst all other galaxies within the halo are satellites. The dynamical friction and tidal stripping which are present in such a system are modelled analytically. These models are based on NFW density profiles for the central halo as well as the satellite halo, while galaxies are modelled as a disc plus spheroid. 

\subsection{Munich models}

The Munich models are based on the original merger trees by~\citet{Springel:2005nw}. One of the key differences between these merger trees and the ones used in the Durham models is that the Munich models explicitly follow dark matter halos even after they are accreted onto larger systems, allowing the dynamics of satellite galaxies residing in the in-falling halos to be followed until the dark matter substructure is destroyed. The galaxy is than assigned to the most bound particle of the sub-halo at the last time the sub-halo could be identified. 

In the Munich models, a two-mode formalism is adopted for active galactic nuclei (AGNs), wherein a high-energy, or $"$quasar$"$ mode occurs subsequent to mergers, and a constant low-energy $"$radio$"$ mode suppresses cooling flows due to the interaction between the gas and the central black hole~\citep{Croton:2005fe}. In the quasar model, accretion of gas onto the black hole peaks at $z\sim 3$, while the radio mode reaches a plateau at $z\sim 2$. AGN feedback is assumed to be efficient only in massive halos, with supernova feedback being more dominant in lower-mass halos.

The galaxies in the~\citet{Croton:2005fe} model can be of three different types: central galaxies (type $0$), satellites of type $1$ and satellites of type $2$. Satellites of type $1$ are associated with dark matter substructures, which usually refers to recently merged halos. Satellites of type $2$ are instead galaxies whose dark matter halo has completely merged with a bigger halo and are not associated with a substructure. We treat both type $2$ and type $1$ as satellite galaxies.

\subsection{Testing semi-analytic models}

In their original paper~\citet{Bower:2005vb} compare the K-band luminosity function, galaxy stellar mass function, and cosmic star formation rate with high-redshift observations. They find that their model matches the observed mass and luminosity functions reasonably well up to $z \approx 1$. \citet{Kitzbichler:2006ec} compare the magnitude counts in the $b_J$, $r_F$, $I$ and $K$-bands, redshift distributions for $K$-band selected samples, $b_J$- and $K$-band luminosity functions, and galaxy stellar mass function from~\citet{Croton:2005fe} and the very similar model by~\citet{De Lucia:2006vua} with high redshift measurements. They find that the agreement of these models with high-redshift observations is slightly worse than that found for the Durham models. In particular, they find that the Munich models tend to systematically overestimate the abundance of relatively massive galaxies at high redshift. 

\citet{Snaith:2011fy} compared four different semi-analytic models~\citep{De Lucia:2006vua,Bower:2005vb,Bertone:2007sj,Font:2008pc}, with observations and found that all models show a shallower, wider magnitude gap, between the brightest group galaxy and the second brightest, compared to observations.

\citet{delaTorre:2010nm} compared measurements of VVDS with the model by~\citet{De Lucia:2006vua}. They found that the model reproduced the galaxy clustering at $z > 0.8$ as well as the magnitude counts in most bands. However the model failed in reproducing the clustering strength of red galaxies and the $b_J$ - $I$ colour distribution. The model tends to produce too many relatively bright red satellites galaxies, a fact that has been reported in other studies as well (over-quenching problem:~\citealt{Weinmann:2006cq,Kimm:2008rp,Liu:2009rm}).

\section{Discussion}
\label{sec:results}

\subsection{Effective halo mass and satellite fraction}
\label{sec:Meff}

As is evident from Table~\ref{tab:results}, the 6dFGS galaxies sit in massive central dark matter halos and most of our galaxies are central galaxies in these halos with only a small fraction being satellite galaxies. Because of the way 6dFGS galaxies are selected, the large majority are red elliptical galaxies and hence our findings agree very well with previous studies, which also found that such galaxies are strongly clustered and therefore must reside in high density regions.

\begin{figure}[tb]
\begin{center}
\epsfig{file=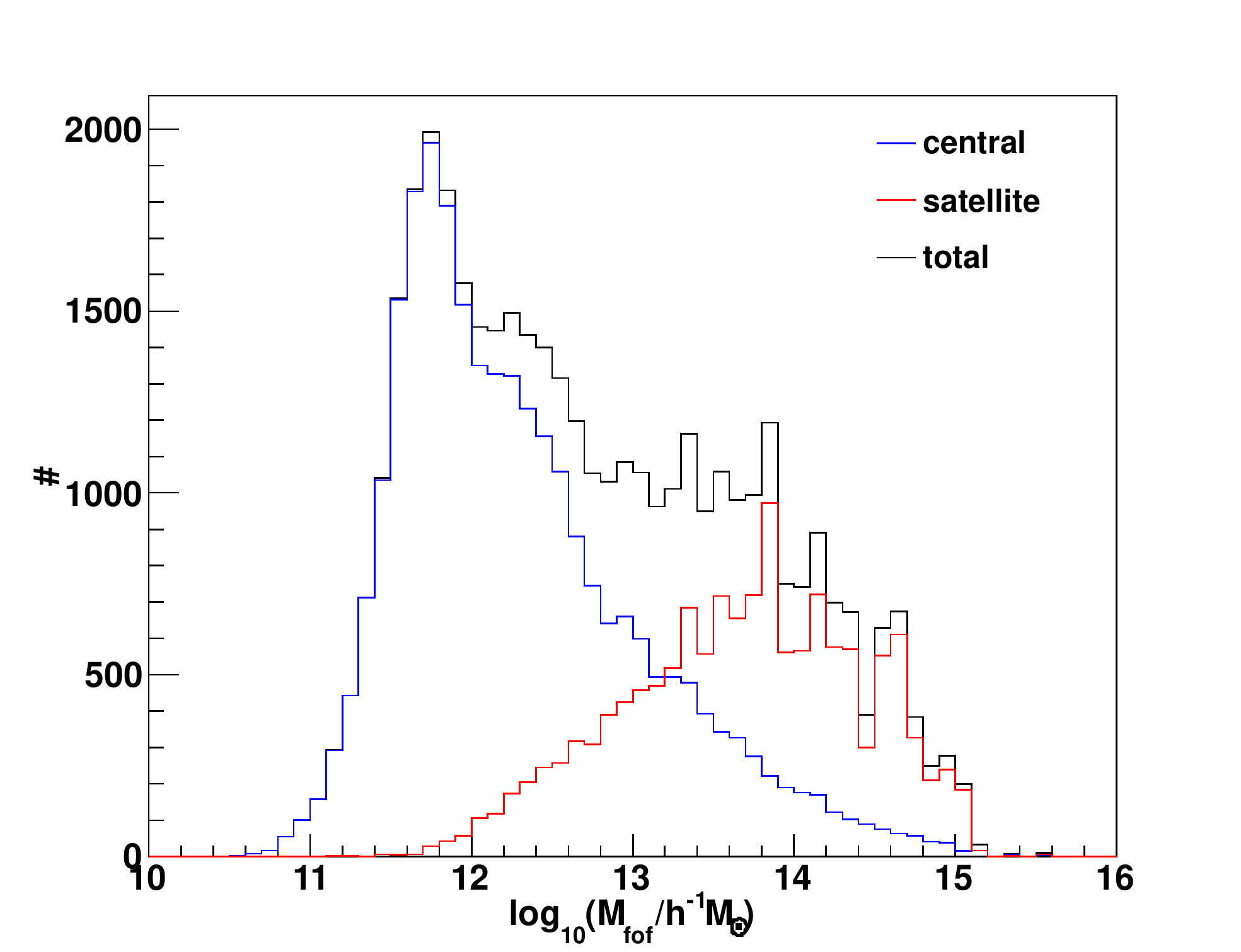,width=10cm}
\caption{This plot shows the friends-of-friends halo mass $\log_{10}( M_{\rm fof})$ of the satellite and central galaxies from the~\citet{Bower:2005vb} semi-analytic catalogue with the 6dFGS selection. The bimodal distribution is caused by a dominant fraction of halos, in which the 6dFGS galaxies are the central galaxy and a smaller fraction with larger halo mass, in which the 6dFGS galaxies are satellite galaxies.}
\label{fig:Meff_mock}
\end{center}
\end{figure}

First we will discuss the effective dark matter halo mass, which represents the effective group or cluster mass for the distribution of galaxies (not the mass of the individual halos which host the galaxies). This quantity appears almost constant for our four different sub-samples. The increasing central galaxy mass ($M_{\rm min}$) with stellar mass is offset by the decreasing satellite fraction, resulting in a fairly constant $M_{\rm eff}$. While the HOD model allows us to derive only averaged parameters for each sample, semi-analytic models directly connect dark matter halo masses with single galaxies. Figure~\ref{fig:Meff_mock} shows the distribution of central and satellite galaxies as a function of friends-of-friends (fof) halo mass derived from the~\citet{Bower:2005vb} semi-analytic catalogue together with the 6dFGS selection criteria. The catalogue contains fewer satellites than central, but the satellites sit in very massive dark matter halos and hence have a significant impact on $M_{\rm eff}$. This plot shows that most 6dFGS galaxies sit in $10^{11}$ - $10^{12}h^{-1}\,M_{\odot}$ halos, while the satellites sit in very massive groups and clusters of up to $10^{15}h^{-1}\,M_{\odot}$. While the median would be around $10^{12}h^{-1}\,M_{\odot}$, the effective mass $M_{\rm eff}$ from the HOD model is the mean of this distribution, which is pushed to very large values by the satellite fraction.

\begin{table}
\begin{center}
\caption{Summary of parameters derived from different semi-analytic models. We impose the 6dFGS $K$-band apparent magnitude selection ($J$-band is not available for these semi-analytic models) as well as correct for incompleteness. The stellar masses are re-calculated using the technique described in section~\ref{sec:stellar}. To calculate the effective dark matter halo mass we used the friends-of-friends halo mass. The stellar mass and the effective halo mass are calculated as the mean of the distribution.}
\vspace{0.4cm}
\begin{small}
\begin{tabular}{cccc}
\hline
sample & $f_s$ & $\log_{10}\left(\frac{\langle M_{\rm stellar}\rangle}{h^{-2}M_{\odot}}\right)$ & $\log_{10}\left(\frac{M_{\rm eff}}{h^{-1}M_{\odot}}\right)$\\
\hline
\citep{Croton:2005fe}\\
\hline
S1 & $0.138$ & $10.38$ & $13.400$\\
S2 & $0.099$ & $10.48$ & $13.389$\\
S3 & $0.085$ & $10.57$ & $13.507$\\
S4 & $0.090$ & $10.67$ & $14.013$\\
\hline
\citep{Bower:2005vb}\\
\hline
S1 & $0.28$ & $10.36$ & $13.695$\\
S2 & $0.25$ & $10.47$ & $13.661$\\
S3 & $0.22$ & $10.58$ & $13.674$\\
S4 & $0.18$ & $10.68$ & $13.777$\\
\hline
\hline
\end{tabular}
\end{small}
\label{tab:sams}
\end{center}
\end{table}

\begin{figure}[tb]
\begin{center}
\epsfig{file=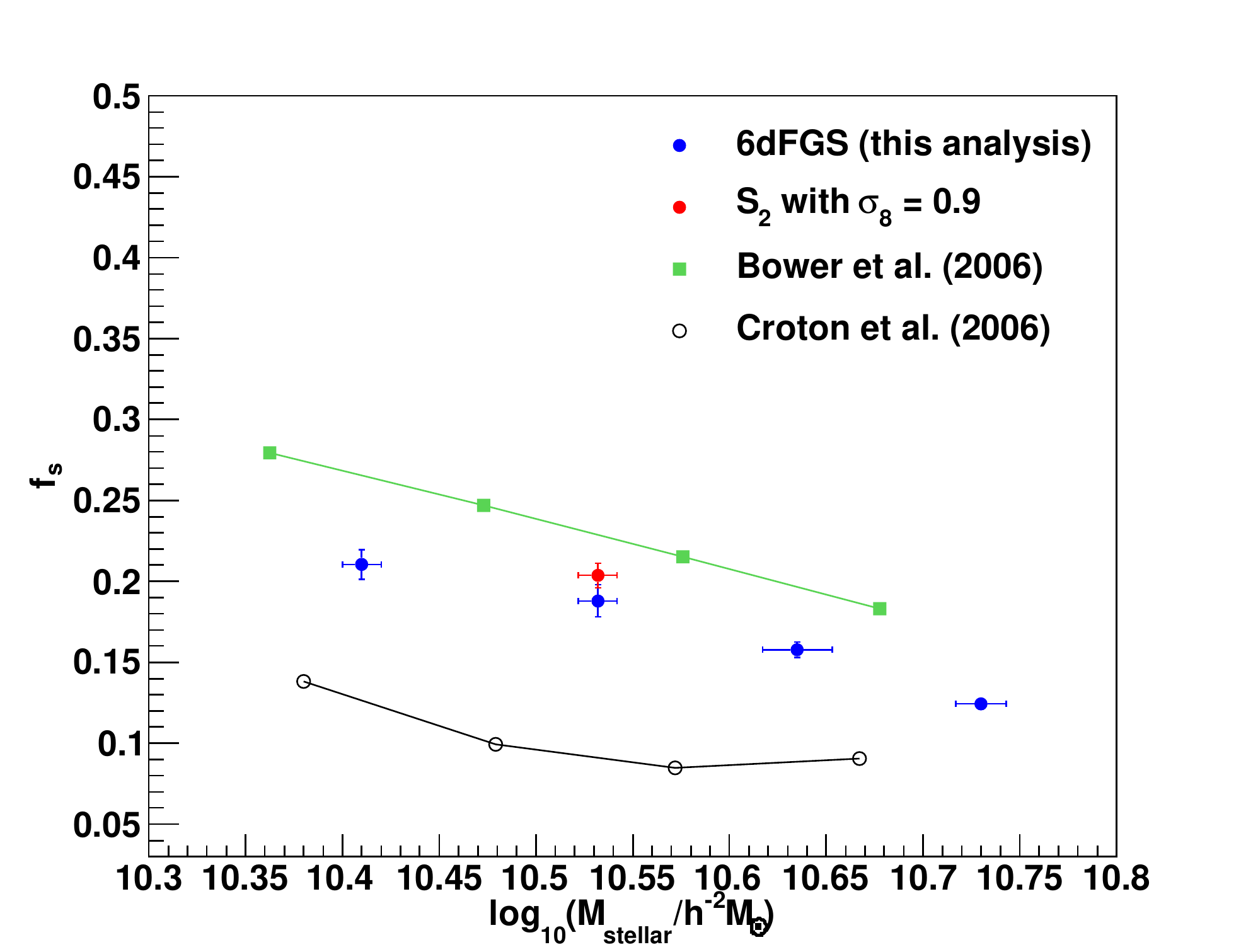,width=10cm}
\caption{Satellite fraction of the 6dFGS stellar mass sub-samples (blue) and the corresponding semi-analytic sub-samples based on the Millennium simulation~\citep{Croton:2005fe, Bower:2005vb}. The red data point represents a fit to sample $S_2$ which assumes $\sigma_8=0.9$, while the blue data points assume $\sigma_8=0.8$.}
\label{fig:cfsat}
\end{center}
\end{figure}

We calculated the effective halo mass $M_{\rm eff} = \langle M_{\rm fof}\rangle$ for the four volume-limited sub-samples derived from the two semi-analytic models and summarise these results in Table~\ref{tab:sams}. While the effective halo mass appears constant in case of the~\citet{Bower:2005vb} model, the~\citet{Croton:2005fe} model shows an increase with increasing stellar mass. These different behaviours for the two different semi-analytic models are most likely connected to the different trend in the satellite fraction shown in Figure~\ref{fig:cfsat}. While the~\citet{Bower:2005vb} model shows a constant decrease in the satellite fraction, the~\citet{Croton:2005fe} model reaches a constant at large stellar mass which causes the effective halo mass to rise. A lower satellite fraction at a fixed stellar mass means that at fixed halo mass the satellite galaxies are less massive~\citep{Weinmann:2006cq,Kimm:2008rp,Liu:2009rm}.

From Figure~\ref{fig:cfsat} we can see that both the~\citet{Croton:2005fe} and the~\citet{Bower:2005vb} model give slightly smaller stellar masses than measured in 6dFGS. This indicates that the semi-analytic catalogues contain too few bright galaxies at low redshift.

Figure~\ref{fig:cfsat} also includes the fit to sample $S_2$ which assumes $\sigma_8=0.9$ (red data point), which agrees with the assumptions of the Millennium Simulation, while the blue data points assume $\sigma_8=0.8$. A larger $\sigma_8$ causes a larger satellite fraction which than causes a larger effective halo mass. This effect seem to bring our results closer to the~\citet{Bower:2005vb} model.

\subsection{The $M_1 - M_{\rm min}$ scaling relation}

\begin{figure*}[tb]
\begin{center}
\epsfig{file=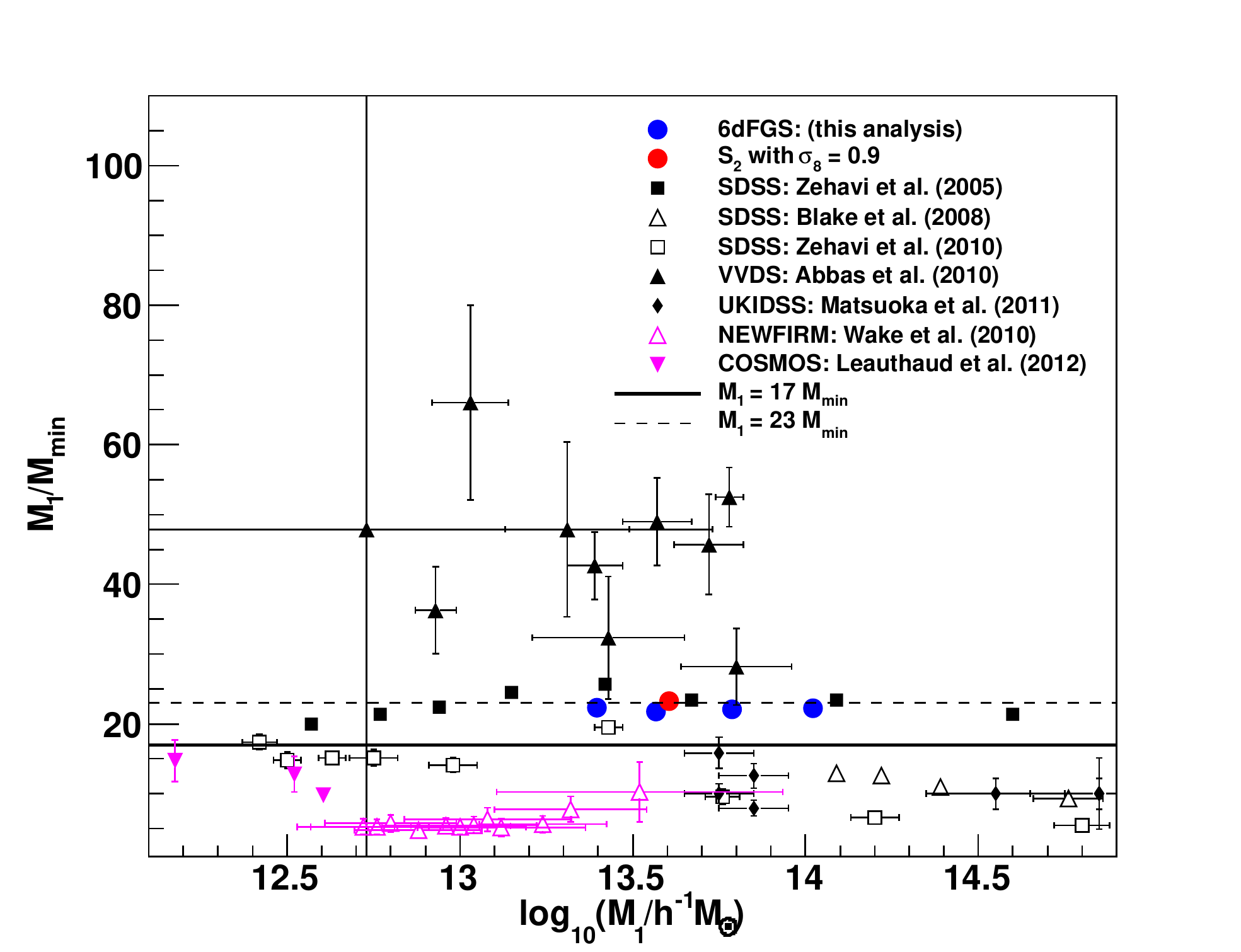,width=10cm}
\caption{The relation between $M_1$ and $M_{\rm min}$ for the four different volume-limited 6dFGS samples (blue data points), compared to~\citet{Zehavi:2004ii} (black solid squares),~\citet{Blake:2007xp} (black open triangles),~\citet{Zehavi:2010bh} (black open squares),~\citet{Abbas:2010hr} (black solid triangles),~\citet{Matsuoka:2010ba} (black open points),~\citet{Wake:2010um} (magenta open triangles) and \citet{Leauthaud:2011rj} (magenta solid triangles). \citet{Zehavi:2004ii} don't report errors on their parameters and the 6dFGS errors are smaller than the data points. All studies which investigate HOD as a function of stellar mass are coloured (including our analysis), while luminosity defined studies are in black.}
\label{fig:M0_Mmin2}
\end{center}
\end{figure*}

The HOD analysis of early data from SDSS by~\citet{Zehavi:2004ii} showed the relation, $M_1 \approx 23\,M_{\rm min}$, between the mass of halos that on average host one additional satellite galaxy, $M_1$ and the minimum dark matter halo mass to host a central galaxy, $M_{\rm min}$. This has been confirmed in subsequent studies ($M_1 \approx 18\,M_{\rm min}$ in~\citealt{Zheng:2007zg} and $M_1 \approx 17\,M_{\rm min}$~\citealt{Zehavi:2010bh}). This relation implies that on average a halo hosting two galaxies of the type studied in their analysis has a mass $\approx 23$ times the mass of a halo hosting only one galaxy of the same type.

\citet{Zehavi:2010bh} also found that this scaling factor is somewhat smaller at the high luminosity end, corresponding to massive halos that host rich groups or clusters. This latter trend likely reflects the relatively late formation of these massive halos, which leaves less time for satellites to merge onto central galaxies and thus lowers the satellite threshold $M_1$.

Theoretical studies of HODs in dark matter simulations~\citep{Kravtsov:2003sg} and those predicted by Smoothed-particle hydrodynamics (SPH) and semi-analytic galaxy formation models~\citep{Zheng:2004id} reveal a similar relation with a scaling factor of $\approx 20$. The large gap between $M_1$ and $M_{\rm min}$ arises because in the low occupation regime, a more massive halo tends to host a more massive central galaxy, rather than multiple smaller galaxies~\citep{Berlind:2002rn}.

\citet{Abbas:2010hr} did a similar study using data from the $I$-band selected VIMOS-VLT Deep Survey (VVDS) and found the ratio $M_1/M_{\rm min}$ to be $\approx 40$ - $50$. The galaxies in this study are at much higher redshift ($z \approx 0.83$) compared to the SDSS galaxies. This result means that in order to begin hosting satellite galaxies, halos sampled by the VVDS survey need to accrete a larger amount of mass compared to SDSS halos.

\citet{Wake:2010um} also studied high redshift galaxies ($1.1 < z < 1.9$) in the NEWFIRM Medium Band Survey (NMBS) and found the ratio $M_1/M_{\rm min}$ to be $\approx 4-10$. This, together with the result by~\citet{Abbas:2010hr} shows, that this relation is not as fundamental as originally thought, but strongly dependent on the type of halos probed in each analysis.

\citet{Leauthaud:2011rj} studied galaxy clustering in the COSMOS survey using threshold stellar mass samples. Using the equations discussed in~\citep{Leauthaud:2011zt} we can derive $M_1$ and $M_{\rm min}$ from their HOD parametrisation, which is included in Figure~\ref{fig:M0_Mmin2}.

\citet{Matsuoka:2010ba} analysed $\sim60\,000$ massive ($\log_{10}(M_{\rm stellar}/h^{-2}M_{\odot}) > 10.7$) galaxies from the UKIRT Infrared Deep Sky Survey (UKIDSS) and the SDSS II Supernova Survey. This analysis shows a very different clustering amplitude depending on whether the observed and theoretical number densities are matched up or not. This dependency makes a comparison of our results with their derived parameters very difficult. Nevertheless, the ratio $M_1/M_{\rm min}$ in their analysis does not depend significantly on their initial assumptions and hence we included their results in Figure~\ref{fig:M0_Mmin2}\footnote{We use their results, in which they match the observed and theoretical number densities, since this agrees with our method.}. Their results indicate a lower $M_1/M_{\rm min}$ ratio at large $M_1$ consistent with~\citet{Zehavi:2010bh} and~\citet{Blake:2007xp}.

In our analysis we found a scaling relation of $M_1 \approx 22\,M_{\rm min}$ (the exact values are $22.29\pm0.39, 21.81\pm0.30, 22.08\pm0.28$ and $22.22\pm0.25$ for $S_1$-$S_4$, respectively). In Figure~\ref{fig:M0_Mmin2} we compare our results with other stellar mass selected samples (coloured data points) and with luminosity threshold selected samples (black data points). The ratio $M_1/M_{\rm min}$ found in 6dFGS is in agreement with~\citet{Zehavi:2004ii} and slightly larger than~\citet{Zehavi:2010bh}. While~\citet{Zehavi:2004ii} and~\citet{Zehavi:2010bh} study the same type of galaxies,~\citet{Zehavi:2004ii} uses an HOD parameterisation very similar to ours, while~\citet{Zehavi:2010bh} uses a parameterisation based on $5$ free parameters, suggesting that the differences might be related to the parameterisation. 

\subsection{Comparison to~\citet{Mandelbaum:2005nx}}

An alternative method for probing the connection between stellar mass and halo mass is galaxy-galaxy weak lensing. Here we will compare our findings to~\citet{Mandelbaum:2005nx}, who used weak lensing of $\sim 350\,000$ galaxies from SDSS and looked at the dependence of the amplitude of the lensing signature as a function of galaxy type and stellar mass. They used stellar masses derived from the $z$-band magnitude and the ratio $M/L_z$ from~\citet{Kauffmann:2002pn} with the assumption of a~\citet{Kroupa:2000iv} IMF. 

The HOD model employed in their analysis is given by
\begin{equation}
N_t(M) = N_c(M) + N_s(M)
\end{equation}
with the central galaxy number given by eq.~\ref{eq:hod} while the satellite galaxies number is modelled by a step-like function
\begin{equation}
N_s(M) = \begin{cases}	kM & \text{ if }\;M\geq 3M_{\rm min}\\
				    	\frac{kM^2}{3M_{\rm min}} & \text{ if }\;M_{\rm min} \leq M < 3M_{\rm min}\\
					0 & \text{ if }\;M < M_{\rm min}.
	        \end{cases}
	        \label{eq:mandel1}
\end{equation}
The normalisation constant $k$ can be determined by matching the measured satellite fractions $f_s^{\rm ma}$ and is given in appendix~\ref{ap:mandel}. Here we have adjusted the nomenclature to the one used in our analysis. We also note that~\citet{Mandelbaum:2005nx} assumed $\sigma_8 = 0.9$, while we assumed $\sigma_8 = 0.8$ for most of our fits.

From the equation above we can derive the effective halo mass using eq.~\ref{eq:Meff} together with the dark matter halo mass function.

\begin{figure}[tb]
\begin{center}
\epsfig{file=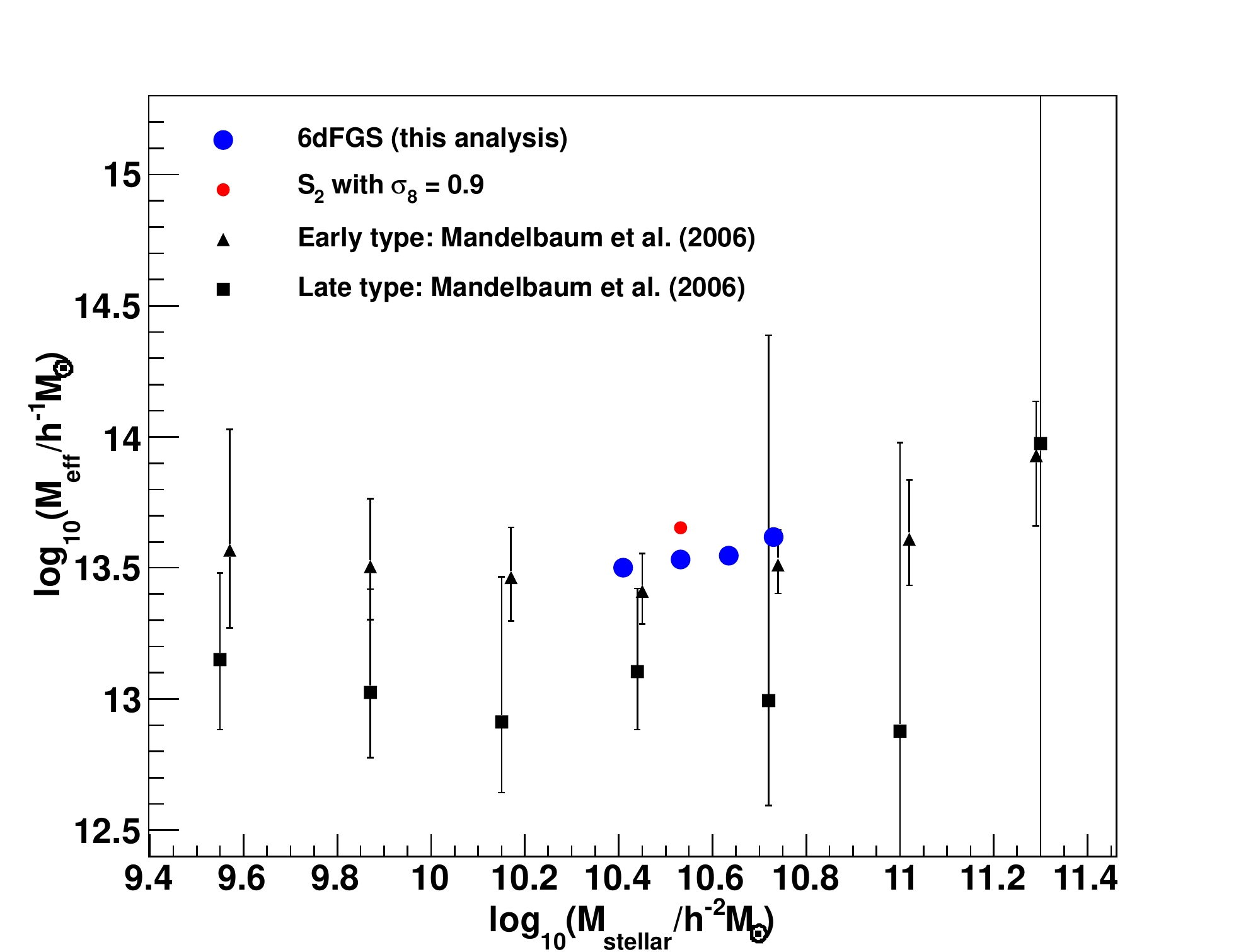,width=10cm}
\caption{The effective dark matter halo mass of the four 6dFGS sub-samples (blue data points) compared to early type (black triangles) and late type (black squares) galaxies from~\citet{Mandelbaum:2005nx}. In red we also include the fit to sample $S_2$ which assumes $\sigma_8=0.9$ in agreement with the assumptions of~\citet{Mandelbaum:2005nx}, while the blue data of assume $\sigma_8=0.8$.}
\label{fig:cMeff_sm}
\end{center}
\end{figure}

\begin{figure}[tb]
\begin{center}
\epsfig{file=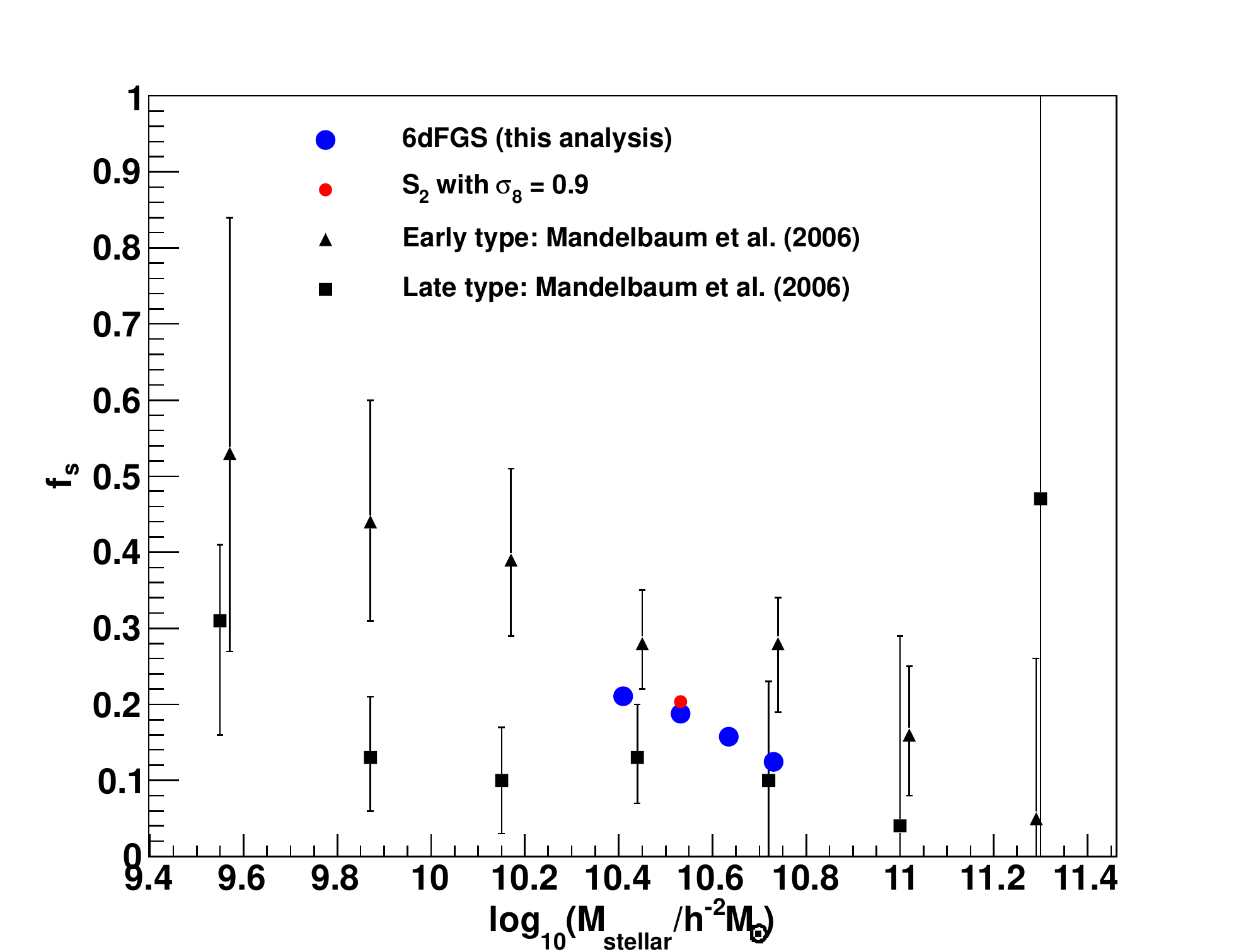,width=10cm}
\caption{The satellite fraction of the four 6dFGS sub-samples (blue data points) compared to early type (black triangles) and late type (black squares) galaxies from~\citet{Mandelbaum:2005nx}. In red we also include the fit to sample $S_2$ which assumes $\sigma_8=0.9$ in agreement with the assumptions of~\citet{Mandelbaum:2005nx}, while the blue data points assume $\sigma_8=0.8$.}
\label{fig:calpha_sm}
\end{center}
\end{figure}

We compare the reported satellite fraction, $f_s^{\rm ma}$, of~\citet{Mandelbaum:2005nx} and the derived effective halo mass, $M_{\rm eff}$, with our results in Figure~\ref{fig:cMeff_sm} and Figure~\ref{fig:calpha_sm}. Figure~\ref{fig:cMeff_sm} shows that the galaxies sampled by~\citet{Mandelbaum:2005nx} have a very similar effective halo mass compared to 6dFGS galaxies. The 6dFGS measurements follow the trend of early-type galaxies in these measurements, while the late types have a slightly smaller effective halo mass. However, the errors in case of the results by~\citet{Mandelbaum:2005nx} don't allow one to distinguish between the early and late type galaxies. Since effective halo mass and satellite fraction are strongly linked, we also compare the satellite fraction of~\citet{Mandelbaum:2005nx} with our results in Figure~\ref{fig:calpha_sm}. Both figures also include the fit to sample $S_2$ where we assumed $\sigma_8=0.9$, instead of the standard $\sigma_8=0.8$, since this agrees with the assumption in~\citet{Mandelbaum:2005nx}. A larger $\sigma_8$ increases the effective halo mass by a small amount and hence does not impact this comparison significantly.

Overall we see very good agreement between our results and the results of~\citet{Mandelbaum:2005nx}, which is reassuring since the two techniques are subject to different systematic uncertainties. We can also emphasise that the 6dFGS results are extremely precise compared to the lensing results.

\subsection{Comparison to other studies}

\citet{Meneux:2007jr} study galaxy cluster mass dependence on stellar mass in the VIMOS-VLT Deep Survey (VVDS) at redshift $0.5 < z < 1.2$. The stellar masses in this sample cover the range $\log_{10}(M_{\rm stellar}/h^{-2}M_{\odot}) = 8.7$ - $10.7$. To quantify the clustering, they use power laws fits to the projected correlation function $w_p(r_p)$ and found an evolution in $r_0$ from $2.76\pm0.17 h^{-1}\,$Mpc at $\log_{10}(M_{\rm stellar}/h^{-2}M_{\odot}) > 8.7$ to $r_0 = 4.28\pm0.45h^{-1}\,$Mpc at $\log_{10}(M_{\rm stellar}/h^{-2}M_{\odot}) > 10.2$. The slope changes over the same range from $\gamma = 1.67\pm 0.08$ to $\gamma = 2.28\pm 0.28$. Comparing the results by~\citet{Meneux:2007jr} with~\citet{Li:2005er} who used the clustering of $200\,000$ SDSS galaxies at $z = 0.15$, showed that the evolution of the amplitude and shape of the correlation function $w_p(r_p)$ with redshift is faster for low stellar mass objects than for high stellar mass objects. At low stellar mass, the amplitude of $w_p(r_p)$ increases by a factor $\sim2$ - $3$ from high to low redshift, while at the high stellar mass range $\log_{10}(M_{\rm stellar}/h^{-2}M_{\odot}) = 10.2$ - $10.7$, the amplitude at $z \sim 0.85$ and $z \sim 0.15$ are very similar, within the error bars. 
In 6dFGS we found a significantly larger clustering amplitude, ranging from $r_0 = 5.14\pm0.23h^{-1}$Mpc at $\log_{10}(M_{\rm stellar}/h^{-2}M_{\odot}) = 10.41$ to $r_0 = 6.21\pm0.17h^{-1}$Mpc at $\log_{10}(M_{\rm stellar}/h^{-2}M_{\odot}) = 10.73$, while our power law index $\gamma$ stays constant at $\gamma \approx 1.84$. This shows that the trend observed in~\citet{Meneux:2007jr} and~\citet{Li:2005er} continues to the 6dFGS redshifts, even for high stellar mass galaxies.

\section{Conclusions}
\label{sec:con}

We present in this paper an analysis of the clustering properties of four stellar mass selected volume-limited sub-samples of galaxies from the 6dF Galaxy Survey. The stellar masses are calculated using the $J$-band magnitude and the $b_J-r_F$ colour following the technique by~\citet{Bell:2000jt}. The average log-stellar mass for the different sub-samples ranges from $\log_{10}(M_{\rm stellar}/h^{-2}\,M_{\odot}) = 10.41-10.73$. Our analysis has the following main results:

\begin{itemize}
\item The projected correlation function, $w_p(r_p)$, for the two low stellar mass sub-samples ($S_1$ and $S_2$) can be described by a power law with an acceptable $\chi^2$, while the two high stellar mass sub-samples ($S_3$ and $S_4$) have a reduced $\chi^2$ of $1.48$ and $1.63$, respectively. We also find patterns in the deviations between the best power law fit and the data, which can naturally be explained within the halo model. This is in agreement with theoretical studies~\citep{Watson:2011cz}, predicting that the disagreement of $w_p(r_p)$ with a power law fit should grow with clustering amplitude.
\item We used an HOD parameterisation with three free parameters ($M_{1}$, $\alpha$, $M_{\rm min}$), representing the typical halo mass, satellite power law index and minimum host halo mass, respectively. The minimum halo mass is fixed by the galaxy number density for all our parameter fits. We performed fits to the projected correlation functions of the four different volume-limited sub-samples. We tested alternative HOD parameterisations, but found that our data does not justify more free parameters. $M_{1}$ and $\alpha$ show increasing trends for increasing stellar mass, with $\log_{10}(M_{1}/h^{-1}\,M_{\odot})$ ranging from $13.4 - 14$ and $\alpha$ from $1.21-1.4$. This means that galaxies with larger stellar mass populate larger dark matter halos, which accrete satellites faster with increasing mass, compared to dark matter halos populated by galaxies with lower stellar mass.
\item From the halo model, we can derive averaged parameters for the four samples such as the satellite fraction, effective dark matter halo mass and the effective galaxy bias. We found that the satellite fraction decreases with stellar mass from $21\%$ at $\log_{10}(M_{\rm stellar}/h^{-2}\,M_{\odot}) = 10.41$ to $12\%$ at $\log_{10}(M_{\rm stellar}/h^{-2}\,M_{\odot}) = 10.73$. The effective dark matter halo mass stays constant at $\log_{10}(M_{\rm eff}/h^{-1}M_{\odot}) \approx 13.55$ for all four sub-samples. The effective galaxy bias increases with stellar mass indicating that galaxies with higher stellar mass reside in denser environments. The increase in the effective galaxy bias can be described by $(1.05 + M_{\rm stellar}/M_*)\times(\sigma_8/0.8)$ with $M_* = 1.18\times 10^{11}h^{-2}M_{\odot}$.
\item We use two semi-analytic models, based on the Millennium Simulation (\citealt{Croton:2005fe} and~\citealt{Bower:2005vb}) to derive 6dFGS mock surveys. We compare the results of these semi-analytic models with our measurements. The~\citet{Croton:2005fe} model under-predicts the satellite fraction, while the~\citet{Bower:2005vb} model is in better agreement with our observations, although it slightly over-predicts the satellite fraction. Since the effective dark matter halo mass is strongly linked to the satellite fraction, the~\citet{Bower:2005vb} model prediction of $M_{\rm eff}$ is again in better agreement with our observations. 6dFGS allows a powerful test of semi-analytic models, because of (1) the robust stellar mass estimates and (2) the focus on $``$red satellites$"$, which semi-analytic models struggled to reproduce in the past. Our results can be used as a new constraint on semi-analytic models in the future. 
\item For the four volume-limited samples we find a constant scaling relation between $M_1$ and $M_{\rm min}$ of $M_1 \approx 22\,M_{\rm min}$, in agreement with studies of SDSS galaxies~\citep{Zehavi:2004ii,Zehavi:2010bh}. This indicates that 6dFGS galaxies populate dark matter halos in a similar way to SDSS galaxies. However, we see a wide spectrum of this ratio ranging from $> 40$ in~\citet{Abbas:2010hr} to $< 10$ in~\citep{Matsuoka:2010ba,Wake:2010um} indicating that this ratio is not universal for all types of galaxies.
\item We compare our results with the results of~\citet{Mandelbaum:2005nx} from galaxy-galaxy weak lensing. We find overall good agreement which represents a valuable crosscheck for these two different clustering measurements. Although our analysis depends on slightly more assumptions and covers a smaller range in stellar mass, the 6dFGS results are extremely precise compared to the lensing results.
\end{itemize}

\chapter{Summary and Conclusion}
\label{ch:conc}

In this thesis we analyse data from the 6-degree Field Galaxy Survey. The three main results of this thesis are:
\begin{enumerate}
\item We used the baryon acoustic oscillation signal in 6dFGS at the effective redshift of $z_{\rm eff} = 0.106$ to constrain the Hubble constant $H_0$ with a small dependence on other cosmological parameters. We found $H_0 = 67.0\pm 3.2\,$km\,s$^{-1}\,$Mpc$^{-1}$, which represents one of the best constraints on this parameter ever obtained, with very different systematic uncertainties compared to more standard methods like the distance ladder technique. 
\item The imprint of redshift-space distortions on the two-dimensional correlation function measures the amplitude of the gravitational force acting between galaxies and therefore allows us to test General Relativity. We combine our measurement of the parameter combination $f\sigma_8 = 0.423\pm0.055$ with WMAP-7 to constrain the gravitational growth index $\gamma = 0.547\pm 0.088$. Our result is in very good agreement with the prediction of General Relativity ($\gamma \simeq 0.55$).
\item We use the clustering properties of 6dFGS galaxies as a function of stellar mass to explore how they populate dark matter haloes and hence place new constraints on semi-analytic models.
\end{enumerate}
In chapter~\ref{ch:BAO} we use the baryon acoustic oscillation signal to make an absolute distance measurement to the effective redshift of 6dFGS at $z_{\rm eff} = 0.106$. We measure a distance of $D_V(z_{\rm eff}) = 456\pm 27\,$Mpc and a distance ratio of $r_s(z_d)/D_V(z_{\rm eff}) = 0.336\pm 0.015$, where $r_s(z_d)$ is the sound horizon at the drag epoch $z_d$. An absolute distance measurement at such a low redshift constraints mainly the Hubble constant $H_0$, almost independently of other cosmological parameters. 
Our measurement of the Hubble constant is $H_0 = 67.0\pm 3.2\,$km\,s$^{-1}\,$Mpc$^{-1}$ ($4.8\%$ precision), which is $1.7\sigma$ smaller than the latest measurement reported using the distance ladder technique ($H_0 = 73.8\pm2.4\,$km/s/Mpc;~\citealt{Riess:2011yx}). This tension has been confirmed by several other publications (e.g.~\citealt{Mehta:2012hh}). While this is an interesting potential indication of systematic errors in either technique, we have to keep in mind that a $1.7\sigma$ deviation corresponds to a probability of $\approx 10\%$ and hence is statistically not significant. Our result represents a competitive and independent alternative to Cepheids and low-z supernovae in measuring the Hubble constant, which depends on very different systematic errors and hence allows a very interesting test of the current standard cosmological model. 

Instead of $r_s(z_d)/D_V(z_{\rm eff})$ we can constrain the parameter $A(z_{\rm eff}) = 100D_V(z_{\rm eff}) \sqrt{\Omega_mh^2}/cz$ which can be used to directly constrain the matter density $\Omega_m$. We found $\Omega_m = 0.287 + 0.039(1+w) + 0.039\Omega_k\pm0.027$ where the weak dependence on the dark energy equation of state parameter $w$ and the curvature parameter $\Omega_k$ occurs because our measurement is not exactly at redshift zero. 

The 6dFGS BAO detection occupies a redshift very different to other galaxy surveys which report such a detection, and hence represents a new data point on the BAO Hubble diagram. While at low redshift we are not directly sensitive to dark energy, 6dFGS can help to break degeneracies arising when fitting the CMB data alone which then leads to improved constraints on dark energy. We show that combining the SDSS DR7 BAO measurement from~\citet{Percival:2009xn} with 6dFGS and WMAP7~\citep{Komatsu:2010fb} results in a constraint on the dark energy equation of state parameter of $w = -0.97\pm0.13$, which represents a $24\%$ improvement compared to the measurement without 6dFGS. 

The 6dFGS BAO result has been combined with other datasets in several subsequent publications (e.g.~\citealt{Blake:2011en,Mehta:2012hh,Anderson:2012sa,Sanchez:2012sg}).\\
 
In chapter~\ref{ch:RSD} we use the redshift-space distortions signal in the 2D correlation function of 6dFGS, $\xi(r_p,\pi)$, to measure the parameter combination $f(z_{\rm eff})\sigma_8(z_{\rm eff}) = 0.423\pm 0.055$, where $f\sim \Omega_m^{\gamma}(z)$ is the growth rate of cosmic structure and $\sigma_8$ is the r.m.s. deviation of matter fluctuations in $8h^{-1}\,$Mpc spheres. The parameter $\gamma$ incorporates the gravitational force and is predicted to be $\gamma \simeq 0.55$ in General Relativity with a $\Lambda$CDM background cosmology. Measuring this parameter allows us to test models of Gravity. Since only the growth rate $f$ is directly related to $\gamma$, we have to combine our measurement of $f\sigma_8$ with another dataset, to break the degeneracy between $f$ and $\sigma_8$. We combine our result with WMAP7~\citep{Komatsu:2010fb}, which yields $\gamma = 0.547\pm0.088$, consistent with the prediction of General Relativity. Because of the low redshift of 6dFGS, this measurement is independent of the fiducial cosmological model (Alcock-Paczynski effect). We also show that our result is not sensitive to the model adopted for non-linear redshift-space distortions or the fitting range.

Furthermore we discuss a method to directly constrain $f$ and $\sigma_8$ with low redshift data only, where we combine 6dFGS with a prior on the Hubble constant from~\citet{Riess:2011yx}. With this technique we find $f = 0.58\pm0.11$, $\sigma_8 = 0.76\pm0.11$ and $\Omega_m = 0.250\pm0.022$, again consistent with the standard model of cosmology.

The 6dFGS measurement of $f\sigma_8$ has been combined with other datasets in several subsequent publications (e.g.~\citealt{Rapetti:2012bu,Hudson:2012gt,Samushia:2012iq}).\\

In chapter~\ref{ch:HOD} we calculate the 6dFGS projected correlation function $w_p(r_p)$ for four volume limited sub-samples with thresholds in redshift and stellar mass in order to investigate how galaxies populate dark matter halos. We found that the halo model fits the data better than empirical power-law fits yielding smaller values of $\chi^2$, especially in the case of our high stellar mass samples for which, the power-law fits give a reduced $\chi^2 \gtrsim 1.5$. 

We find that the typical halo mass $M_1$, where the average dark matter halo has its first satellite, as well as the satellite power law index $\alpha$, increase with stellar mass. This indicates that galaxies with higher stellar mass sit in more massive dark matter halos, which accumulate satellites faster with growing mass compared to low stellar mass galaxies. Furthermore we found a relation between $M_1$ and the minimum dark matter halo mass $M_{\rm min}$ of $M_1 \approx 22M_{\rm min}$, in good agreement with studies of SDSS galaxies~\citep{Zehavi:2004ii,Zehavi:2010bh}. The satellite fraction declines with increasing stellar mass from $21\%$ at $M_{\rm stellar} = 2.6\times10^{10}h^{-2}M_{\odot}$ to $12\%$ at $M_{\rm stellar} = 5.4\times 10^{10}h^{-2}M_{\odot}$ indicating that higher stellar mass galaxies are more likely to be central galaxies. The effective galaxy bias increases with stellar mass and can be approximated by a power law given by $[\log_{10}(M_{\rm stellar}/h^{-2}M_{\odot})/10]^{5.74}\times(0.8/\sigma_8)$.

We compare these results with two mock catalogues, based on the~\citet{Croton:2005fe} and~\citet{Bower:2005vb} semi-analytic models. We derive these mock catalogues from the Millennium Simulation including the 6dFGS selection effects. The~\citet{Croton:2005fe} model under-predicts the satellite fraction, while the~\citet{Bower:2005vb} model is in better agreement with our observations. Since the satellite fraction and the effective dark matter halo mass, $M_{\rm eff}$ are strongly linked, the~\citet{Bower:2005vb} model predictions of $M_{\rm eff}$ are again in better agreement with our observations. 6dFGS allows a powerful test of semi-analytic models, because of (1) the robust stellar mass estimates and (2) the focus on $``$red satellites$"$, which semi-analytic models struggled to reproduce in the past. Our results can be used as a new constraint on semi-analytic models in the future. 

Finally we compare our results with~\citet{Mandelbaum:2005nx} who studied halo occupation as a function of stellar mass using galaxy-galaxy weak lensing. We find overall good agreement, which represents a valuable crosscheck of these two different techniques for studying matter clustering. Although our analysis depends on slightly more assumptions and covers a smaller range in stellar mass, the 6dFGS results are extremely precise compared with the lensing results.\\

The BAO analysis in chapter~\ref{ch:BAO} as well as the RSD analysis in chapter~\ref{ch:RSD} include predictions for the two future galaxy surveys TAIPAN\footnote{TAIPAN: Transforming Astronomical Imaging surveys through Polychromatic Analysis of Nebulae} and WALLABY\footnote{http://www.atnf.csiro.au/research/WALLABY}. Our analysis shows that these two very different surveys will be able to improve upon the 6dFGS result discussed in this thesis. The TAIPAN survey has an effective volume three times as big as 6dFGS and will be able to constrain the Hubble Constant to $3\%$ precision using the BAO technique. The WALLABY survey, with its very small galaxy bias, is much better set-up to measure the growth of structure, where it will be able to constrain $f\sigma_8$ to about $4$ - $10\%$ precision, depending on the modelling of non-linear structure formation.\\

In summary, in this thesis we have presented tests of the current standard model of cosmology using baryon acoustic oscillations as a geometrical probe of the expansion of the Universe as well as redshift-space distortions to test the growth of structure. Since geometric probes are able to measure the dark energy equation of state parameter $w$, they can test different dark energy models, which predict deviations from a cosmological constant at $w=-1$. The measurement of the growth of structure, however, is able to test the laws of Gravity. A weaker gravitational force on cosmic scales could be an alternative explanation for the accelerated expansion measured by geometrical probes. 

All our measurements are in agreement with the current standard model of cosmology, which treats dark energy as a cosmological constant. We discussed the theoretical difficulties of this model in our introduction, but unfortunately our data is not able to solve these problems. To find a physical explanation for dark energy and its behaviour as a cosmological constant remains one of the biggest challenges in cosmology and will call for better observational data and more innovative theoretical models.

\subsection{Future galaxy surveys}

The tremendous success of galaxy redshift surveys, in complementing the CMB to constrain cosmological models, sparked a large interest in the field. We already discussed extensively the two future galaxy surveys WALLABY and TAIPAN. To conclude this thesis we will give a short overview of some other projects and discuss their scientific goals. 

Future galaxy surveys require dedicated larger telescopes, like the ongoing Baryon Oscillation Spectroscopic Survey (BOSS)~\citep{Schlegel:2009hj}, which will run until 2014, or the Dark Energy Survey (DES)~\citep{Abbott:2005bi} which commenced data taking in 2012. The main motivation for these surveys is to study the physics behind the acceleration of the expansion of the Universe. However, the same data will allow us to learn more about many open questions in astrophysics and cosmology like galaxy formation, inflation and neutrino properties. 

The Dark Energy Survey is a photometric galaxy survey using the 4m Victor M. Blanco Telescope located at Cerro Tololo Inter-American Observatory (CTIO) in Chile. The main innovation of that project consists in the development of a new camera called DECam. The survey will image $5000\,\deg^2$ of the southern sky and will take five years to complete. 

Looking even further into the future the proposed BigBOSS experiment~\citep{Schlegel:2011wb} will study Baryon Acoustic Oscillations in the redshift range  $0.2 < z < 3.5$ using a dedicated 4m telescope. A new $5000$-fibre spectrograph covering a 3-degree diameter field will detect the redshifts of about $15.3$ million Emission Line Galaxies (ELGs), $3.4$ million Luminous Red Galaxies (LRGs), and $630\,000$ Quasars (QSOs).

The ESA space mission Euclid~\citep{Laureijs:2009} is a 1.2m telescope together with a CCD and a spectrograph and will map about $2/3$ of the sky ($15\,000\, \deg^2$). The mission is scheduled to launch in 2020, measuring the redshifts of billions of galaxies through the detection of the H$\alpha$ emission line.

On a similar time scale there is the Large Synoptic Survey Telescope (LSST), which is a planned wide-field ``survey" telescope, capable of covering $20\,000\,\deg^2$ of the southern sky using an $8.4$m ground-based telescope. Each patch of sky will be visited about $1000$ times in ten years detecting roughly $250\,000$ supernovae per year. First light is scheduled for 2020 and full operations for a ten-year survey will commence in January 2022.

Finally, the Square Kilometre Array (SKA) will use thousands of radio telescopes covering a collecting area of about one square kilometre to detect the signal of hydrogen in the Universe, up to the epoch of re-ionisation~\citep{Blake:2004pb}. Radio galaxy surveys have the advantage that it is relatively simple to build interferometers, which means that the physical limitations of building one big structure can be bypassed, by building many small telescopes and linking them together. 

Galaxy redshift surveys are amongst the most successful projects in cosmology/astronomy. The Sloan Digital Sky Survey (SDSS) lead to currently $\approx 5000$ refereed papers and $\approx 200\,000$ citations (Michael Strauss, private communication, see also \citealt{Madrid:2009rf,Savaglio:2012cw}). For comparison, WMAP lead to $\approx 1700$ refereed papers and $100\,000$ citations. 
Galaxy redshift surveys are fantastically equipped to address the big open questions in astronomy and cosmology. With the future survey projects outlined above the coming years will be very exciting for cosmology and who knows what we will find.

\singlespacing
\setlength{\bibhang}{2em}
\setlength{\labelwidth}{0pt}

\newpage

\doublespacing
\appendixpage
\begin{subappendices}

\section{Generating log-normal mock catalogues}
\label{ap:log}

Here we explain in detail the different steps used to derive a log-normal mock catalogue, as a useful guide for researchers in the field. We start with an input power spectrum, (which is determined as explained in Section~\ref{sec:log}) in units of $h^{-3}$Mpc$^{3}$. We set up a 3D grid with the dimensions $L_x\times L_y\times L_z = 1000\times 1000\times 1000h^{-1}\;$Mpc with $200^3$ sub-cells. We then distribute the quantity $P(\vec{k})/V$ over this grid, where $V$ is the volume of the grid and $\vec{k} = \sqrt{k_x^2 + k_y^2 + k_z^2}$ with $k_x = n_x2\pi/L_x$ and $n_x$ being an integer value specifying the $x$ coordinates of the grid cells.

Performing a complex-to-real Fourier transform (FT) of this grid will produce a 3D correlation function. Since the power spectrum has the property $P(-\vec{k}) = P(\vec{k})^*$ the result will be real.

The next step is to replace the correlation function $\xi(r)$ at each point in the 3D grid by $\ln[1 + \xi(r)]$, where $\ln$ is the natural logarithm. This step prepares the input model for the inverse step, which we later use to produce the log-normal density field. 

Using a real-to-complex FT we can revert to $k$-space where we now have a modified power spectrum, $P_{\ln}(\vec{k})$. At this point we divide by the number of sub-cells $N_c$. The precise normalisation  depends on the definition of the discrete Fourier transform. We use the FFTW library~\citep{Frigo:2005}, where the discrete FT is defined as
\begin{equation}
Y_i = \sum^{N_c-1}_{j=0} X_j \exp\left[\pm 2\pi ij\sqrt{-1}/N_c\right].
\end{equation}
The modified power spectrum $P_{\ln}(\vec{k})$ is not guarantied to be neither positive defined nor a real function, which contradicts the definition of a power spectrum. \citet{Weinberg:1991qe} suggested to construct a well defined power spectrum from $P_{\ln}(\vec{k})$ by
\begin{equation}
P'_{\ln}(\vec{k}) = \text{max}\left[0,\text{Re}[P_{\ln}(\vec{k})]\right].
\end{equation}
We now generate a real and an imaginary Fourier amplitude $\delta(\vec{k})$ for each point on the grid by randomly sampling from a Gaussian distribution with r.m.s. $\sqrt{P'_{\ln}(\vec{k})/2}$. However, to ensure that the final over-density field is real, we have to manipulate the grid, so that all sub-cells follow the condition $\delta(-\vec{k}) = \delta(\vec{k})^*$.

Performing another FT results in an over-density field $\delta(\vec{x})$ from which we calculate the variance $\sigma_G^2$. The mean of $\delta(\vec{x})$ should be zero. The log-normal density field is then given by
\begin{equation}
\mu_L(\vec{x}) = \exp\left[\delta(\vec{x}) - \sigma_G^2/2\right],
\end{equation}
which is now a quantity defined on $[0,\infty[$ only, while $\delta(\vec{x})$ is defined on $]-\infty,\infty[$.

Since we want to calculate a mock catalogue for a particular survey we have to incorporate the survey selection function. If $W(\vec{x})$ is the selection function with the normalisation $\sum W(\vec{x}) = 1$, we calculate the mean number of galaxies in each grid cell as 
\begin{equation}
n_g(\vec{x}) = N\;W(\vec{x})\;\mu_L(\vec{x}),
\end{equation}
where $N$ is the total number of galaxies in our sample. The galaxy catalogue itself is than generated by Poisson sampling $n_g(\vec{x})$.

The galaxy position is not defined within the sub-cell, and we place the galaxy in a random position within the box. This means that the correlation function calculated from such a distribution is smooth at scales smaller than the sub-cell. It is therefore important to make sure that the grid cells are smaller than the size of the bins in the correlation function calculation. In the 6dFGS calculations presented in this paper the grid cells have a size of $5h^{-1}\;$Mpc, while the correlation function bins are $10h^{-1}\;$Mpc in size.

\section{Comparison of log-normal and jack-knife error estimates}
\label{ap:jk_comp}

\begin{figure}[tb]
\begin{center}
\epsfig{file=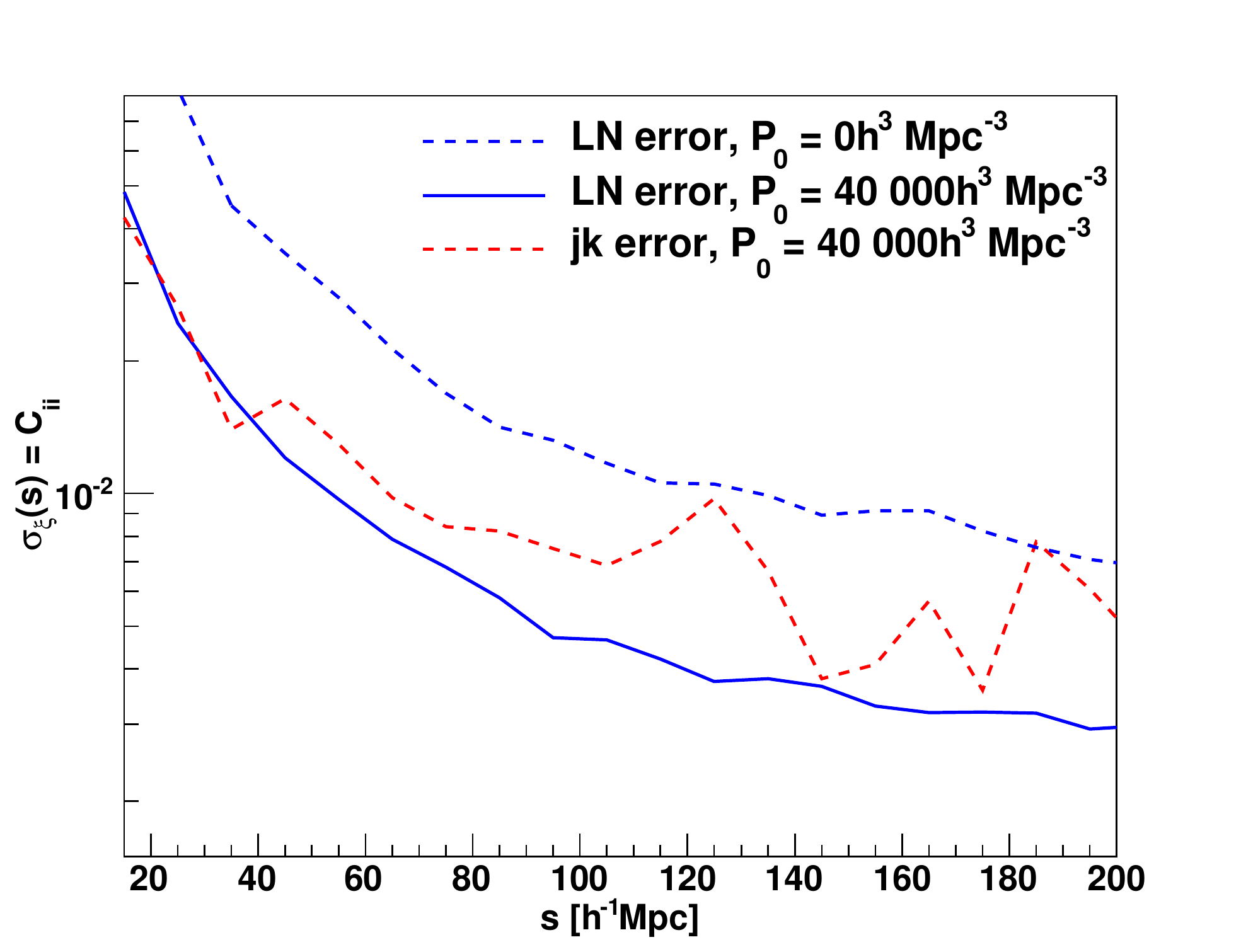, width=10cm}
\caption{Correlation function error for different values of $P_0$. The weighting with $P_0 = 40\,000h^{3}\;$Mpc$^{-3}$ reduces the error at the BAO scale by almost a factor of four compared to the case without weighting.  The red dashed line indicates the jack-knife error.}
\label{fig:sigP0}
\end{center}
\end{figure}

\begin{figure}[tb]
\begin{center}
\epsfig{file=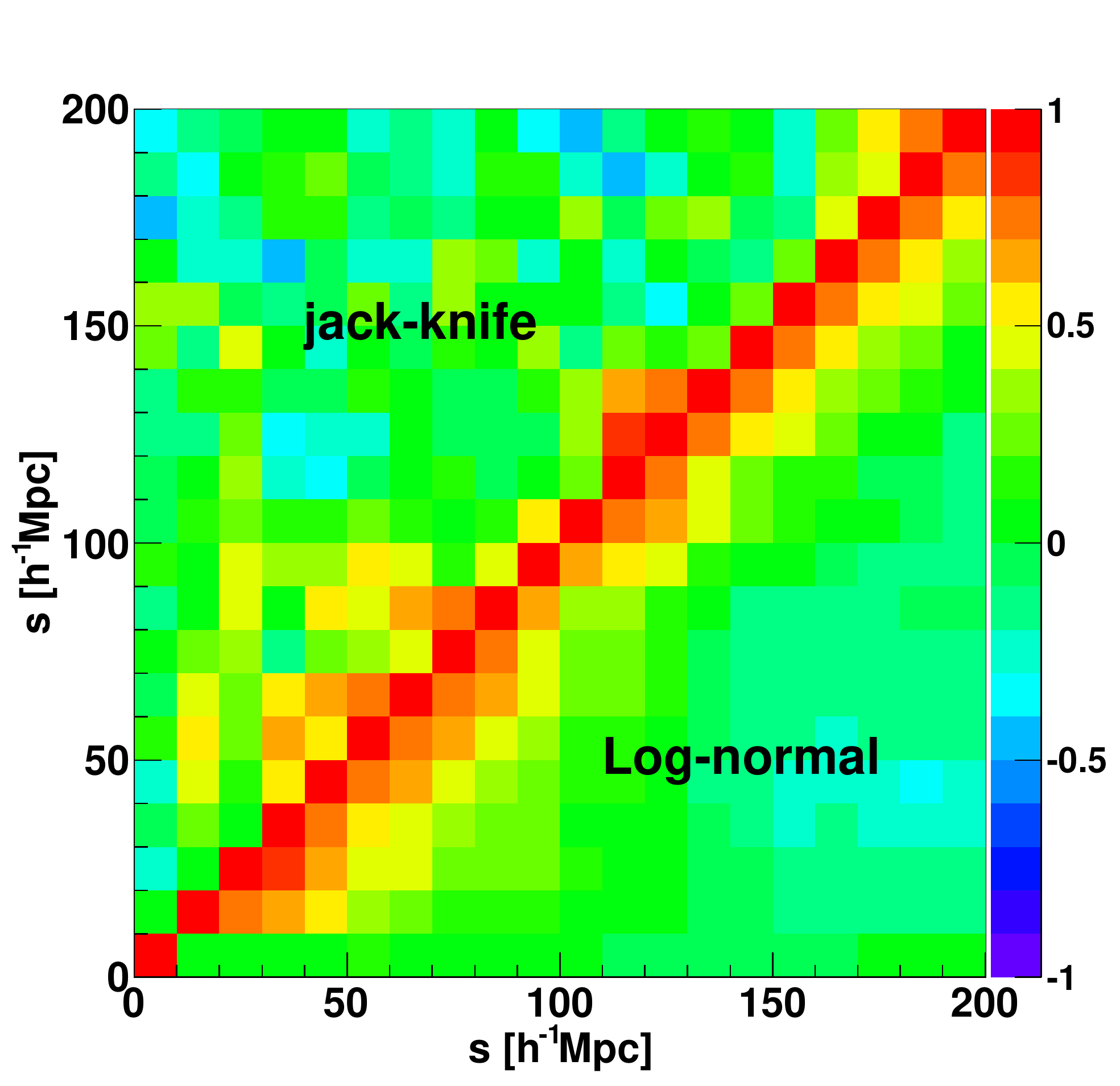,width=10cm}
\caption{Correlation matrix of the jack-knife errors (upper left triangle) and log-normal errors (lower right triangle).}
\label{fig:matrix2}
\end{center}
\end{figure}

We have also estimated jack-knife errors for the correlation function, by way of comparison. We divided the survey into $18$ regions and calculated the correlation function by excluding one region at a time. We found that the size of the error-bars around the BAO peak varies by around $20\%$ in some bins, when we increase the number of jack-knife regions from $18$ to $32$. Furthermore the covariance matrix derived from jack-knife resampling is very noisy and hard to invert.

We show the jack-knife errors in Figure~\ref{fig:sigP0}. The jack-knife error shows more noise and is larger in most bins compared to the log-normal error. The error shown in Figure~\ref{fig:sigP0} is only the diagonal term of the covariance matrix and does not include any correlation between bins.

The full error matrix is shown in Figure~\ref{fig:matrix2}, where we plot the correlation matrix of the jack-knife error estimate compared to the log-normal error. The jack-knife correlation matrix looks much more noisy and seems to have less correlation in neighbouring bins.

The number of jack-knife regions can not be chosen arbitrarily. Each jack-knife region must be at least as big as the maximum scale under investigation. Since we want to test scales up to almost $200h^{-1}\;$Mpc our jack-knife regions must be very large. On the other hand we need at least as many jack-knife regions as we have bins in our correlation function, otherwise the covariance matrix is singular. These requirements can contradict each other, especially if large scales are analysed. Furthermore the small number of jack-knife regions is the main source of noise (for a more detailed study of jack-knife errors see e.g.~\citealt{Norberg:2008tg}).

Given these limitations in the jack-knife error approach, correlation function studies on large scales usually employ simulations or log-normal realisations to derive the covariance matrix. We decided to use the log-normal error in our analysis. We showed that the jack-knife errors tend to be larger than the log-normal error at larger scales and carry less correlation. These differences might be connected to the much higher noise level in the jack-knife errors, which is clearly visible in all our data. It could be, however, that our jack-knife regions are too small to deliver reliable errors on large scales. We use the minimum number of jack-knife regions to make the covariance matrix non-singular (the correlation function is measured in $18$ bins). The mean distance of the jack-knife regions to each other is about $200h^{-1}\;$Mpc at the mean redshift of the survey, but smaller at low redshift.

\section{Wide-angle formalism}
\label{ap:wide}

\begin{figure}[tb]
\begin{center}
\epsfig{file=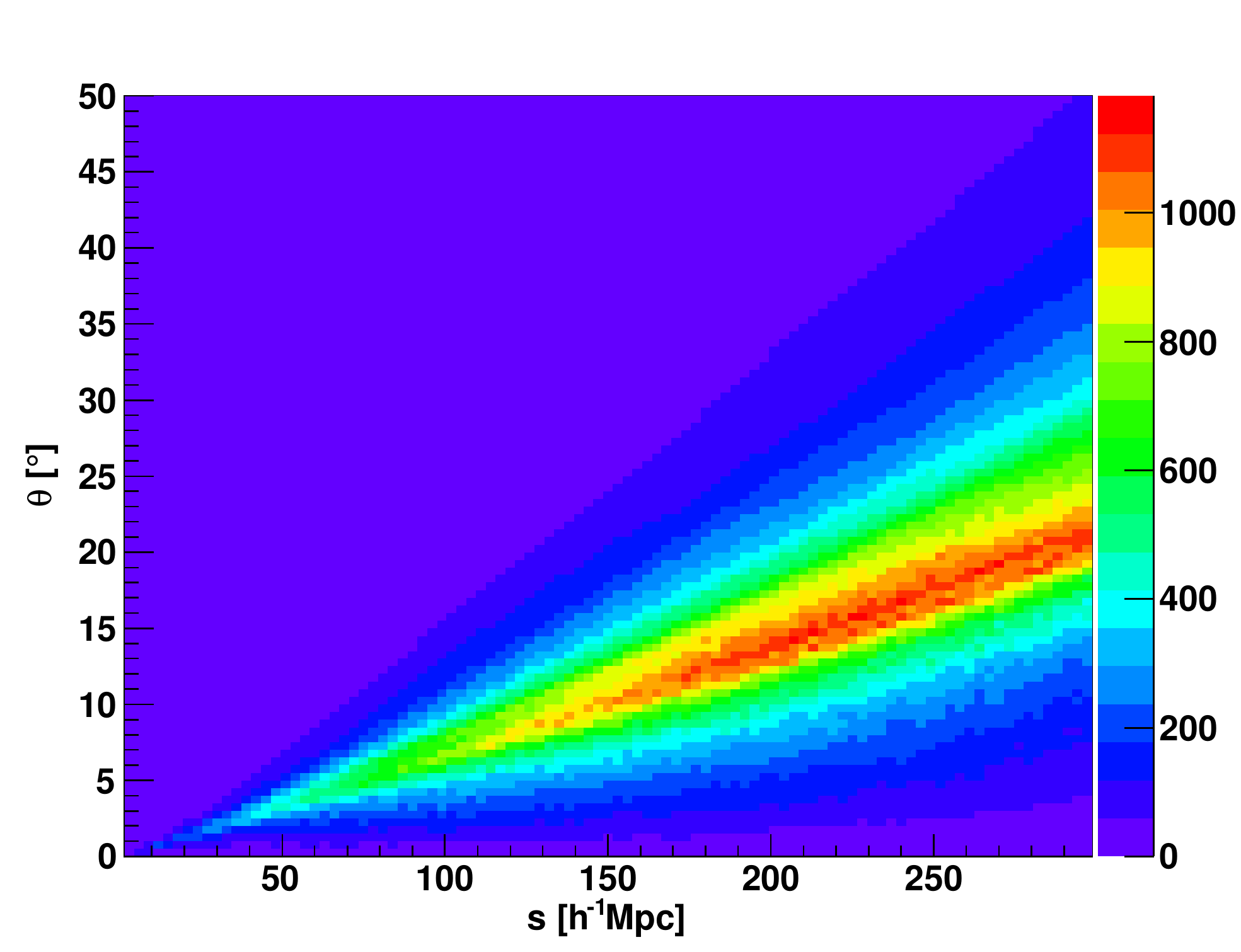, width=10cm}
\caption{The half opening angle $\theta$ as a function of separation $s$ of the 6dFGS weighted catalogue. The plane parallel approximation assumes $\theta = 0$. The mean half opening angle at the BAO scale is $\lesssim 10^{\circ}$. The colour bar gives the number of pairs in each bin.}
\label{fig:theta_s}
\end{center}
\end{figure}

The general redshift space correlation function (ignoring the plane parallel approximation) depends on $\phi$, $\theta$ and $s$. Here, $s$ is the separation between the galaxy pair, $\theta$ is the half opening angle, and $\phi$ is the angle of $s$ to the line of sight (see Figure~$1$ in~\citealt{Raccanelli:2010hk}). For the following calculations it must be considered that in this parametrisation, $\phi$ and $\theta$ are not independent.

The total correlation function model, including $O(\theta^2)$ correction terms, is then given by~\cite{Papai:2008bd},
\begin{equation}
\begin{split}
\xi(\phi,\theta,s) &= a_{00} + 2a_{02}\cos(2\phi) + a_{22}\cos(2\phi) + b_{22}\sin^2(2\phi)\\
& +\Big[ - 4a_{02}\cos(2\phi) - 4a_{22} - 4b_{22} - 4a_{10}\cot^2(\phi)\\
& + 4a_{11}\cot^2(\phi) - 4a_{12}\cot^2(\phi)\cos(2\phi) + 4b_{11}\\
& - 8b_{12}\cos^2(\phi)\Big]\theta^2 + O(\theta^4)
\end{split} 
\label{eq:wide1}
\end{equation}
This equation reduces to the plane parallel approximation if $\theta = 0$. The factors $a_{xy}$ and $b_{xy}$ in this equation are given by
\begin{equation}
\begin{split}
a_{00} &= \left[1 + \frac{2\beta}{3} + \frac{2\beta^2}{15}\right]\xi_0^2(r)\\
&- \left[\frac{\beta}{3} + \frac{2\beta^2}{21}\right]\xi^2_2(r) + \frac{3\beta^2}{140}\xi^2_4(r)\\
a_{02} &= -\left[\frac{\beta}{2} + \frac{3\beta^2}{14}\right]\xi^2_2(r) + \frac{\beta^2}{28}\xi^2_4(r)\\
a_{22} &= \frac{\beta^2}{15}\xi^2_0(r) - \frac{\beta^2}{21}\xi^2_2(r) + \frac{19\beta^2}{140}\xi^2_4(r)\\
b_{22} &= \frac{\beta^2}{15}\xi_0^2(r) - \frac{\beta^2}{21}\xi^2_2(r) - \frac{4\beta^2}{35}\xi^2_4(r)\\
a_{10} &= \left[2\beta + \frac{4\beta^2}{5}\right]\frac{1}{r}\xi^1_1(r) - \frac{\beta^2}{5r}\xi^1_3(r)\\
a_{11} &= \frac{4\beta^2}{3r^2}\left[\xi^0_0(r) - 2\xi^0_2(r)\right]\\
a_{12} &= \frac{\beta^2}{5r}\left[2\xi^1_1(r) - 3\xi^1_3(r)\right]\\
b_{11} &= \frac{4\beta^2}{3r^2}\left[\xi^0_0(r) + \xi^0_2(r)\right]\\
b_{12} &= \frac{2\beta^2}{5r}\left[\xi^1_1(r) + \xi^1_3(r)\right] ,
\end{split}
\end{equation}
where $\beta = \Omega_m(z)^{0.545}/b$, with $b$ being the linear bias. The correlation function moments are given by 
\begin{equation}
\xi^m_l(r) = \frac{1}{2\pi^2}\int^{\infty}_0 dk\;k^mP_{\rm lin}(k)j_l(rk)
\end{equation}
with $j_l(x)$ being the spherical Bessel function of order $l$.

\begin{figure}[tb]
\begin{center}
\epsfig{file=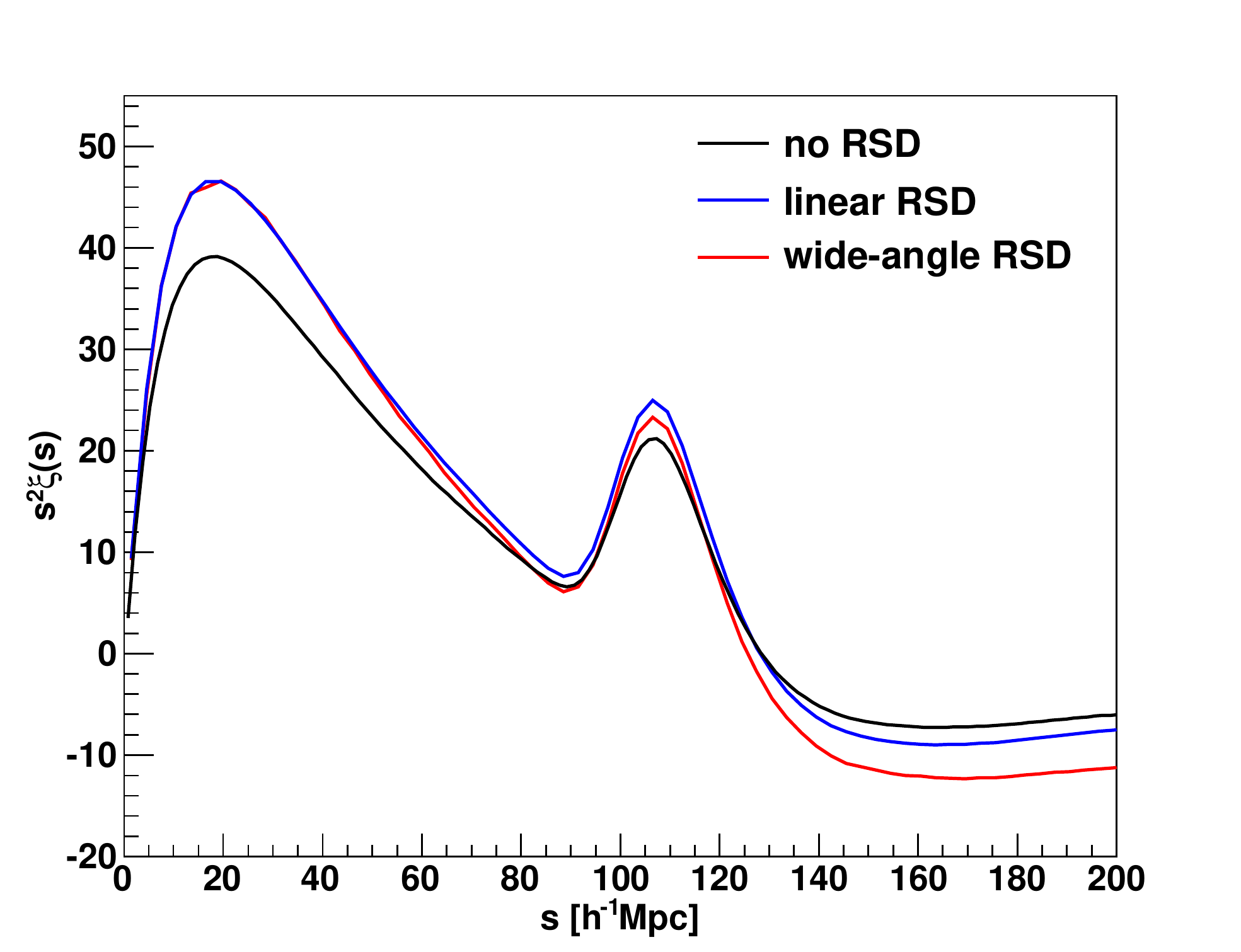,width=10cm}
\caption{The black line represents the plain correlation function without redshift space distortions (RSD), $\xi(r)$, obtained by a Hankel transform of our fiducial $\Lambda$CDM power spectrum. The blue line includes the linear model for redshift space distortions (linear Kaiser factor) using $\beta = 0.27$. The red line uses the same value of $\beta$ but includes all correction terms outlined in eq.~\ref{eq:wide1} using the $N(\phi,\theta,s)$ distribution of the weighted 6dFGS sample employed in this analysis.}
\label{fig:wide2}
\end{center}
\end{figure}

The final spherically averaged correlation function is given by
\begin{equation}
\xi(s) = \int^{\pi}_0\int_{0}^{\pi/2}\xi(\phi,\theta,s)N(\phi,\theta,s)\;d\theta d\phi,
\label{eq:mom}
\end{equation}
where the function $N(\phi,\theta,s)$ is obtained from the data. $N(\phi,\theta,s)$ counts the number of galaxy pairs at different $\phi$, $\theta$ and $s$ and includes the areal weighting $\sin(\phi)$ which usually has to be included in an integral over $\phi$. It is normalised such that 
\begin{equation}
\int^{\pi}_0\int^{\pi/2}_0 N(\phi,\theta,s)\;d\theta d\phi = 1 .
\end{equation}
If the angle $\theta$ is of order $1\;$rad, higher order terms become dominant and eq.~\ref{eq:wide1} is no longer sufficient. Our weighted sample has only small values of $\theta$, but growing with $s$ (see figure~\ref{fig:theta_s}). In our case the correction terms contribute only mildly at the BAO scale (red line in figure~\ref{fig:wide2}). However these corrections behave like a scale dependent bias and hence can introduce systematic errors if not modelled correctly. 

\section{Partial analytical solution for the correlation function moments integral}
\label{sec:ana}

The double integrals in eqs.~\ref{eq:xilm1} and \ref{eq:xilm2} are difficult to solve numerically. Here we show an analytical solution for the integral over $\mu$ which allows for a faster numerical solution of the full integral. First we re-write eq.~\ref{eq:xilm2} as
\begin{equation}
\begin{split}
\xi^m_{\ell,xy}(r) &= \int^{\infty}_{0}\int^1_{-1}\frac{k^mdkd\mu}{(2\pi)^2}e^{-(k\mu\sigma_v)^2}\cr
&\times \cos(kr\mu)Q_{xy}(k)\mathcal{P}_{\ell}(\mu)\cr
& = Re\left[ \int^{\infty}_{0}\frac{k^mdk}{(2\pi)^2}Q_{xy}(k) \int^1_{-1}d\mu\; e^{ikr\mu-(k\mu\sigma_v)^2}\mathcal{P}_{\ell}(\mu)\right],\notag
\end{split}
\end{equation}
where $i$ is the complex number. More generally the integral over $\mu$ can be written as
\begin{equation}
\begin{split}
F_{n}(\mu)=&\int^1_{-1}d\mu\; e^{ikr\mu-(k\mu\sigma_v)^2}\mu^n\cr
=&\left(-i\frac{\partial}{\partial (kr)}\right)^{n}\int^1_{-1}d\mu\; e^{ikr\mu-(k\mu\sigma_v)^2}\cr
=&\left(-\frac{\partial}{\partial (k^2\sigma_v^2)}\right)^{n/2}\int^1_{-1}d\mu\; e^{ikr\mu-(k\mu\sigma_v)^2}.
\end{split}
\end{equation}
The integral on the right can be solved analytically. If we set $a = kr$ and $b = (k\sigma_v)^2$ we obtain
\begin{equation}
\begin{split}
&\int^1_{-1}d\mu\; e^{ikr\mu-(k\mu\sigma_v)^2} \cr
=& \frac{i}{\sqrt{b}}e^{-ia-b}\left[\text{Daw}\left(\frac{a - 2ib}{2\sqrt{b}}\right) - e^{2ia}\text{Daw}\left(\frac{a + 2ib}{2\sqrt{b}}\right)\right].
\end{split}
\end{equation}
Taking the n-th derivatives of the real part of the term above gives $F_n(\mu)$. The Dawson integral $\text{Daw}(x)$ can be calculated using the imaginary error function $\text{erfi}(x)$:
\begin{equation}
\begin{split}
\text{Daw}(x) &= e^{-x^2}\int^x_0e^{y^2}dy\cr
&= \frac{\sqrt{\pi}}{2}e^{-x^2}\text{erfi}(x).
\end{split}
\end{equation}
We can than construct eq.~\ref{eq:xilm2} for the different correlation function moments:
\begin{equation}
\begin{split}
\xi^m_{\ell=0,xy}(r) &= \int^{\infty}_{0}\frac{k^mdk}{(2\pi)^2}Q_{xy}(k) F_0(\mu)\cr
\xi^m_{\ell=1,xy}(r) &= \int^{\infty}_{0}\frac{k^mdk}{(2\pi)^2}Q_{xy}(k) F_1(\mu)\cr
\xi^m_{\ell=2,xy}(r) &= \int^{\infty}_{0}\frac{k^mdk}{(2\pi)^2}Q_{xy}(k) \frac{1}{2}\left[3F_2(\mu) - F_0(\mu)\right]\cr
\xi^m_{\ell=3,xy}(r) &= \int^{\infty}_{0}\frac{k^mdk}{(2\pi)^2}Q_{xy}(k) \frac{1}{2}\left[5F_3(\mu) - 3F_1(\mu)\right]\cr
\xi^m_{\ell=4,xy}(r) &= \int^{\infty}_{0}\frac{k^mdk}{(2\pi)^2}Q_{xy}(k) \frac{1}{8}\left[35F_4(\mu) - 30F_2(\mu) + 3F_0(\mu)\right]\cr
\dots&
\end{split}
\end{equation}
and so on.

\section{Normalisation constant in M\lowercase{andelbaum et al. 2006}}
\label{ap:mandel}

We can determine the normalisation constant $k$ of eq.~\ref{eq:mandel1} by matching to the satellite fraction $f_s^{\rm ma}$, which in terms of the mass function $dn(M)/dM$ is given by
\begin{align}
f_s^{\rm ma} &= \frac{\int^{\infty}_0 dM\frac{dn(M)}{dM}N_s(M)}{\int^{\infty}_0dM\frac{dn(M)}{dM} \left[N_c(M) + N_s(M)\right]}\\
&=\frac{kH}{ \int_{M_{\rm min}}^{\infty} dM \frac{dn(M)}{dM}  + kH}
\end{align}
with 
\begin{equation}
\begin{split}
H =& \frac{1}{3M_{\rm min}}\int^{3M_{\rm min}}_{M_{\rm min}}dM\frac{dn(M)}{dM}M^2 +\\ 
&\int^{\infty}_{3M_{\rm min}}dM\frac{dn(M)}{dM}M,
\end{split}
\end{equation}
which leads to
\begin{equation}
k = \frac{f_s^{\rm ma}\int_{M_{\rm min}}^{\infty} dM \frac{dn(M)}{dM}}{(1-f_s^{\rm ma})H},
\end{equation}
where $k$ has units of $1/h^{-1}M_{\odot}$ and is constant for a given set of $f_s^{\rm ma}$ and $M_{\rm min}$.

\end{subappendices}


\bibliographystyle{plainnat}

\bibliography{sample}            


\end{document}